\documentclass[12pt, twoside, openright]{Thesis} % The default font size and one-sided printing (no margin offsets)

\usepackage{lipsum}
\usepackage{notoccite}

\hypersetup{urlcolor=black, citecolor=black, linkcolor=black, colorlinks=true}

\title{\ttitle} % Defines the thesis title - don't touch this

\begin{document}

\frontmatter % Use roman page numbering style (i, ii, iii, iv...) for the pre-content pages

\setstretch{1.5} % Line spacing of 1.3

% Define the page headers using the FancyHdr package and set up for one-sided printing
\fancyhead{} % Clears all page headers and footers
\rhead{\thepage} % Sets the right side header to show the page number
\lhead{} % Clears the left side page header

\pagestyle{fancy} % Finally, use the "fancy" page style to implement the FancyHdr headers

\newcommand\blfootnote[1]{%
  \begingroup
  \renewcommand\thefootnote{}\footnote{#1}%
  \addtocounter{footnote}{-1}%
  \endgroup
}

\newcommand{\HRule}{\rule{\linewidth}{0.5mm}} % New command to make the lines in the title page

\newcommand{\bs}{\boldsymbol}
\newcommand{\bt}{\mathbf}

\newcommand{\rojo}{\textcolor{red}}
\newcommand{\red}{\textcolor{red}}
\newcommand{\azul}{\textcolor{blue}}
\newcommand{\blue}{\textcolor{blue}}

\newcommand{\vicente}[1]{\textcolor{red}{ #1}}
\newcommand{\vicentebis}[1]{\textcolor{green}{ #1}}
\newcommand{\fran}[1]{\textcolor{black}{ #1}}

\newcommand\beq{\begin{equation}}
\newcommand\eeq{\end{equation}}
\newcommand\beqa{\begin{eqnarray}}
\newcommand\eeqa{\end{eqnarray}}
\newcommand{\dd}{\text{d}}
\newcommand{\nn}{\nonumber\\}
\newcommand{\qq}{r}
\newcommand{\al}{\alpha}

\newcommand{\at}{\frac{1+\alpha}{2}}
\newcommand{\atp}{\left(\frac{1+\alpha}{2}\right)}
\newcommand{\sg}{(\widehat{\bm{\sigma}}\cdot\mathbf{g})}
\newcommand{\si}{\widehat{\sigma}}
\newcommand{\nutz}{\nu_{2|0}}
\newcommand{\nuzt}{\nu_{0|2}}
\newcommand{\nuto}{\nu_{2|1}}
\newcommand{\nuzth}{\nu_{0|3}}
\newcommand{\nufz}{\nu_{4|0}}
\newcommand{\nutt}{\nu_{2|2}}
\newcommand{\nuzf}{\nu_{0|4}}
\newcommand{\omegatz}{\omega_{2|0}}
\newcommand{\omegazt}{\omega_{0|2}}
\newcommand{\omegato}{\omega_{2|1}}
\newcommand{\omegazth}{\omega_{0|3}}
\newcommand{\omegafz}{\omega_{4|0}}
\newcommand{\omegatt}{\omega_{2|2}}
\newcommand{\omegazf}{\omega_{0|4}}
\newcommand{\nubar}{\omega_{0|2}}
\newcommand{\omeganu}{\nu}

% PDF meta-data
\hypersetup{pdftitle={\ttitle}}
\hypersetup{pdfsubject=\subjectname}
\hypersetup{pdfauthor=\authornames}
\hypersetup{pdfkeywords=\keywordnames}

%----------------------------------------------------------------------------------------
%	TITLE PAGE
%----------------------------------------------------------------------------------------

\begin{titlepage}
\begin{center}

\textsc{\LARGE \univname}\\[1.5cm] % University name
\textsc{\Large Doctoral Thesis}\\[0.5cm] % Thesis type

\HRule \\[0.4cm] % Horizontal line
{\huge \bfseries \ttitle}\\[0.4cm] % Thesis title
\HRule \\[1.5cm] % Horizontal line
 
\begin{minipage}{0.4\textwidth}
\begin{flushleft} \large
\emph{Author:}\\
%\href{http://}{\authornames} % Author name - remove the \href bracket to remove the link
\end{flushleft}
\end{minipage}
\begin{minipage}{0.4\textwidth}
\begin{flushright} \large
\emph{Supervisors:} \\
%\href{http://}{\supname} % Supervisor name - remove the \href bracket to remove the link  
\end{flushright}
\end{minipage}\\[3cm]
 
\large \textit{A thesis submitted in fulfilment of the requirements\\ for the degree of \degreename}\\[0.3cm] % University requirement text
\textit{in the}\\[0.4cm]
%\groupname\\
\deptname\\[2cm] % Research group name and department name
 
{\large 2017}\\[4cm] % Date
%{\large \today}\\[4cm] % Date
%\includegraphics{Logo} % University/department logo - uncomment to place it
 
\vfill
\end{center}

\end{titlepage}

\pagebreak

\thispagestyle{empty} % The page style headers have been "empty" all this time, now use the "fancy" headers as defined before to bring them back
{\color{white}.}

\thispagestyle{empty} % The page style headers have been "empty" all this time, now use the "fancy" headers as defined before to bring them back
{\color{white}.}
\pagebreak

\addtotoc{Abstract} % Add the "Abstract" page entry to the Contents
\abstract{\addtocontents{toc}{\vspace{1em}} % Add a gap in the Contents, for aesthetics
Granular matter under rapid flow conditions can be modeled as a granular gas, namely, a gas of hard spheres dissipating part of their kinetic energy during binary collisions (inelastic hard spheres, IHS). On the other hand, given that collisions are inelastic one has to inject energy into the system to compensate for the inelastic cooling and maintain it in rapid conditions. Although in real experiments the external energy is supplied to the system by the boundaries, it is quite usual in computer simulations to heat the system by the action of an external driving force or \emph{thermostat}. Despite thermostats have been widely employed in the past, their influence on the dynamic properties of the system (for elastic and granular fluids) is not yet completely understood. In this work, we determine the transport properties of driven granular systems by using two independent and complementary routes, one of them being analytic (Chapman-Enskog method, BGK solution and Grad's moments method) and the other one being computational (Monte Carlo simulations).
}

\pagebreak

\thispagestyle{empty} % The page style headers have been "empty" all this time, now use the "fancy" headers as defined before to bring them back
{\color{white}.}
\pagebreak

%-------------------
%	ACKNOWLEDGEMENTS
%-------------------

\acknowledgements{
\addtocontents{toc}{\vspace{1em}} % Add a gap in the Contents, for aesthetics

Firstly, I would like to express my sincere gratitude to my advisors Prof. Vicente Garz\'o and Prof. Francisco Vega Reyes for the continuous support of my Ph.D study and related research. They offerred me their immense patience, motivation, and knowledge. Their constant guidance helped me in all the time of research and writing of this thesis. I could not have imagined having a better advisor and mentor for my Ph.D study.

Besides my advisors, I would like to thank my thesis external review committee: Prof. Mariano L\'opez de Haro, and Prof. Rodrigo Soto, for their insightful reading and comments, and also for the hard questions which incented me to widen my research from various perspectives.

My sincere thanks also goes to Prof. Emmanuel Trizac, who provided me an opportunity to join their team as intern, and who gave access to the laboratory and research facilities. %Without they precious support it would not be possible to conduct this research.
 
I thank my department fellows and labmates, Andr\'es Santos, Juan Jes\'us Ruiz, Santos Bravo, Enrique Abad, Miguel \'Angel Gonz\'alez, Antonio Gordillo and Antonio Astillero for the stimulating coffee discussions, help, and for all the fun we have had in the last five years.

I would like to thank to my old friends and colleagues who always listened and helped me when I most needed to: Elena Su\'arez, Mar\'ia Su\'arez, Zazu Os\'es, Gloria P\'erez, Espe Mor\'on, Miguel Blanco, Fran Naranjo, Joaqu\'in Aparicio, Arcadio Guerra, Javi S\'aez and of course all the wonderful members of the Choir \emph{Vocalis} with special mention to Jos\'e Mar\'ia y Asun.

I want to thank specially to my parents Timoteo and M\'axima and my sister Gema for supporting me spiritually throughout writing this thesis and my life in general.

I am grateful to Ministerio de Ciencia en Innovaci\'on from the Spanish Government (Spain) for the finantial support through grant No. FIS2010-16587 and EEBB-I-2014-08635 .

}

%------------------------------------------
%	LIST OF CONTENTS/FIGURES/TABLES PAGES
%------------------------------------------

\pagestyle{fancy} % The page style headers have been "empty" all this time, now use the "fancy" headers as defined before to bring them back

\lhead{\emph{Contents}} % Set the left side page header to "Contents"
\tableofcontents % Write out the Table of Contents

\lhead{\emph{List of Figures}} % Set the left side page header to "List of Figures"
\listoffigures % Write out the List of Figures

%------------------
%	ABBREVIATIONS
%------------------

\setstretch{1.5} % Set the line spacing to 1.5, this makes the following tables easier to read

\lhead{\emph{Abbreviations}} % Set the left side page header to "Abbreviations"
\listofsymbols{ll} % Include a list of Abbreviations (a table of two columns)
{
\textbf{BGK}  & \textbf{B}hatnagar-\textbf{G}ross-\textbf{K}rook  (114)\footnote{The numbers in parenthesis correspond to the page where the abbreviation appears for the first time.}                    \\
\textbf{CE}   & \textbf{C}hapman-\textbf{E}nskog (9)                                          \\
\textbf{DSMC} & \textbf{D}irect \textbf{S}imulation of \textbf{M}onte \textbf{C}arlo (8)      \\
%\textbf{EB}   & \textbf{E}nskog-\textbf{B}oltzmann (kinetic equation)                         \\
%\textbf{EHS}  & \textbf{E}lastic \textbf{H}ard \textbf{S}pheres (model)                       \\
\textbf{HCS}  & \textbf{H}omogeneous \textbf{C}ooling \textbf{S}tate (9)                      \\
\textbf{IHS}  & \textbf{I}nelastic \textbf{H}ard \textbf{S}pheres (9)                         \\
\textbf{IMM}  & \textbf{I}nelastic \textbf{M}axwell \textbf{M}odel   (9)                      \\
\textbf{MD}   & \textbf{M}olecular \textbf{D}ynamics (67)                                     \\
\textbf{ME}   & \textbf{M}aximum \textbf{E}ntropy (121)                                       \\
\textbf{USF}  & \textbf{U}niform \textbf{S}hear \textbf{F}low  (10)                           \\
\textbf{NS}   & \textbf{N}avier-\textbf{S}tokes  (4)                                          \\
}

%------------------
%	SYMBOLS
%------------------

\lhead{\emph{Symbols}} % Set the left side page header to "Symbols"
\listofnomenclature{lllll} % Include a list of Symbols (a three column table)
{
$a_{ij}$                  & shear rate,                                               \\
$a_p$                     & Sonine expansion coefficient,                             \\
$\alpha$                  & coefficient of normal restitution,                        \\ 
% \quad                    & \quad                                                     \\
$\bt{c}$                  & scaled velocity,                                          \\
$\langle c^p \rangle $    & $p$-moment of the scaled velocity distribution function,    \\
$\chi$; $\chi(\phi)$      & pair correlation function,                                \\
% \quad                    & \quad                                                     \\
$d$                       & dimensionality,                                           \\
$D_\text{T}$              & thermal diffusivity,                                      \\
% \quad                    & \quad                                                     \\
$\eta$                    & shear viscosity,                                          \\
% \quad                    & \quad                                                     \\
$f(\bt{r},\bt{v},t)$      & velocity distribution function,                           \\
$f^{(k)}(\bt{r},\bt{v},t)$& $k$-order distribution function,                        \\
%$f^{(0)}(\bt{r},\bt{v},t)$& distribution function in zeroth order,                    \\
%$f^{(1)}(\bt{r},\bt{v},t)$& first-order distribution function,                        \\
% \quad                    & \quad                                                     \\
$\bt{g}_{12}$             & relative velocity,                                        \\
$\bt{G}_{12}$             & center of mass velocity,                                  \\
$\gamma_\text{b}$         & drag or friction coefficient,                             \\
% \quad                    & \quad                                                     \\
$J[f,f]$                  & Boltzmann collision operator,                             \\
$J_\text{E}[f,f]$         & Enskog collision operator,                                \\
$J_\text{IMM}[f,f]$       & Inelastic Maxwell Model collision operator,               \\
$J^*[f,f]$                & scaled collision operator,                                \\
% \quad                    & \quad                                                     \\
Kn                        & Knudsen number,                                           \\
$\kappa$                  & thermal conductivity,                                     \\
% \quad                    & \quad                                                     \\
$\ell$                    & mean free path,                                           \\
$\lambda$                 & bulk viscosity,                                           \\
% \quad                    & \quad                                                     \\
$m$                       & mass of particles,                                        \\
$\mu$                     & diffusive heat conductivity or Dufour-like coefficient,   \\
% \quad                    & \quad                                                     \\
$n$                       & number density,                                           \\
$\nu_0$                   & collision frequency,                                      \\
%$\nu_l$                   & longitudinal viscosity,                                   \\
% \quad                    & \quad                                                     \\
% \quad                    & \quad                                                     \\
$\Omega_d$                & total solid angle in $d$-dimensions                         \\
$P_{ij}$                  & pressure tensor,                                          \\
$\Pi_{ij}$                & traceless pressure tensor,                                \\
$\phi$                    & solid volume fraction,                                    \\
$\varphi(\bt{c})$         & scaled velocity distribution function,                    \\  
% \quad                    & \quad                                                     \\
$\rho$                    & mass density,                                             \\
% \quad                    & \quad                                                     \\
$T$                       & granular temperature,                                     \\
$\Theta(x)$               & Heaviside step function,                                  \\
% \quad                    & \quad                                                     \\
$\sigma$                  & diameter of particles,                                    \\
$\widehat{\bs{\sigma}}$   & unit vector joining the centers of colliding particles,   \\
% \quad                    & \quad                                                     \\
$\bt{U}$                  & mean flow velocity of solid particles,                    \\
$\bt{U}_g$                & mean flow velocity of interstitial gas,                   \\
% \quad                    & \quad                                                     \\
$\bt{V}$                  & peculiar velocity,                                        \\
$v_0$                     & thermal velocity,                                         \\
% \quad                    & \quad                                                     \\
$\xi_b^2$                 & noise intensity,
}

%------------------
%	DEDICATION
%------------------

\setstretch{1.3} % Return the line spacing back to 1.3
\pagestyle{empty} % Page style needs to be empty for this page
\dedicatory{To M\'axima, Timoteo and Gema, who always believed in me, even when I didn't.} % Dedication text
\addtocontents{toc}{\vspace{2em}} % Add a gap in the Contents, for aesthetics

%------------------------------
%	THESIS CONTENT - CHAPTERS
%------------------------------

\mainmatter % Begin numeric (1,2,3...) page numbering

\pagestyle{fancy} % Return the page headers back to the "fancy" style

% Include the chapters of the thesis as separate files from the Chapters folder
% Uncomment the lines as you write the chapters
% Chapter Template

\chapter{Introduction to granular gases} 

\label{Chapter1} 

\lhead{Chapter 1. \emph{Introduction to granular gases}} 

\section{Introduction}
%mezcla ANDREOTTI, Serero 2005
Granular matter is a vast and diverse family of materials with the common property of being composed of a large number of macroscopic grains with very different shapes and range of sizes \cite{ANDREOTTI13}. This kind of material is quite commonplace in Nature and industry in examples like cereals, salt, sand, etc., Their study is of great interest in a wide variety of the industry and technology sectors as well as in different fields of fundamental and applied science such as biophysics, astrophysics, fluid mechanics, statistical physics and even in optics applications.

The knowledge of physical properties of granular matter has also great practical importance in Engineering for %the characterization of the particulate materials and 
the design of many industrial processes such as  conveying, handling and storage. This is important because it might prevent  malfunctions of the devices due to phenomena of obstructions, irreversible stuck of grains %due to huge internal chain forces, 
and potentially dramatic events such as the collapse of a silo. It is also usual to deal with processes of separating or mixing several substances in the form of powder for the manufacturing of pharmaceutical or chemical products.\cite{LEVY2001}

Some authors estimate that, nowadays, granular matter is involved in more than $50\%$ of trade in the world \cite{B06}. Many of the products we daily use have been made using granular matter in a stage of their fabrication process. In fact, granular media are the second most used type of material in industry after water. Rough estimates of the losses suffered in the world economy due to \textit{granular ignorance} amount to billions of dollars a year \cite{ANDREOTTI13}.

But not only in Earth one can find this kind of material. Out of our planet, granular matter also abounds in space in the form of dust and grains. This is illustrated for instance by Martian dunes or astronomic-range features as planetary rings, asteroids, comets clouds (as the Kuiper's Belt and the Oort's Cloud) and interstellar dust that reaches vital importance for the proper functioning of communication satellites, probes and man ships such as the  International Space Station \cite{POSCHEL01, SSPS00, BKB15, SAS04}. %\rojo{más citas aquí}

% mezcla Chamorro PRE92-2015 Suspensiones y Serero JFM 2005 y ANDREOTTI 2013
On the other hand, although in nature one can find granular systems in vacuum (as for instance the previous mentioned interstellar dust), in most cases of interest granular particles are immersed in a fluid (air, water, etc.). This kind of mixtures (granular suspensions) is widely used in many industrial processes, for instance, in civil engineering works with concretes, asphalt, bitumen or in the chemical industry in fuel or catalysts deployed in the form of grain to maximize the active surfaces. % \rojo{más citas aquí}

Another interesting problem where granular theories are useful is the diffusion of fluids through densely packed cobblestone and rocks that is vital for the industry of natural combustibles and subterranean water finding.%\rojo{más citas aquí}
 The comprehension of the coupling between fluid and solid phase is essential in geological problems as soil stability and water controls, surface modelling by soil erosion, sand dunes movement and the dangerous ripples formation in the sand under shallow sea water. Furthermore, understanding of the dynamics of many natural disasters such as avalanches, landslides, mud flows, pyroclastic flows, etc. can be achieved by means of models of granular media \cite{BAGNOLD54}. In particular, an important target of the research in granular matter concerns the description and prediction of natural hazards that the above events suppose to the human activities in order to avoid or minimize their impact on lives and economy around the world \cite{JNB96}. 

% Mezcla Serero JFM 2005, ANDREOTTI  %Apart of the environmental importance and practical interest, 
Apart from their industrial and geophysical applications, there exist many important scientific reasons to study the laws underlying the behaviour of granular materials. On the other hand, despite its practical importance our understanding of granular media remains still incomplete. No theoretical framework is available to describe the different phenomena observed in nature for granular flows.

It is well-known that granular media can behave like a solid, liquid or a gas \cite{JNB96b}.  
Grains can create static structures sustaining great stresses but they can also flow as liquids or even gases when strongly excited. In addition the three states can coexist in a single system. 
In spite of this, it is not so easy to accept that such flow can be described by hydrodynamic equations \cite{TG98, K99}. Notwithstanding, when the system is externally driven (rapid flow regime), fluidized granular media may exhibit most of the known hydrodynamic flows and instabilities such as Taylor-Couette \cite{LBLG00} and Couette-Fourier \cite{TTMGSD01} flows, Bénard convection \cite{KM03}, etc. Furthermore, they may present a complex rheology exhibiting different non-Newtonian features as nonzero anisotropic normal stresses differences and non-linear relation between the shear stress and the shear rate much like in other non-Newtonian materials. The normal stress in these fluids is often anisotropic, like in other non-Newtonian materials. In addition, they show features that do have not their counterpart in ordinary fluids. For instance, in vertically vibrated shallow layers of grains stable geyser-like excitations called \textit{oscillons} can be observed \cite{UMS96}.

This intriguing behaviour intermediate between solid and fluid is a basic characteristic of granular matter. Above certain density threshold the system becomes a compact solid because of the dissipating character of grain interactions but if the system is externaly excited or its density is decreased, then it can flow.

For all these reasons, in the last years a vast bibliography on granular dynamics has been reported \cite{dG99, G03}.

\section{Granular Gases}

Generally, we may differentiate the high and low density regimes in granular matter. The latter regime is essentially characterized by binary particle collisions whereas the former presents multiparticle contacts. As a consequence, the theoretical modeling and mathematical treatment to obtain their physical properties are quite different in each regime \cite{G03}. In this work we will focus on the binary collision regime where the system is usually called a \textit{granular gas} \cite{H83} whose physical realization can be observed in rings of planets, small planets, suspended particles in fluidized beds, aerosols, rapid granular flows, etc. \cite{POSCHEL00}. 

One of the fundamental properties of the grains in granular matter is the inelastic character of their collisions. When two particles collide, part of their kinetic energy is irreversibly transformed into internal degrees of freedom (temperature rising of particles, plastic deformations, etc.). This provokes a persistent loss of mechanical energy in the whole system. Dissipative interactions between particles in unforced granular gases is the reason for which these systems are inherently out of equilibrium. %From the point of view of thermodynamics, granular gases can be considered as \rojo{open systems} and 
Granular gases also reveal self-organized spatio-temporal structures and instabilities. When a granular gas has no energy input, then it become unstable to density perturbations and if they freely evolve, they will eventually collapse by a mechanism of clustering instabilities (which is increasingly stronger with increasing inelasticity) that will destroy the homogeneity of the system \cite{MY92, GT96, MIMA08, GZ93, POSCHEL04}. % M93, WK00, One of the most remarkable features is the formation of clusters or clustering. \cite{POSCHEL04}
This tendency to collapse into clusters occurs even for \emph{initially prepared} homogeneous mass distributions \cite{G91, HL91, GZ93}. The clustering instability can be easily understood from a qualitative argument. Fluctuations of density in granular gases generate relatively denser domains where the rate of collisions (proportional to the number density) is higher than in dilute domains and hence the kinetic energy loss due to inelasticity increases in these regions. As a result, the grains tend to move from dilute into dense domains driven by the granular pressure difference between them thereby further increasing the density of the latter and giving rise to bigger and denser clusters. This mechanism allows for the growing of the clusters which may further coagulate by coarsening with other into larger clusters \cite{G03} or collide thereby destroying each other \cite{TG97,GZ93, DB97, BRC99}. %This feature is exhibited even in very dilute systems in form of linear instabilities that render them non-homogeneous \cite{GZ93, DB97, BRC99}. 
The critical length scale for the onset of instability can be determined via stability analysis of the linearized \emph{Navier-Stokes} (NS) hydrodynamic equations \cite{BDKS98, G05, NE00, MDCPH11, MGHEH12, GMD06, BM13, MGH14, G15}

%% SERERO et al. JFM554 (2006) SGNT06
%Granular gases possess a tendency to coagulate into clusters even for “initially pre-
%pared” homogeneous mass distributions (Goldhirsch 1991; Hopkins & Louge 1991; Gold-
%hirsch & Zanetti 1993). This property can be understood on the basis of the following
%simple considerations, as well as detailed calculations (Goldhirsch & Zanetti 1993). A
%granular gas, like any other many-body system, experiences fluctuations, in particular
%of the density. Since in relatively dense domains the rate of collisions (proportional to
%the square of the number density) is higher than in dilute domains, the kinetic energy
%in dense domains decays at a higher rate than in dilute ones, due to inelasticity. The
%ensuing (grain) pressure difference leads grains from dilute into dense domains, thereby
%further increasing the density of the latter, and giving rise to dense clusters. 
%The clusters may further coagulate into larger clusters by coarsening (Goldhirsch 2003 and references
%therein) and they may collide thereby destroying each other (Tan & Goldhirsch 1997).
%
%Other structures or microstructures, such as plugs in sheared flows can be attributed to
%similar mechanisms. 
%Sufficiently small granular systems do not exhibit clustering, but
%even then there are linear instabilities that render them non-homogeneous (Goldhirsch &
%Zanetti 1993; Deltour & Barrat 1997; Brey & al 1999). Clustering is one of the reasons
%for multistability of granular gases, since they can be rather stable once formed.

In order to maintain the system in rapid flow conditions, it is neccesary to compensate the loss of energy due to the inelastic dissipation with the introduction of external non-conservative forces acting over the whole system. This is commonly done either by driving through the boundaries, (e.g., shearing the system or vibrating its walls \cite{POSCHEL04,YHCMW02,YHCMW04}) or alternatively by bulk driving (as in air-fluidized beds \cite{AD06}), gravity (as in a chute) or other techniques. On the other hand, this way of supplying energy causes in most of the cases strong spatial gradients in the system. 
% Garzo-Chamorro 2013. PRE87
To avoid the difficulties associated with non-homogeneous states, it is quite usual in computer simulations to homogeneously heat the system by the action of an external driving force 
\cite{PLMV98,PLMV99,CLH00,PEU02,PBL02,VPBTW06,FAZ09,SVGP10,KSZ10,PGGSV12,NETP99,VAZ11,GSVP11-1}. 
Borrowing a terminology often used in nonequilibrium molecular dynamics of ordinary fluids \cite{EM90}, this type of external forces are usually called \emph{thermostats}. Nevertheless, in spite of its practical importance, the effect of the external driving force on the dynamical properties of the system (such as the transport coefficients) is still not completely understood \cite{DSBR86, GSB90, GARZO03}. In particular, recent computer simulations \cite{VAZ11,GSVP11-1} have obtained some transport coefficients by measuring the static and dynamical structure factors for shear and longitudinal modes in a driven granular fluid. Given that the expressions for the transport coefficients were not known in this driven problem, the simulation data were compared with their corresponding elastic system. Thus, it would be desirable to provide simulators with the appropriate theoretical tools to work when studying problems in granular fluids driven by thermostats.

When externally excited, a granular system can become sufficiently fluidized so that the grain interactions are mostly nearly-instantaneous binary collisions and a steady non-equilibrium state is achieved \cite{G03}. In this regime each grain moves freely and \emph{independently} instead of moving joined in clusters. Hence, the velocity of each particle may be decomposed into a sum of the mean or bulk velocity of the whole system and an apparently random component to describe the motion of the particle relative to the mean flux, usually named \emph{peculiar velocity}. Such random motion resembles the thermal motion of atoms or molecules in ordinary gases where the collision time is much smaller than the mean free time between collisions. This analogy between granular and ordinary gases allows one to manage with a kinetic-theory picture of such systems. In that context, the mean-square value of the random velocites is commonly referred to as the \emph{granular temperature}. This term, first coined by Ogawa \cite{O78}, has nothing to do with the usual thermal temperature, which plays no role in the dynamics of granular flows, despite such name. %However, by using this kinetic analogy it is possible to generate pressure and governs the internal transport rates of mass, momentum and energy.

% SERERO JFM 2005
%Instead of moving in many-particle blocks, each particle moves freely and “independently”. In the
%rapid-flow regime, the velocity of each particle may be decomposed into a sum of the mean velocity of
%the bulk material and an apparently random component to describe the motion of the particle relative
%to the mean. The analogy between the random motion of the granular particles and the thermal motion
%of molecules in the kinetic-theory picture of gases is so strong that the mean-square value of the random
%velocities is commonly referred to as the “granular temperature” - a term first used by Ogawa [133].

Under these conditions, kinetic theory together with numerical simulations are the best tools to describe the behaviour and provide constitutive equations for rapid and diluted granular flows which gives insight into the physical origin of the transport properties. The analogy between granular and ordinary gases was first introduced by Maxwell in 1859 to describe Saturn's planetary rings \cite{MAXWELL27} and constitutes one of the most remarkable applications of the kinetic theory of granular media. 
% Garzo-Chamorro 2013. PRE87
Thus, from the point of view of kinetic theory, the study of granular gases is an interesting and fundamental challenge since it involves the generalization of classical kinetic equations (such as the Boltzmann, Enskog or Boltzmann-Lorentz equations, for instance) to dissipative dynamics.

On the other hand, driven granular gases can be seen as a prototype model of a suspension of solid grains inmersed in a fluid in the dilute limit \cite{GTSH12}. In those cases, %the material may still be considered to be granular if 
the stress due to the grains exceeds that due to the fluid (the ratio of the two is known as the Bagnold number), so that the effects of the fluid can be ignored \cite{SGNT06}. This condition is accomplished for example in aerosols or suspensions in wich the gravity is balanced with bouyancy \cite{FRIEDLANDER1997, KTT98, KSTT00}. In these systems, the influence of the interstitial fluid on the dynamic properties of the solid phase is neglected in most theoretical and computational works. On the other hand, the effects of the interstitial fluid turns out to be significant for a wide range of practical applications and physical phenomena like for instance species segregation \cite{MLNJ01, NSK03, Yan03, SSK04, MCEKNJ05, WZXS08, ZHK08, Idler09, CPSK10, PGM14} or in biophysics where active matter may be considered as a driven granular suspension \cite{KSRS14}. For this reason, the study of gas-solid flows has atracted the attention of engineering and physicist communities in the last few years \cite{KH01}.

The description of gas-solid suspensions, whose dynamics is very complex, is a long-standing branch of classical fluid mechanics \cite{B74}. For instance, particles suspended in a fluid feel a lubrication force, transmitted by the surrounding fluid but originated by the presence of another nearby particle. It is known that this kind of interaction (usually called \emph{hydrodynamic interaction}) depends also on the global configuration of the set of grains \cite{BB88}, giving rise to tensor-rank force equations. The modeling of these lubrication forces is rather involved and several approaches can be used. For this reason, there is a large bibliography that extends for decades and that is devoted to the study of this kind of interactions (Stokesian or Stokes dynamics) \cite{BB88, BGP11, LXWZJ14}.

Nevertheless, in the dilute suspension limit, these hydrodynamic interactions become less relevant \cite{BB88,B74} and only the isolated body resistance is retained, usually in the form of a simple drag force. On the other hand, due to the inherent complexity of the interaction between the interstitial fluid and the granular particles, early kinetic theory studies have neglected in most cases the effect of inelasticity in suspended particle collisions \cite{TMS84,LMJ91,MS94,SM94,TK95}. This kind of approach is not entirely accurate since of course in most real cases the sizes of suspended particles are big enough to render particle collisions inelastic (bigger than $1~\mu\mathrm{m}$, otherwise particles may be considered as colloids, for which collisions are elastic \cite{LXWZJ14, GHP15}). Inelasticity in the collisions can play a major role in the dynamics of granular (as opposed to colloidal) suspensions, specially in the dilute limit at high Stokes number, where grain-grain collisions effects dominate over many particle hydrodynamic interactions \cite{E11}. However, only more recent works have dealt with inelastic collisions in the case of dilute \cite{YPTKT01,TFAJ07} and moderately dense \cite{SMTK96} suspensions.

%In dense suspensions the hydrodynamic interaction among particles are important \cite{IH95}, but effects of the interstitial fluid may be regarded as a thermostat driven by the Langevin forces in dilute suspensions.
%\rojo{ATENCIÓN. EN CONSTRUCCIÓN $\rightarrow$}
%Most researches for granular gases are interested in undriven systems which are difficult to be achieved in actual experiences. We believe that more systematic studies for driven granular gases are required.

Despite the apparent similarity between granular and molecular gases, there are, however, fundamental differences to take into account. The first one is related with the size of the grains in a granular gas. Due to the macroscopic dimensions of the granular particles, the typical number of them in laboratory conditions is much smaller than Avogadro's number and hence, the fluctuations of their hydrodynamic fields are much bigger than in molecular gases \cite{TG98}. However, their number is sufficiently large to admit a statistical description.
%This difficulty could be solved by sufficently large sized granular gas exhibiting properties similar to those of a molecular gas. 

Granular gases present, nevertheless, a deeper difference with molecular gases. This difference comes from the inelastic character of collisions, which gives rise to a loss of kinetic energy. Thus, in order to keep the granular gas in rapid flow conditions, energy must be externally injected into the system to compensate for the energy dissipated by collisions. Therefore, granular matter can be considered as a good example of a system that inherently is a non-equilibrium state.

Apart from the collisional cooling, there is another fundamental open question in granular gases: the posible lack of separation between microscopic and  macroscopic length or time scales. To apply a continuum hydrodynamic approach it is necessary that there exists a clear separation between macroscopic and microscopic scales, that is, spatial variations of hydrodynamic fields must occur on a length scale larger than the mean free path of the gas molecules. Correspondingly, the typical macroscopic time scale should be larger than the mean free time between two collisions. 

However, several authors claim that the above scale separation does not exist for finite dissipation \cite{GS96, SGN96, SG98, TG98, DLK95, G99} and the granular hydrodynamic description only applies in the quasi-elastic limit. The reason for this concern resides in the fact that the inverse of the cooling rate (which measures the rate of energy loss due to collisional dissipation) introduces a new time scale not present for elastic collisions. The variation of the (granular) temperature $T$ over this new time scale is faster than over the usual hydrodynamic time scale and hence, as inelasticity increases, it could be possible that $T$ were not considered a \emph{slow} variable as in the usual hydrodynamic description. 

%To put the above controversial issue in a proper context, let us consider a granular gas (with a coeffcient of normal restitution $\alpha\leq1$. The spatial variation of the mean velocity over the mean free path in the spanwise direction is $\Delta u \sim a\,\ell$. This macroscopic velocity gradient must be small compared with the microscopic velocity fluctuations of the particles to ensure the validity of continuum approach. That is $a\,\ell \ll \sqrt{T}$, where $T$ is the granular temperature and has dimensions of squared velocity as follows from dimensional analysis. For shear flows this temperature is $T\sim a^2\ell^2/(1-\alpha^2)$ and, therefore, continuum approach requires the condition $\sqrt{1-\alpha^2}\ll1$. That is, the assumptions of a small velocity gradient implies small inelasticity. Similarly, the ratio between microscopic mean free time $\tau=\ell/\sqrt{T}$ and the macroscopic time $a^{-1}$ is also proportional to $\sqrt{1-\alpha^2}$. However, strictly speaking, the above arguments do not invalidate the validity conditions for hydrodynamics; they are only a signal of the failure of the NS approximation (small shear rates). In this regime, one needs constitutive relations more complex than the NS ones to obtain the rheological properties of the system as happens for molecular gases under simpleor uniform shear flow \cite{GARZO03}. The need for more complex constitutive equations does not  mean a breakdown of hydrodynamics \cite{TG98}, only shows the limitations of the NS hydrodynamic equations \cite{DB99}

Despite the above difficulties, in recent years it has been proved that it is possible to apply a hydrodynamic description for the study of granular gases. The main condition for a flow to be considered as a candidate for a hydrodynamic description is a state of continual collisions. This implies that all particles within each small cell are moving randomly relative to the mean flow velocity of the cell \cite{D09}.

%CHAMORRO PRE92 2015
Nevertheless, the ranges of interest of the physics of granular gases fall frequently beyond Newtonian hydrodynamics since the strength of the spatial gradients is large in most situations of practical interest (for example, in steady states). This is essentially due to %the coupling between collisional cooling and spatial gradients that under steady states
the balance between viscous heating and collisional cooling and usually moderately large spatial gradients can appear \cite{G03,SGD04,VU09}. As said before, in these steady states, a hydrodynamic description is still valid but with complex constitutive equations \cite{VSG10,VSG13}. 
%A very neat example of this is the simple or uniform shear flow (USF) \cite{C90}, that except in the quasi-elastic limit, is essentially non-newtonian. It is characterized by a linear velocity field (that is $\partial U_x/\partial y\equiv a =\text{const}$), constant density $n$ and constant temperature $T$. In particular, in the USF state the presence of shearing induces anisotropies in the pressure tensor $P_{ij}$, namely, nonzero shear stress $P_{xy}$ and normal stress differences $P_{xx}-P_{yy}$ and $P_{yy}-P_{zz}$.

%@@@@@@   for moderate inelasticities by choosing a suitable base state \cite{SGD04} or taking into account higher-order terms in the expansion of the distribution function in powers of the Knudsen number $Kn=\ell/L$ \cite{G99}

\section{Structure of the Thesis}

We have organized this thesis as follows.

In Chapter 2 we present the details of the model for driven granular gases previously explained as a paradigm of dilute gas-solid suspensions. In addition, we display the general mathematical and numerical tools to be employed in the present work. 

Homogeneous steady states of a driven granular fluid are analyzed in Chapter 3. After a transient regime, the gas reaches a steady state characterized by a \emph{scaled} distribution function $\varphi$ that does not only depend on the dimensionless velocity $\mathbf{c}\equiv \mathbf{v}/v_0$ ($v_0$ being the thermal velocity) but also on the dimensionless driving force parameters characterizing the external driving forces. The dependence of $\varphi$ and its first relevant velocity moments $a_2$ and $a_3$ (which measure non-Gaussian properties of $\varphi$) on both the coefficient of restitution $\al$ and the driving parameters is widely investigated by means of the Direct Simulation Monte Carlo (DSMC) method. In addition, approximate forms for $a_2$ and $a_3$ are also derived from an expansion of $\varphi$ in Sonine polynomials. The theoretical expressions of the above Sonine coefficients agree  well with simulation data, even for quite small values of $\alpha$.  Moreover, the third order expansion of the distribution function makes a significant accuracy improvement for larger velocities and inelasticities over theoretical predictions made by considering only the second order expansion. Results also show that the non-Gaussian corrections to the distribution function $\varphi$ are smaller than those observed for undriven granular gases.
%, we perform direct Monte Carlo simulations \cite{BIRD94} of the Enskog equation to confirm that indeed the scaled distribution $\varphi$ presents this universal character for arbitrary values of the coefficient of restitution $\al$ and the external driving parameters. As a second goal, we shall characterize the behavior of $\varphi(c,\xi^*)$ in the domain of thermal velocities by evaluating the two first non-trivial coefficients ($a_2$ and $a_3$) of an expansion of $\varphi$ in Sonine polynomials. Given that both coefficients cannot be exactly obtained, we will propose two different approximation to estimate $a_2$ and $a_3$. In particular, we provide expressions for the coefficient $a_2$ in a more accurate calculation method than in our previous work \cite{GCV13}. Therefore, we give an analytical expression for the distribution function with one more term, and in a more refined approximation.  As we will see, the comparison with the direct simulation Monte Carlo (DSMC) results obtained specifically for this work shows that the analytical expression of the distribution function derived here describes very well the system in a wide range of velocities. %A preliminary report of part of the results offered in this Chapter has been published elsewhere \cite{M12}.

The aim of Chapter 4 is to determine the NS transport coefficients of a \emph{dense} driven granular gas of inelastic hard spheres in the framework of the Enskog kinetic equation. Like in the undriven case \cite{GD99,L05}, the transport coefficients are obtained by solving the Enskog equation by means of the Chapman-Enskog (CE) expansion \cite{CHAPMAN70} around a certain reference state $f^{(0)}$ (zeroth-order approximation). While in the undriven case the distribution $f^{(0)}$ is chosen to be the local version of the \emph{Homogeneous Cooling State} (HCS), there is some flexibility in the choice of $f^{(0)}$ for a driven gas. For simplicity, one possibility is to take a local thermostat such that the distribution $f^{(0)}$ is still stationary at any point of the system. This was the choice assumed in previous works \cite{GM02,G11} to compute the transport coefficients of a heated granular gas. On the other hand, for \emph{general} small deviations from the steady reference state, the zeroth-order distribution $f^{(0)}$ is not in general a stationary distribution since the collisional cooling cannot be compensated \emph{locally} by the heat injected by the driving force. This fact introduces additional difficulties not present in previous studies \cite{GM02,G11}. In this Chapter, we will adopt this point of view and will consider this kind of thermostat that seems to be closer to the one used in computer simulations. 

The determination of the transport coefficients involves, like in the undriven case \cite{BDKS98,GD02}, the evaluation of certain collision integrals that cannot be \emph{exactly} computed due to the complex mathematical structure of the (linearized) Enskog collision operator for Inelastic Hard Spheres (IHS). Thus, in order to obtain explicit expressions for the above coefficients one has to consider additional approximations. In Chapter 5 we propose a possible way of circumventing these technical difficulties inherent to IHS, by considering instead the so-called Inelastic Maxwell Models (IMM) for dilute granular gases. As for ordinary gases, the collision rate for these models is independent of the relative velocity of the two colliding particles. In the case of elastic collisions (conventional molecular gases), Maxwell models are characterized by a repulsive potential that (in three dimensions) is proportional to the inverse fourth power of distance between particles. On the other hand, for inelastic collisions, Maxwell models can be introduced in the framework of the Boltzmann equation at the level of the cross section, without any reference to a specific interaction potential \cite{E81}. In addition, apart from its academic interest, it is worthwhile remarking that experiments \cite{KSSAOB05} for magnetic grains with dipolar interactions are well described by IMM.
Therefore, the motivation of the Chapter is twofold. On the one hand, the knowledge of the first collisional moments for IMM allows one to re-examine the problem studied in the previous Chapter in the context of the (inelastic) Boltzmann equation and without taking any additional and sometimes uncontrolled approximations. On the other hand, the comparison between the results obtained from IMM with those derived from IHS \cite{GCV13, GCV13err} can be used again as a test to assess the reliability of IMM as a prototype model for characterizing real granular flows. Previous comparisons have shown a mild qualitative agreement in the freely cooling case  \cite{S03,GA05} while the agreement between IMM and IHS significantly increases for low order velocity moments in the case of driven states (for instance, the simple shear flow problem) \cite{G03,G07,SG07}.
The main advantage of using IMM instead of IHS is that a velocity moment of order $k$ of the Boltzmann collision operator only involves moments of order less than or equal to $k$. This allows to evaluate the Boltzmann collision moments without the explicit knowledge of the distribution function \cite{GS07}. This property opens up the search of exact solutions to the Boltzmann equation and justifies the interest of physicists and mathematicians in IMM in the last few years \cite{BCG00, CCG00, BK00, C01, EB02-1, EB02-2, EB02-3, BMP02, BK02-1, KB02, BK02-2, MP02, BC03, BCT03, SE03, BG06, ETB06-1, ETB06-2, BTE07, SGV09, GS11, GT11, GT12-1, SG12}. Thus, in this Chapter, we determine in the steady state the exact forms of the shear viscosity $\eta$, the thermal conductivity $\kappa$ and the transport coefficient $\mu$ (that relates the heat flux with the density gradient) as a function of the coefficient of restitution $\alpha$ and the thermostat forces. As for IHS \cite{GCV13}, the expressions of $\eta$, $\kappa$ and $\mu$ are obtained by solving the Boltzmann equation for IMM up to first order in the spatial gradients by means of the CE expansion \cite{CHAPMAN70}.

In Chapter 6 we study a steady laminar shear flow with null heat flux, usually called \emph{Uniform Shear Flow} (USF), in a gas-solid suspension at low density. The solid particles are modeled as a gas of smooth hard spheres with inelastic collisions while the influence of the surrounding interstitial fluid on the dynamics of grains is modeled by means of a volume drag force, in the context of a rheological model for suspensions. The model is solved by means of three different but complementary routes, two of them being theoretical (Grad's moment method applied to the corresponding Boltzmann equation \cite{G49} and an exact solution of a kinetic model adapted to granular suspensions \cite{BDS99}) and the other being computational (Monte Carlo simulations of the Boltzmann equation \cite{MGSB99}). Unlike in previous studies on granular sheared suspensions \cite{TK95, SMTK96}, the collisional moment associated with the momentum transfer is determined in Grad's solution by including \emph{all} the quadratic terms in the stress tensor. This theoretical enhancement allows us for the detection and evaluation of the normal stress differences in the plane normal to the laminar flow. In addition, the exact solution of the kinetic model gives the explicit form of the velocity moments of the velocity distribution function. Comparison between our theoretical and numerical results shows in general a good agreement for the non-Newtonian rheological properties, the kurtosis (fourth velocity moment of the distribution function) and the velocity distribution of the kinetic model for quite strong inelasticity and not too large values of the (scaled) friction coefficient characterizing the viscous drag force. This shows the accuracy of our analytical results that allows us to describe in detail the flow dynamics of the granular sheared suspension.

\chapter{Kinetic Theory of driven granular gases} 

\label{Chapter2} 

\lhead{Chapter 2. \emph{Kinetic Theory of driven granular gases}} 

\section{Introduction}

In this Chapter we describe in detail the model of driven granular gases studied in this work and the theoretical background and numerical tools that will be used throughout the next Chapters.

As discussed previously, granular matter in rapid flow regime obeys a hydrodynamic description that is different, and more general, than the hydrodynamics of ordinary gases. This is due to the absence of energy conservation which introduces modifications in the kinetic and its corresponding momentum balance equations. %conventional NS equations of states with small spatial gradients in hydrodynamic fields.
%\azul{
The energy loss in the inelastic collisions makes neccesary the introduction of externals forces in order to avoid instabilities and keep the system in rapid flow conditions. Thus, as we said before, granular gases may be regarded as prototypes of non-equilibrium systems and kinetic theory is an appropriate tool to study their properties \cite{FERZIGER72}. 
%From a fundamental point of view, Enskog and Boltzmann equations \cite{POSCHEL04,GoSh95, BDS97} (for dense and dilute systems, respectively) with the appropriate modifications to account for the effect of inelastic binary collisions have been used in the past few years as starting points to derive the transport properties of granular fluids.
% }

%\rojo{algo más...?}

%\section{Basic concepts}
Kinetic Theory is based on the assumption that the macroscopic properties of a collection of gas molecules can be obtained from the \emph{one-particle velocity distribution function} $f(\textbf{r},\textbf{v},t)$, where $\textbf{r}$ and $\textbf{v}$ are the position and velocity of one particle, respectively. In other words, $f(\textbf{r},\textbf{v},t)$ provides all of the relevant information about the state of the system. The distribution function $f(\textbf{r},\textbf{v},t)$ is defined as the average number of particles having velocity between $\textbf{v}$ and $\textbf{v}+d\textbf{v}$ in a volume $d\textbf{r}$ centered at point $\textbf{r}$ in the instant $t$.

\section{The model for Driven Granular Gases}

We consider a system of smooth inelastic hard spheres (or disks) in $d$ dimensions ($d=2$ for disks and $d=3$ for spheres) with mass $m$ and diameter $\sigma$ driven by external non-conservative forces that act homogeneously over the system. We assume in this work that inelastic collisions are characterized by a constant coefficient of normal restitution $0 < \alpha < 1$, where $\alpha=1$ corresponds to elastic collisions, and $\alpha=0$ to completely inelastic collisions (all the kinetic energy contained in the velocity components in the direction of contact line at collision is lost).
% A more realistic model considers that $\alpha$ depends on the relative velocity of the colliding particles \cite{STRONGE2000, JOHNSON1985, Alfadepend}. Some effects of such dependence of $\alpha$ on the transport coefficients of the system can be seen in Ref. \cite{BP03}.

%\rojo{
Although at moderate densities correlations between the velocities of two particles that are about to collide could not be negligible \cite{BK01, S08}, in this work we have still assumed the molecular chaos hypothesis \cite{GoSh95} and therefore the two-body distribution function can be factorized into the product of the one-particle velocity distribution functions $f(\bt{r},\bt{v},t)$.%}
 
%At moderate densities, one can still assume that there are no correlations between the velocities of two particles that are about to collide, that is, we assume the molecular chaos hypothesis \cite{GoSh95} and therefore the two-body distribution function can be factorized into the product of the one-particle velocity distribution functions $f(\bt{r},\bt{v},t)$. 
 
%\rojo{\emph{FRAN: cuidado. se sabe, porque se ha medido, que esas correlaciones existen y son importantes. otra cosa es decir (en una redaccion mas humilde), que es lo que yo haria, que esas correlaciones no son el efecto mas importante en nuestro estudio, y que las vamos a despreciar. ademas, daria referencias sobre mediciones experimentales de esas correlaciones. \cite{BK01, S08}}}

As a result of the action of the external volume forces, the system reaches a non-equilibrium stationary fluidized state. We can model the forces $\bt{F}^{\text{th}}(t)$ that the surrounding fluid exerts on the granular gas. Thus, the equation of motion for a particle $i$ with velocity $\bt{v}_i$ can be written as \cite{SVGP10, KSZ10, PGGSV12, VAZ11, GSVP11-1}
%\cite{PLMV98,PLMV99,CLH00,PEU02,PBL02,VPBTW06,FAZ09,SVGP10,KSZ10,PGGSV12,NETP99,VAZ11,GSVP11-1} %\cite{thermostat}
%This granular gas of inelastic hard particles is maintained in a stationary fluidized state by mean of external volume forces that act locally on each particle and try to mimic the interaction with a surrounding molecular fluid \cite{WM96,PLM98,PLM99,PLM07,VPV08,PV09}.
\begin{equation}
  \label{2.1}
  m\dot{\bt{v}}_i=\bt{F}^{\text{th}}_i(t)+\bt{F}^{\text{coll}}_i,
\end{equation}
where $\bt{F}^{\text{th}}_i(t)$ \fran{stands for the forces coming from the surrounding fluid} and $\bt{F}^{\text{coll}}_i$ is the force due to inelastic collisions. 

\fran{We will model ${\bf F}_i^{\text{th}}$ as a force composed} by two different terms: (i) a stochastic force where the particles are randomly kicked between collisions \cite{WM96} and (ii) a viscous drag force which mimics the interaction of the particles with an effective viscous \emph{bath} at temperature $T_\text{b}$. Under the above conditions one can consider the following generalized Langevin model for the instantaneous acceleration of a grain:
%More explicitly,
%In this work we have considered a thermostat force composed by two independent terms:
\begin{equation}
 \label{2.2}
 \bt{F}_i^{th}(t)=\bt{F}_i^{\text{st}}(t) + \bt{F}_i^{\text{drag}}(t).
\end{equation}

The first term $\bt{F}_i^{\text{st}}(t)$ attempts to simulate the kinetic energy gain due to eventual collisions with the rapidly moving particles of the surrounding fluid. \fran{This effect is specially important for small granular particles.} 
%The first term corresponds to a Gaussian white noise force that %tries to 
%simulates the kinetic energy gain due to eventual collision with the molecules of the surrounding fluid. It does this by adding a \emph{random} velocity to each particle. 
\fran{The additional velocity is drafted} from a Maxwellian distribution with a characteristic variance determined by the \emph{noise intensity} $\xi^2_b$ \cite{WM96}. The stochastic force is assumed to have the form of a Gaussian white noise and satisfies the conditions \cite{MS00}:
\begin{equation}
  \label{noise}
  \langle\bt{F}_i^{\text{st}}(t)\rangle = \bt{0}, \qquad   \langle\bt{F}_i^{\text{st}}(t)\bt{F}_j^{\text{st}}(t')\rangle = \mathbb{1} m^2 \xi_b^2 \delta_{ij} \delta(t-t'),
\end{equation}
where $\mathbb{1}$ is the $d\times d$ unit matrix and $\delta_{ij}$ is the Kronecker delta function. Here the subindexes $i$ and $j$ refer to particles $i$ and $j$, respectively.

For homogeneous states the drag force $\bt{F}_i^{\text{drag}}(t)$ is proportional to the instantaneous particle velocity $\textbf{v}_i$. The generalization of $\bt{F}_i^{\text{drag}}(t)$ to non-homogeneous situations is a matter of choice. Here, since our model attempts to incorporate the effect of the interstitial viscous fluid into the dynamics of grains, we define $\bt{F}_i^{\text{drag}}(t)$ as
\begin{equation}
\label{eqFdrag}
%  \bt{F}_i^{\text{drag}}=-\gamma_b \left(\bt{v}_i(t)-\bt{U}_g\right),
   \bt{F}_i^{\text{drag}}=-\gamma_b \left(\bt{V}_i+\Delta\bt{U}\right),
\end{equation}  
%where $\gamma_b$ is a drag or friction coefficient and $\bt{U}_g$ can be interpreted as the mean velocity of \fran{the interstitial gas}. It is assumed to be a known quantity of the model. 
%Eq. \eqref{eqFdrag} defines the characteristic interaction time with the gas phase $\tau_b^{-1}=\gamma_b/m$. 
where $\gamma_b$ is a drag or friction coefficient, $\bt{V}_i=\bt{v}_i-\bt{U}$ is the particle fluctuation or \emph{peculiar} velocity, $\Delta \bt{U}=\bt{U}-\bt{U}_g$ is the difference between the mean velocity of the interstitial gas $\bt{U}_g$ (assumed to be a known quantity of the model) and the mean flow velocity of grains $\bt{U}$ defined by
\begin{equation}
\label{U}
 \bt{U}(\textbf{r},t) \equiv \frac{1}{n(\textbf{r},t)}\int d\textbf{v} \; \textbf{v} f(\textbf{r},\textbf{v},t).
\end{equation}

%In addition, since our model attempts to mimic the effect of the interstitial gas phase on grains, in the right hand side of Eq.\ \eqref{eqFdrag} we have considered the \emph{peculiar velocity} redefinited as $\mathbf{v}-\mathbf{U}$ (rather the instantaneous velocity $\mathbf{v}$ of particle) in the drag force expression. Here, $\mathbf{U}_g$ can be interpreted as the mean velocity of gas surrounding the solid particles that is assumed to be a known quantity of the model and $\Delta \bt{U}=\bt{U}-\bt{U}_g$ the difference between mean flux velocity of grains and the mean velocity of the insterstitial gas.

This kind of thermostat composed by two different forces has been widely employed in \fran{the regime of Stokesian dynamics for which the many-body hydrodynamic forces are weak.}
%\cite{PLM98,PLM99,PLM07,VPV08,PV09}. One of the advantages of using the model \eqref{2.1} instead of other kind of thermostats is that the temperature of the thermostat $T_\text{b}$ (different from the temperature of the granular fluid $T<T_\text{b}$) is always well defined. In particular, this thermostat is able to equilibrate the system when collisions are elastic. 
Moreover, a similar external driving force to that of Eq.\ \eqref{2.2} has been recently proposed to model the effect of the interstitial fluid on grains in monodisperse gas-solid suspensions \cite{GTSH12}.

The corresponding term in the Enskog kinetic equation associated with the stochastic forces is represented by the Fokker-Plank operator $-\frac{1}{2}\xi_b^2\partial^2/\partial v^2$ \cite{NE98}. 
%Therefore, the contribution of the stochastic and drag forces to the kinetic equation is through terms of the form
%\begin{equation}
%  \label{eqForcesTh0}
%  \mathcal{F}\,f = \mathcal{F}_{st}\,f + \mathcal{F}_{drag}\,f, 
%\end{equation}
%where 
%\begin{equation}
%  \label{eqForcesTh}
%  \mathcal{F}_{st}\,f = \frac{1}{2}\xi_b^2\frac{\partial^2}{\partial v^2}\,f  , \quad \mathcal{F}_{drag}\,f = \frac{\gamma_b}{m}\frac{\partial}{\partial\bt{v}}\cdot \bt{v}\,f.
%\end{equation}

For moderately dense gases, the Enskog kinetic equation for the one-particle velocity distribution function $f(\bt{r},\bt{v},t)$ adapted to dissipative collisions reads \cite{GTSH12}:
%\begin{equation}
%  \label{eqEB00}
%  \partial_t f(\bt{r},\bt{v},t) + \bt{v}\cdot\nabla f(\bt{r},\bt{v},t) - {\cal{F}} f(\bt{r},\bt{v},t) = J_E[\bt{r},\bt{v}|f,f]
%%  \partial_t f + \bt{v}\cdot\nabla f - \frac{\gamma_b}{m}\frac{\partial}{\partial \bt{v}}\cdot \bt{V} f - \frac{1}{2}\xi_b^2\frac{\partial^2}{\partial v^2} f = J_E[\bt{r},\bt{v}|f,f]
%\end{equation}
%where $\cal{F}$ is an operator representing the effect of an external force.
%Thus, the corresponding Enskog kinetic equation for the one-particle velocity distribution function $f({\bf r},{\bf v},t)$ reads
\begin{equation}
\label{eqEB}
\partial_{t}f + \mathbf{v} \cdot \mathbf{\nabla}f 
- \frac{\gamma_\text{b}}{m}\frac{\partial }{\partial{\bf v}}\cdot{\bf V}f 
- \frac{\gamma_\text{b}}{m} \Delta \mathbf{U} \cdot \frac{\partial }{\partial{\bf v}}f 
- \frac{1}{2}\xi_\text{b}^2\frac{\partial^2}{\partial v^2}f
= J_{\text{E}}\left[f,f\right],
%\partial_{t}f+\mathbf{v}\cdot \mathbf{\nabla}f -\frac{\gamma_\text{b}}{m} \Delta \mathbf{U}\cdot\frac{\partial f}{\partial{\bf v}} -\frac{\gamma_\text{b}}{m} \frac{\partial}{\partial {\bf v}}\cdot {\bf V} f-\frac{1}{2}\xi_\text{b}^2\frac{\partial^2}{\partial v^2}f=J_{\text{E}}\left[f,f\right],
\end{equation}
where
\begin{eqnarray}
  \label{eqJE}
  J_E[f,f]&=&\sigma^{d-1}\int\;d\bt{v}_2 \int\;d\widehat{\bs{\sigma}}\;\Theta(\widehat{\bs{\sigma}}\cdot\bt{g}_{12})(\widehat{\bs{\sigma}}\cdot\bt{g}_{12})  \nonumber\\
&\times&[\alpha^{-2}\chi(\bt{r},\bt{r}-\bs{\sigma}) f(\bt{r},\bt{v}'_1;t) f(\bt{r}-\bs{\sigma},\bt{v}'_2;t) \nonumber\\
&-& \chi(\bt{r},\bt{r}+\bs{\sigma}) f(\bt{r},\bt{v}_1;t) f(\bt{r}+\bs{\sigma},\bt{v}_2;t)]\nonumber\\
\end{eqnarray}
is the Enskog collision operator. 
%In Eq. \eqref{eqEB} $\bt{V}=\bt{v}-\bt{U}$ is the particle fluctuation or \emph{peculiar} velocity and $\Delta \bt{U}=\bt{U}-\bt{U}_g$ the difference between the mean velocity of the interstitial gas and the mean flow velocity of grains defined by
%\begin{equation}
%\label{U}
% \bt{U}(\textbf{r},t) \equiv \frac{1}{n(\textbf{r},t)}\int d\textbf{v} \; \textbf{v} f(\textbf{r},\textbf{v},t).
%\end{equation}

In Eq.\ \eqref{eqJE} $\bt{g}_{12}=\bt{v}_1-\bt{v}_2$ is the relative velocity of two colliding particles, $\Theta$ is the Heaviside step function, $\bs{\sigma}=\sigma\widehat{\bs{\sigma}}$ with $\widehat{\bs{\sigma}}$ the unit vector along the line of centers of the colliding particles, that is, the apsidal vector defined by $(\mathbf{g}'_{12}-\mathbf{g}_{12})/|\mathbf{g}'_{12}-\mathbf{g}_{12}|$ with $\mathbf{g}'_{12}=\bt{v}'_1-\bt{v}'_2$, and $\chi[\bt{r},\bt{r}+\bs{\sigma}|n(\bt{r},t)]$ is the equilibrium pair correlation function at contact as a functional of the nonequilibrium density field defined by
\begin{equation}
\label{n}
n(\bt{r},t)\equiv\int d\textbf{v} f(\textbf{r},\textbf{v},t).
\end{equation}
The quantity $\chi$ accounts for the increase of the collision frequency due to excluded volume effects by the finite size of particles. For spheres ($d=3$), we consider the Carnahan-Starling \cite{CS69} approximation for $\chi$ given by
\begin{equation}
\label{chi3d}
  \chi(\phi)=\frac{1-\frac{1}{2}\phi}{(1-\phi)^3}.
\end{equation}
In the case of disks ($d=2$), $\chi$ is approximately given by \cite{JM87}
\begin{equation}
\label{chi2d}
  \chi(\phi)=\frac{1-\frac{7}{16}\phi}{(1-\phi)^2}.
\end{equation}
In Eqs.\ \eqref{chi3d} and \eqref{chi2d}, $\phi$ is the solid volume fraction. For a $d$-dimensional system it is defined as
\begin{equation}
  \phi=\frac{\pi^{d/2}}{2^{d-1}d\,\Gamma\left(\frac{d}{2}\right)}n\sigma^d.
\end{equation}

Notice that the introduction of the two thermostat terms in the kinetic equation \fran{involves} the emergence of two new and independent time scales given by $\tau_{st}=v_0^2/\xi_b^2$ and $\tau_{drag}=m/\gamma_b$, respectively. 

%\rojo{\emph{FRAN: aqui si merece la pena extenderse algo más.}}
%
%\rojo{\emph{MOY: ¿Cómo?}}

For uniform states, the collision operator \eqref{eqJE} is identical to the Boltzmann collision operator for a low-density gas except for the presence of the factor $\chi$, i.e. $J_E[\bt{v}_1|f,f)]=\chi J[\bt{v}_1|f,f)]$ where% wich accounts for the increase of the collision frequency due to excluded volume effects.
\begin{eqnarray}
  \label{eqJE2}
    J[\bt{v}_1|f,f)]= \sigma^{d-1}\int\;d\bt{v}_2 \int\;d\widehat{\bs{\sigma}}\;\Theta(\widehat{\bs{\sigma}}\cdot\bt{g}_{12})(\widehat{\bs{\sigma}}\cdot\bt{g}_{12})[\alpha^{-2}f(\bt{v}'_1)f(\bt{v}'_2)]-f(\bt{v}_1)f(\bt{v}_2)].\nonumber\\
\end{eqnarray}
In the dilute limit ($\phi\rightarrow0$) the size of the particles is negligible compared with the mean free path $\ell$ and then $\chi\rightarrow 1$. In this case there are no collisional contributions to the fluxes. It is important to recall that the assumption of molecular chaos is maintained in the Enskog equation. This means that the two-body function factorizes into the product of one-particle distribution functions \cite{BDS99, GoSh95}.

The primes in Eq.\eqref{eqJE} denote the the initial values of velocities {${\bt{v}'_1,\bt{v}'_2}$} that lead to {${\bt{v}_1,\bt{v}_2}$} following binary collisions:
\begin{equation}
  \label{eqBinCol}
  \bt{v}'_{1,2}=\bt{v}_{1,2} \mp \frac{1}{2}(1+\alpha^{-1})(\widehat{\bs{\sigma}}\cdot\bt{g}_{12})\widehat{\bs{\sigma}}.
\end{equation}

%\rojo{AQUI VIENE LA FIGURA DE COLISION}

The macroscopic balance equations for the system are obtained by multiplying the Enskog equation  \eqref{eqEB} by $\{1, m\bt{v},\frac{1}{2}m v^2\}$ and integrating over velocity. After some algebra one gets \cite{GD99, NE98}
\begin{equation}
  \label{eqBaln01}
  D_tn + n\nabla\cdot\bt{U}=0,
\end{equation}
\begin{equation}
  \label{eqBalU01}
  D_t \bt{U} + \rho^{-1}\nabla\cdot\mathsf{P}=-\frac{\gamma_b}{m}\Delta\bt{U},
\end{equation}
\begin{equation}
  \label{eqBalT01}
  D_tT+\frac{2}{dn}(\nabla\cdot\bt{q}+\mathsf{P}:\nabla\bt{U})=-\frac{2T}{m}\gamma_b+m\xi^2-\zeta T,
\end{equation}
where
\begin{equation}
\label{T}
  T(\textbf{r},t) \equiv \frac{m}{d\, n(\textbf{r},t)}\int d\textbf{v} \; V^2 f(\textbf{r},\textbf{v},t),
\end{equation}
is the \emph{granular temperature}\footnote{Here $T\equiv k_B T$ has units of energy.}.
In the above equations, $D_t=\partial_t+\bt{U}\cdot\nabla$ is the material time derivative and $\rho=mn$ is the mass density. The term $\zeta$ in the right hand of Eq.\eqref{eqBalT01} is the so-called \emph{cooling rate} given by
\begin{eqnarray}
\label{eqCool}
\zeta &=& -\frac{m}{dnT}\int d\mathbf{v}_1 V^2 J_E[\bt{r},\bt{v}_1|f,f] \nonumber\\
&=&\frac{\left(1-\alpha^{2}\right)}{4dnT} m \sigma^{d-1}\int d\mathbf{v}_{1}\int d\mathbf{v}_{2}\int d\widehat{\boldsymbol {\sigma }} \Theta (\widehat{\boldsymbol {\sigma }}\cdot \mathbf{g}_{12})(\widehat{ \boldsymbol {\sigma }}\cdot
\mathbf{g}_{12})^{3}f^{(2)}(\mathbf{r}, \mathbf{r}+\boldsymbol {\sigma} ,\mathbf{v}_{1},\mathbf{v}_{2};t), \label{14}\nonumber\\
\end{eqnarray}
where
\begin{equation}
f^{(2)}(\bt{r}_1,\bt{r}_2,\bt{v}_1,\bt{v}_2;t)=\chi(\bt{r}_1,\bt{r}_2|n(t)) f(\bt{r}_1,\bt{v}_1,t)f(\bt{r}_2,\bt{v}_2,t).
\end{equation}
The cooling rate is proportional to $1-\alpha^2$ and characterizes the rate of energy dissipated due to collisions \cite{VSG13}. The pressure tensor $\mathsf{P}(\bt{r},t)$ and the heat flux $\bt{q}(\bt{r},t)$ have both \emph{kinetic} and \emph{collisional transfer} contributions, $\mathsf{P}=\mathsf{P}^k+\mathsf{P}^c$ and $\bt{q}=\bt{q}^k+\bt{q}^c$. The kinetic contributions are given by
\begin{equation}
  \label{eqPk}
  \mathsf{P}^k(\bt{r},t)=\int d\bt{v}m\bt{V}\bt{V}f(\bt{r},\bt{v},t),
\end{equation}
\begin{equation}
  \label{eqqk}
  \bt{q}^k(\bt{r},t)=\int d\bt{v}\frac{m}{2}V^2\bt{V}f(\bt{r},\bt{v},t),
\end{equation}
%\begin{equation}
%  \label{eqKinCon}
%  \mathsf{P}^k=\int d\bt{v}m\bt{V}\bt{V}f(\bt{r},\bt{v},t) \qquad \bt{q}^k=\int d\bt{v}\frac{m}{2}V^2\bt{V}f(\bt{r},\bt{v},t),
%\end{equation}
and the collisional transfer contributions are \cite{GD99}
\begin{eqnarray}
  \label{eqColConP}
  \mathsf{P}^c(\bt{r},t) &=& \frac{1+\alpha}{4} m\sigma^d \int d\bt{v}_1\int d\bt{v}_2\int d\widehat{\bs{\sigma}}\Theta(\widehat{\bs{\sigma}}\cdot\bt{g}_{12})(\widehat{\bs{\sigma}}\cdot\bt{g}_{12})^2   \nonumber\\
  &&\times\widehat{\bs{\sigma}}\widehat{\bs{\sigma}} \int_0^1dx f^{(2)}[\bt{r}-x\bs{\sigma}, \bt{r}+(1-x)\bs{\sigma},\bt{v}_1,\bt{v}_2;t],
\end{eqnarray}
\begin{eqnarray}
  \label{eqColConq}
  \bt{q}^c(\bt{r},t) &=& \frac{1+\alpha}{4} m\sigma^d \int d\bt{v}_1\int d\bt{v}_2\int d\widehat{\bs{\sigma}}\Theta(\widehat{\bs{\sigma}}\cdot\bt{g}_{12})(\widehat{\bs{\sigma}}\cdot\bt{g}_{12})^2 \nonumber\\
  &&\times(\bt{G}_{12}\cdot\widehat{\bs{\sigma}})\widehat{\bs{\sigma}} \int_0^1dx f^{(2)}[\bt{r}-x\bs{\sigma}, \bt{r}+(1-x)\bs{\sigma},\bt{v}_1,\bt{v}_2;t],
\end{eqnarray}
where $\bt{G}_{12}=\frac{1}{2}(\bt{V}_1+\bt{V}_2)$ is the velocity of the center of mass.
%\begin{equation}
%f^{(2)}(\bt{r}_1,\bt{r}_2,\bt{v}_1,\bt{v}_2;t)=\chi(\bt{r}_1,\bt{r}_2|n(t)) f(\bt{r}_1,\bt{v}_1,t)f(\bt{r}_2,\bt{v}_2,t).
%\end{equation}

%GTSH 2012. Final de la sección 4.
\fran{Let us point out that the macroscopic equations for a granular suspension (or a driven granular gas), given by \eqref{eqBaln01}--\eqref{eqBalT01} \fran{have three additional terms} respect to the freely cooling granular gas} \cite{GD99} by the inclusion of three terms arising from the action of the surrounding fluid on the dynamic of grains. The term on the right hand side of Eq. \eqref{eqBalU01} gives the mean drag force between the two phases (fluid and solid). The other two are included in the granular energy balance equation \eqref{eqBalT01}. %The term that contains $\gamma_b$ represents the sink of energy due to viscous drag while the term containing $\xi_b^2$ represents the source arising from the change in particle momentum due to neighbour particles.
%CHAMORRO PRE 92. Suspensiones
%Furthermore, the form of the Enskog collision operator \eqref{eqJE}, (or Boltzmann operator in the dilute limit) is the same as for a dry granular gas and, hence, the collision dynamics does not contain any gas-phase parameter. Such important assumption requieres that the mean-free time between collisions is much less than the time taken by the fluid forces (viscous relaxation time) to significantly affect the motion of solid particles. Thus, the suspension model \eqref{eqEB} is expected to describe situations where the stresses exested by the intestitial fluid on particles are suffiently small that they have a weak influence on the dynamics of grains. However, as the density of fluid increases (liquid flows), the above assumption could be not reliable and hence one should take into account the presence of fluid into the binary collisions event.

\fran{The model \eqref{eqEB} can be seen as the Fokker-Planck model studied previously by Hayakawa for homogeneous systems \cite{H03} but with $\gamma_\text{b}$ and $\xi_\text{b}^2$ related by $\xi_\text{b}^2=2 \gamma_\text{b} T_\text{b}/m^2$}.

\section{The Chapman-Enskog method}
%%%%%%%%%%%%%%%%%%%%%%%%%%%%%%%%%%%%%%%%%%%%%%%%%%
% LO siguente ESTABA ANTES EN EL chap4
%%%%%%%%%%%%%%%%%%%%%%%%%%%%%%%%%%%%%%%%%%%%%%%%%%

%Chapman-Enskog method is a perturbative procedure by which one extracts the long time behavior of non-homogeneous solutions of the Enskog Eq. \eqref{eqEB} and computes the macroscopic description of the gas in terms of the hydrodynamical quantities of pressure, temperature, etc., for slightly non-uniform gases, when the field gradients are weak around a reference state. \cite{CHAPMAN70}
%The basic assumption is that for times long compared to the mean time between collisions, the distribution function $f(\bt{r},\bt{v},t)$ should have a universal functional form, so that the individual $f$ depends in its details only upon the local hydrodynamical quantities density, mean flow velocity and granular temperature.
%The main idea of this procedure is search of a parameter $\epsilon$ that measures the nonuniformity of the system. 

The macroscopic balance Eqs. \eqref{eqBaln01}--\eqref{eqBalT01} are not entirely expressed in terms of the hydrodynamic fields due to the presence of the pressure tensor $\mathsf{P}$, the heat flux $\bt{q}$ and the cooling rate $\zeta$ which are given in terms of the one-particle velocity distribution function $f(\bt{r},\bt{v},t)$. On the other hand, if this distribution function is expressed as a \fran{functional} of the fields, then $\mathsf{P}$, $\bt{q}$ and $\zeta$ will become functionals of the hydrodynamic fields through Eqs. \eqref{eqPk}--\eqref{eqColConq} and \eqref{eqCool}. \fran{The relations obtained after integration of \eqref{eqPk}--\eqref{eqColConq} are the constitutive hydrodynamic relations. They yield a} closed set of equations for \fran{the} hydrodynamic fields $n$, $\bt{U}$ and $T$.
The above hydrodynamic description can be derived by looking for a \emph{normal} solution to the Enskog equation. A normal solution is one whose all space and time dependence of the distribution $f(\bt{r},\bt{v},t)$ occurs through a functional dependence on the hydrodynamic fields:
\begin{equation}
  \label{eqfnormal}
  f(\bt{r},\bt{v},t) = f[\bt{v}| n(\bt{r},t), \bt{U}(\bt{r},t), T(\bt{r},t) ]
\end{equation}
That can be achieved by studing the order of magnitude of the various terms appearing in the Enskog equation \eqref{eqEB}. If we denote by $t_0$ a typical time, $L$ a typical length and $v_0$ a typical velocity, then:
\begin{equation}
  \frac{\partial f}{\partial t} = \mathcal{O}(t_0^{-1}f); \qquad \bt{v}\cdot\frac{\partial f}{\partial \bt{r}}=\mathcal{O}(v_0\mathit{L}^{-1}f); \qquad J[f,f]=\mathcal{O}(n\sigma^{d-1}v_0 f)
\end{equation} 
We can relate the quantity $n\sigma^{d-1}$ with the mean free path $\ell$, that is, the lenght of the free flight of particles between two successive collisions. For hard spheres, 
\begin{equation}
  \ell\approx(n\sigma^{d-1})^{-1} \, .
\end{equation}
The combination $(v_0\ell^{-1})$ can be considered as defining naturally a mean free time $\tau$ and its inverse as a measure of the collision frequency $\nu$. 

It seems clear the existence of two basic nondimensional numbers in the Enskog equation, $\tau/t_0$ and $\ell/\mathit{L}$. In a first approximation, we can take time and length scales to be comparable and, so, we can express the relative magnitudes of both sides of the Enskog equation by a single non-dimensional number
\begin{equation}
  \text{Kn}=\frac{\ell}{\mathit{L}},
\end{equation}
where Kn is called Knudsen number \cite{CERCIGNANI68}. The main assumption of the CE method is that the mean free path $\ell$ is small compared with the linear size of gradients $L$ which is of the order of the linear size of the experiment. In this case, $\text{Kn}\rightarrow0$ and there is a clear separation between the microscopic length scale $\ell$ and its macroscopic counterpart $L$.

\fran{The small Knudsen number condition is equivalent to small spatial gradients %(and time derivatives) 
of the hydrodynamic fields, if referred to the microscopic length scale of the mean free path (the collision frequency). For ordinary (elastic) gases this can be controlled by the initial or the boundary conditions. However, in granular gases inelasticity generates an independent macroscopic time derivative \cite{GS96,SG98,SGNT06} and as a consequence, the steady granular flows created by energy injection from the boundaries may be intrinsically non-Newtonian.}
%Due to this coupling, some authors  have given the CE solution in powers of both the Knudsen number (or equivalently, the spatial hydrodynamic gradients) and the degree of dissipation $\delta=1-\alpha^2$. 
%There, as for ordinary gases, we also assume that the spatial gradients are independent of the coefficient of restitution and hence, the correspondign NS order hydrodynamic equations apply for small spatial gradients but they are not limited a priori to weak inelasticity.

For small spatial variations, the functional dependence \eqref{eqfnormal} can be made local in space through an expansion in spatial gradients of the hydrodynamic fields. To generate it, $f$ is expanded in powers of the non-uniformity parameter $\epsilon$:
\begin{equation}
  \label{eqfexpand}
  f=f^{(0)}+\epsilon f^{(1)}+\epsilon^2 f^{(2)}+\cdots\, ,
\end{equation}
where each factor $\epsilon$ means an implicit gradient. Thus $f^{(0)}$ denotes the solution in the absence of spatial gradients, $f^{(1)}$ the solution obtained in the linear-order approximation with respect to the hydrodynamic gradients, $f^{(2)}$ the solution is the second-order approximation, etc. With these approximations we finally construct  a system  of equations where the first one contains only $f^{(0)}$, the second one $f^{(1)}$ and $f^{(0)}$, the third one $f^{(2)}$, $f^{(1)}$ and $f^{(0)}$, etc.

Since $f$ qualifies as a normal solution, then its time derivative can be obtained as
\begin{equation}
  \label{eqdf}
  \frac{\partial f}{\partial t}=\frac{\partial f}{\partial n}\frac{\partial n}{\partial t} + \frac{\partial f}{\partial \bt{U}}\cdot\frac{\partial \bt{U}}{\partial t} + \frac{\partial f}{\partial T}\frac{\partial T}{\partial t},
\end{equation}
where the time derivatives $\partial_t n$, $\partial_t \bt{U}$, and $\partial_t T$ can be determined from the hydrodynamic balance Eqs. \eqref{eqBaln01}--\eqref{eqBalT01}
% PRE87 2013 Garzó-Chamorro-Vega 
%Note that while the strength of the gradients can be controlled by the initial or boundary conditions in the case of elastic collisions, the problem becomes harder for systems of inelastic particles since in some cases there exists an intrinsic relation between collisional dissipation and some hydrodynamic gradients (e.g, steady states such as Uniform Shear Flow). In such situations the usual NS approximation only applies for quasi-elastic systems \cite{G03, SGD04}. Here, however, we consider situations where the spatial gradients are sufficently small (low Knudsen number).

\fran{In order to establish an appropriate order in the different levels of approximation in the kinetic equation it is neccesary to charaterize the magnitude of the external forces (thermostats) in relation with the gradient, as well. }

%Here, since both drag and stochastic forces do not induce any flux in the system then the parameters $\gamma_b$ and $\xi_b^2$ are taken to be of zeroth order in gradients.

A different treatment must be given to the relative difference $\Delta \bt{U}=\bt{U}-\bt{U}_g$. According to the momentum balance Eq. \eqref{eqBalU01}, in the absence of spatial gradients $\bt{U}$ relaxes towards $\bt{U}_g$ after a transient period. As a consequence, the term $\Delta \bt{U}$ must be considered to be at least of first order in spatial gradients.

Following the form of the expansion \eqref{eqfexpand}, the Enskog collision operator and time derivative can be also expanded in powers of $\epsilon$:
\begin{equation}
  \label{eqJexpand}
  J_E=J_E^{(0)} + \epsilon J_E^{(1)} + \epsilon^2J_E^{(2)} + \cdots, \quad 
    \partial_t=\partial_t^{(0)} + \epsilon \partial_t^{(1)} + \epsilon^2\partial_t^{(2)} + \cdots
\end{equation}
The action of the operators $\partial_t^{(k)}$ over the hydrodynamic fields can be obtained from the balance equations \eqref{eqBaln01}--\eqref{eqBalT01} when one takes into account the corresponding expansions for the fluxes and the cooling rate.

%hysically, the expansion of the time derivatives is based on the requirement that the higher the order of the space gradient, the slower the time variations it causes, This means that the fastest variations are caused by the zeroth-order gradient term, then follows the first-order term, etc.

The expansions \eqref{eqJexpand} lead to similar expansions for the heat and momentum fluxes when one replaces the expansion \eqref{eqfexpand} for $f$ into Eqs. \eqref{eqPk}-\eqref{eqColConq}:
\begin{equation}
  \label{eqPexpand}
  P_{ij}=P_{ij}^{(0)} + \epsilon P_{ij}^{(1)} + \epsilon^2 P_{ij}^{(2)} + \cdots, \quad \bt{q}=\bt{q}^{(0)} + \epsilon\bt{q}^{(1)} + \epsilon^2\bt{q}^{(2)} + \cdots
\end{equation}

These expansions introduced into the Enskog equation lead to a set of integral equations at different order which can be separately solved. Each equation governs the evolution of the distribution function on different space and time scales. 
%The coefficients in the time derivative expansion are identified by a representation of the fluxes and the cooling rate in the macroscopic balance equations as a similar series through their definitions as functionals of $f$. 
This is the usual CE method \cite{CHAPMAN70,GARZO03} for solving kinetic equations. The main difference in this work with respect to previous ones carried out by Brey \emph{et al.} \cite{BDKS98} and Garz\'o and Dufty \cite{GD99} is the new time dependence of the reference state $f^{(0)}$ through the parameters of the thermostat. 

%In the zeroth order approximation the distribution function is locally Maxwellian providing the Euler equations of change of hydrodynamic fields. 
\fran{In contrast to ordinary gases, the form of the zeroth-order solution $f^{(0)}$ is not Maxwellian, yielding a slightly different functional form \cite{NE98}}. The first order solution results in the NS equations, while the second and third order expansion give the so-called Burnett and super-Burnett hydrodynamic equations. 
%The zeroth order approximation to $f$ is valid when the system is at equilibrium and the gas properties contain no or very small macroscopic gradients. In particular, when the system is at equilibrium the heat fluxes and the viscous stresses vanish. On the other hand, the NS equations are valid whenever the typical lenght of the gradients of $n(\bt{r},t)$, $\bt{U}(\bt{r},t)$ and $T(\bt{r},t)$ are small compared to the mean free path.

In this work we shall restrict our calculations to the first order in the parameter $\epsilon$ (Navier-Stokes approximation).

\section{Grad's moment method}
Apart from the CE method, we will use here the classical Grad's moment method \cite{G49}. This method is based on the assumption that the velocity distribution function can be expanded in a complete set of orthogonal polynomials (generalized Hermite polynomials) of the velocity around a local Maxwellian distribution:
\begin{equation}
  \label{eqGrad01}
  f(\bt{r},\bt{v},t) \rightarrow f_M(\bt{r},\bt{v},t)\sum^{N-1}_{k=0} C_k(\bt{r},t)H_k(\bt{v}),
\end{equation}
where
\begin{equation}
  \label{eqfM}
  f_M(\bt{r},\bt{v},t) = n(\bt{r},t)\left(\frac{m}{2\pi T(\bt{r},t)}\right)^{d/2}\exp\left(-\frac{m \bt{V}^2}{2T}\right).
\end{equation}
%and $\bt{V}=\bt{v}-\bt{U}$ is the peculiar velocity.

This method was originally devised to solve the Boltzmann equation for monodisperse dilute systems although it has been easily extended to determine the NS transport coefficients of a dense granular fluid described by the Enskog equation \cite{G13, JR85, JR85b, KM11} since only kinetic contributions to the fluxes are considered in the trial expansion \cite{G03}.

The coefficients $C_k(\bt{r},t)$ appearing in Eq. \eqref{eqGrad01} in each of the velocity polynomials $H_k(\bt{v})$ are chosen by requiring that the corresponding velocity moments of Grad's solution be the same as those of the exact velocity distribution function. The infinite hierarchy of moment equations is not a closed set of equations and furthermore some truncation is neccesary in the above expansion after a certain order. After this truncation, the hierarchy of moment equations becomes a closed set of coupled equations which can be recursively solved.

The standard application of Grad's moment method implies that the retained moments are the hydrodynamic fields ($n$, $\bt{U}$, and $T$) plus the kinetic contributions to the irreversible momentum and heat fluxes ($P_{ij}^k-nT \delta_{ij}$ and $\bt{q}_k$). These moments have to be determined by recursively solving the corresponding transfer equations. In the three-dimensional case there are 13 moments involved in the form of the velocity distribution function $f$ and, consequently, this method is usually referred to as the 13-moment method. The explicit form of the non-equilibrium distribution function $f(\bt{r},\bt{v},t)$ in this approximation is
\begin{equation}
  \label{eqGrad02}
  f \rightarrow f_M \left[ 1 + \frac{m}{2nT^2}V_iV_j\Pi_{ij} + \frac{2}{d+2}\frac{m}{nT^2}\bt{S} \cdot \bt{q}^k \right],
\end{equation}
where
\begin{equation}
  \bt{S}(\bt{V})=\left( \frac{mV^2}{2T} -\frac{d+2}{2}\right)\bt{V}
\end{equation}
and $\Pi_{ij}=P_{ij}-nT\delta_{ij}$ is the traceless part of the kinetic contribution to the pressure tensor.

%\azul{000000000}

\section{Direct Simulation Monte Carlo method}

A different but complementary method used here is the Direct Simulation Monte Carlo (DSMC) method. This numerical method, first propossed by G.A. Bird \cite{B63,B70,BIRD94}, is based on the implementation of a probabilistic Monte Carlo method \cite{MU49} in a simulation to solve the Boltzmann equation for rarefied gases with finite Knudsen number. In these simulations the mean free path of particles is at least of the same order of a representative physical length scale of the system. %\azul{
%This condition implies that DMSC method is only applicable to systems not too far from equilibrium in which hydrodynamical gradients are smaller than the mean free path and local equilibrium states can be considered.
%}

In the classical DSMC simulations fluids are modeled using numerical $particles$ which represent a large number of real particles. Particles are moved through a simulation of physical space in a realistic manner, that is, following the trajectories given by Newton's equations for ballistic particles under the action of external forces if they exist. Collisions between particles are computed using probabilistic, phenomenological models. Common collision models include the Hard Sphere model, the Variable Hard Sphere model, and the Variable Soft Sphere model.

%The solution obtained by means DSMC simulation has the advantage that it can also determine homogeneous non-steady states and no \emph{a priori} assumption of normal solution neither specific scaling form of the distribution function must be introduced. Therefore, 
Although the DSMC method does not avoid the assumptions inherent to kinetic theory (molecular chaos hypothesis), it gives the possibility of obtaining solutions to the Botlzmann or Enskog equations in non-equilibrium situations without making any assumption on the validity of a normal or hydrodynamic solution. In this context, a comparison between numerical and analytical solutions is a direct way of validating the reliability of kinetic theory for describing granular flows.

\subsection{Description of DSMC method}
The underlying assumption of the DSMC method is that the free movement and collision phases can be decoupled over time periods that are smaller than the mean collision time. This basic condition is, in fact, inherent to the Boltzmann equation and allows us to separately deal with convective and collisional terms in the time evolution of the velocity distribution function. Thus,
\begin{equation}
  \partial_t f = -D[f] + J[f,f],
\end{equation}
where $D$ is the convective operator defined as:
\begin{equation}
  D[f]=\bt{v}\cdot \frac{\partial}{\partial\bt{r}} f + \frac{\partial}{\partial\bt{v}} \cdot (m^{-1}\bt{F}_{ext}) f,
\end{equation}
being $\bt{F}_{ext}$ the external forces acting on particles.

DSMC simulation is based on spatial and time discretization. Space is divided into $d$-dimensional cells with a typical length $l_c$ less than a mean free path $\ell$ and time is divided in intervals $\delta t$ that are taken much smaller than mean free time $\tau$.

The basic DSMC algorithm is composed by two steps computed in each time interval: advection and collision stage. The order of those stages has no importance. In the present work, we will only perform simulations in homogeneous states and hence, only one cell is used and no boundary conditions are needed.

The homogeneous velocity distribution function is represented only by the velocities $\bt{v}_i$ of the $N$ simulated particles:
\begin{equation}
  \label{eqfDSMC}
  f(\bt{v},t) \rightarrow \frac{n}{N} \sum_{i=0}^N \delta(\bt{v}_i(t)-\bt{v}).
\end{equation}
In the advection phase, velocities of every particle are updated following the corresponding Newton's equations of movement under the action of the external forces: %When it is neccesary, boundary conditions has to be applied in order to maintain the particles inside the volume of integration. Various types of boundary conditions are used in DSMC simulation and its choice depend on the nature of the problem we are studied. The more common boundary conditions are periodic boundaries, specular surfaces, thermal walls, etc. When a particle reaches a boundary, velocity and positions is processed according to the characteristic of it.
\begin{equation}
  \label{eqFreeStream}
  \bt{v}_i\rightarrow\bt{v}_i+\bt{w}_i,
\end{equation}
As we pointed out before, our thermostat is composed by two differents terms: a deterministic external force porportional to the velocity of the particle plus a stochastic force. Thus, $\bt{w}_i=\bt{w}_i^{drag}+\bt{w}_i^{st}$, where $\bt{w}_i^{drag}$ and $\bt{w}_i^{st}$ denote the velocity increments due to the drag and stochastic forces, respectively. In the case of the drag force the velocity increment is given by
\begin{equation}
  \label{eqdrag}
  \bt{w}_i^{drag}=(1-e^{\gamma_b\delta t})\bt{v}_i,
\end{equation}
while $\bt{w}_i^{st}$ is randomly drawn from the Gaussian probability distribution with a variance characterized by the noise intensity $\xi_b^2$ fulfilling the conditions:
\begin{equation}
  \label{eqstoch01}
  \langle\bt{w}_i\rangle=\bt{0}, \qquad   \langle \bt{w}_i \bt{w}_j\rangle = \xi_b^2\delta t \delta_{ij},
\end{equation}
where
\begin{equation}
  \label{eqstoch02}
  P(\bt{w})=(2\pi\xi_b^2\delta t)^{-d/2}e^{-w^2/(2\xi_b^2\delta t)}
\end{equation}
is a Gaussian probability distribution \cite{MS00}.

Intrinsic time scales produced by the inclusion of the two thermostat forces in our system ($\tau_{drag}=m/\gamma_b$ and $\tau_{st}=v_0^2/\xi_b^2$) must be not too fast compared to the algorithm time step $\delta t$ (which is small compared to the characteristic collision time) in order to describe properly the collision integral of the Enskog equation \cite{BIRD94}. Thus, $\tau_{drag}\leq \nu^{-1}$ and $\tau_{st}\leq \nu^{-1}$ where $\nu=v_0/\ell$.

In the collision stage a sample of $\frac{1}{2}N\omega_{max}\delta t$ pairs of particles ($i$,$j$) are choosen at random with equiprobability where $N$ is the number of particles and $\omega_{max}$ is an upper bound estimate of the probability that a particle collides per unit of time. For each pair ($i,j$) belonging to this sample a given direction $\widehat{\bs{\sigma}}_{ij}$ is chosen at random with equiprobability and the collision between particles $i$ and $j$ is accepted with a probability equal to $\Theta(\bt{g}_{ij} \cdot \widehat{\bs{\sigma}}_{ij})\omega_{ij}/\omega_{max}$ where $\omega_{ij}=(4\pi n\sigma^2\chi)|\bt{g}_{ij} \cdot \widehat{\bs{\sigma}}_{ij}|$ for hard spheres and $\omega_{ij}=(2\pi n\sigma\chi)|\bt{g}_{ij} \cdot \widehat{\bs{\sigma}}_{ij}|$ for hard disks. Here, $\bt{g}_{ij}=\bt{v}_{i}-\bt{v}_{j}$ is the relative velocity. If the collision is accepted, post-collisional velocities are assigned according to the scatering rule \eqref{eqBinCol}. If the frequency $\omega_{ij}>\omega_{max}$ for any collision, the estimate of $\omega_{max}$ is updated as $\omega_{max}=\omega_{ij}$ \cite{NE98}. It is worthwhile to remark here that the acceptance probability $\Theta(\bt{g}_{ij} \cdot \widehat{\bs{\sigma}}_{ij})\omega_{ij}/\omega_{max}$ is independent of the pair correlation function and, thus, the DSMC algorithm is formally identical for both Boltzmann and Enskog equations when they describe homogeneous systems.

%\rojo{000000000}
         % Kinetic theory
% Chapter 1

\chapter{Homogeneous state} 
\label{Chapter3} 

\blfootnote{The results obtained in this Chapter have been published in M.G. Chamorro, F. Vega Reyes, and V. Garzó, \emph{ AIP Conf. Proc.} \textbf{1501}:1024-1030, (2012) and M.G. Chamorro, F. Vega Reyes and V. Garz\'o, \emph{J. Stat. Mech.}, P07013 (2013)  \cite{CVG12,CVG13}}

\lhead{Chapter 3. \emph{Homogeneous state}}
%----------------------------------------------------------------------------------------
%	SECTION 1
%----------------------------------------------------------------------------------------
\section{Introduction}
%As explained in previous Chapters granular systems are constituted by macroscopic particles that collide inelastically so that, if not excited, the total energy decreases with time. In this context, granular matter can be considered as a prototype of system in an inherenty non-equilibrium state. 
Under rapid flow conditions (state which can be reached by externally vibrating or shearing the system \cite{AMSV02}), the grains interact by nearly instantaneous collisions (compared with the mean free time) and their motion resembles the clasical picture of a molecular gas. In these conditions, kinetic theory can be a quite useful tool to study these systems \cite{G03}.
%\rojo{\emph{FRAN: yo quitaria esa frase. decir que es rapid flow y por eso es un gas granular es redundante.}}
%the motion of grains resembles a granular gas where binary collisions prevail and kinetic theory can be a quite useful tool to study these systems. 
%In all of these cases the grains interact by nearly instantaneous collisions (compared with the mean free time) in a way that is reminiscent of the classical picture of a molecular gas. \cite{G03}
%This is perhaps one of the main reasons why these systems have received considerable attention in the past few years, apart from its practical interest. %
% FRAN:es subjetivo pero realmente creo que el interes viene de las aplicaciones industriales y tecnologicas.  (OK) 
Thus, in order to maintain the granular medium in a fluidized state, an external energy input is needed for collisional cooling compensation and for attaining a steady state. In most experiments, energy is supplied through the boundaries. This creates spatial gradients in the system. To avoid the difficulties associated with non-homogeneous states, it is usual in computer simulations to homogeneously heat the system by the action of an external driving force (thermostat). 
% FRAN: a mi me gusta mas otro tipo de planteamiento. es decir, la introduccion de un termostato uniforme yo la veo mas 

Nevertheless, in spite of its practical importance, little is known about the influence of the external force (or thermostat) on the properties of the system \cite{GARZO03,GSB90}.

%Before considering non-homogeneous systems, it is conventient to analyze first the homogeneous state. In this situation, the density and mean flux velocity are $constant$ and the granular temperature is spatially uniform. 
%FRAN: es redundante. si es homogeneo es porque sus propiedades son uniformes. di una frase o la otra. las dos juntas no es necesario. pero yo mejor pasaría directamente a la siguiente frase borrando estas dos.

The goal of this Chapter is to analyze the homogeneous steady state of a driven granular fluid described by the Enskog kinetic equation. % In order to reach a steady state, %FRAN: para llegar a un estacionario no hacen falta dos fuerzas. solo el ruido blanco es necesario. tal como esta escrito lleva a confusion. sugiero re-redactar.
The particles are assumed to be under the action of an external thermostat composed by two different forces: ($i$) a stochastic force where the particles are randomly kicked between collisions \cite{WM96} and ($ii$) a viscous drag force which mimics the interaction of the particles with an effective viscous \emph{bath} at temperature $T_b$. 
The viscous drag force allows us to model the friction from a surrounding fluid over a moderately dense set of spheres \cite{GTSH12} while the stochastic force would model the energy transfer from the surrounding fluid molecules to the granular particles due to molecular thermal motions in a similar way as in a Brownian particle.
There exists, thus, a balance in the system between the injection of energy due to the stochastic thermostat and the loss of it due to the friction and particle collisions. %inelastic dissipation. 
Given an arbitrary initial state, the system will evolve towards a steady state characterized by a constant and homogeneous granular temperature.

Under these conditions, our kinetic equation has the structure of a Fokker-Planck equation \cite{KAMPEN81} plus the corresponding inelastic collisional operator of the Enskog equation.

One of the main advantages of using this kind of thermostat \cite{GSVP11-1} with respect to others present in the literature \cite{WM96} is that the temperature of the thermostat $T_b$ (different from the temperature $T < T_b$ of the granular fluid) is always well defined. In particular, for elastic collisions, the fluid equilibrates to the \emph{bath temperature} ($T=T_b$). 
%\rojo{\emph{FRAN: el argumento de los de Roma, como ya he sugerido anteriormente, no tiene mucho significado ni profundidad fisica. temperatura del baño? termodinamica? granular? se puede medir? a mi no me gusta y solo confunde. es una cuestion de apreciacion no obstante.}}
This happens because, in addition to the random driving, the thermostat acts on the grains also through a finite drag. Moreover, some recent results \cite{GSVP11-2} suggest that this thermostat is the most appropriate to model some experiments. 

%\rojo{\emph{FRAN: yo andaria con cautela a la hora de defender el realismo del termostato. siendo realistas, es solo un modelo teorico aun no contrastado con medidas experimentales.}}

%-----------------------------------
%	SECTION 2
%-----------------------------------

\section{Enskog Kinetic Theory for \\ Homogeneous Driven States}
For a spatially uniform system, the Enskog kinetic equation for $f(\bt{v},t)$ reads:
\begin{equation}
  \label{eqEB02}
  \partial_t f - \frac{\gamma_b}{m}\frac{\partial}{\partial \bt{v}}\cdot \bt{v} f - \frac{1}{2}\xi_b^2\frac{\partial^2}{\partial v^2} f = \chi J[f,f].
\end{equation}
%The first $d-2$ moments of the velocity distribution funtions are defined as
%\begin{equation}
%    n(t)=\int d{\bf v} f({\bf v},t)
%\end{equation}
%\begin{equation}
%    \bt{U}(t) = \frac{1}{n(t)} \int d{\bf v} \; {\bf v} f({\bf v,t})
%\end{equation}
%\begin{equation}
%    T(t) = \frac{m}{d \; n(t)} \int d{\bf v} \; V^2 f({\bf v})
%\end{equation}	

In the homogeneous state the mean flux velocity is constant and uniform, and so, one can choose an \fran{appropriate} frame of reference where this velocity vanishes without lost of generality ($\bt{U}=\bt{0}$ and hence $\bt{V}=\bt{v}$). Furthermore, \fran{the energy balance is the only relevant equation. Energy balance is} obtained from Eq.\ \eqref{eqEB02} by multiplying it by $V^2$ and integrating over velocity. The result is
\begin{equation}
  \label{eqBalHom}
  \partial_t T = - \frac{2T}{m}\gamma_b + m\xi_b^2 - \zeta T,
\end{equation}
where
\begin{equation}
  \label{eqzeta}
  \zeta=-\frac{m}{dnT}\int d\bt{v}\,V^2\,J[f,f]
\end{equation}
is the cooling rate proportional to $1-\alpha^2$. This term \fran{describes} the loss of kinetic energy due to the inelastic character of collisions.

In the hydrodynamic regime, the distribution function $f$ qualifies as a $normal$ solution for the Boltzmann equation and therefore its time dependence only occurs through the hydrodynamic fields. %In the homogeneous state, the density and mean \fran{flow} velocity are homogeneous and time-independent. \fran{Granular temperature is homogeneous too.} 
Therefore, given that the only relevant hydrodynamic field is $T(t)$ then the time dependence of $f$ is through $T(t)$: %the only relevant hydrodynamic function is $T(t)$; i.e., the time dependence of the homogeneous granular temperature.
\begin{equation}
  \label{eqT}
  \partial_t f = \frac{\partial f}{\partial T} \partial_t T = -\left( \frac{2}{m}\gamma_b - \frac{m}{T}\xi_b^2 + \zeta \right) T\, \frac{\partial f}{\partial T}.
\end{equation}

Substitution of Eq.\ \eqref{eqT} into Eq.\ \eqref{eqEB02} yields
\begin{equation}
  \label{eqEB03}
   -\left( \frac{2}{m}\gamma_b - \frac{m}{T}\xi_b^2 + \zeta \right) T\, \frac{\partial f}{\partial T} 
- \frac{\gamma_b}{m}\frac{\partial}{\partial \bt{v}}\cdot \bt{v} f - \frac{1}{2}\xi_b^2\frac{\partial^2}{\partial v^2} f = \chi J[f,f].
\end{equation}
In the elastic limit ($\alpha=1$), the cooling rate vanishes and the solution to Eq.\ \eqref{eqEB03} is the Maxwellian distribution
\begin{equation}
  \label{eqfMaxwell}
  f_M(v)=n\left(\frac{m}{2\pi T_b}\right)^{d/2}\exp\left(-\frac{mv^2}{2T_b}\right),
\end{equation}
where $T_b$ is defined as
\begin{equation}
  \label{eqTb}
  T_b=\frac{m^2\xi_b^2}{2\gamma_b}.
\end{equation}

%As in the work of Gradenigo \emph{et al} \cite{GSVP11-1}, Eq.\ \eqref{eqTb} defines a \emph{bath temperature}. Its name may be justified since it is determined by the two thermostat parameters ($\gamma_\text{b}$ and $\xi_\text{b}^2$ ), and thus it can be considered as a remnant of the physical equilibrium temperature of the surrounding ordinary (\emph{elastic}) fluid. 

%\rojo{\emph{FRAN: no estoy de acuerdo}}

%In this sense, for elastic collisions $T=T_b$ and the energy equipartition is fulfilled in accordance to equilibrium statistical mechanics principles. The relation \eqref{eqTb} is a direct consequence of the fluctuation-dissipation theorem \cite{KAMPEN81} relating the \fran{energy loss} due to the action of an external non-conservative drag force to the spontaneous fluctuations at thermal equilibrium. Conversely, for inelastic granular gases ($\alpha<1$, $\zeta_s\neq 0$) the fluctuation-dissipation theorem does not strictly apply hence, the bath is not at equilibrium and $T_s<T_b$ according Eq.\ \eqref{eqTs01}. From a physical point of view, this can be seen as the granular gas of inelastic particles is cooler than  the surrounding ordinary fluid. In that case, the drag coefficient $\gamma_b$ and the amplitude of the stochastic force $\xi_b^2$ are generally not related and can be chosen as independent parameters.

%\rojo{\emph{FRAN yo no le daria mucha vuelta a las ideas del parrafo de arriba. uno, que no son relevantes para tu tesis, dos que no hay consenso en la comunidad al respecto. esto es subjetivo, obviamente. asi que lo dejo a tu eleccion.}}

In the long time limit, the system reaches a steady state in which the energy lost due to inelastic collision and drag force \fran{is compensated} by the heating due to the stochastic force. Let us call $T_s$ the value of the granular temperature at the stationary state \eqref{eqBalHom}.
\begin{equation}
  \label{eqTs01}
\fran{  \zeta_s T_s + \frac{2\gamma_b}{m}T_s = m\xi_b^2, }
\end{equation}
where $\zeta_s$ is the cooling rate evaluated in the steady state. This equation establishes a relation between the two driving parameters $\gamma_b$ and $\xi_b^2$ and the inelasticity of the particles, and hence, only one of them will be considered as independent at a fixed $\alpha$. 

Using Eq.\ \eqref{eqTs01} in Eq.\ \eqref{eqEB03} we obtain the kinetic equation of the steady distribution function $f_s$:
\begin{equation}
  \label{eqEB04}
   \frac{1}{2}\left(\zeta_s - \frac{m\xi_b^2}{T_s}\right) \frac{\partial}{\partial \bt{v}}\cdot \bt{v} f_s 
    - \frac{1}{2}\xi_b^2\frac{\partial^2}{\partial v^2} f_s = \chi J[f_s,f_s].
\end{equation}

Note that the solution of Eq.\ \eqref{eqEB04} must depend on the model parameter $\xi_b^2$, the steady granular temperature $T_s$ and the coefficient of restitution $\alpha$. %The steady cooling rate is defined in terms of the steady distribution $f_s$
%As for \fran{the HCS} \cite{EP97, POSCHEL04} where the granular temperature monotonically decreases in time, we assume that all the other characteristic parameters of the granular gas (particle mass, density, mean flux velocity, etc) remain constant in time. 
%\rojo{\emph{FRAN esa frase no estoy seguro si la entiendo... te refieres a que la masa de las particulas, la densidad, etc se consideran constantes como en el HCS? si es asi, la frase creo que es inadecuada. asi que si quieres puedes borrarla. (se puede tener HCS sin tener gas monocomponente por ejemplo).}} 

This allows us to consider that all the time dependence of the velocity distribution function occurs only through the time-dependent \emph{square velocity}, that is, the temperature. Consequently it is expected that Eq.\ \eqref{eqEB04}  admits a scaling solution of the form:
\begin{equation}
  \label{eqfscaled}
  f_s(\bt{v},\xi_b^2)\rightarrow n\,v_{0,s}^{-d}\varphi(\bt{c},\xi^*_s),
\end{equation}
where $\bt{c}=\bt{v}/v_{0,s}$ is the scaled velocity, $v_{0,s}=\sqrt{2T_s/m}$ is the steady thermal velocity and $\xi^*_s$ is the scaled model parameter defined as
\begin{equation}
  \label{eqxistar}
  \xi^*_s = \frac{m\ell} {\chi T_s v_{0,s}} \xi_b^2.
\end{equation}

Here, $\ell=(n\sigma^{d-1})^{-1}$ is the mean free path for hard spheres (or disks). 

In terms of the scaled distribution function, the Eq.\ \eqref{eqEB04} can be rewritten as:
\begin{equation}
  \label{eqEBscaled}
  \frac{1}{2} (\zeta_s^*-\xi_s^*) \frac{\partial}{\partial \bt{c}} \cdot \bt{c} \varphi - \frac{1}{4}\xi_s^* \frac{\partial^2}{\partial c^2} \varphi = J^*[\varphi,\varphi ].
\end{equation}

Here, we have introduced the dimensionless quantities:
\begin{equation}
  \zeta_s^*\equiv\frac{\zeta_s}{\chi \nu_s}\;,  \qquad   J^*[\varphi,\varphi]\equiv \frac{v_{0,s}^d}{n\nu_s}J[f,f],
\end{equation}
where 
\begin{equation}
  \label{eqfreqcol}
  \nu_s=\frac{v_{s,0}}{\ell}=\sqrt{\frac{2T_s}{m}}n\sigma^{d-1}
\end{equation}
is the collision frequency \cite{CHAPMAN70}. %FRAN poner referencia. Chapman-Cowling por ejemplo.
%\rojo{ATENCIÓN A LO SIGUIENTE. QUIZÁ VAYA MEJOR EN OTRO SITIO}
%The steady cooling rate $\zeta_\text{s}$ can be written up to the first order in the Sonine expansion in terms of $a_{2,\text{s}}$ as:
%\begin{equation}
%\label{eqZetas} 
%\zeta_\text{s}=\frac{2}{d}\frac{\pi^{\left( d-1\right) /2}} {\Gamma \left( \frac{d}{2}\right)}(1-\alpha^2)\chi \left(1+\frac{3}{16}a_{2,\text{s}}\right) n_\text{s}\sigma^{d-1}\sqrt{\frac{T_\text{s}}{m}},
%\end{equation}
%where the steady granular temperature $T_\text{s}$ obeys the equation
%\begin{equation}
%\label{eqTs02}
%T_\text{s}=\frac{m^2\xi_\text{b}^2}{2\gamma_\text{b}} - \frac{2^{d-1}}{\sigma}\sqrt{\frac{m}{\pi}}\frac{\chi\phi} {\gamma_\text{b}}(1-\alpha^2)\left(1+\frac{3}{16}a_{2,\text{s}}\right)T_\text{s}^{3/2}.
%\end{equation}
%Here,
%\begin{equation}
%\label{eqphi}
%\phi=\frac{\pi^{d/2}}{2^{d-1}d\Gamma \left(\frac{d}{2}\right)}n_\text{s}\sigma^d
%\end{equation}
%is the solid volume fraction. Eq.\ \eqref{eqTs02} gives the granular temperature $T_\text{s}$ in the non-equilibrium stationary state.
%\rojo{ATENCION ¿mencionar a Hayakawa aquí?} \rojo{******************************************}

It is easy to see from Eq.\ \eqref{eqEBscaled} that the dependence on the temperature of the scaled distribution function $\varphi$ is through the dimensionless velocity $\bt{c}$ and the reduced noise strength $\xi^*$. This differs from the HCS and for homogeneously gases heated by a single thermostat \cite{BRC96,NE98, MS00}, where only one parameter (scaled velocity) was neccesary to characterize the stationary distribution function $\varphi$. A similar scaling solution to the form \eqref{eqfscaled} has been recently used in the transient time-dependent regime for a granular gas driven by a stochastic force \cite{GMT12}.

In the elastic limit ($\alpha=1$), the cooling rate vanishes and the solution of Eq.\ \eqref{eqEBscaled} is a Maxwellian distribution \cite{CHAPMAN70, FERZIGER72, CERCIGNANI88}
\begin{equation}
  \label{eqphiM}
  \varphi_M=\pi^{-d/2}e^{-c^2}.
\end{equation}

Note that in this case the scaled distribution function does not depend on the thermostat forces. However when the particles collide inelastically ($\alpha < 1$) the cooling rate has a non-zero value and the exact form of the distribution function is unknown. In that case we can find an approximate solution in the regions of thermal velocities ($c \sim 1$) by measuring the departure of $\varphi(\bt{c},\xi^*)$ from the Maxwellian function \eqref{eqphiM} through the first nontrivial coefficients of an expansion in Sonine polynomials \cite{POSCHEL04}.

The steady-state condition \eqref{eqTs01} can be rewritten using reduced units as:
\begin{equation}
  \label{eqBalRed}
  2\gamma^*_s=\xi^*_s-\zeta^*_s,
\end{equation}
where
\begin{equation}
  \label{eqGamRed}
  \gamma^*_s=\frac{\gamma_b}{\chi m \nu_s}
\end{equation}
is the reduced drag coefficient.

Since $\gamma_s^*\geq 0$\fran{, equation} \eqref{eqBalRed} requires $\xi_s^* \geq \zeta_s^*$. Thus, there exists a minimum threshold value of $\xi_{th}^*(\alpha)$ of the noise intensity for a given restitution coefficient $\alpha$ needed to reach a steady state. This minimum value occurs when $\gamma_s^*=0$ and coincides with the reduced cooling rate $\zeta_s^*(\alpha)$. Given that the latter cannot be exactly determined, a good estimate of it is obtained when one replaces the true scaled distribution function $\varphi$ by its Maxwellian form $\varphi_M$ \cite{GoSh95}. In this case:
\begin{equation}
  \label{eqzetaM}
  \zeta_s^*\rightarrow\zeta_M^*=\frac{\sqrt{2}}{d}\frac{\pi^{(d-1)/2}}{\Gamma(d/2)}(1-\alpha^2). 
\end{equation}

\section{Analytical solution of the scaled distribution function}

We provide in this Section a perturbative analytical solution of the scaled distribution function as a Sonine polynomials expansion
% in the range of thermal velocities % no se exactamente que significa esa frase. si te refieres a que no representamos velocidades mucho mayores que la velocidad termica, la frase no es precisa puesto que el limite v->0 esta incluido y obviamente alli la diferencia con la velocidad termica es infinita. creo por otro lado que especificar mas la frase o la frase en si misma no es necesaria. se puede borrar.
\begin{equation}
  \label{eqSonine01}
  \varphi(\bt{c},\xi_s^*)=\varphi_M(c)\left(1+\sum^\infty_{p=1}a_p(\xi_s^*)S_p(c^2)  \right),
\end{equation}
where $S_p$ are generalized Laguerre or Sonine polynomials defined as \cite{ABRAMOWITZ72}:
% FRAN citar para todo el libro de Poschel no da buena imagen. es como si hubieras basado tu tesis en este trabajo. por ejemplo aqui sugiero poner algo mas apropiado. como el abramowitz o la libreria del NIST (http://dlmf.nist.gov/18.27#E15).
\begin{equation}
  \label{eqSoniPoly}
  S_p(x)=\sum^p_{k=0}\frac{(-1)^k(\frac{d}{2}-1+p)!}{(\frac{d}{2}-1+k)!(p-k)!k!}x^k.
\end{equation}

They satisfy the orthogonality relations:
\begin{equation}
  \int d\bt{c} \; \varphi_M(c) \; S_p(c^2) \; S_{p'}(c^2) = \mathcal{N}_p\;\delta_{pp'},
\end{equation}
where $\mathcal{N}_p=\frac{(\frac{d}{2}-1+p)!}{2p!}$ is a normalization constant \cite{POSCHEL04}.

The first relevant Sonine polynomials in our case are
\begin{equation}
  S_0(x)=1,
\end{equation}
\begin{equation}
  S_1(x)=-x+\frac{d}{2},
\end{equation}
\begin{equation}
  S_2(x)=\frac{1}{2}x^2-\frac{d+2}{2}x+\frac{d(d+2)}{8},
\end{equation}
\begin{equation}
  S_3(x)=-\frac{1}{6}x^3 + \frac{d+4}{4}x^2 - \frac{(d+2)(d+4)}{8}x + \frac{d(d+2)d+4)}{48}.
\end{equation}

The \fran{$a_p$ coefficients} in Eq.\ \eqref{eqSonine01} \fran{correspond to the different velocity moments of the Sonine polynomial.} They are defined as
\begin{equation}
  \label{eqap}
  a_p(\xi_s^*)=\frac{1}{\mathcal{N}_p}\int\;d\bt{c}S_p(c^2)\varphi(\bt{c},\xi_s^*).
\end{equation}

Since $\langle c^2\rangle = \frac{d}{2}$, then the coefficient $a_1=0$ by definition.
The first two nontrivial coefficients are $a_2$ and $a_3$ and they are related with the fourth and sixth velocity moments as
\begin{equation}
  \label{eqc4}
  \langle c^4\rangle = \frac{d(d+2)}{4}(1+a_2),
\end{equation}
\begin{equation}
  \label{eqc6}
  \langle c^6\rangle = \frac{d(d+2)(d+4)}{8}(1+3a_2-a_3),
\end{equation}
where
\begin{equation}
  \label{eqcaverage}
  \langle c^p\rangle = \int\;d\bt{c}\;c^p\;\varphi(\bt{c}).
\end{equation}

Multiplying Eq.\ \eqref{eqEBscaled} by $c^{2p}$ and integrating over velocities we can easily derive the hierarchy of equations for the moments and determine the coefficients $a_k$:
\begin{equation}
  \label{eqJerarquia}
  p(\zeta_s^*-\xi_s^*)\langle c^{2p}\rangle + \frac{p(2p+d-2)}{2}\xi_s^*\langle c^{2p-2}\rangle = \mu_{2p},
\end{equation}
where
\begin{equation}
  \label{eqmu2p}
  \mu_{2p}=-\int\;d\bt{c}\;c^{2p} J^*[\varphi,\varphi]
\end{equation}
are the velocity moments of the collisional operator. Notice that $p=1$ yields the cooling rate since $\mu_2=\frac{2}{d}\zeta_s^*$. To obtain Eq.\ \eqref{eqJerarquia} use has been made of the results:
\begin{equation}
  \int\;d\bt{c}\;c^{2p} \frac{\partial}{\partial\bt{c}}\cdot\bt{c}\varphi(\bt{c}) = -2p\langle c^{2p}\rangle,
\end{equation}
\begin{equation}
  \int\;d\bt{c}\;c^{2p} \frac{\partial^2}{\partial c^2}\varphi(\bt{c}) = 2p(2p+d-2)\langle c^{2p-2}\rangle.
\end{equation}

In Eq.\ \eqref{eqJerarquia} the collisional moments $\mu_{2p}$ are functionals of the distribution function $\varphi$ and we obtain an infinite hierarchy of moment equations where all the Sonine coefficients $a_p$ are coupled. In order \fran{to get an explicit form} of them one has to make some kind of truncation. This truncation is based \fran{on the expectation} that the Sonine coefficients \fran{will be small} enough and, consequently, high-order and nonlinear terms can be neglected \cite{NE98}. \fran{In particular,} the first three collisional moments with $p=1,2$ and $3$ are given by:
\begin{equation}
  \label{eqmu2}
  \mu_2\rightarrow A_0 + A_2\;a_2  + A_3\;a_3, \\
\end{equation}
\begin{equation}
  \label{eqmu4}
  \mu_4\rightarrow B_0 + B_2\;a_2  + B_3\;a_3, \\
\end{equation}
\begin{equation}
  \label{eqmu6}
  \mu_6\rightarrow C_0 + C_2\;a_2  + C_3\;a_3,
\end{equation}
where the coefficents $A_i$, $B_i$ and $C_i$ are known functions of the coefficients of restitution $\alpha$ and dimensionality $d$. These coefficients were independently obtained by van Noije and Ernst \cite{NE98} and Brilliantov and P\"oschel \cite{BP06} and their complete expressions are displayed in Appendix \ref{AppendixA}. As it was said before, the Sonine coefficients are expected to be small and, for this reason, the coefficients $a_p$ with $p\geq 4$ and nonlinear terms (like $a_2^2$, $a_2 a_3$ and $a_3^2$) have been neglected in Eqs.\ \eqref{eqmu2},\eqref{eqmu4} and \eqref{eqmu6}.

By introducing the expression for the collisional moments \eqref{eqmu2}--\eqref{eqmu6} and the velocity averages \eqref{eqc4} and \eqref{eqc6} in the exact moment equation \eqref{eqJerarquia} and retaining only linear terms in $a_k$ for $p=2$, one gets
\begin{equation}
\label{eqJerarquia01}
\left[B_2-(d+2)(A_0+A_2)+\frac{d(d+2)}{2}\xi_s^*\right]a_2+\left[B_3-(d+2)A_3\right]a_3=(d+2)A_0-B_0,
\end{equation}
while the result for $p=3$ is
\begin{equation}
  \label{eqJerarquia02}
  \left[\widehat{C}_2+(d\xi_s^*-3A_0-A_2)\right]a_2 + \left[\widehat{C}_3-\left(A_3-A_0+\frac{d}{2}\xi_s^*\right)\right]a_3 = A_0 - \widehat{C}_0,
\end{equation}
where
\begin{equation}
  \widehat{C}_i= \frac{4}{3(d+2)(d+4)}C_i.
\end{equation}

%In order to obtain the later equations we have taken into account that the reduced thermostat parameter $\xi_s^*$ is function of the Sonine coefficients through it dependence on steady temperature granular, $\xi_s^*(a_2,a_3)$. Since it is expected that both coefficients are quite small, we evaluate $\xi_s^*$ by assuming $a_2=a_3=0$. Consequently,  

Eqs.\ \eqref{eqJerarquia01} and \eqref{eqJerarquia02} \fran{become a linear algebraic} set of equations that can be easily solved to give $a_2$ and $a_3$ in terms of $d$, $\alpha$ and $\xi^*$.
As noted previously by Montanero and Santos \cite{MS00,SM09}, there exists a certain degree of ambiguity in the approximations used in the determination of $a_2$ and $a_3$. Here, two kinds of approximations will be used to solve the set of equations. In the first one (Approximation I) we consider $a_3 \ll a_2$ in the collisional moments $\mu_2$ and $\mu_4$ but not in $\mu_6$ given that the latter is expected to be smaller than $\mu_4$. With this approach, $a_2$ can be independently calculated of $a_3$ from Eq. \eqref{eqJerarquia01} with $a_3=0$. 
% We reffer this estimation as $a_2^{(I)}$ and it was obtained in previous works \cite{CVG12, GCV13}
Its explicit expression is given by
\begin{equation}
\label{eqa2I}
a_2^{(I)}(\alpha,\xi^*) = \frac{(d+2)A_0-B_0}{B_2-(d+2)(A_0+A_2)+\frac{d(d+2)}{2}\xi^*},
\end{equation}
while $a_3^{(I)}$ is
\begin{equation}
\label{eqa3I}
a_3^{(I)}(\alpha,\xi^*)=F\left(\alpha,a_2^{(I)}(\alpha),\xi^*\right),
\end{equation}
where the function $F(\alpha, a_2, \xi_s^*)$ is given by Eq.\ \eqref{apA13}. 

In Approximation II, both Sonine coefficients $a_2$ and $a_3$ are considered as being of the same order in Eqs.\ \eqref{eqmu2} and \eqref{eqmu4} giving rise to the linear set of Eqs.\ \eqref{eqJerarquia01}, \eqref{eqJerarquia02} for the Sonine coefficients. Their expressions in Approximation II have then the following forms:
\begin{equation}
\label{eqa2II}
a_2^{(II)}(\alpha,\xi^*)=\frac{M(\alpha,\xi^*)}{N(\alpha,\xi^*)},
\end{equation}
\begin{equation}
\label{eqa3II}
a_3^{(II)}(\alpha,\xi^*)=F\left(\alpha,a_2^{(II)}(\alpha),\xi^*\right),
\end{equation}
where the explicit expressions of $M(\alpha,\xi^*)$ and $N(\alpha,\xi^*)$ are given in Appendix \ref{AppendixA} by Eqs.\ \eqref{apA14} and \eqref{apA15}, respectively.

%Here, in order to solve the set of Eqs.\ \eqref{3.18} and \eqref{3.19}, we consider two basic classes of approximations. \azul{ In Approximation I, we first assume that $a_3\ll a_2$ so that, $a_3$ can be neglected versus $a_2$ in Eq. \eqref{3.18} but not in Eq. \eqref{3.19}. This is equivalent to neglect $a_3$ in Eqs. \eqref{3.15} and \eqref{3.16} for $\mu_2$ and $\mu_4$, respectively. Given that $\mu_6$ is expected to be smaller than $\mu_4$, it seems to be more accurate to neglect $a_3$ in Eq. \eqref{3.18} rather than in Eq. \eqref{3.19}. The comparison with computer simulations confirm this expectation.} In Approximation II, both Sonine coefficients $a_2$ and $a_3$ are considered as being of the same order of magnitude. Since the latter does not assume negligible contributions of $a_3$ to the expression of $a_2$, this approximation should be more accurate.

\section{Numerical solutions of the BE equation}

\fran{In the previous sections, we have used analytical tools to obtain a solution to the Enskog equation} as an expansion of the velocity distribution function in Sonine polynomials. 
As it was before mentioned, and given the complex structure of Enskog equation, it is very difficult to get an exact solution for far from equilibrium situations. 
For this reason, it is necessary to resort to perturbative methods only valid in situations near equilibrium or kinetic models that, given the simplifications which they are based on, may involve results quite different to those directly obtained from the Boltzmann equation \cite{GaSa95}. In our case, several kinds of truncations have been neccesary to get an approximate expresion for the Sonine expansion for the Enskog equation. 

%\rojo{\emph{FRAN creo que seria necesaria una referencia a cualquiera de los trabajos de los que se ha tomado directamente la base del codigo. por ejemplo (por poner un ejemplo, PRE 85, p. 021308 (2012))}}%\cite{GV12}
In this Section we obtain the exact solution of Eq.\ \eqref{eqEB02} by means of the DSMC method.
The DSMC solution has the advantage that it can also determine transport states and neither an \emph{a priori} assumption of a normal solution nor a specific scaling form of the distribution function must be introduced. Therefore, a comparison of both numerical and analytical solutions is a direct way of validating, for steady states, the hypothesis of existence of a normal solution and of the special scaling form of the distribution function used in Eq.\ \eqref{eqEBscaled}. The simulation code used here was based on the one previously used in Ref.\ \cite{GV12} for studying a segregation criterion based on the thermal diffusion factor of an intruder in a heated granular gas described by the inelastic Enskog equation.

For practical reasons we have introduced in our simulations the following dimensionless quantitites ($\gamma_{sim}^*$ and $\xi_{sim}^*$) characterizing the \fran{driving} parameters:
\begin{equation}
  \label{eqGammaSim}
  \gamma_{sim}^*=\frac{\gamma_b}{\chi m \nu_0}=\left(\frac{T_s}{T_0}\right)^{1/2}\gamma^*,
\end{equation}
\begin{equation}
  \label{eqXiSim}
  \xi_{sim}^*=\frac{m\xi_b^2}{\chi T_0 \nu_0}=\left(\frac{T_s}{T_0}\right)^{3/2}\xi^*,
\end{equation}
where the last equality in Eqs.\ \eqref{eqGammaSim} and \eqref{eqXiSim} provides the relation between the simulation reduced quantities $\gamma_{sim}^*$ and $\xi_{sim}^*$ and their corresponding theoretical ones $\gamma^*$ and $\xi^*$, respectively.

In the simulations carried out, the system is always initialized with a Maxwellian velocity distribution with temperature $T_0$. A number of particles $N=2\times 10^6$ and a time step $\delta t=5\times 10^{-2}\nu_0$ has been used, where $\nu_0=(2T_0/m)^{1/2}n\sigma^{d-1}$ and $T_0$ is the initial temperature.

%%%%%%%%%%%%%%%%%%%
\subsection{Comparison between theory and simulations}
In this Section we compare numerical results of DSMC simulations with analytical ones obtained in previous sections. In particular, we are interested in the scaled distribution function and the coefficients that measure its deviation from the Maxwellian distribution function corresponding to a state in equilibrium.

\subsubsection{Transient regime}

\begin{figure}[h]%tbp]
\centering
\subfigure[]{\includegraphics[width=95mm]{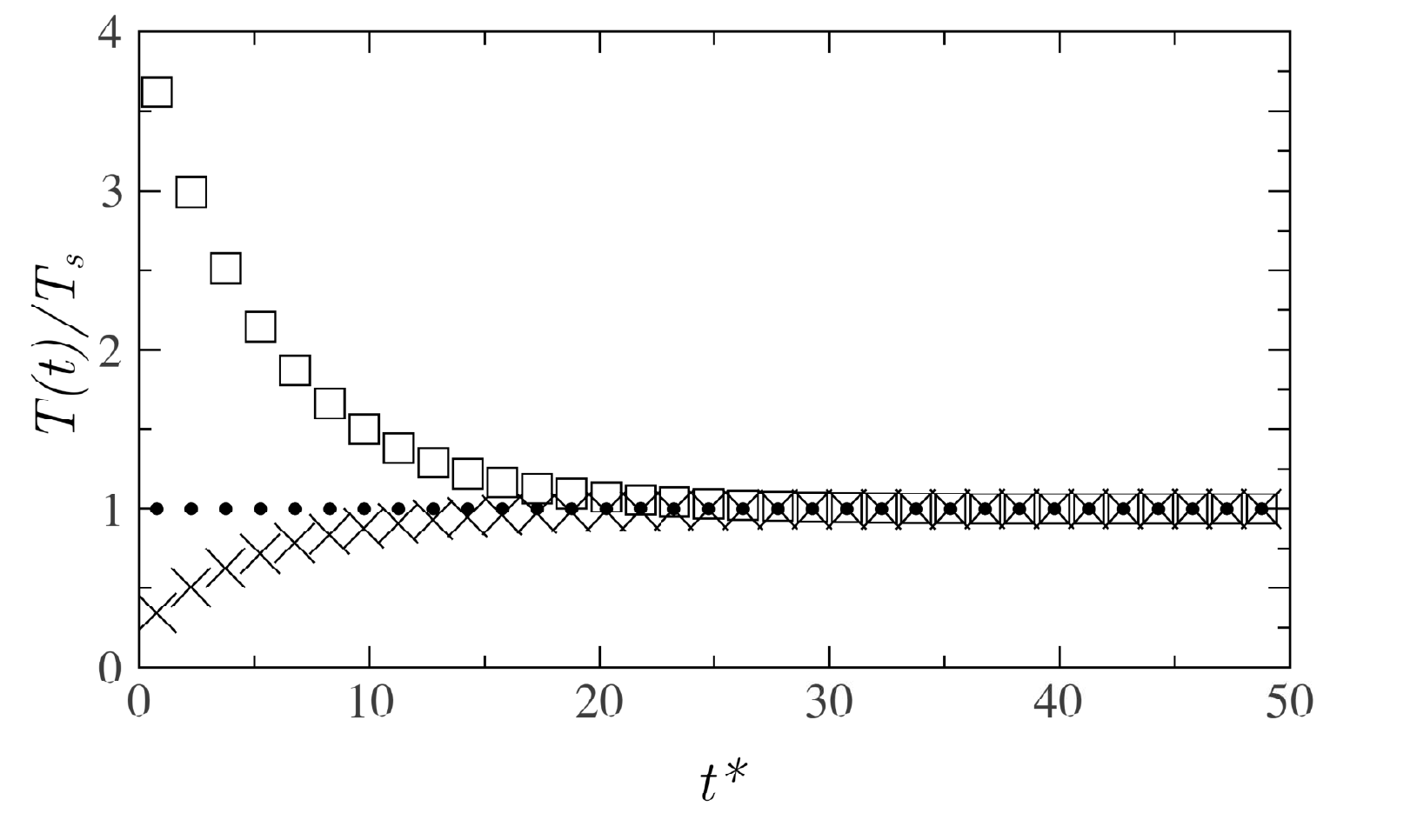}}\hspace{5mm}%\vspace{10mm} 
\subfigure[]{\includegraphics[width=100mm]{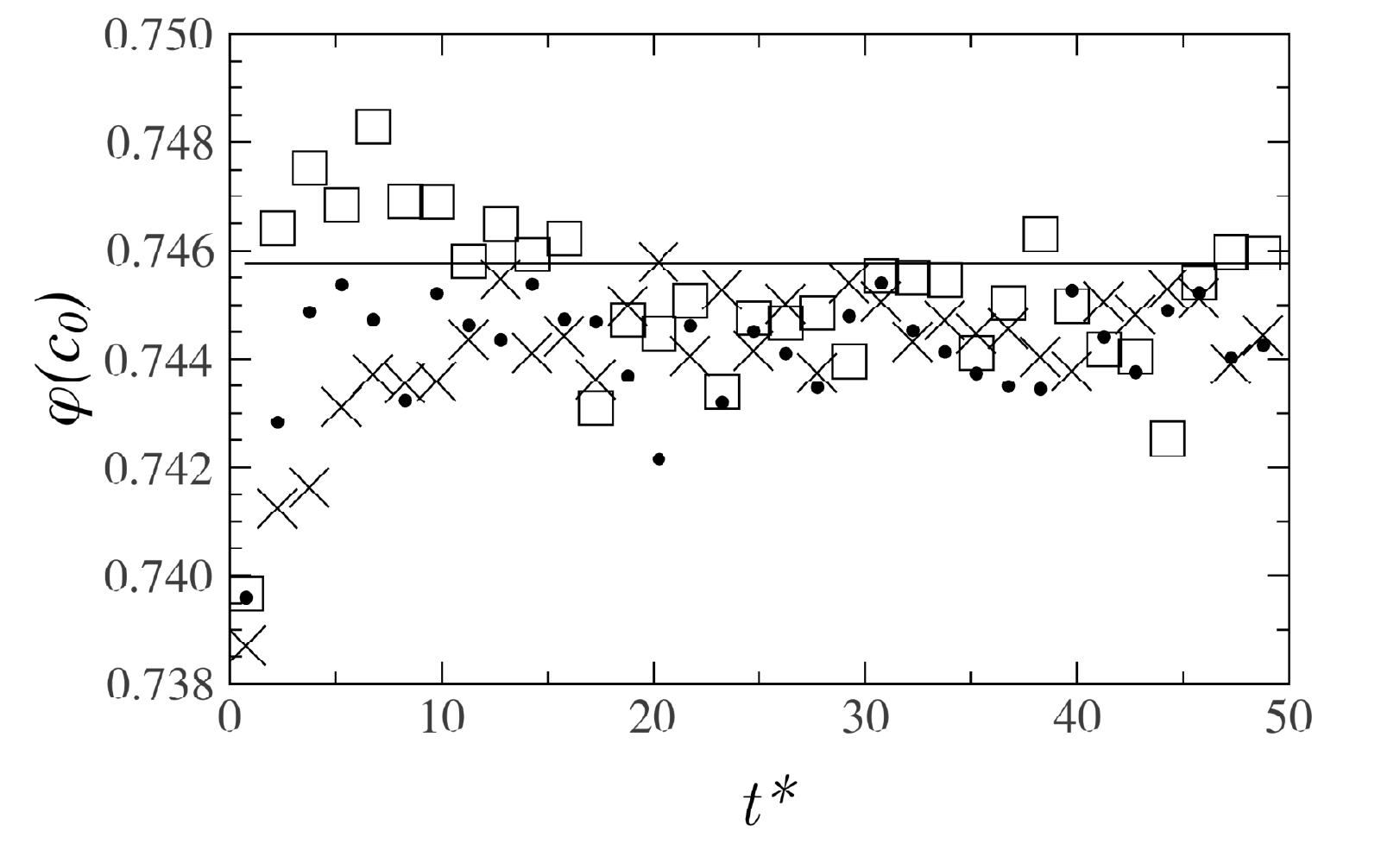}}%\vspace{10mm}

\caption{Time evolution for hard disks of the reduced temperature $T(t)/T_\text{s}$ (a) and the scaled distribution function $\varphi(c_0)$ (b) for $\xi^*=0.478$, $\gamma^*=0.014$, and $\alpha=0.9$. Three different initial temperatures have been considered: $T(0)/T_\text{s}=0.25 (\times), 1 (\cdots),$ and $4 (\square)$. Here, $T_\text{s}$ is the steady value of the temperature and $c_0(t)=v_{0,\text{s}}/v_0(t)$, $v_{0,\text{s}}=\sqrt{2T_\text{s}/m}$ being the steady value of the thermal speed.The symbols correspond to the simulation results while the horizontal lines refer to the theoretical predictions for $T_\text{s}$ and  $\varphi(c_0)$. The latter has been obtained by retaining the three first Sonine polynomials (see Eq.\ \eqref{4.1}) and evaluating  $a_2$ and $a_3$ with Approximation II. Time is measured in units of $\nu^{-1}$ ($t^*=t \nu^{-1}$).}
\label{fig01chap3}
\end{figure}

Although the main target in this work is the evaluation of all the relevant quantities of the problem ($a_2$, $a_3$ and $\varphi$) in the steady state, it is worthwhile to analyze the approach of some of these quantities towards the steady state. Fig.\ \ref{fig01chap3} shows the time evolution of a) the reduced temperature $T(t)/T_\text{s}$  and b) the distribution function $\varphi(\bt{c}_0)$  for the dimensionless velocity $c_0=v_{0,\text{s}}/v_0(t)$. Here, $T_\text{s}$ and $v_{0,\text{s}}=\sqrt{2T_\text{s}/m}$ refer to the theoretical steady values of the granular temperature and thermal velocity, respectively. The solid horizontal lines correspond to the theoretical predictions by considering the first two non-Gaussian corrections (third Sonine approximation) to the distribution $\varphi$ [see Eq.\ \eqref{eqSonine01}]. We have made runs of identical systems except that they are initialized with different temperatures. After a transient regime, as expected we observe that all simulation data tend to collapse to the same steady values for sufficiently long times. In addition, the corresponding steady values obtained from the simulation for both temperature and distribution function practically coincide with those predicted by the Sonine solution. It is also to be noticed that the convergence to the steady values occurs approximately at the same time for both $T(t)/T_\text{s}$ and $\varphi(c_0)$ (thermal fluctuations make difficult to determine the exact point for steady state convergence for the distribution function). This is another and indirect way of checking that indeed the normal solution exists for simulations, since its existence implies, from Eq.\ \eqref{eqT}, that we reach the scaled form \eqref{eqfscaled} when the temperature is stationary. %Although this Chapter is devoted only on the study of an homogeneous granular gas in the asymptotic steady state and all the analytical expressions for the relevant quantities ($a_2$, $a_3$ and $\varphi$) have been obtained for this regime, DSMC simulations allow us to analyze the approach of some of these quantites towards the steady state

Some previous works on a granular gas heated by the stochastic thermostat \cite{GMT12} and on the simple shear flow \cite{AS07} have shown that before reaching the steady state the system evolves towards a universal \emph{unsteady} state that depends on a new parameter measuring the distance to the steady state. A similar behavior is expected here where the different solutions to the Enskog equation \eqref{eqEB02} would be attracted by the universal distribution function $f(v,t)\rightarrow n v_0(t)^{-1}\varphi(\bt{c}(t),\tilde{\gamma}(t),\tilde{\xi}(t))$, where $\bt{c}(t)=\bt{v}/v_0(t)$ and
\begin{equation}
  \tilde{\gamma}(t)\equiv\frac{\ell \gamma_b}{\chi m v_0(t)}, \qquad   \tilde{\xi}(t)\equiv\frac{\ell \xi_b^2}{\chi T(t) v_0(t)}.
\end{equation}

The dimensionless driving parameters $\tilde{\gamma}(t)$ and $\tilde{\xi}(t)$ measure the distance to the steady state. For asymptotically long times, the steady state is eventually reached, i.e., $\varphi(\bt{c}(t),\tilde{\gamma}(t),\tilde{\xi}(t)) \rightarrow \varphi_s(\bt{c},\xi^*)$. The above unsteady hydrodynamic regime (for which the system has forgotten its initial condition) is expected to be achieved after a certain number of collisions per particle.% On the other hand, although the characterization of this unsteady state is a very interesting problem, its study lies beyond the goal of the present paper.

\subsubsection{Steady regime}

\begin{figure}[h]%tbp]
\centering
\subfigure[]{\includegraphics[width=90mm]{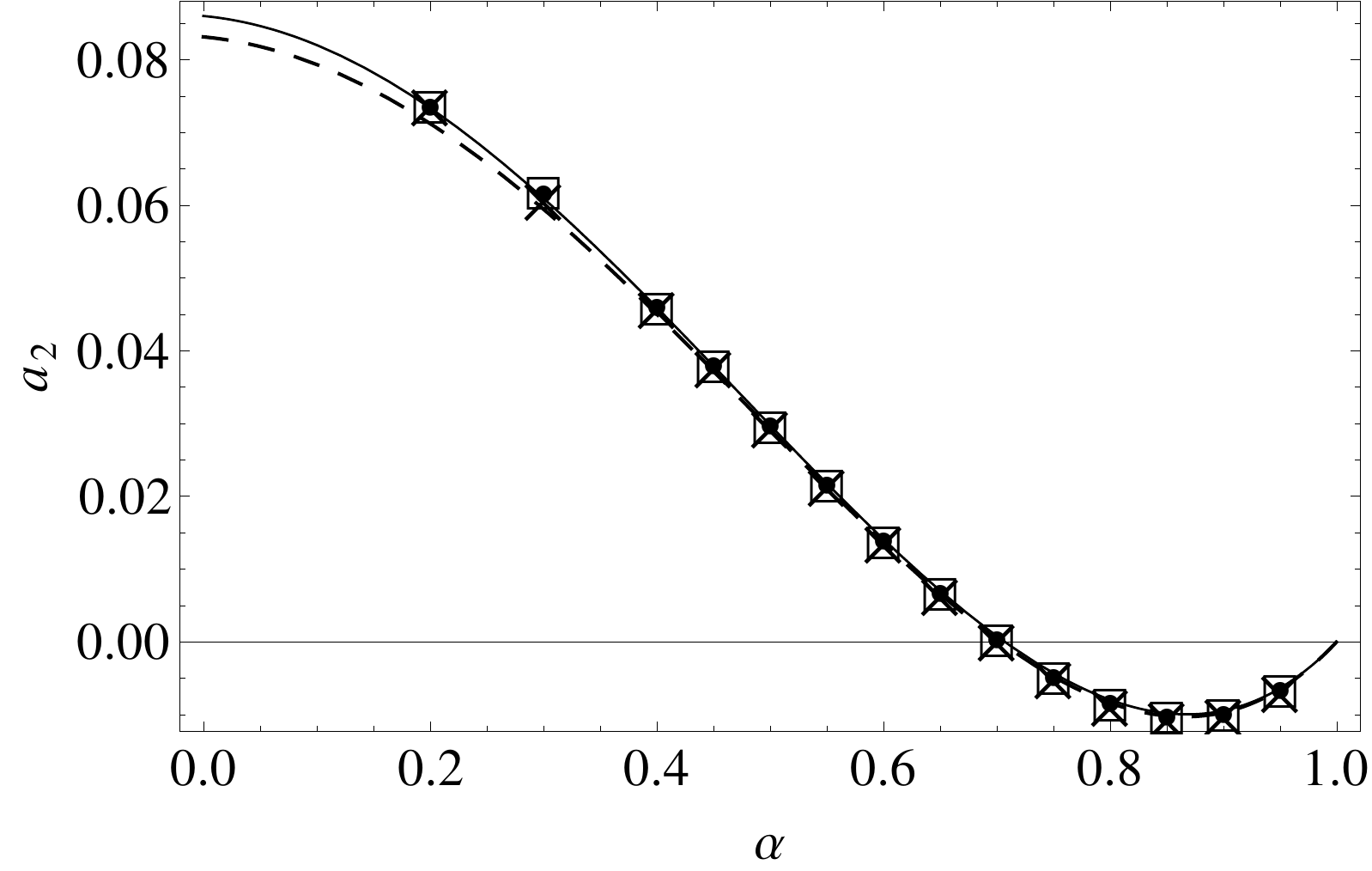}} \hspace{5mm} %\vspace{10mm}
\subfigure[]{\includegraphics[width=90mm]{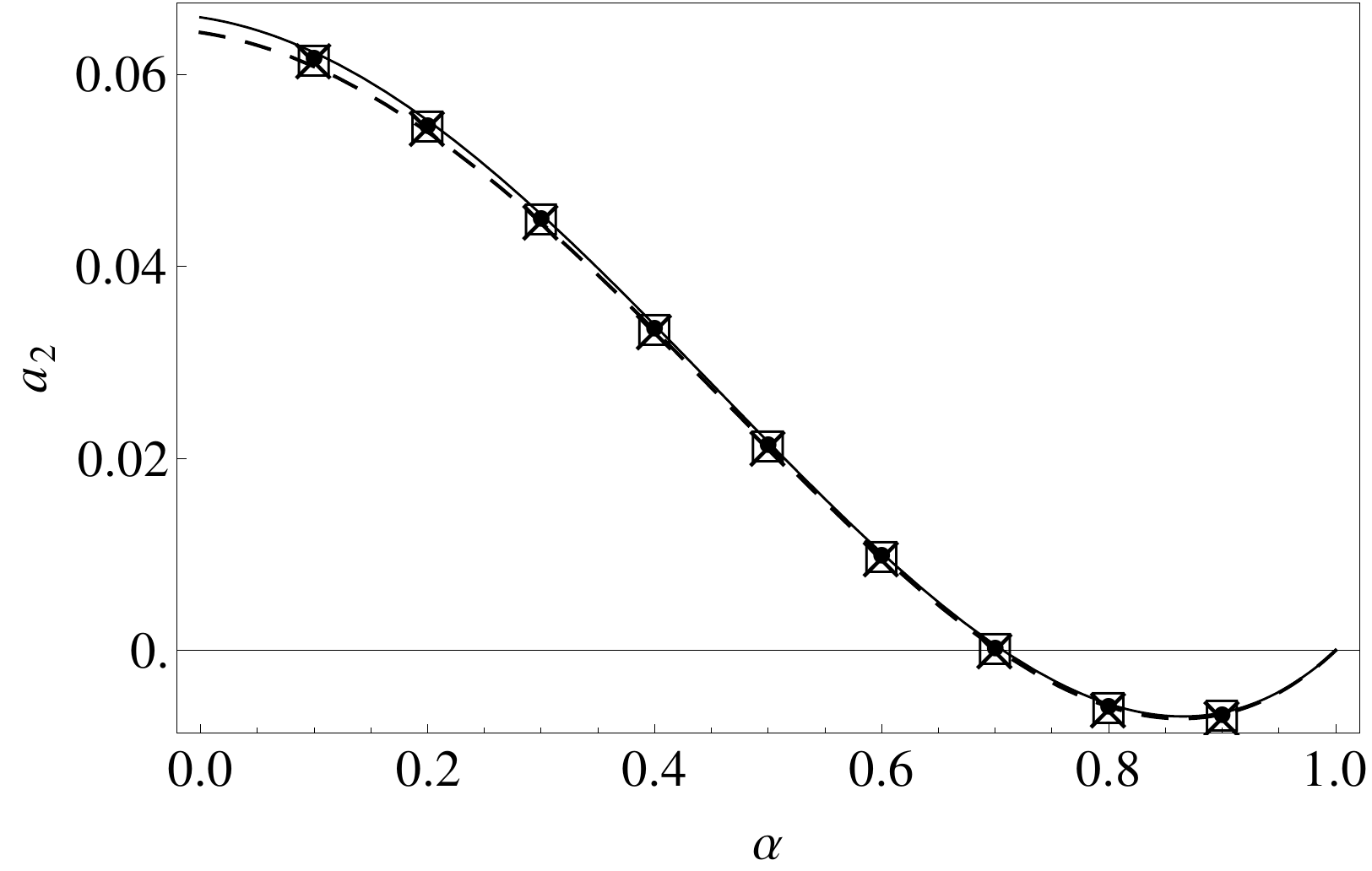}}%\vspace{10mm}
\caption{Plot of the second Sonine coefficient $a_2$ versus the coefficient of restitution $\alpha$ for hard disks (a) and hard spheres (b). The symbols refer to three different systems with different values of the simulation parameters $\gamma_{sim}^*$ and $\xi_{sim}^*$ but with the same value of $\xi^*$ ($\xi^*=1.26$ for disks and $\xi^*=1.68$ for spheres). The solid and dashed lines are the values obtained for $a_2$ by means of Approximation I and Approximation II, respectively. \label{fig02chap3}} 
\end{figure}

\begin{figure}[h]%tbp]
\centering
\subfigure[]{\includegraphics[width=90mm]{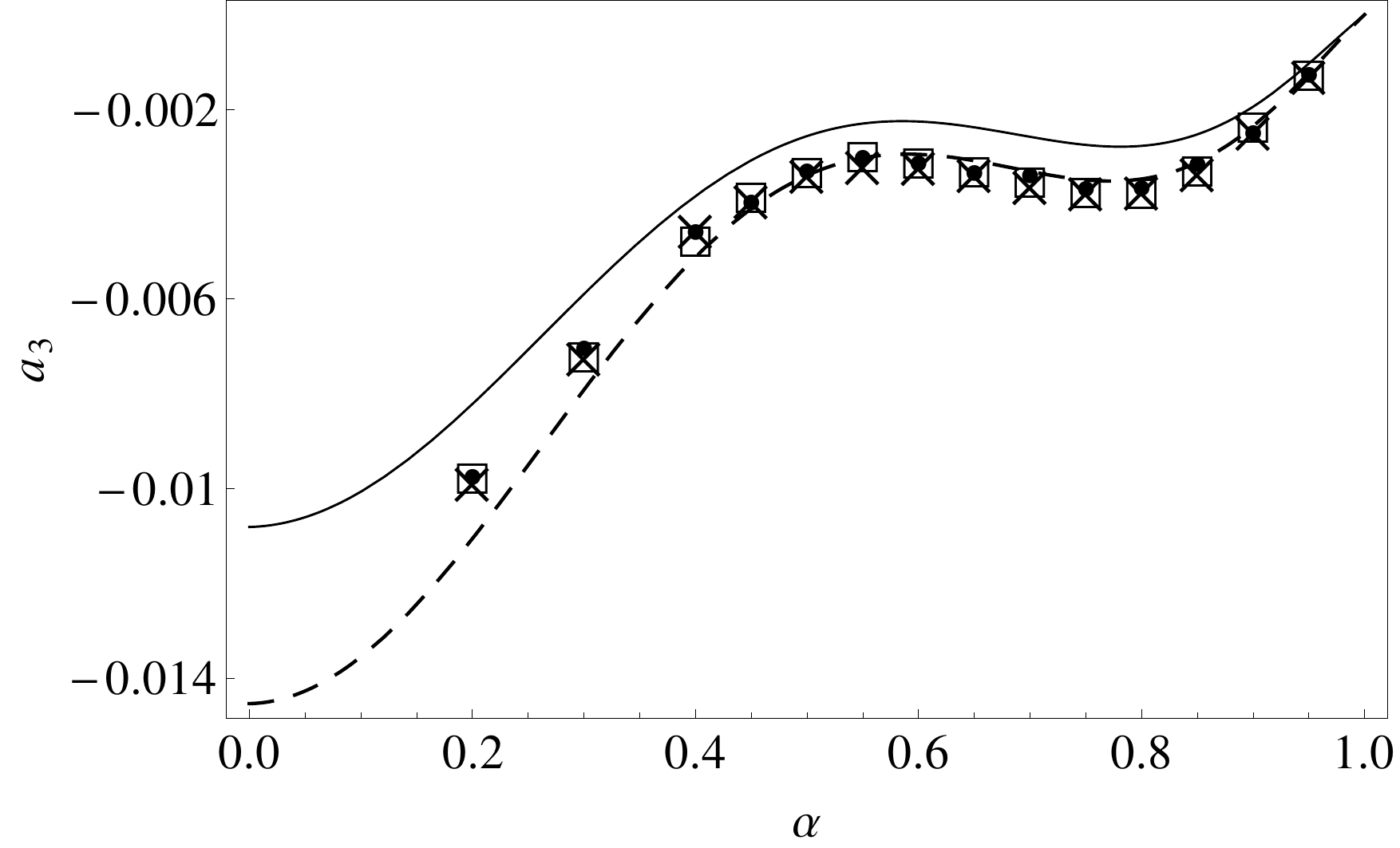}}\vspace{10mm} %\hspace{5mm}
\subfigure[]{\includegraphics[width=90mm]{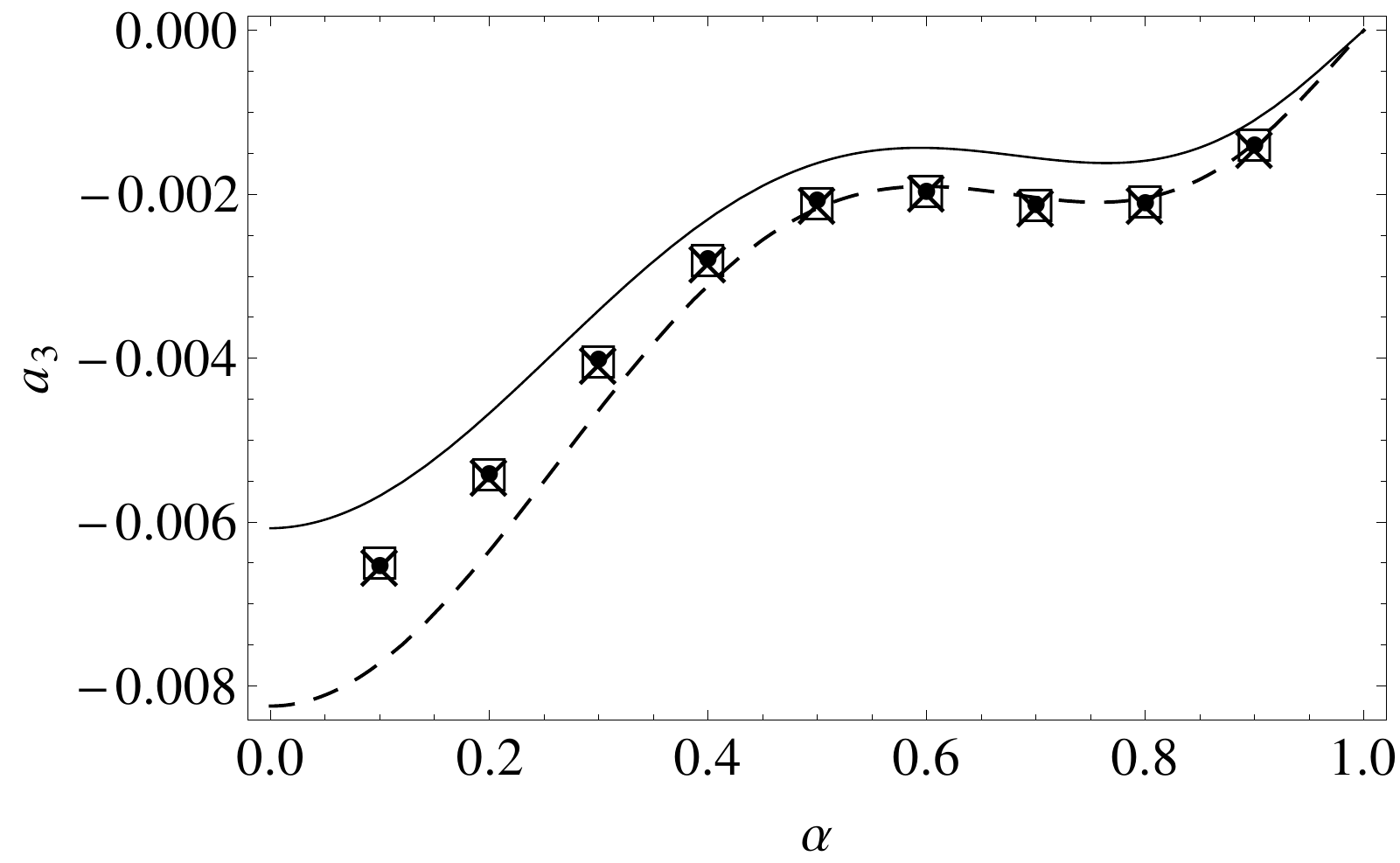}}%\vspace{10mm}
\caption{Plot of the third Sonine coefficient $a_3$ versus the coefficient of restitution $\alpha$ for hard disks (a) and hard spheres (b). The symbols refer to three different systems with different values of the simulation parameters $\gamma_\text{sim}^*$ and $\xi_\text{sim}^*$ but with the same value of $\xi^*$ ($\xi^*=1.26$ for disks and $\xi^*=1.68$ for spheres). The solid and dashed lines are the values obtained for $a_3$ by means of Approximation I and Approximation II, respectively.} 
\label{fig03chap3}
\end{figure}

Now, we will focus on the steady state values of the relevant quantities of the problem. In particular,
the basic quantities measuring the deviation of the distribution function from its Maxwellian form are the second and third Sonine coefficients $a_2$ and $a_3$, respectively. The dependence of $a_2$ and $a_3$ on the coefficient of restitution $\al$ is shown in Figs.\ \ref{fig02chap3} and \ref{fig03chap3}, respectively, for hard disks (a) and spheres (b). Three different systems with different values of the simulation parameters $\gamma_\text{sim}^*$ and $\xi_\text{sim}^*$ but with the same value of $\xi^*$ ($\xi^*=1.263$ for disks and $\xi^*=1.688$ for spheres) have been considered. We observe that, at a given $\al$, the corresponding three simulation data collapse in a common curve, showing that indeed both Sonine coefficients are always of the form $a_i(\alpha, \xi^*)$. Regarding the comparison between theory and simulation, it is quite apparent that while both Approximations I and II compare quantitatively quite well with simulations in the case of $a_2$, Approximation II has a better performance than Approximation I in the case of $a_3$, specially at very strong dissipation. This is the expected result since Approximation II is in principle more accurate that Approximation I, although the latter is simpler than the former. In this sense and with respect to the $\al$-dependence of $a_2$ and $a_3$, Approximation I could be perhaps preferable to Approximation II since it has an optimal compromise between simplicity and accuracy.

%\begin{figure}[hl]%tbp]

\begin{figure}[h]%tbp]
\centering
\subfigure[]{\includegraphics[width=90mm]{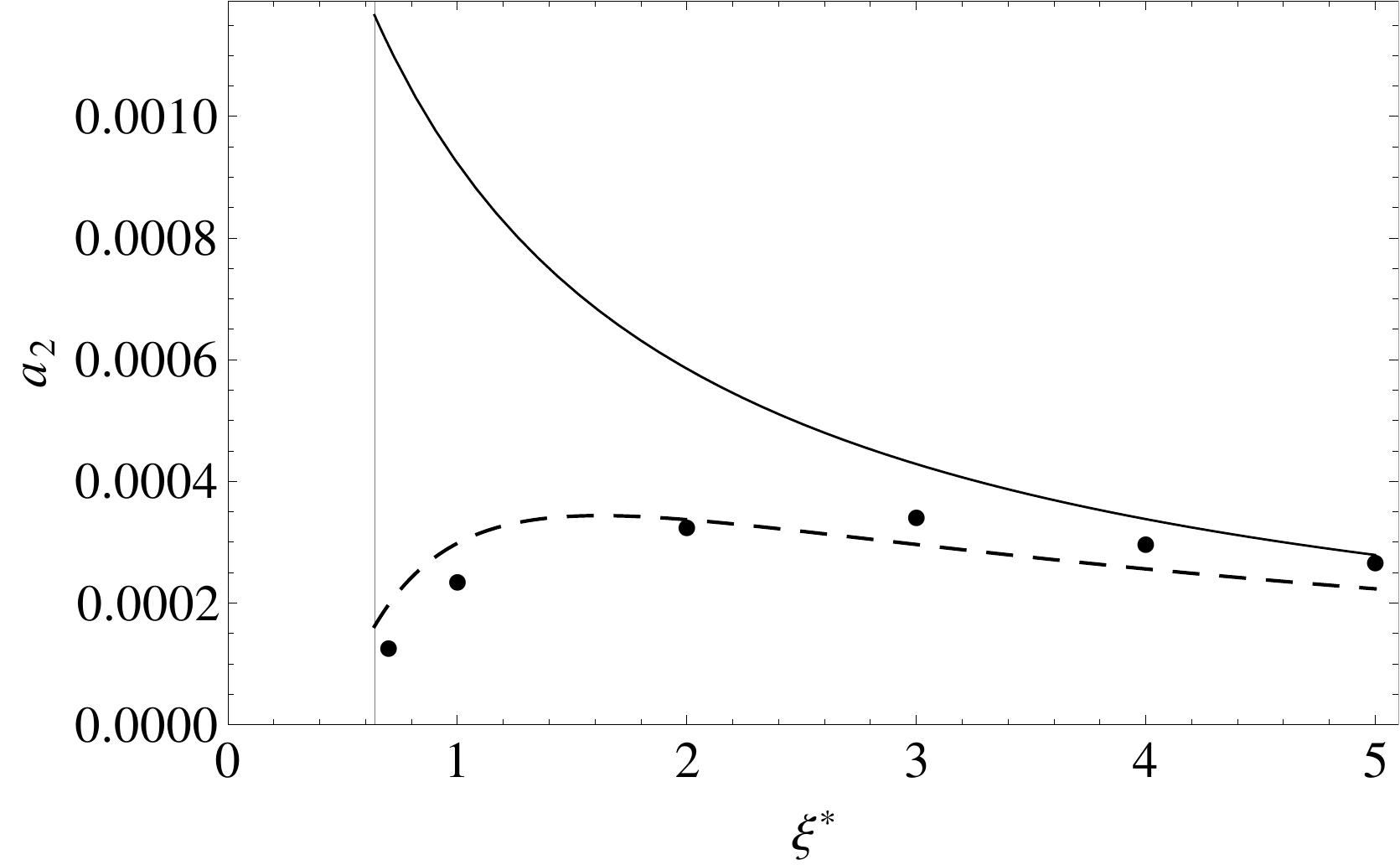}}\vspace{10mm} %\hspace{5mm}
\subfigure[]{\includegraphics[width=90mm]{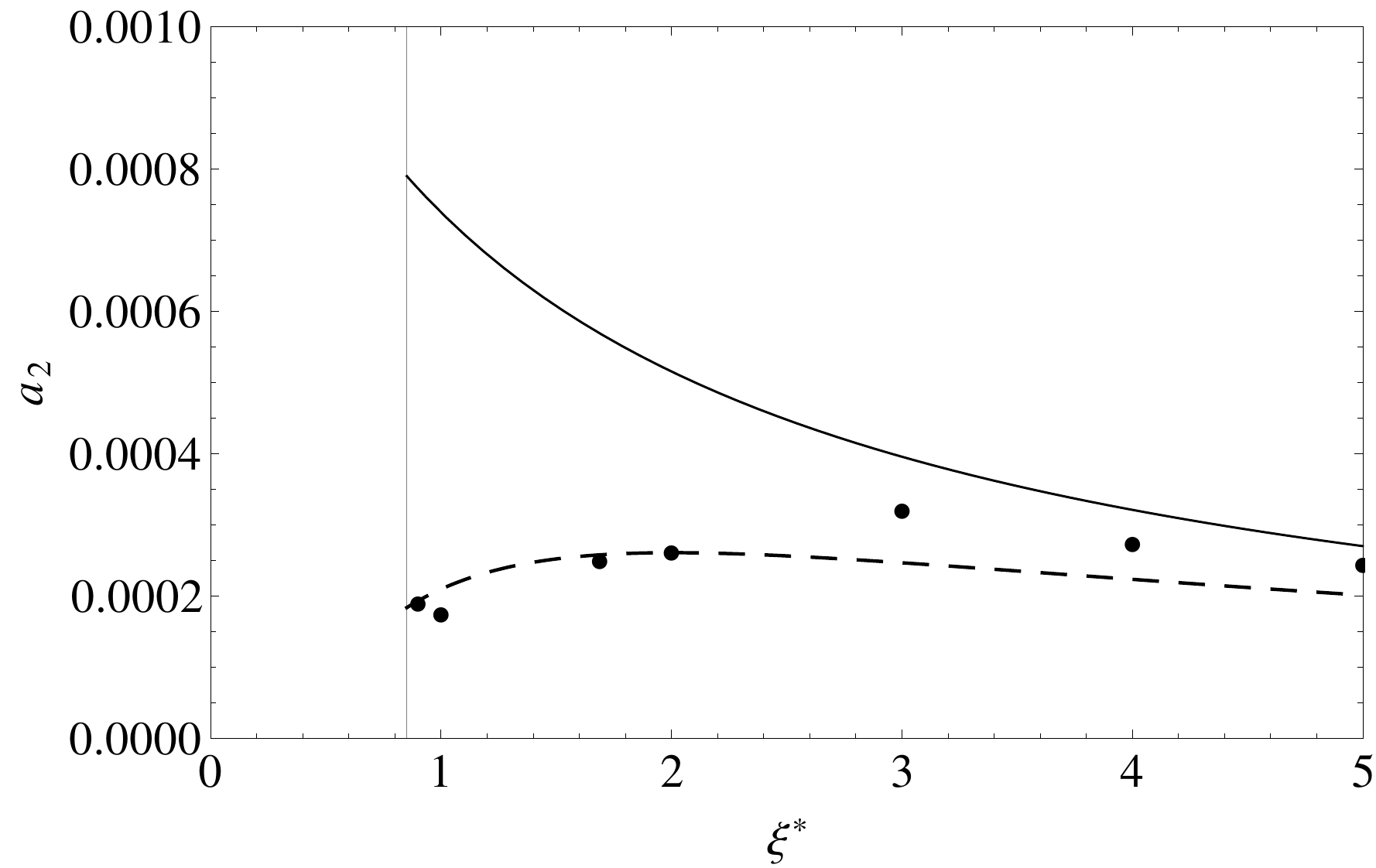}}%\vspace{10mm}
\caption{Plot of the second Sonine coefficient $a_2$ versus the (reduced) noise strength $\xi^*$ for $\alpha=0.7$ in the case of hard disks (a) and hard spheres (b). The symbols refer to simulation results while the solid and dashed lines are the values obtained for $a_2$ by means of Approximation I and Approximation II, respectively. The vertical lines indicate the threshold values $\xi^*_\text{th}$.} 
\label{fig04chap3}
\end{figure}

\begin{figure}[h]%tbp]
\centering
\subfigure[]{\includegraphics[width=90mm]{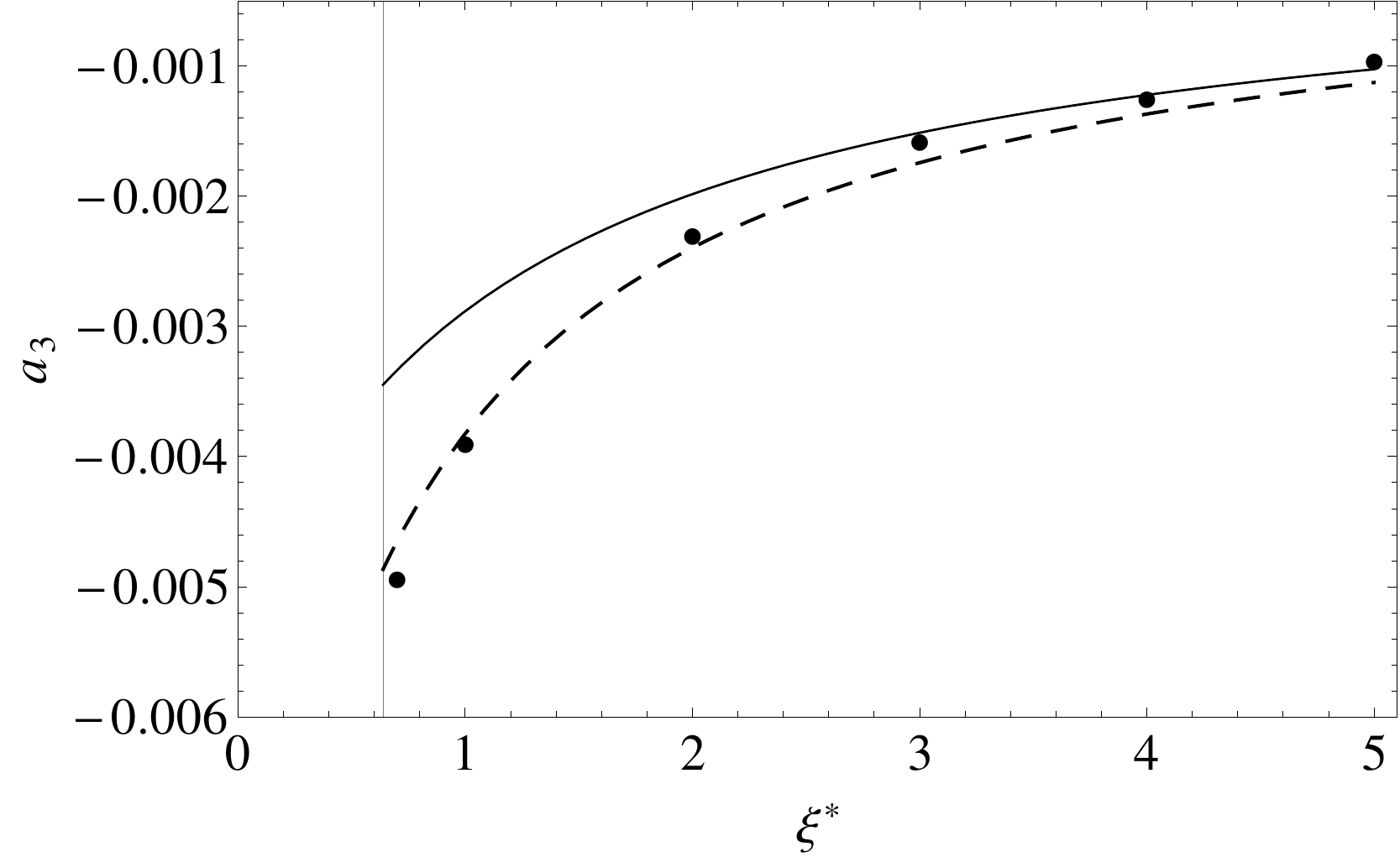}}\vspace{10mm} %\hspace{5mm}
\subfigure[]{\includegraphics[width=90mm]{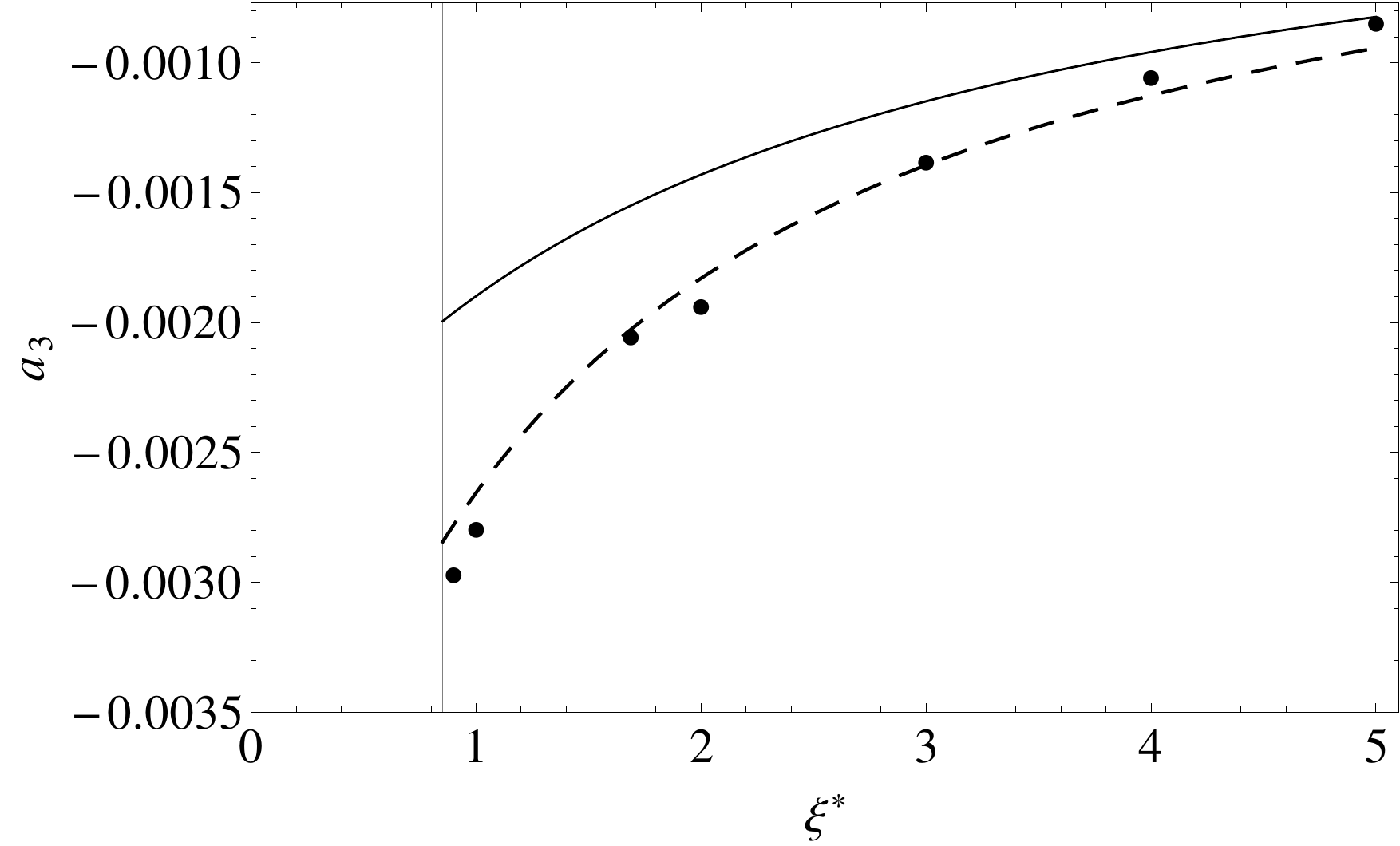}}%\vspace{10mm}
\caption{Plot of the third Sonine coefficient $a_3$ versus the (reduced) noise strength $\xi^*$ for $\alpha=0.7$ in the case of hard disks (a) and hard spheres (b). The symbols refer to simulation results while the solid and dashed lines are the values obtained for $a_3$ by means of Approximation I and Approximation II, respectively. The vertical lines indicate the threshold values $\xi^*_\text{th}$.} 
\label{fig05chap3}
\end{figure}

On the other hand, more quantitative discrepancies between both Approximations are found when one analyzes both Sonine coefficients vs. $\xi^*$ with constant $\al$. Figs.\ \ref{fig04chap3} and \ref{fig05chap3} show $a_2$ and $a_3$, respectively, versus $\xi^*$ at $\al=0.7$. We see that Approximation I exhibits a poor agreement with simulations since it predicts a dependence on the noise strength opposite to the one found in the simulations. On the other hand, Approximation II agrees very well with simulation data in all the range of values of $\xi^*$  (note that $\xi^*\gtrsim 0.639$ for $d=2$ and $\xi^*\gtrsim 0.852$ for $d=3$ to achieve a steady state for $\al=0.7$). It must be also noted that for the systems studied in Figs.\ \ref{fig04chap3} and \ref{fig05chap3}, although the magnitudes of both Sonine coefficients are very small, $|a_2|$ is of the order of ten times smaller than $|a_3|$. This may indicate a poor convergence of the Sonine polynomial expansion \cite{BP06} for high inelasticity.
%\begin{figure}[hl]%tbp]
%\begin{figure}[hl]%tbp]

\begin{figure}[h]%tbp]
\centering
\subfigure[]{\includegraphics[width=90mm]{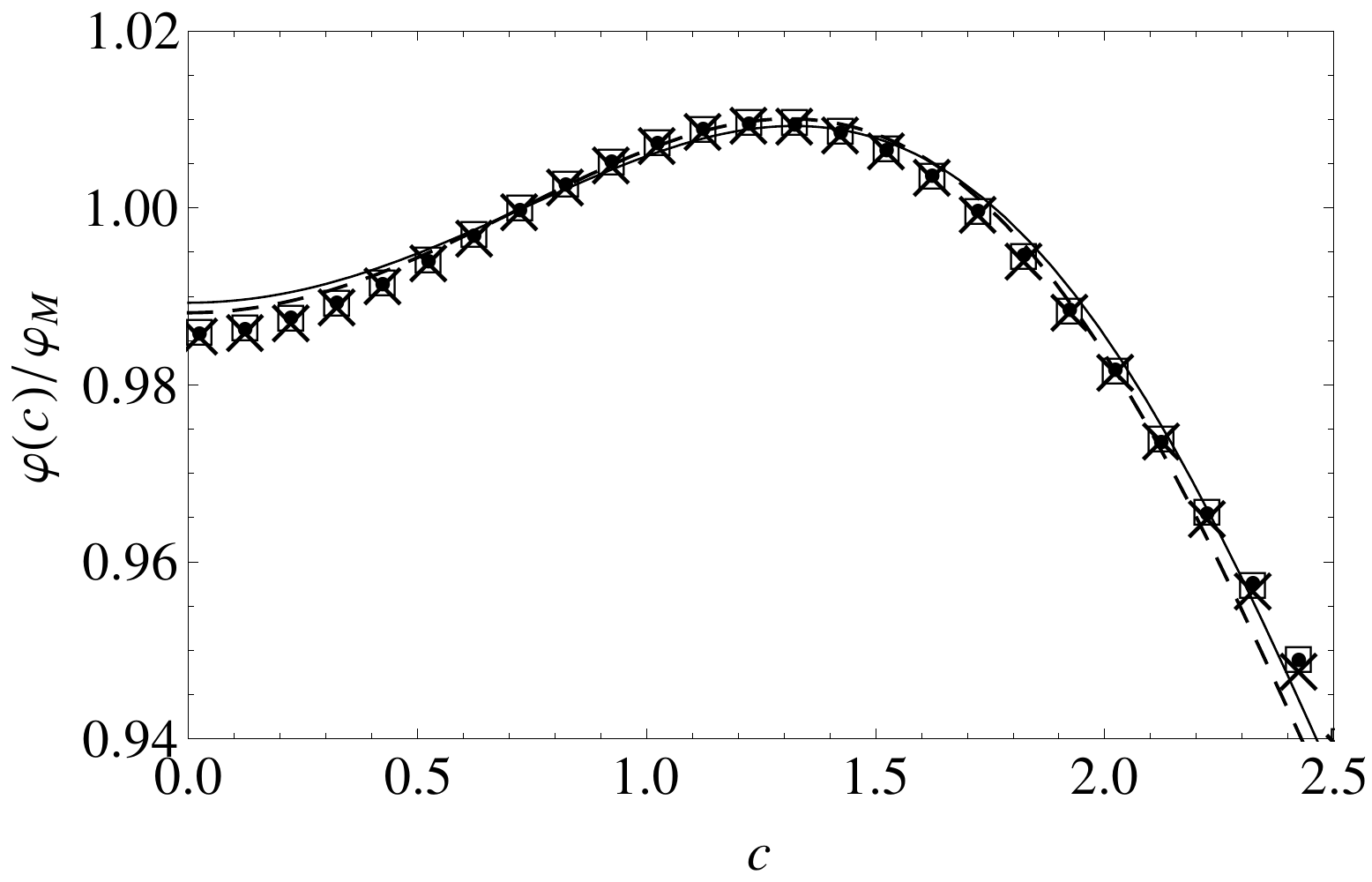}}\vspace{10mm} %\hspace{5mm}
\subfigure[]{\includegraphics[width=90mm]{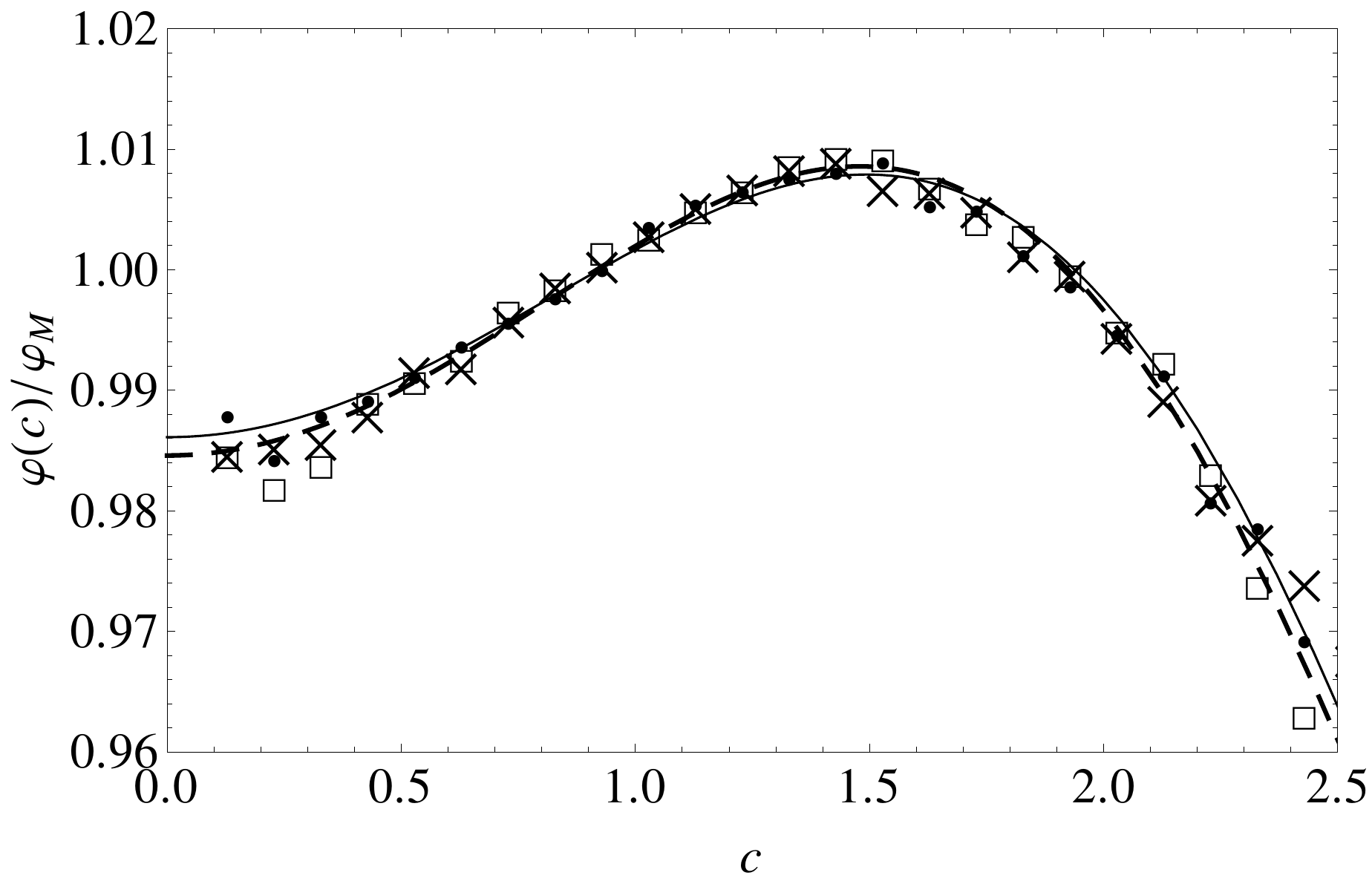}}%\vspace{10mm}
\caption{Plot of the scaled distribution function $\varphi(\bt{c},\xi^*)/\varphi_\text{M}(c)$ in the steady state for $\alpha=0.8$. The symbols refer to DSMC data obtained for three different systems with parameters: $\{\gamma_\text{sim}^*, \xi_\text{sim}^*\}=\{(1.4\times 10^{-2}, 5.2\times 10^{-5}), (9.8\times 10^{-3}, 1.8\times 10^{-5}), (7\times 10^{-3}, 6.5\times 10^{-6})\}$ for $d=2$ and $\{\gamma_\text{sim}^*, \xi_\text{sim}^*\}=\{(7.1\times 10^{-3}, 2.9\times 10^{-6}), (5\times 10^{-3}, 9.8\times 10^{-7}), (3.6\times 10^{-3}, 3.6\times 10^{-7})\}$ for $d=3$. These values yield a common value of $\xi^*$: $\xi^*=1.263$ for $d=2$ and $\xi^*=1.688$ for $d=3$. The lines correspond to Eq.\ \eqref{4.1} with expressions for the cumulants given by Approximation I (solid lines) and Approximation II (dashed lines).} 
\label{fig06chap3}
\end{figure}

\begin{figure}[h]%tbp]
\centering
\subfigure[]{\includegraphics[width=90mm]{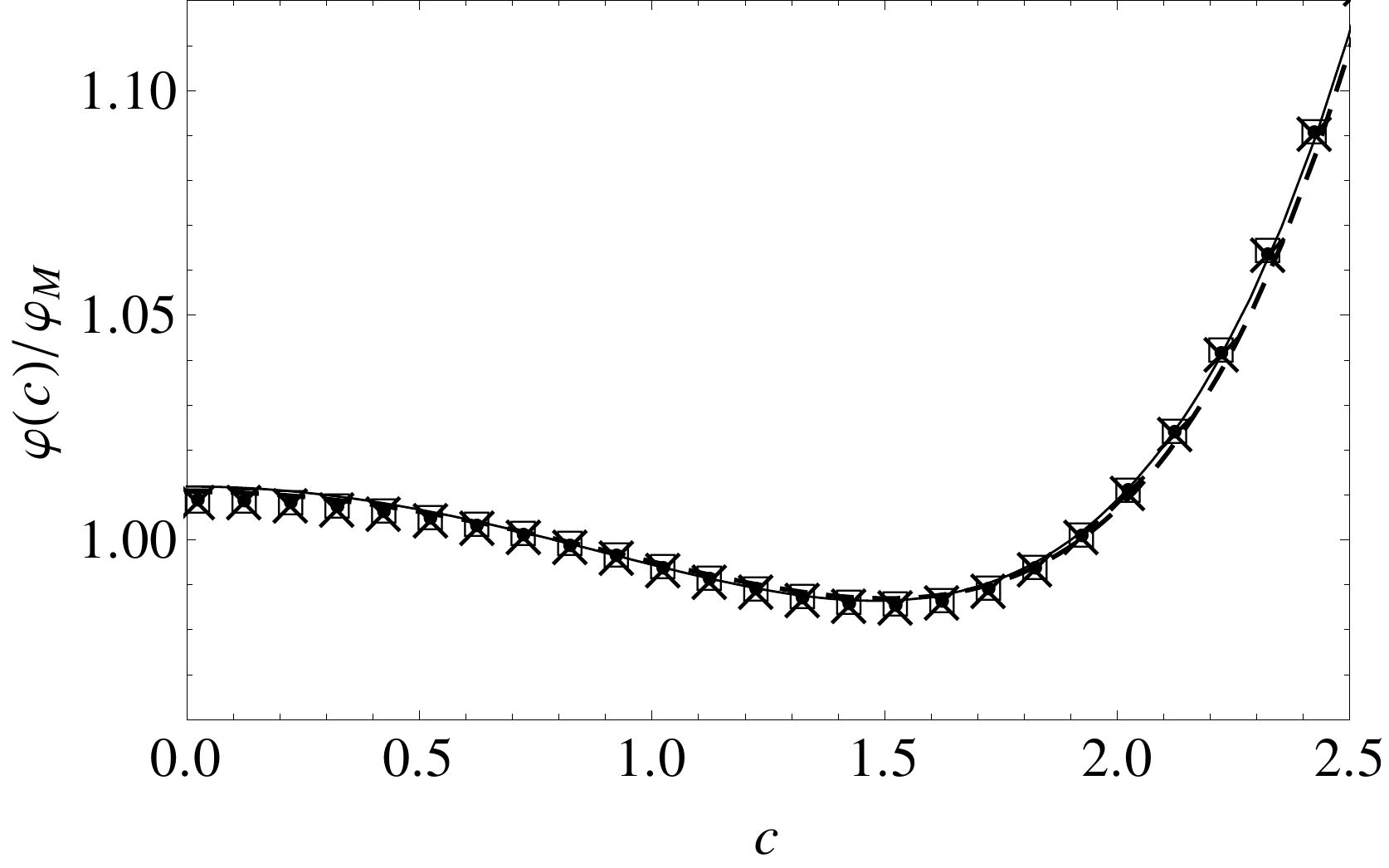}}\vspace{10mm} %\hspace{5mm}
\subfigure[]{\includegraphics[width=90mm]{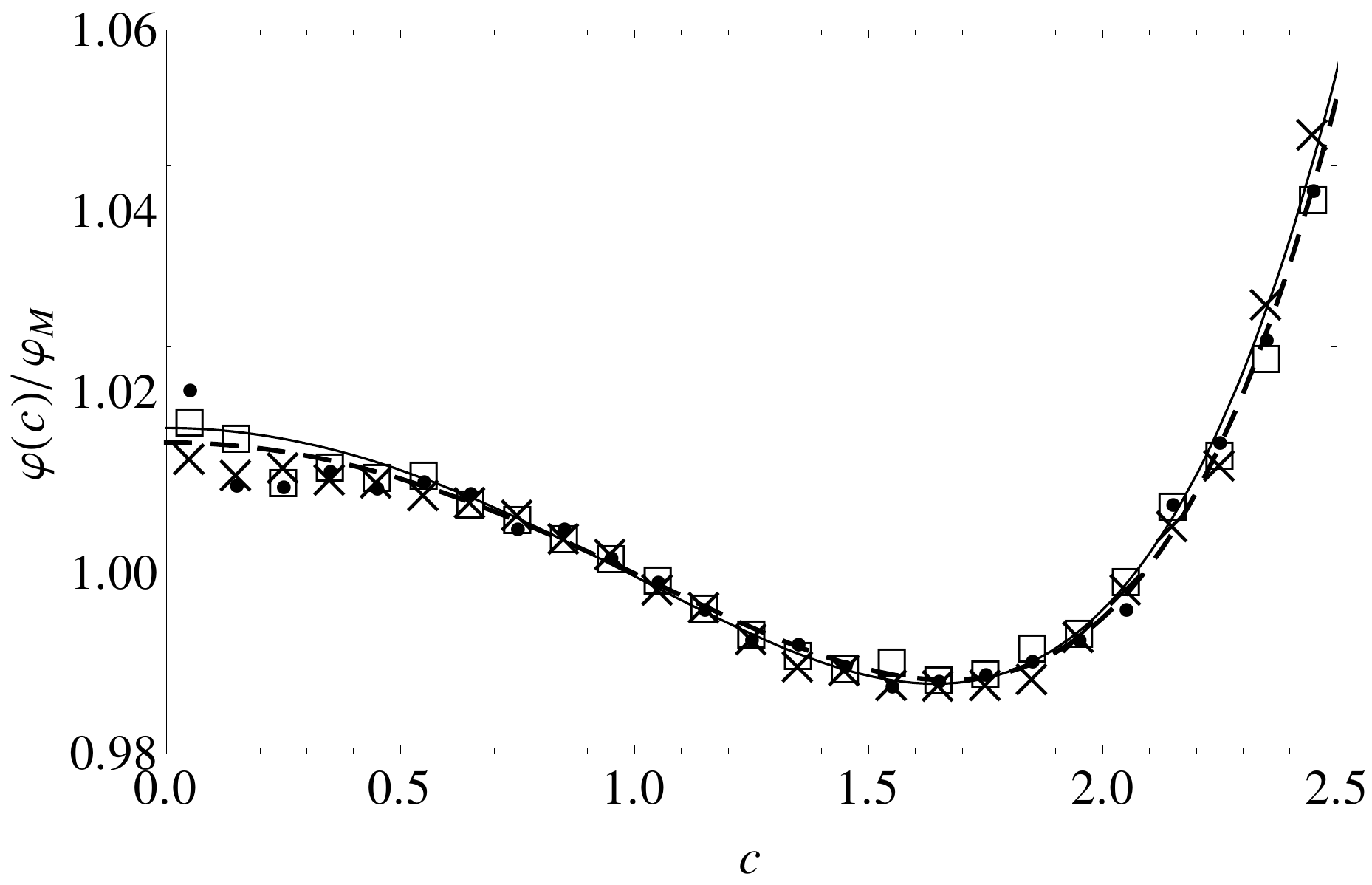}}%\vspace{10mm}
\caption{Plot of the scaled distribution function $\varphi(\bt{c},\xi^*)/\varphi_\text{M}(c)$ in the steady state for $\alpha=0.6$. The symbols refer to DSMC data obtained for three different systems with parameters: $\{\gamma_\text{sim}^*, \xi_\text{sim}^*\}=\{(1.4\times 10^{-2}, 2.9\times 10^{-4}), (9.8\times 10^{-3}, 10^{-4}), (7\times 10^{-3}, 3.6\times 10^{-5})\}$ for $d=2$ and $\{\gamma_\text{sim}^*, \xi_\text{sim}^*\}=\{(7.1\times 10^{-3}, 1.5\times 10^{-5}), (5\times 10^{-3}, 5.4\times 10^{-6}), (3.6\times 10^{-3}, 1.9\times 10^{-6})\}$ for $d=3$. These values yield a common value of $\xi^*$: $\xi^*=1.263$ for $d=2$ and $\xi^*=1.688$ for $d=3$. The lines correspond to Eq.\ \eqref{4.1} with expressions for the cumulants given by Approximation I (solid lines) and Approximation II (dashed lines).} 
\label{fig07chap3}
\end{figure}

The small values of the coefficients $a_2$ and $a_3$ support the assumption of a low-order truncation in polynomial expansion and suggests that the scaled distribution function $\varphi(\bt{c},\xi^*)$ for thermal velocities can be well represented by the three first contributions (note that $a_1=0$) in the Sonine polynomial expansion \eqref{eqSonine01}. To confirm it, we have measured the deviation of $\varphi(\bt{c},\xi^*)$ from its Maxwellian form $\varphi_\text{M}(c)$. In Figs.\ \ref{fig06chap3} an \ref{fig07chap3} we plot the ratio $\varphi(\bt{c},\xi^*)/\varphi_\text{M}(c)$ versus the reduced velocity $c$ in the steady state for two values of the coefficient of restitution ($\al=0.8$ and $\al=0.6$). As before, we have considered a system of inelastic hard disks (Fig.\ \ref{fig06chap3} with $\xi^*=1.26$) and inelastic hard spheres (Fig.\ \ref{fig07chap3} with $\xi^*=1.69$). As in Figs.\ \ref{fig02chap3}--\ref{fig05chap3}, symbols correspond to simulation results obtained for different values of $\gamma_\text{sim}^*$ and $\xi_\text{sim}^*$. The solid and dashed lines are obtained from Eq.\ \eqref{eqSonine01} with the series truncated at $p=3$, i.e.,
\beqa
\label{4.1}
\frac{\varphi(\bt{c},\xi^*)}{\varphi_\text{M}(c)}&\to& 1+a_2(\xi^*)\left(\frac{1}{2}c^4-\frac{d+2}{2}c^2+\frac{d(d+2)}{8}\right)\nonumber\\
& & -a_3(\xi^*)\left(
\frac{1}{6}c^6-\frac{d+4}{4}c^4+\frac{(d+2)(d+4)}{8}c^2\right.\nonumber\\
& & -\left.\frac{d(d+2)(d+4)}{48}\right).
\eeqa

The coefficients $a_2$ and $a_3$ in Eq.\ \eqref{4.1} are determined by using Approximation I (solid lines) and Approximation II( dashed lines). First, it is quite apparent that simulations confirm that the reduced distribution function $\varphi(\bt{c},\xi^*)$ is a universal function of $\xi^*$ since all simulation series at constant $\xi^*$ collapse to the same curve (within non-measurable marginal error). We also see that the simulation curves agree very well with the corresponding third-order Sonine polynomial in this range of velocities, especially in the two-dimensional case. Surprisingly, in the high velocity region, the curves obtained from Approximation I fit the simulation data slightly better than those obtained by using the improved Approximation II. In any case, the agreement between theory and simulation is again excellent, especially taking into account the very small discrepancies we are measuring.

%\begin{figure}[hl]%tbp]

%\begin{figure}[hl]%tbp]

%\azul{ATENCION: Hay que terminar esta parte y explicar los resultados anteriores}

\section{Summary and Discussion}
%\label{discussion}

In this Chapter we have analyzed the homogeneous steady state of a granular gas driven by a stochastic bath with friction described \fran{with the} Enskog kinetic equation. One of the primary objectives of this work has been to check the velocity scaling and the functional form assumed for the distribution function in the steady state. As Eq.\ \eqref{eqfscaled} indicates, the new feature of the scaled distribution $\varphi$ is the dependence on both the granular temperature $T$ through the scaled velocity $\bt{c}$ and also through the reduced noise strength $\xi^*$ [defined in Eq.\ \eqref{eqxistar}]. The simulation results reported here (see Figs.\ \ref{fig06chap3} and \ref{fig07chap3}) have confirmed the above dependence since different systems sharing the same values of $\xi^*$ and $\al$ lead to the same distribution function $\varphi$. This is consistent with the existence of a \emph{normal} solution in the long-time limit.

We have also characterized the distribution $\varphi$ through its first velocity moments. More specifically, we have obtained the second $a_2$ and third $a_3$ Sonine coefficients. While the coefficient $a_2$ measures the fourth-degree velocity moment of $\varphi$, the coefficient $a_3$ is defined in terms of the sixth-degree velocity moment of $\varphi$. Both Sonine coefficients provide information on the deviation of $\varphi$ from its Maxwellian form $\varphi_\text{M}$. Moreover, the knowledge of those coefficients \fran{is important}; for instance, in the precise determination of the transport coefficients \cite{GCV13}. On the other hand, given that the Sonine coefficients cannot be \emph{exactly} determined (they obey an infinite hierarchy of moments), one has to truncate the corresponding Sonine polynomial expansion in order to estimate them. Here, we have considered two different approaches (Approximation I and II) to get explicit expressions of $a_2$ and $a_3$ in terms of the dimensionality of the system $d$, the coefficient of restitution $\al$ and the driving parameter $\xi^*$. Approximation II is more involved than Approximation I since it considers both Sonine coefficients as being of the same order of magnitude.  The comparison between the analytical solution and DSMC results shows in general a good agreement, even for high-inelasticity for both approaches. Moreover, the improved Approximation II for $a_2$ and $a_3$ shows only a slightly better agreement with simulations than Approximation I (see Figs.\ \ref{fig02chap3}--\ref{fig05chap3}). Thus, taking into account all the above comparisons, we can conclude that a good compromise between accuracy and simplicity is represented by  Approximation I.

The results derived in this Chapter show clearly that the combination of analytical and computational tools (based on the DSMC method) turns out to be \fran{a useful} way to characterize properties in granular flows. On the other hand, given that most of the Sonine coefficients can be directly calculated by DSMC, one could in principle make a least-square fit to obtain explicit forms for those coefficients. However, this procedure would not be satisfactory from a more fundamental point of view, especially if one is interested in capturing the behavior of $\varphi(\bt{c})$ and its Sonine polynomial expansion. In this context, our analytical solution of the distribution function (redundant as it may seem) has the advantage of providing a rational description of the physical properties of the kinetic equation of the system. This is not accomplished by the numerical solution. \fran{Nevertheless}, the fact that the DSMC method gives an accurate numerical solution of the Enskog equation makes it complementary to the theoretical one and thus both conform a complete description of the kinetic equation of our system.

         % Homogeneous
\chapter{Transport properties for driven granular fluids in situations close to homogeneous steady states}
\label{Chapter4}
\lhead{Chapter 4. \emph{Transport properties for driven granular fluids in situations close to homogeneous steady states}}
\blfootnote{The results obtained in this Chapter have been published in V. Garz\'o, M.G. Chamorro and F. Vega Reyes, \emph{Phys. Rev. E}, \textbf{87}:032201 (2013) and V. Garz\'o, M.G. Chamorro and F. Vega Reyes, \emph{Phys. Rev. E}, \textbf{87}:059906 (2013) [\emph{erratum}] \cite{GCV13, GCV13err}} 
 
\section{Introduction}
\label{sec1chap4}
%The results of the homogeneous steady state obtained in previous Chapter is the starting point to study the transport properties of a (dense) granular gas driven by a stochastic bath with friction. 

%Dissipative character of the collisions between particles of unforced granular gases is the reason that, conversely it happens in ordinary gases, the only equilibrium state is one of zero kinetic energy, or zero granular temperature. Furthermore, such gases are unstable to density perturbations and give rise to the formation of clusters that destroy the homogeneity of the system \cite{MY92,GT96}. In order to avoid such problems it is ussual to compensate the lost of energy due to inelastic collisions with the introduction of external forces acting over the whole system that allow it to reach a steady \emph{nonequilibrium} state. This can be done either by driving through the boundaries, e. g., shearing the system or vibrating its walls \cite{YHCMW02,YHCMW04} or alternatively by bulk driving, as in air-fluidized beds \cite{AD06}

The aim of the present Chapter is to study the transport properties of a moderately dense granular gas driven by a stochastic bath with friction in the frame of the Enskog kinetic theory. The homogeneous steady state described in Chapter \ref{Chapter3} is now perturbed by small spatial gradients in the hydrodynamical fields and the response of the system to these perturbations will give rise to nonzero contributions to the heat and momentum fluxes which are characterized by the transport coefficients. In order to obtain them, states that deviate from steady homogeneous states by small spatial gradientes are considered and, therefore, the Enskog kinetic equation \eqref{eqEB} is solved by means of the CE method \cite{CHAPMAN70} conveniently adapted to dissipative dynamics.

\section{Small spatial perturbations around the Homogeneous Steady State}
\label{sec2chap4}
%In order to obtain the transport properties of a driven granular gas, 
In the previous Chapter, the velocity distribution function of a driven granular gas in homogenous steady conditions was characterized by means of the coefficients of the Sonine expansion. In particular the first two non-zero coefficients $a_2$ and $a_3$ were calculated by using two different approximations. On the other hand, Figs. \ref{fig02chap3} and \ref{fig03chap3} show that, in general, $a_3$ is much smaller than $a_2$ and thus in further calculations of the present Chapter the coefficient $a_3$ will be neglected. In that case, the steady cooling rate $\zeta_\text{s}$ can be written up to the first order in the Sonine expansion in terms of $a_{2,\text{s}}$ as
\begin{equation}
\label{eqZetas} 
\zeta_\text{s}=\frac{2}{d}\frac{\pi^{\left( d-1\right) /2}}
{\Gamma \left( \frac{d}{2}\right)}(1-\alpha^2)\chi \left(1+\frac{3}{16}a_{2,\text{s}}\right)
n_\text{s}\sigma^{d-1}\sqrt{\frac{T_\text{s}}{m}},
\end{equation}
where the steady granular temperature $T_\text{s}$ obeys the equation
\begin{equation}
\label{eqTs02}
T_\text{s}=\frac{m^2\xi_\text{b}^2}{2\gamma_\text{b}} -\frac{2^{d-1}}{\sigma}\sqrt{\frac{m}{\pi}}\frac{\chi\phi}{\gamma_\text{b}}(1-\alpha^2)\left(1+\frac{3}{16}a_{2,\text{s}}\right)T_\text{s}^{3/2}.
\end{equation}
%and $\phi$ is the solid volume fraction defined by Eq.\ \eqref{eqphi}. 
Equation \eqref{eqTs02} gives the granular temperature $T_\text{s}$ in the non-equilibrium stationary state.

\begin{figure}[h]
\centering
  \includegraphics[width=0.75 \columnwidth,angle=0]{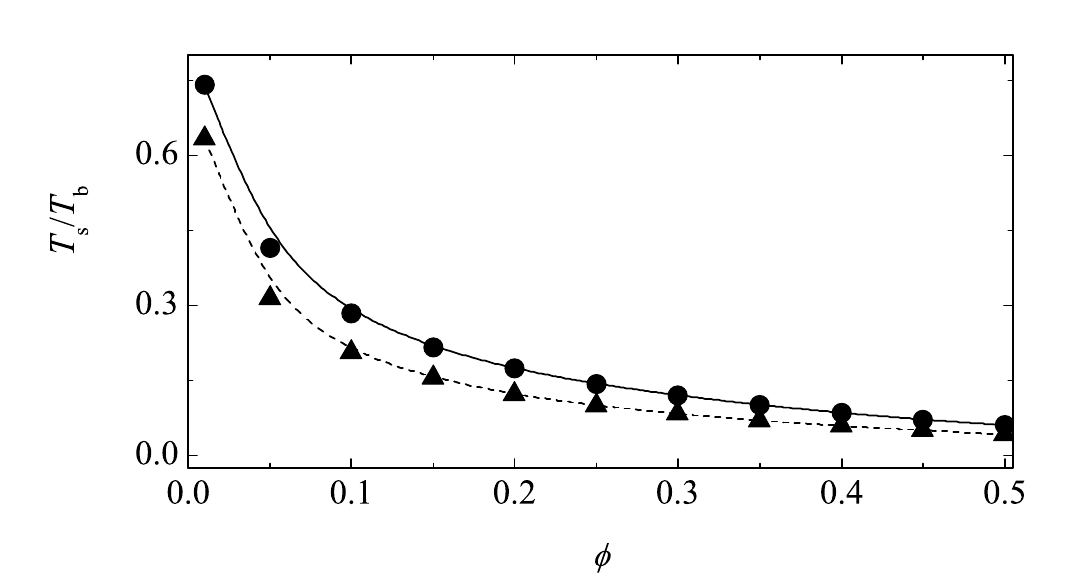}
  \caption{Plot of the reduced granular temperature $T_\text{s}/T_\text{b}$ versus the volume fraction $\phi$ for a two-dimensional ($d=2$) granular fluid and two different values of the coefficient of restitution: $\alpha=0.8$ (solid line) and $\alpha=0.6$ (dashed line). The symbols are the Monte Carlo simulation results (circles for $\alpha=0.8$ and triangles for $\alpha=0.6$).}
   \label{fig01chap4}
\end{figure}

To illustrate the dependence of $T_s$ on both $\alpha$ and $\phi$, Fig. \ref{fig01chap4} shows the (reduced) steady temperature $T_\text{s}/T_\text{b}$ versus the volume fraction $\phi$ for two different values of the coefficient of restitution $\alpha$. The theoretical results obtained from Eq.\ \eqref{eqTs02} for hard disks ($d=2$) are compared with those obtained by numerically solving the Enskog equation from the DSMC method \cite{BIRD94}. As in Ref.\ \cite{GSVP11-1}, the fixed parameters of the simulations are $m=1$, $\sigma=0.01$, $\gamma_\text{b}=1$, $\xi_\text{b}^2=2$, and $T_\text{b}=1$.

We observe an excellent agreement between theory and simulation in the complete range of values of $\phi$ considered. As expected, at a given value of the solid volume fraction, the steady granular temperature $T_\text{s}$ decreases as the gas becomes more inelastic.

Once the homogeneous steady state is characterized, our goal now is to solve the Enskog equation by means of the CE expansion. We consider here the zeroth and the first-order approximations.

\subsection{Zeroth-order approximation}

Inserting Eqs. \eqref{eqfexpand} and \eqref{eqJexpand} into Eq.\ \eqref{eqEB} and arranging terms of the same order in the parameter $\epsilon$, we can determine the different approximations to $f$ separately. To zeroth-order in the expansion, the distribution $f^{(0)}$ obeys the kinetic equation

\begin{equation}
  \label{eqf0}
  \partial_t^{(0)}f^{(0)}-\frac{\gamma_\text{b}}{m}\frac{\partial}{\partial \bt{v}}\cdot \bt{V}f^{(0)} -\frac{1}{2}\xi_\text{b}^2\frac{\partial^2}{\partial v^2}f^{(0)}=J_E^{(0)}[f^{(0)},f^{(0)}],
\end{equation}
where $J_E^{(0)}[f^{(0)},f^{(0)}]$ is given by Eq.\ \eqref{eqJE2} with the replacements of the steady state density $n_\text{s}$ and distribution function $f_\text{s}$ by their unsteady non-homogeneous versions $n(\bt{r},t)$ and $f^{(0)}(\bt{r},\bt{v},t)$, respectively.  In addition, as mentioned in Chapter \ref{Chapter2}, the term $\Delta\bt{U}=\bt{U}-\bt{U}_g$ is considered to be at least of first order in spatial gradients.

%Multiplying Eq.\ \eqref{eqf0} by $\{1, \bt{v}, v^2\}$ and integrating over velocity we obtain 
The conservation laws of the hydrodynamic fields at this order are given by
\begin{equation}
  \label{eqBal0}
  \partial_t^{(0)}n=0, \quad   \partial_t^{(0)}\bt{U}= \mathbf{0},
\end{equation}
\begin{equation}
  \label{eqBalT0}
  \partial_t^{(0)}T = - \frac{2T}{m}\gamma_\text{b}  +m\xi_\text{b}^2 - \zeta^{(0)}T,
\end{equation}
where $\zeta^{(0)}$ is determined by Eq.\ \eqref{eqZetas} to zeroth-order.
Thus, in the zeroth-order approximation $\zeta^{(0)}$ is given by Eq.\ \eqref{eqZetas} with $n_s\rightarrow n(\bt{r},t)$ and $T_s\rightarrow T(\bt{r},t)$.

%\vicente{Upon writing \eqref{eqBal0} we have taken $\Delta \mathbf{U}$ as a zeroth-order term in gradients.} 
The time derivative $\partial_t^{(0)}f^{(0)}$ can be rewritten more explicity in terms of the variation of the hydrodynamic fields as
\begin{eqnarray}
  \label{eqdtf0}
  \partial_t^{(0)}f^{(0)} &=& \frac{\partial f^{(0)}}{\partial n} \partial_t^{(0)}n + 
  \frac{\partial f^{(0)}}{\partial U_i} \partial_t^{(0)} U_i + 
  \frac{\partial f^{(0)}}{\partial T} \partial_t^{(0)} T \nonumber\\ 
  &=& -\left(\frac{2\gamma_\text{b}}{m} -\frac{m\xi_\text{b}^2}{T} + \zeta^{(0)}\right) T \frac{\partial f^{(0)}}{\partial T}.
%\vicente{+\frac{\gamma_\text{b}}{m} \Delta \mathbf{U}\cdot \frac{\partial f^{(0)}}{\partial \mathbf{V}}}
\end{eqnarray}
With this result, Eq.\ \eqref{eqf0} reduces to
\begin{equation}
  \label{eqf0bis}
  -\left(\frac{2\gamma_\text{b}}{m} -\frac{m\xi_\text{b}^2}{T} + \zeta^{(0)}\right) T \frac{\partial f^{(0)}}{\partial T}
  -\frac{\gamma_\text{b}}{m}\frac{\partial}{\partial \bt{v}}\cdot \bt{V}f^{(0)} - 
  \frac{1}{2}\xi_\text{b}^2\frac{\partial^2}{\partial v^2}f^{(0)} 
  = J_E^{(0)}[f^{(0)},f^{(0)}].
\end{equation}

%At this level of the derivation it is worthwhile noting that for given values of the thermostat parameters $\gamma_\text{b}$ and $\xi_\text{b}^2$ and the inelasticity $\alpha$ there exists a mapping between the density $n(\bt{r},t)$ and temperature $T(\bt{r},t)$ in such way that every density corresponds to one and only one temperature. 

At this level of the derivation, it is worthwhile noting that for given values of the thermostat parameters $\gamma_b$ and $\xi^2_b$ and the coefficient of restitution $\alpha$, the steady state condition \eqref{eqTs01} establishes a mapping between the density $n(\bt{r},t)$ and temperature $T(\bt{r},t)$ and hence, every density corresponds to one and only one temperature.
Since the density and temperature are given separately in the reference $local$ state $f^{(0)}$, the collisional cooling is only partially compensated for the energy injected by the thermostat forces and so, $\partial_t^{(0)}T \neq 0$. Consequently, the zeroth-order distribution function $f^{(0)}$ depends on time through its functional dependence on the temperature. %In previous works \cite{GM02, G11} for the case of stochastic thermostat ($\gamma_\text{b}=0$) where , steady-state conditions has been impose at any point of the system and so $\partial_t^{(0)} T = 0$S
On the other hand, for simplicity, one could impose the steady-state condition \eqref{eqTs01} at any point of the system and so, $\partial_t^{(0)} T = 0$. This was the choice used in previous theoretical works \cite{GM02, G11} in the case of the stochastic thermostat  ($\gamma_\text{b}=0$) where the relation $m\xi_\text{b}^2 = \zeta^{(0)} T$ was assumed to apply also in the non-homogeneous state. 

As we will see below, while the expressions of the shear and bulk viscosities are the same in both choices ($\partial_t^{(0)} T \neq 0$ and $\partial_t^{(0)} T = 0$), the transport coefficients of the heat flux are different. The former choice of thermostat ($\partial_t^{(0)} T \neq 0$) will be referred here to as choice $A$ while the latter ($\partial_t^{(0)} T = 0$) will be referred as to choice $B$. 
Although the choice $A$ has the advantage of a simpler implementation in computer simulations at the level of kinetic theory, the fact that $\partial_t^{(0)} T \neq 0$ gives rise to conceptual and practical difficulties not present in the previous analysis \cite{GM02, G11} carried out by using the choice $B$. The above difficulties are also present in a CE-like method proposed to analyze rheological properties around the steady shear flow state \cite{L06,G06}.

Although for granular gases the drag parameter $\gamma_\text{b}$ and the white noise $\xi_\text{b}^2$ can be considered in general as independent parameters, to make contact here with previous results obtained for granular fluids \cite{GSVP11-1,GSVP11-2}, we assume that both parameters are related by
\begin{equation}  
  \label{eqgamG}
  \gamma_\text{b}=\beta\frac{m^2\xi_\text{b}^2}{T_\text{b}},
\end{equation}
where $\beta$ is an \emph{arbitrary} constant. Thus, when $\beta=0$, $\gamma_\text{b}=0$ and our thermostat reduces to the stochastic thermostat \cite{GMT13} while the choice $\beta=1/2$ leads to the conventional Fokker-Plank model for ordinary gases \cite{H03}. According to Eq.\ \eqref{eqgamG}, the reduced parameter $\gamma^*\equiv \gamma_\text{b}\ell/mv_0$ can be expressed in terms of $\xi^*$ as
\begin{equation}
\label{3.6.1}
\gamma^*=\theta \xi^{*1/3}, \quad \theta\equiv \beta \left(\frac{m\xi_\text{b}^2}{n\sigma^{d-1}T_\text{b}\sqrt{2T_\text{b}/m}}\right)^{2/3}.
\end{equation}
Note that here the reduced model parameter $\xi^*$ is defined as in Eq.\ \eqref{eqxistar} with the replacement $T_s\rightarrow T(\bt{r},t)$. Upon writing Eq.\ \eqref{3.6.1}, use has been made of the identity $\beta T^*=\theta \xi^{*-2/3}$, where $T^*=T(t)/T_b$

In the \emph{unsteady} state, the zeroth-order distribution function obeys Eq. \eqref{eqEB03}. Dimensional analysis requires that $f^{(0)}$ is also given by the scaled form \eqref{eqfscaled} [once one uses the relation \eqref{eqgamG}], namely,
\begin{equation}
  \label{eqf0scaled}
  f^{(0)}(\bt{r},\bt{v},t)\rightarrow n(\bt{r},t)\,v_{0}(\bt{r},t)^{-d}\varphi(\bt{c},\theta,\xi^*),
\end{equation}
where now $\bt{c}=\bt{V}/v_0$. 
In this case, the thermal velocity $v_0$ and the reduced parameter $\xi^*$ are local quantities according to their definitions %\eqref{eqvter} and \eqref{eqxistar} respectively 
 with the change $T_\text{s}\rightarrow T(\bt{r}, t)$. The scaling \eqref{eqf0scaled} gives rise to the presence of new terms in the kinetic equations. As in the steady state, the temperature dependence of $f^{(0)}$ is not only through $v_0$ and $c$ but also through $\xi^*$. Thus,
\begin{equation}
  \label{eqTtime}
  T\frac{\partial f^{(0)}}{\partial T} = - \frac{1}{2}\frac{\partial}{\partial\bt{V}}\cdot \bt{V} f^{(0)} - \frac{3}{2}\xi^*\frac{\partial f^{(0)}}{\partial \xi^*},
\end{equation}

In reduced units, the equation for $\varphi$ is
%The Enskog equation can be written now in reduced units as:
\begin{equation}
  \label{eqBEphi}
  \frac{3}{2}\left[(2\beta T^* -1)\xi^* + \zeta^*_0\right]\xi^*\frac{\partial\varphi}{\partial \xi^*} 
  + \frac{1}{2}(\zeta^*_0-\xi^*)\frac{\partial}{\partial \bt{c}}\cdot \bt{c}\varphi 
  - \frac{1}{4}\xi^*\frac{\partial^2}{\partial c^2}\varphi 
  = J_E^*[\varphi,\varphi],
\end{equation}
where
\begin{equation}
  \zeta_0^*\equiv\frac{\zeta^{(0)}}{n\sigma^{d-1}\sqrt{2T/m}},
\end{equation}
and
\begin{equation}
  J_E^*\equiv\frac{\ell \; v_0^{d-1}}{n}J_E^{(0)}.
\end{equation}

As in the previous Chapter, the explicit form of the solution of Eq. \eqref{eqBEphi} is not known. However, it is possible to obtain certain information on the scaled distribution function $\varphi$ through its fourth cumulant $a_2(\alpha,\xi^*)$. % defined in Eq.\ \eqref{eqc4}. 
To obtain this cumulant, we multiply both sides of Eq.\ \eqref{eqBEphi} by $c^4$ and integrate over velocity. The result is
\begin{equation}
  \label{eqa2inhomo}
  - \frac{3d(d+2)}{8}\left[(2\beta T^*-1)\xi^* + \zeta_0^* \right]\xi^*\frac{\partial a_2}{\partial \xi^*} 
  + \frac{d(d+2)}{2}\left[ \zeta_0^*(1+a_2) - \xi^*a_2 \right] = \mu_4,
\end{equation}
where
\begin{equation}
  \mu_p=-\int d\bt{c}c^p J_E^*[\varphi,\varphi].
\end{equation}

In the steady state, Eq.\ \eqref{eqTs01} applies and the first term on the left hand side of Eq.\ \eqref{eqa2inhomo} vanishes. In this case, the solution to Eq.\ \eqref{eqa2inhomo} is given by Eq.\ \eqref{eqa2I}. In general, due to its complexity, Eq.\ \eqref{eqa2inhomo} must be solved numerically to get the dependence of $a_2$ on the thermostat reduced parameter $\xi^*$ (or on the reduced temperature $T^*$). However, in the vicinity of the steady state, it is possible to give an analytical expression for $\partial a_2 / \partial \xi^*$.  This derivative appears in the expressions of the heat flux transport coefficients and the first order contribution to the cooling rate $\zeta_U$. 
In order to determine $\partial a_2 / \partial \xi^*$ from Eq.\ \eqref{eqa2inhomo}, we first assume that $\varphi$ can be well described by the lowest Sonine approximation \eqref{eqSonine01}. Then, approximate forms for $\zeta_0^*=(2/d)\mu_2$ and $\mu_4$ are obtained when one uses the distribution \eqref{eqSonine01} and neglects nonlinear terms in $a_2$ . The results are:
\begin{equation}
  \mu_2\rightarrow \mu_2^{(0)} + \mu_2^{(1)} a_2; \qquad   \mu_4\rightarrow \mu_4^{(0)} + \mu_4^{(1)} a_2,
\end{equation}
where
\begin{equation}
  \mu_2^{(0)}=\frac{\pi^{(d-1)/2}}{\sqrt{2}\Gamma\left(\frac{d}{2}\right)}\chi(1-\alpha^2), \qquad \mu_2^{(1)}=\frac{3}{16}\mu_2^{(0)},
\end{equation}  
\begin{equation}
  \mu_4^{(0)}=\left(d+\frac{3}{2}+\alpha^2\right)\mu_2^{(0)}  ,
\end{equation}  
\begin{equation}
 \qquad \mu_4^{(1)}=\left[\frac{3}{32}(10d+39+10\alpha^2) + \frac{d-1}{1-\alpha}\right]\mu_2^{(0)}.
\end{equation}  

Substituting these terms in Eq.\ \eqref{eqa2inhomo} and retaining only linear terms in $a_2$, one obtains the relation
\begin{equation}
\label{a6}
  \frac{\partial a_2}{\partial \xi^*} = - \frac{\mu_4^{(0)}-(d+2)\mu_2^{(0)} - \left[ \frac{19}{16}(d+2)\mu_2^{(0)} - \mu_4^{(1)} - \frac{d(d+2)}{2}\xi^* \right]a_2 }{ \frac{3d(d+2)}{8}\xi^*\left[ (2\beta T^*-1)\xi^* +\frac{2}{d}(\mu_2^{(0)} + \mu_2^{(1)} a_2 )  \right]  }.
\end{equation}

However, some care must be taken in Eq.\ \eqref{a6} at the steady state, since the numerator and denominator of Eq.\ \eqref{a6} vanish. Thus, the corresponding expression for the derivative $\partial a_{2}/\partial \xi^*$ in the steady state becomes indeterminate. This difficulty can be solved by means of l'Hopital's rule. After some algebra, it is straightforward to see that the steady-state value of the derivative $\Delta \equiv (\partial a_2/\partial \xi^*)_\text{s}$ obeys the quadratic equation:
\begin{eqnarray}
\label{a7}
& &\frac{3}{4}(d+2)\mu_2^{(1)}\xi_\text{s}^* \Delta^2 + \left[\frac{d(d+2)}{8}\left(1+2\beta T_\text{s}^*\right)\xi_\text{s}^* - \frac{19}{16}(d+2)\mu_2^{(0)}+\mu_4^{(1)}\right] \Delta \nonumber \\
& & + \frac{d(d+2)}{2}a_{2,s} = 0,
\end{eqnarray}
where $T_\text{s}^*=T_\text{s}/T_\text{b}$. Since $a_{2,\text{s}}$ is in general very small, it is expected that the magnitude of $\Delta$ be also quite small. An analysis of the solutions to Eq.\ \eqref{a7} shows that in general one of its roots is much larger than $a_{2,s}$ while the other is of the order of $a_{2,s}$. We take the latter one as the physical root of the quadratic Eq.\ \eqref{a7}.

Since $\Delta$ is in general very small, one may neglect the term proportional to $ \left(\Delta\right)^2$ in Eq.\ \eqref{a7}. In this case, the derivative $\Delta$ can be explicitly written as
\begin{equation}
\label{a8}
\Delta=\frac{a_{2,\text{s}}}{\frac{19}{8d}\mu_2^{(0)}-\frac{1+2\beta T_\text{s}^*}{4}\xi_\text{s}^*-\frac{2}{d(d+2)}\mu_4^{(1)}}.
\end{equation}

As we will see later, we also need to know the derivative $(\partial a_2/\partial \theta)_\text{s}$ where $\theta$ is defined by Eq.\ \eqref{3.6.1}. It can be directly obtained from Eq.\ \eqref{a6} with the result
\begin{equation}
\label{a9}
\left(\frac{\partial a_2}{\partial \theta}\right)_\text{s}=\frac{\xi_\text{s}^{*4/3}\Delta}
{\frac{19}{12d}\mu_2^{(0)}-\frac{2}{3}\xi_\text{s}^*-\frac{4}{3d(d+2)}\mu_4^{(1)}-\frac{\mu_{2}^{(1)}}{d}\xi_\text{s}^*\Delta}.
\end{equation}

\begin{figure}
  \centering
  \includegraphics[width=0.75 \columnwidth,angle=0]{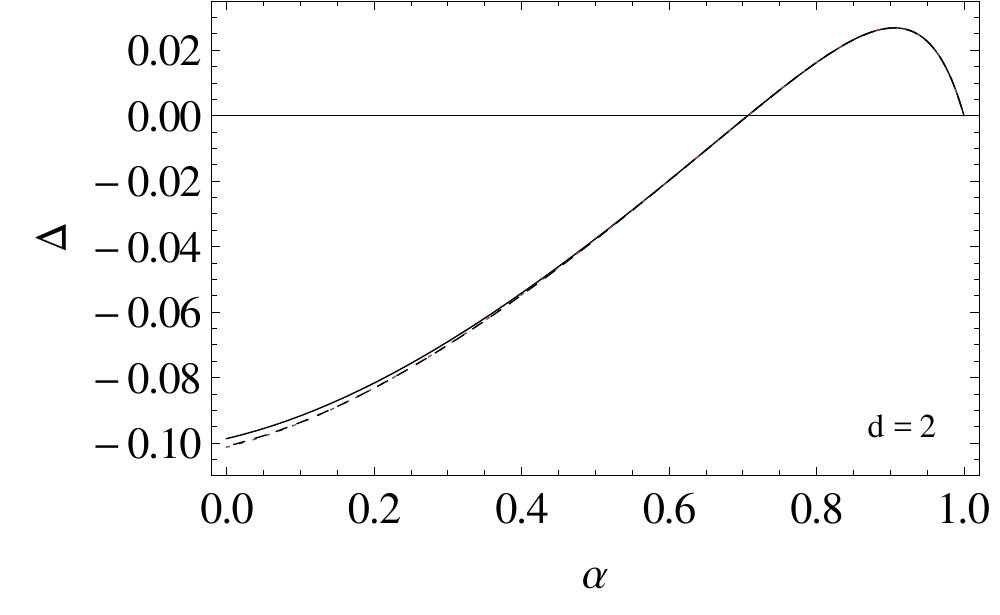}
  \includegraphics[width=0.75 \columnwidth,angle=0]{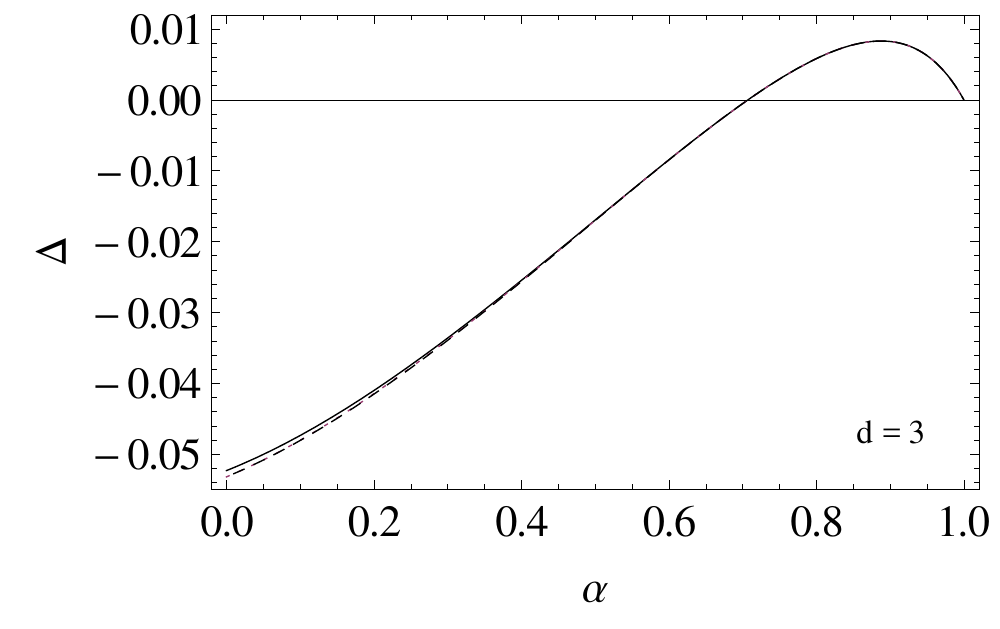}
  \caption{Plot of the derivative versus the coefficient of restitution $\alpha$ for the stochastic thermostat $\xi_\text{s}^*=\zeta_\text{s}^*$  for disks ($d=2$) and spheres ($d=3$). The solid line is the result given by Eq.\ (4.22) while the dashed line is the result obtained by Garc\'ia de Soria \emph{et al} \cite{GMT12}.  \label{fig02chap4}}
\end{figure}

Fig.\ \ref{fig02chap4} shows the dependence of $\Delta$ on the coefficient of restitution $\alpha$ when the gas is heated by the stochastic thermostat ($\beta=0$ and $\xi_\text{s}^*=\zeta_\text{s}^*$). 
We plot here the physical solution root to Eq.\ \eqref{a7} and its simpler form \eqref{a8}. Moreover, for the sake of comparison, the result obtained by Garc\'ia de Soria \emph{et al.} \cite{GMT12} by using a different method is also shown.

It is quite apparent that the results obtained here yield identical results for the derivative $\Delta$ with those obtained in Ref.\ \cite{GMT12}. In addition, given that the solution to Eq.\ \eqref{a7} and the form \eqref{a8} are indistinguishible, henceforth we will take the last form \eqref{a8}.

\subsection{First-order approximation}
The analysis to first order in the spatial gradients is similar to the one worked out in the undriven case \cite{GD99,L05}. Some technical details on the determination of the transport coefficients and the cooling rate are provided in Appendices \ref{AppendixB} and \ref{AppendixC}.
The form of the first-order velocity distribution function $f^{(1)}$ is given by
\begin{eqnarray}
\label{eq4.1}
f^{(1)} &=& \boldsymbol{\mathcal{A}}\left(\mathbf{V}\right)\cdot  \nabla \ln T+\boldsymbol{\mathcal{B}}\left( \mathbf{V}\right) \cdot \nabla \ln n \nonumber\\
&&+ \mathcal{C}_{ij}\left( \mathbf{V} \right)\frac{1}{2}\left( \partial _{i}U_{j}+\partial _{j}U_{i}-\frac{2}{d}\delta _{ij}\nabla \cdot \mathbf{U} \right)
+\mathcal{D}\left( \mathbf{V} \right) \nabla \cdot \mathbf{U},
  \end{eqnarray}
where the quantities $\boldsymbol{\mathcal{A}}\left(\mathbf{V}\right)$, $\boldsymbol{\mathcal{B}}\left(
\mathbf{V}\right)$, $\mathcal{C}_{ij}\left( \mathbf{V} \right)$ and $\mathcal{D}\left( \mathbf{V} \right)$ are the solutions of the linear integral Eqs.\ \eqref{apB15}--\eqref{apB18}, respectively. However, the evaluation of the transport coefficients from the above integral equations requires to know the complete time dependence of the first order contributions to the pressure tensor and the heat flux vector. This is quite an intricate problem. On the other hand, some simplifications occur if attention is restricted to linear deviations from the steady state. In particular, since the kinetic and collisional contributions to the heat and momentum fluxes are already of first order in the deviations from the steady state, one only needs to know the transport coefficients to zeroth order in the deviations. This means that we can evaluate the transport coefficients under steady state conditions, namely, when the condition \eqref{eqTs01} applies.
%In order to obtain the transport coefficients from the integral Eqs.\ \eqref{apB15}-\eqref{apB18} it is necessary to know the complete time dependence of the first-order contributions to the pressure tensor and the heat flux vector but this becomes a fairly complex problem. Nevertheless it is possible to introduce some simplifications if attention is only restricted to linear deviations from the steady state described in Chapter \ref{Chapter2}. In particular, since the kinetic and collisional contributions to the heat and momentum fluxes are already of first order in the deviations from the steady state, one only needs to know the transport coefficients to zeroth order in the deviations. This means we can evaluate the transport coefficients in the steady-state conditions when the term $\frac{2\gamma_\text{b}}{m}-\frac{m\xi_\text{b}^2}{T}+\zeta^{(0)}$ vanishes. The integral equations for the transport coefficients thus become simple coupled algebraic equations that can be easily solved. In that case, the quantities $\bf{A}(\bf{V})$, $\bf{B}(\bf{V})$, $C_{ij}(\bf{V})$ and $D(\bf{V})$ are evaluated in the steady state and all the  transport coeffcients are given in terms of the steady granular temperature $T_\text{s}$.
%The form of the collisional contributions to the momentum and heat fluxes are the same as those previously obtained in the undriven case \cite{GD99, L05} with the difference that now the fourth cumulant $a_2$ depends on the thermostat parameter $\xi_\text{s}^*$. Here we will focus on the kinetic contributios to the transport coefficients and the cooling rate.

In this case, Eqs.\ \eqref{apB15}--\eqref{apB18} become
\begin{equation}
\label{eq4.2}
\left\{
-\left[\frac{m}{T}\xi_\text{b}^2\left(1-\frac{3}{2}\frac{\partial \zeta_0^*}{\partial \xi^*}\right)+\frac{1}{2}\zeta^{(0)}\right]  
- \frac{\gamma_\text{b}}{m}\frac{\partial}{\partial {\bf v}}\cdot{\bf V}  
- \frac{1}{2}\xi_\text{b}^2 \frac{\partial^2}{\partial v^2}  
+ {\cal L} \right\} \boldsymbol{\mathcal{A}}={\bf A},
\end{equation}
%\begin{eqnarray}
%\label{eq4.2}
%&-&\left[
%\frac{m}{T}\xi_\text{b}^2\left(1-\frac{3}{2}\frac{\partial \zeta_0^*}{\partial \xi^*}\right)+\frac{1}{2}%\zeta^{(0)}\right]
%\boldsymbol{\mathcal{A}}-\frac{\gamma_\text{b}}{m}\frac{\partial}{\partial {\bf v}}\cdot
%{\bf V}\boldsymbol{\mathcal{A}}\nonumber\\
%&-&\frac{1}{2}\xi_\text{b}^2
%\frac{\partial^2}{\partial v^2}\boldsymbol{\mathcal{A}}+{\cal L}\boldsymbol{\mathcal{A}}={\bf A},
%\end{eqnarray}
\begin{eqnarray}
\label{eq4.3}
&&
\left[
-\frac{\gamma_\text{b}}{m}\frac{\partial}{\partial {\bf v}}\cdot {\bf V} 
-\frac{1}{2}\xi_\text{b}^2 \frac{\partial^2}{\partial v^2} 
+ {\cal L} \right] \boldsymbol{\mathcal{B}} = \nonumber\\
&=&{\bf B}+
\left\{\zeta ^{(0)} g(\phi)
+\left[\phi \frac{\partial \chi}{\partial \phi}\frac{\partial}{\partial \chi}\left(\frac{\zeta^{(0)}}{\chi}\right)-\xi^*\frac{\partial \zeta^{(0)}}{\partial \xi^*}
-\frac{2}{3}\theta\frac{\partial \zeta^{(0)}}{\partial \theta}\right]\right\}
\boldsymbol{\mathcal{A}},
\end{eqnarray}
%\begin{eqnarray}
%\label{eq4.3}
%&-&
%\frac{\gamma_\text{b}}{m}\frac{\partial}{\partial {\bf v}}\cdot {\bf V}\boldsymbol{\mathcal{B}}
%-\frac{1}{2}\xi_\text{b}^2
%\frac{\partial^2}{\partial v^2}\boldsymbol{\mathcal{B}}+{\cal L}\boldsymbol{\mathcal{B}}={\bf B}
%\nonumber\\
%&+&
%\zeta ^{(0)}\left(1+\phi\frac{\partial}{\partial \phi}\ln \chi \right)\boldsymbol{\mathcal{A}}\nonumber\\
%&+&\left[\phi \frac{\partial \chi}{\partial \phi}\frac{\partial}{\partial \chi}\left(\frac{\zeta^{(0)}}%{\chi}\right)-\xi^*\frac{\partial \zeta^{(0)}}{\partial \xi^*}
%-\frac{2}{3}\theta\frac{\partial \zeta^{(0)}}{\partial \theta}\right]
%\boldsymbol{\mathcal{A}},
%\end{eqnarray}
\begin{equation}
  \label{eq4.4}
\left[-\frac{\gamma_\text{b}}{m}\frac{\partial}{\partial {\bf v}}\cdot {\bf V}
-\frac{1}{2}\xi_\text{b}^2 \frac{\partial^2}{\partial v^2} + 
{\cal L}\right] \mathcal{C}_{ij} = C_{ij},
\end{equation}
\begin{equation}
\left[-\frac{\gamma_\text{b}}{m}\frac{\partial}{\partial {\bf v}}\cdot {\bf V}
-\frac{1}{2}\xi_\text{b}^2
\frac{\partial^2}{\partial v^2}\mathcal{D}+{\cal L}\right]\mathcal{D}=D,  \label{eq4.5}
\end{equation}
%\end{widetext}
where 
\begin{equation}
\label{eqgphi}
g(\phi)=\left(1+\phi\frac{\partial}{\partial \phi}\ln \chi\right).
\end{equation}

Moreover, in Eqs.\ \eqref{eq4.2}--\eqref{eq4.5}, it is understood that the quantities $\mathbf{A}(\mathbf{V})$, $\mathbf{B}(\mathbf{V})$, $C_{ij}(\mathbf{V})$, and $D(\mathbf{V})$ [defined in Appendix \ref{AppendixB} by Eqs.\ \eqref{apB5}--\eqref{apB8}, respectively] and the derivatives $\partial \zeta^{(0)} / \partial \xi^*$ and $\partial \zeta^{(0)} /\partial \theta$ are evaluated in the steady state. Consequently, \emph{all} the transport coefficients are given in terms of the steady granular temperature $T_\text{s}$.

The forms of the collisional contributions to the momentum and heat fluxes are exactly the same as those previously obtained in the undriven case \cite{GD99,L05} except that $a_{2,\text{s}}$ depends on $\xi_\text{s}^*$ [see Eq.\ \eqref{eqa2I}]. Thus, we will focus here our attention in the evaluation of the kinetic contributions to the transport coefficients and the cooling rate. Technical details on this calculation are given in Appendix \ref{AppendixC}.

\section{Transport coefficients}
\label{sec3chap4}

%In order to close the equations for the hydrodynamic fields $n$, $\bf{U}$ and $T$, the pressure tensor and the heat flux must be expressed in terms of these fields and their gradients by the constituve relations. Here we restrict only to the linear-order terms with respect to the gradients (\emph{NS hydrodynamics}). In such approximation the pressure tensor and heat flux may be written as:

To first-order in the spatial gradients (NS hydrodynamics), the pressure tensor and the heat flux  are given by
\begin{equation}
\label{eq4.6}
P_{ij}^{(1)}=-\eta\left( \partial_{i}U_{j}+\partial _{j}U_{i}-\frac{2}{d}\delta _{ij}\nabla \cdot
\mathbf{U} \right) - \lambda \delta_{ij} \nabla \cdot \mathbf{U},
\end{equation}
%and
\begin{equation}
\label{eq4.11}
{\bf q}^{(1)}=-\kappa \nabla T-\mu \nabla n,
\end{equation}
where $\eta$, $\lambda$ and $\kappa$ are called respectively \emph{shear viscosity}, \emph{bulk viscosity} and \emph{thermal conductivity}. As for ordinary gases, the shear viscosity characterizes  the flux of momentum due to the gradient of the flow velocity whereas the thermal conductivity characterizes the heat flux due to the gradient of temperature. Here, a new coefficient $\mu$ (the diffusive heat or \emph{Dufour-like} coefficient) not present in ordinary gases arises which relates the heat flux with the density gradient \cite{BDKS98, BC01}.% and it is due to the increased collision frequency in regions of high density \cite{BDKS98, BC01}
%While the shear viscosity has kinetic and collisional contributions, the bulk viscosity has only a collisional contribution. The bulk viscosity $\lambda$ is given by

\subsection{Viscosity}
%To first order in the spatial gradients, the pressure tensor is given by where $\eta$ is the shear viscosity and $\lambda$ is the bulk viscosity. 
Here, we provide the final expressions for the shear and bulk viscosities. Technical details of the calculations are given in Appendix \ref{AppendixC}. While the shear viscosity has kinetic and collisional contributions, the bulk viscosity has only a collisional contribution. The bulk viscosity $\lambda$ is given by

\begin{equation}
\label{eq4.7}
\lambda=\frac{2^{2d+1}}{\pi(d+2)}\phi^2 \chi (1+\alpha)\left(1-\frac{a_{2,\text{s}}}{16} \right)\eta_0,
\end{equation}
where
\begin{equation}
\label{eq4.8}
\eta_0=\frac{d+2}{8}\frac{\Gamma \left( \frac{d}{2}\right)}{\pi ^{\left( d-1\right) /2}}\sigma^{1-d}\sqrt{mT_\text{s}}
\end{equation}
is the low density value of the shear viscosity in the elastic limit. The shear viscosity $\eta$ can be written as
\begin{eqnarray}
\label{eq4.9}
\eta&=&\frac{\eta_0\nu_0}{\nu_\eta+\frac{2\beta m}{T_\text{b}}\xi_\text{b}^2}
\left[1-\frac{2^{d-2}}{d+2}(1+\alpha)
(1-3 \alpha)\phi \chi \right]
\left[1+\frac{2^{d-1}}{d+2}(1+\alpha)
\phi \chi \right] \nonumber\\
&&+\frac{d}{d+2}\lambda,
\end{eqnarray}
where $\nu_0=n_\text{s}T_\text{s}/\eta_0$ and the collision frequency $\nu_\eta$ is \cite{GSM07}
\begin{equation}
\label{eq4.10}
\nu_\eta=\frac{3\nu_0}{4d}\chi \left(1-\alpha+\frac{2}{3}d\right)(1+\alpha) \left(1+\frac{7}{16}a_{2,\text{s}}\right).
\end{equation}

\subsection{Thermal conductivity}

The thermal conductivity $\kappa$ is given by
\begin{equation}
\label{eq4.12}
\kappa = \kappa_k\left(1+3\frac{2^{d-2}}{d+2}\phi \chi (1+\alpha)\right) + 
\kappa_0\frac{2^{2d+1}(d-1)}{(d+2)^2\pi}\phi^2 \chi (1+\alpha)\left(1+\frac{7}{16} a_{2,\text{s}} \right),
\end{equation}
where
\begin{equation}
\label{eq4.13}
\kappa_0=\frac{d(d+2)}{2(d-1)}\frac{\eta_0}{m}
\end{equation}
is the thermal conductivity coefficient of an elastic dilute gas. The expression of the kinetic part $\kappa_k$ appearing
in Eq.\ \eqref{eq4.12} is
%\begin{widetext}
\begin{eqnarray}
\label{eq4.14}
\kappa_k&=&\kappa_0\nu_0\frac{d-1}{d}\left[\nu_\kappa+\frac{1}{2}\frac{m\xi_\text{b}^2}{T_\text{s}}\left(1+3 \zeta_\text{M}
\left(\frac{\partial a_2}{\partial \xi^*}\right)_\text{s}\right)-2\zeta_\text{s}^{(0)}
\right]^{-1}
\nonumber\\
& &\left\{1+2a_{2,s}-\frac{3}{2}\xi_\text{s}^*\left(\frac{\partial a_2}{\partial \xi^*}\right)_\text{s}+3\frac{2^{d-3}}{d+2}\phi \chi(1+\alpha)^2 \right.\nonumber\\
& & \left.\left[2\alpha-1+a_{2,\text{s}}(1+\alpha) - \frac{3}{8}
(1+\alpha)\xi_\text{s}^*\left(\frac{\partial a_2}{\partial \xi^*}\right)_\text{s}\right]\right\}.\nonumber\\
\end{eqnarray}
In Eq.\ \eqref{eq4.14}, $\zeta_\text{s}^{(0)}$ is given by Eq.\ \eqref{eqZetas},
\begin{equation}
\label{eq4.14.1}
\zeta_\text{M}=\frac{3\sqrt{2}}{16d}\frac{\pi^{\left( d-1\right) /2}}
{\Gamma \left( \frac{d}{2}\right)}(1-\alpha^2)\chi,
\end{equation}
and the value of the derivative $(\partial a_2/\partial \xi^*)_\text{s}$ in the steady-state is given by Eq.\ \eqref{a8}. Moreover, the collision frequency $\nu_\kappa$ is \cite{GSM07}
\begin{equation}
\label{eq4.15}
\nu_\kappa=\nu_0\frac{1+\alpha}{d}\chi\left[\frac{d-1}{2}+\frac{3}{16}(d+8)(1-\alpha)+\frac{296+217d-3(160+11d)\alpha}{256}a_{2,\text{s}}\right].
\end{equation}
%\end{widetext}

The coefficient $\mu$ is
\begin{equation}
\label{eq4.16}
\mu=\mu_k\left[1+3\frac{2^{d-2}}{d+2}\phi \chi (1+\alpha)\right],
\end{equation}

where its kinetic contribution $\mu_k$ is
\begin{eqnarray}
\label{eq4.17}
\mu_k &=& \frac{\kappa_0\nu_0T_\text{s}}{n_\text{s}}\left[\nu_\kappa-\frac{3}{2}\left(\zeta_\text{s}^{(0)}
-\frac{m\xi_\text{b}^2}{T_\text{s}}\right)\right]^{-1}
\nonumber\\
& & \left\{\frac{\kappa_k}{\kappa_0\nu_0} \left[  \zeta_\text{s}^{(0)}g(\phi) + \frac{\zeta_\text{M}v_0}{\ell}\left(\frac{\phi}{\chi}\frac{\partial \chi}{\partial \phi}
\frac{\partial a_2}{\partial \chi}-\xi_\text{s}^*\frac{\partial a_2}{\partial \xi^*}
-\frac{2}{3}\theta\frac{\partial a_2}{\partial \theta}\right)\right]\right.
\nonumber\\
& & \left.+\frac{d-1}{d}\left(a_{2,\text{s}}+\phi \frac{\partial \chi}{\partial \phi}
\frac{\partial a_2}{\partial \chi}-\xi_\text{s}^*\frac{\partial a_2}{\partial \xi^*}
-\frac{2}{3}\theta\frac{\partial a_2}{\partial \theta}\right)\right.
\nonumber\\
& & +3\frac{2^{d-2}(d-1)}{d(d+2)}\phi \chi (1+\alpha)\left(1+\frac{1}{2}\phi  \frac{\partial}{\partial\phi}    \ln \chi\right)
\nonumber\\
& & \times\left[\alpha(\alpha-1)+\frac{a_{2,\text{s}}}{6}(10+2d-3\alpha+3\alpha^2)\right]
\nonumber\\
& &\left. + 3 \frac{2^{d-4}(d-1)}{d(d+2)}\chi \phi (1+\alpha)^3 \left(  \phi \frac{\partial \chi}{\partial \phi} \frac{\partial a_2}{\partial \chi}-\xi^*\frac{\partial a_2}{\partial \xi^*} -\frac{2}{3}\theta\frac{\partial a_2}{\partial \theta}\right)\right\}.
\end{eqnarray}

\subsection{Cooling rate}
The cooling rate $\zeta$ is given by
\begin{equation}
\label{eq4.18} \zeta=\zeta_\text{s}^{(0)}+\zeta_U \nabla \cdot {\bf U}.
\end{equation}
At first order in spatial gradients, the proportionality constant $\zeta_U$ is a new transport coefficient for granular fluids \cite{GD99,L05}. For a driven gas, $\zeta_U$ can be written as
\begin{equation}
\label{eq4.19} \zeta_U=\zeta_{10}+\zeta_{11},
\end{equation}
where
\begin{equation}
\label{eq4.20} \zeta_{10}= -3\frac{2^{d-2}}{d}\chi \phi (1-\alpha^2),
\end{equation}
\begin{eqnarray}
\label{eq4.21}
\zeta_{11}&=&\frac{9(d+2)\;2^{d-8}}{d^2}\chi(1-\alpha^2)
\left(\nu_\gamma+\frac{2m\xi_\text{b}^2}{T}-2\zeta_\text{s}^{(0)}\right)^{-1}\nonumber\\
&&\left[\frac{\omega \phi \chi}{2(d+2)}-2^{2-d}\frac{d+3}{3}\xi_\text{s}^*\left(\frac{\partial a_2}{\partial \xi^*}\right)_\text{s}\nu_0\right.\nonumber\\
& & \left.
-(1+\alpha)\left(\frac{1}{3}-\alpha\right)\left(2a_{2,\text{s}}-\frac{3}{2}
\xi_\text{s}^*\left(\frac{\partial a_2}{\partial \xi^*}\right)_\text{s}\right)\phi\chi \nu_0\right]
,\nonumber\\
\end{eqnarray}
%\begin{widetext}
and the collision frequencies $\omega$ and $\nu_\gamma$ are
\begin{equation}
\label{eq4.22}
\omega=(1+\alpha)\nu_0\left\{
(1-\alpha^2)(5\alpha-1)\frac{a_{2,\text{s}}}{6}\left[15\alpha^3-3\alpha^2+3(4d+15)\alpha-
(20d+1)\right]\right\},
\end{equation}
\begin{equation}
\label{eq4.23}
\nu_\gamma = -\frac{1+\alpha}{192}\chi \nu_0\left[30\alpha^3-30\alpha^2+(105+24 d) \alpha-56d-73\right].
\end{equation}
%\end{widetext}

In Eq.\ \eqref{eq4.21}, the contributions proportional to the derivatives $\partial a_2/\partial\chi$ and $\partial a_2/\partial\theta$ have been neglected for the sake of simplicity.

Note that the first-order contribution $\zeta_U$ to the cooling rate vanishes for elastic gases ($\alpha=1$, arbitrary solid volume fraction $\phi$). However, for dilute inelastic gases ($\phi=0$, arbitrary values of the coefficient of restitution $\alpha$), at variance with the undriven case \cite{BDKS98} there is here a nonzero contribution to $\zeta_{U}$ proportional to $(\partial a_{2}/\partial \xi^*)_\text{s}$ [see Eq.\ \eqref{eq4.21}]. This result is consistent with those obtained \cite{GMT13} from the Boltzmann equation.

The expressions for the NS transport coefficients obtained by using the choice B [i.e., when the condition \eqref{eqBalT0} holds locally and so, $\partial_t^{(0)}T=0$] are displayed in Appendix \ref{AppendixD}. While the expressions of $\eta$ and $\lambda$ are also given by Eqs.\ \eqref{eq4.7}--\eqref{eq4.9}, the forms of $\kappa$ and $\mu$ are different to those derived from the choice A.

\subsection{Special limits}

It is quite apparent that the expressions of the transport coefficients are rather complicated, given the different parameters (inelasticity, density and the model parameter $\xi_\text{b}^2$) involved in them. Thus, in order to show more clearly the dependence of each parameter on transport, it is instructive to consider some simple cases.

\subsubsection*{Elastic limit.}
%First, in the case $\xi_\text{b}^2=0$, the results derived in this Section agree with those obtained in the \emph{undriven} case \cite{GD99,L05}.
In the elastic limit ($\alpha=1$), $T_\text{s}=m^2\xi_\text{b}^2/2\gamma_\text{b}$, $\zeta_\text{s}^{(0)}=a_{2,\text{s}}=0$, $\nu_\eta=\chi \nu_0$, and $\nu_\kappa=(1-d^{-1})\chi \nu_0$. In this case, $\mu=\zeta_U=0$ and the coefficients $\lambda$, $\eta$ and $\kappa$ become, respectively,
\begin{equation}
\label{eq4.23.1}
\lambda=\frac{2^{2(d+1)}}{\pi(d+2)}\phi^2 \chi \eta_0,
\end{equation}
\begin{equation}
\label{eq4.23.2}
\eta=\frac{\eta_0}{\chi+\frac{2\beta m}{T_\text{b}\nu_0}\xi_\text{b}^2}
\left(1+\frac{2^{d}}{d+2}\phi \chi \right)^2+\frac{d}{d+2}\lambda,
\end{equation}
\begin{equation}
\label{eq4.23.3}
\kappa=\kappa_0\frac{\left(1+3\frac{2^{d-1}}{d+2}\phi \chi \right)^2}{\chi+\frac{d}{d-1}\frac{\gamma_\text{b}}{m\nu_0}}
+\frac{2^{2(d+1)}(d-1)}{(d+2)^2\pi}
\phi^2 \chi\kappa_0.
\end{equation}

Note that the expressions \eqref{eq4.23.2} and \eqref{eq4.23.3} for $\eta$ and $\kappa$ differ from their corresponding elastic counterparts for undriven gases.

\subsubsection*{Low-density regime.}

We consider now a low-density granular gas ($\phi=0$). In this limit case, $\lambda=0$ while $\eta$, $\kappa$ and $\mu$ are given, respectively, by
\begin{equation}
\label{eq4.9.dilute}
\eta=\frac{\eta_0\nu_0}{\nu_\eta+\frac{2\beta m}{T_\text{b}}\xi_\text{b}^2},
\end{equation}

\begin{eqnarray}
\label{eq4.14.dilute}
\kappa &=& \kappa_0\nu_0\frac{d-1}{d}\left[\nu_\kappa+\frac{1}{2}\frac{m\xi_\text{b}^2}{T_\text{s}}\left(1 + 3 \zeta_\text{M} \left(\frac{\partial a_2}{\partial \xi^*}\right)_\text{s}\right)-2\zeta_\text{s}^{(0)}
\right]^{-1} \nonumber\\
&& \left[1+2a_{2,s}-\frac{3}{2}\xi_\text{s}^*\left(\frac{\partial a_2}{\partial \xi^*}\right)_\text{s}
\right],
\end{eqnarray}

\begin{eqnarray}
\label{eq4.17.dilute}
\mu&=&\frac{\kappa_0\nu_0T_\text{s}}{n_\text{s}}\left[\nu_\kappa-\frac{3}{2}\left(\zeta_\text{s}^{(0)}
-\frac{m\xi_\text{b}^2}{T_\text{s}}\right)\right]^{-1} 
\left\{\frac{\kappa}{\kappa_0\nu_0}\left[\zeta_\text{s}^{(0)} -  
\frac{\zeta_\text{M}v_0}{\ell} \left( \xi_\text{s}^*\frac{\partial a_2}{\partial \xi^*}+\frac{2}{3}\theta\frac{\partial a_2}{\partial \theta}\right)\right]\right.\nonumber\\
&&
\left.+\frac{d-1}{d}\left( a_{2,\text{s}} -\xi_\text{s}^*\frac{\partial a_2}{\partial \xi^*}
-\frac{2}{3}\theta\frac{\partial a_2}{\partial \theta}\right)\right\},
\end{eqnarray}
where $\nu_\eta$ and $\nu_\kappa$ are defined by Eqs.\ \eqref{eq4.10} and \eqref{eq4.15}, respectively, with $\chi=1$.  The expressions \eqref{eq4.9.dilute} and \eqref{eq4.14.dilute} agree with recent results \cite{GMT12} derived from the linearized Boltzmann equation for a granular gas heated by the stochastic thermostat ($\beta=0$). In addition, as mentioned before, when $\beta=\frac{1}{2}$ in Eq.\ \eqref{eqgamG}, our model reduces to the Fokker-Planck model studied previously by Hayakawa \cite{H03} for dilute gases. In that paper, Hayakawa determines the transport coefficients $\eta$, $\kappa$, and $\mu$ by neglecting the dependence of the fourth cumulant $a_2$ on the (reduced) model parameters $\gamma^*$ and $\xi^*$. In particular,  in the steady state, Eqs.\ \eqref{eq4.9.dilute}--\eqref{eq4.17.dilute} agree with the results obtained in Ref.\ \cite{H03} when $(\partial a_2/\partial \xi^*)_\text{s}=0$. All the above limit situations confirm the self-consistency of the results derived here for a dense granular fluid.

\section{Comparison with computer simulations}
\label{sec4chap4}
The expressions obtained in Sec.\ \ref{sec3chap4} for the transport coefficients and the cooling rate depend on the (steady) granular temperature $T_\text{s}$, the coefficient of restitution $\alpha$, the solid volume fraction $\phi$ along with the parameter $\xi_\text{b}^2$ characterizing the external energy source. In this Section we will compare our theoretical predictions for the thermostats A and B with recent Langevin dynamics simulations carried out by Gradenigo \emph{et al.} \cite{GSVP11-1} for hard disks ($d=2$). In these simulations, the fluid is also driven by a stochastic bath with friction and the two external parameters $\gamma_\text{b}$ and $\xi_\text{b}^2$ are related by Eq.\ \eqref{eqgamG} with $\beta=\frac{1}{2}$. In the steady state, they measured the static and dynamic structure factors for shear and longitudinal modes for several values of the coefficient of restitution $\alpha$ and volume fraction $\phi$. The corresponding best fit of the simulation results of the above structure factors allows them to identify the kinematic viscosity $\nu=\eta/\rho$, the longitudinal viscosity
\begin{equation}
\label{eq4.1bis}
\nu_l=\frac{1}{\rho}\left(2\frac{d-1}{d}\eta+\lambda\right),
\end{equation}
and the thermal diffusivity
\begin{equation}
\label{eq4.2bis}
D_\text{T}=\frac{2}{dn}\kappa.
\end{equation}

\begin{figure}[h]
  \centering
  \includegraphics[width=0.75 \columnwidth,angle=0]{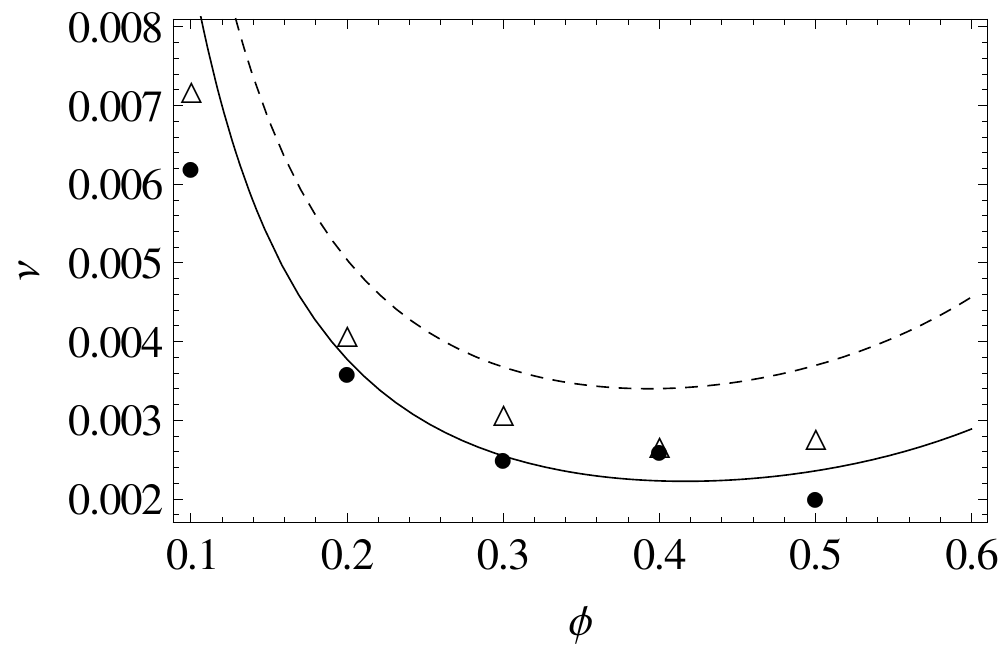}
  \caption{Plot of the kinematic viscosity $\nu=\eta/\rho$ as a function of the volume fraction $\phi$ for $\alpha=0.6$. The solid line is the theoretical prediction given by Eq. (4.33) while the dashed line is the theoretical result obtained by assuming the elastic form of the shear viscosity $\eta$. Symbols are the simulation results obtained by Gradenigo \emph{et al.} \cite{GSVP11-1} from the static (circles) and dynamical (triangle) correlations of transversal shear modes.}
   \label{fig03chap4}
\end{figure}

\begin{figure}[h]
  \centering
  \includegraphics[width=0.75 \columnwidth,angle=0]{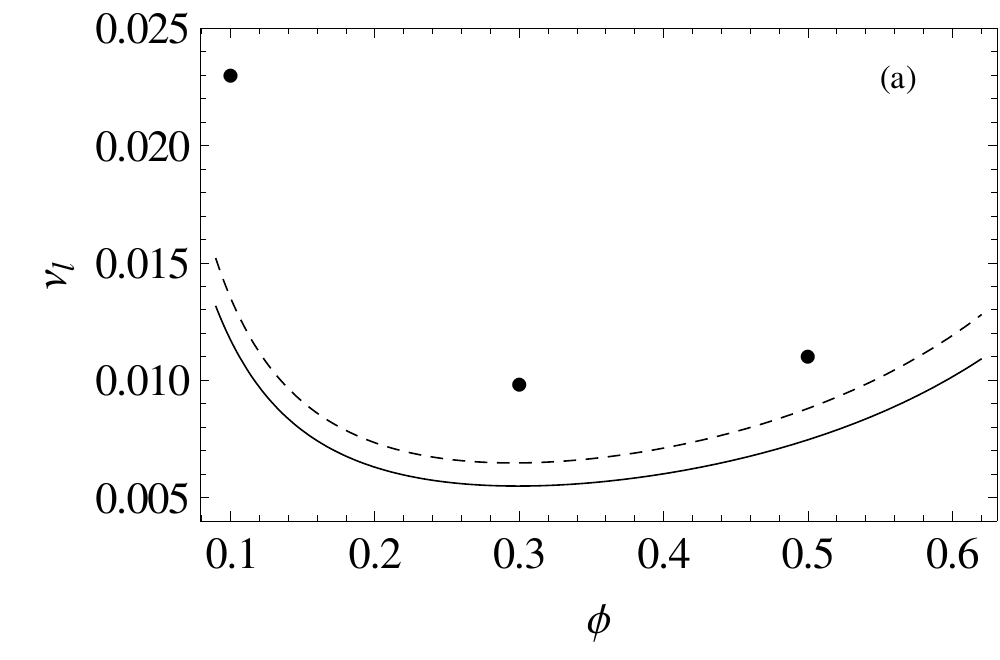}
  \includegraphics[width=0.75 \columnwidth,angle=0]{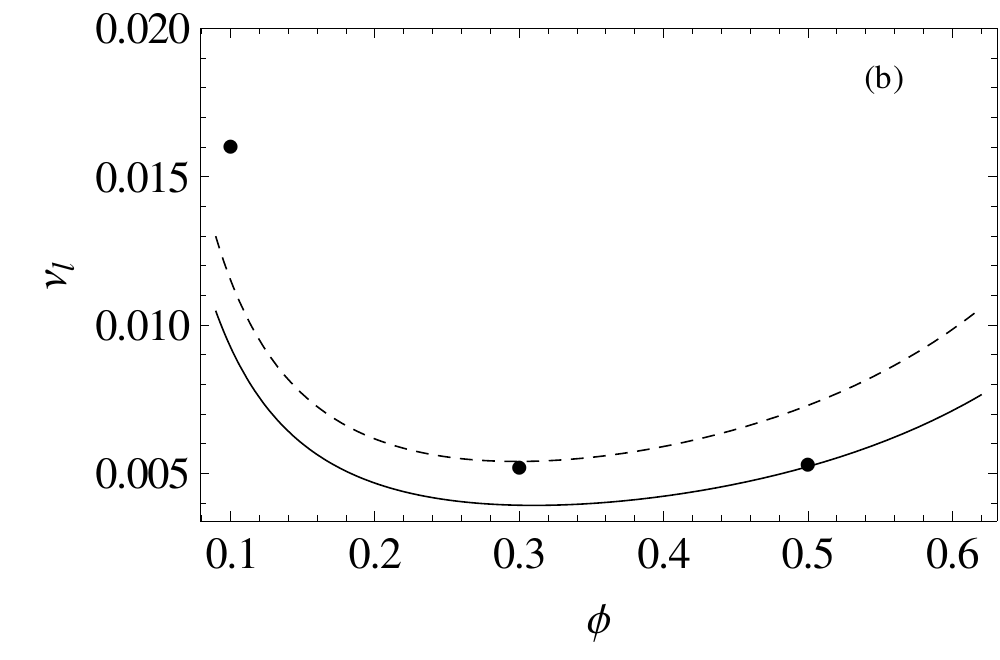}
  \caption{Plot of the longitudinal viscosity $\nu_l$ as a function of the volume fraction $\phi$ for two values of the coefficient of restitution: $\alpha=0.8$ (panel a), and $\alpha=0.6$ (panel b). The solid lines are the theoretical predictions for $\nu_l$ obtained by using Eqs.\ \eqref{eq4.7} and \eqref{eq4.9}  while the dashed lines are the theoretical results obtained by assuming the elastic forms of the shear viscosity $\eta$ and the bulk viscosity $\lambda$. Symbols are the simulation results obtained by Gradenigo {\em et al.} \cite{GSVP11-1} by fitting their numerical data for the dynamical correlations of the longitudinal modes. 
  \label{fig04chap4}}
\end{figure}
\begin{figure}[h]
  \centering
  \includegraphics[width=0.75 \columnwidth,angle=0]{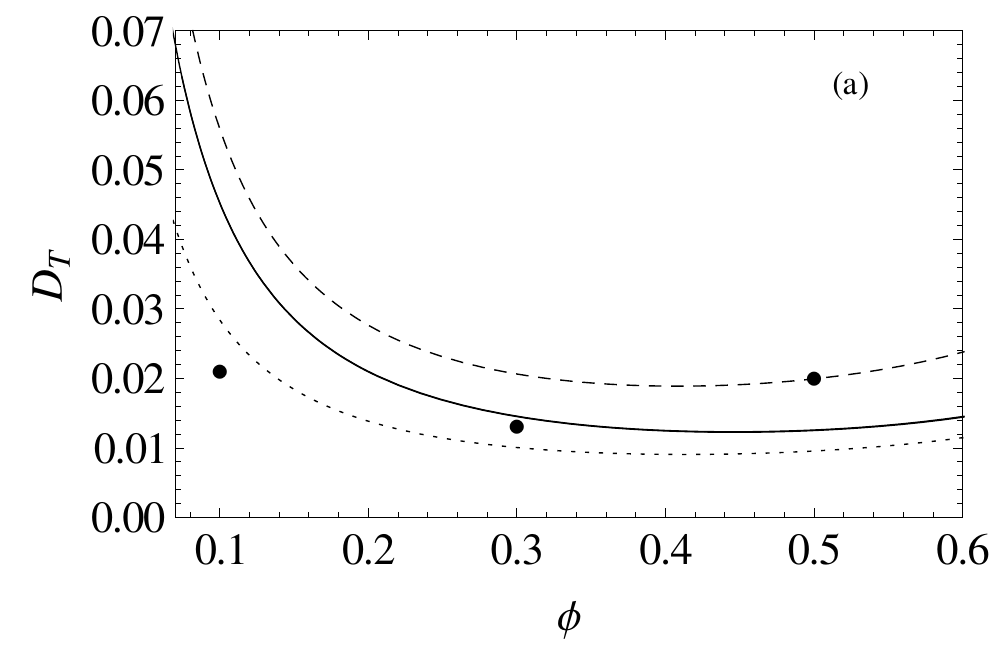}
  \includegraphics[width=0.75 \columnwidth,angle=0]{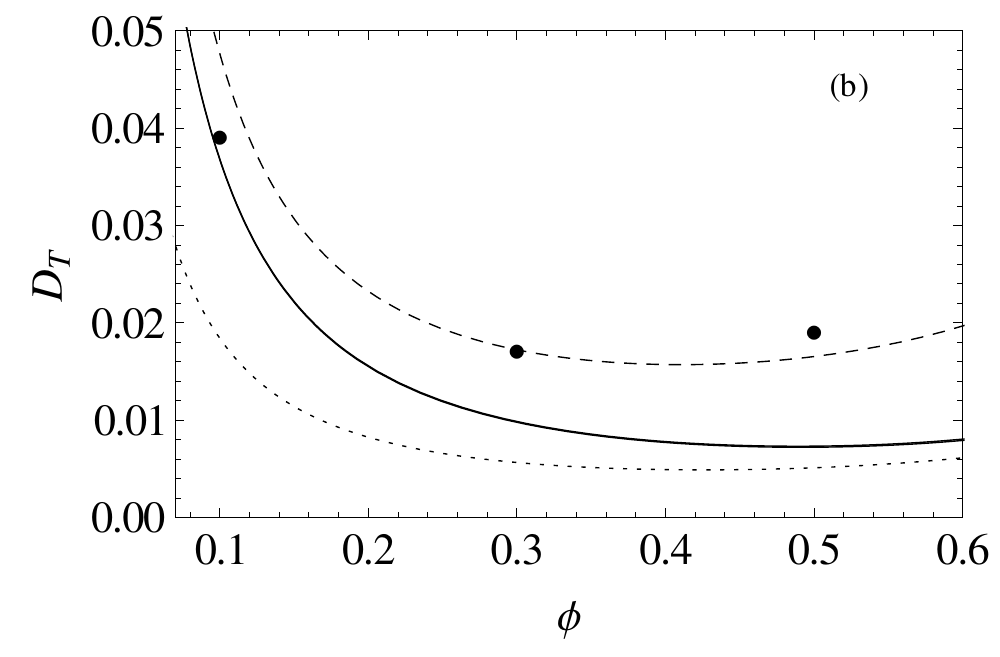}
  \caption{Plot of the thermal diffusivity $D_T=2\kappa/dn$ as a function of the volume fraction $\phi$ for two values of the coefficient of restitution: $\alpha=0.8$ (panel a), and $\alpha=0.6$ (panel b). Symbols are the simulation results obtained by Gradenigo \emph{et al.} \cite{GSVP11-1} by fitting their numerical data for the dynamical correlations of the longitudinal modes. The solid lines are the theoretical predictions for $D_T$ obtained by using Eqs.\ (4.35)--(4.37) (thermostat A),  the dotted lines are the theoretical predictions for $D_T$ obtained by using Eq.\ (D.1) (thermostat B) and the dashed lines are the theoretical results obtained by assuming the elastic form of the thermal conductivity $\kappa$.  
  \label{fig05chap4}}
\end{figure}

Fig.\ \eqref{fig03chap4} shows the kinematic viscosity $\nu$ for disks as a function of the volume fraction $\phi$ for $\alpha=0.6$. Symbols refer to the values of $\nu$ obtained from Langevin dynamics simulations by Gradenigo \emph{et al.} \cite{GSVP11-1} by using two different procedures: (i) via the equal-time correlation of the transversal shear mode (static correlations) and (ii) via the correlation of the transversal shear mode at different times (dynamical correlations). As in Fig.\ \ref{fig01chap4}, the parameters of the simulation are $\gamma_\text{b}=1$, $T_\text{b}=1$, $m=1$ and $\sigma=0.01$. We observe first that the simulation data obtained with the two independent procedures are compatible. Regarding the theoretical results, note that for the kinematic viscosity the results obtained by using both kind of thermostats are the same. The theoretical prediction for $\eta$ in the \emph{elastic} limit [i.e., Eq.\ \eqref{eq4.9} with $\alpha=1$ and $\gamma_\text{b}=\xi_\text{b}=0$] but considering the $\alpha$-dependence of the granular temperature given by Eq.\ \eqref{eqTs02} is also plotted. This was the theoretical expression for $\nu$ used in Ref.\ \cite{GSVP11-1} to compare with simulation data. At a qualitative level, we observe that both theories (the elastic Enskog theory and the one derived here) reproduce the general trends of simulation data. However, at a more quantitative level, it appears that the analytical results obtained here for granular fluids agree much better with simulation data than those obtained in the elastic case, since the latter clearly overestimates the value of $\nu$. This is the expected result since the simulations were carried out for inelastic gases in the presence of a stochastic bath with friction.

The longitudinal viscosity $\nu_l$ is plotted in Fig.\ \ref{fig04chap4} versus the volume fraction $\phi$ for the same systems as in Fig.\ \ref{fig03chap4}. We observe that, in general, the influence of the thermostat on the longitudinal viscosity is less significant than for the kinematic viscosity $\nu$ since both theories agree relatively well. However, the discrepancies with computer simulations are more important than in the case of $\nu$, specially in the low-density limit ($\phi=0.1$). While the elastic theory is closer to the simulation data than the inelastic theory when $\alpha=0.8$ [panel $(a)$ of Fig.\ \ref{fig04chap4}], the opposite happens at $\alpha=0.6$ for denser systems [see the panel $(b)$ of Fig.\ \ref{fig04chap4}]. Since the dependence of the shear viscosity $\eta$ on $\phi$ is well captured by the inelastic Enskog theory (see Fig.\ \ref{fig03chap4}), it is evident that the discrepancies between theory and simulations are essentially due to the bulk viscosity $\lambda$, whose value is specially underestimated at low-density. This is a quite surprising result since one would expect that the influence of $\lambda$ on the value of $\nu_l$ increases with increasing density since $\lambda=0$ for a dilute gas ($\phi=0$).

The thermal diffusivity is shown in Fig.\ \ref{fig05chap4} for the same cases as those considered in Figs.\ \ref{fig03chap4} and \ref{fig04chap4}. Surprisingly, for strong dissipation and quite dense systems [see the panel (b) of Fig.\ \ref{fig05chap4}], the comparison between theory and simulation agrees in general better when one uses the elastic form for $D_\text{T}$ instead of its inelastic expression \eqref{eq4.9}. These results contrast with the ones recently obtained \cite{G11} for the stochastic driving (i.e., when $\gamma_\text{b}\to 0$, keeping $\gamma_\text{b} T_\text{b}$ finite) where it was shown the accuracy of the inelastic Enskog theory (see Fig.\ 1 of Ref.\ \cite{G11}) for moderate densities and finite collisional dissipation. It is important to note that the identification of the transport coefficients from Langevin dynamics simulations requires to fit the simulation results for small but not zero values of the wave number $k$. Given that the expressions for the Enskog transport coefficients are independent of the wave number (since the hydrodynamic regime only strictly holds in the limit $k\to 0$), it is possible that the transport coefficients measured in the simulations are still functions of $k$, specially when the smallest value of $k$ considered to get the fit results is not close to 0. In particular, the simulation data for $\phi=0.3$ and 0.5 in the panel (b) of Fig.\ \ref{fig04chap4} were obtained for $k\sigma=0.4$ and 0.5, respectively. In this sense, if we extrapolate the data shown in Table 3 of Ref.\ \cite{GSVP11-1}, one could conclude that the true value of $D_\text{T}$ is smaller than the one shown in this figure when $k\sigma=0$. More simulations would be needed to clarify this point.

\begin{figure}
  \centering
  \includegraphics[width=0.75 \columnwidth,angle=0]{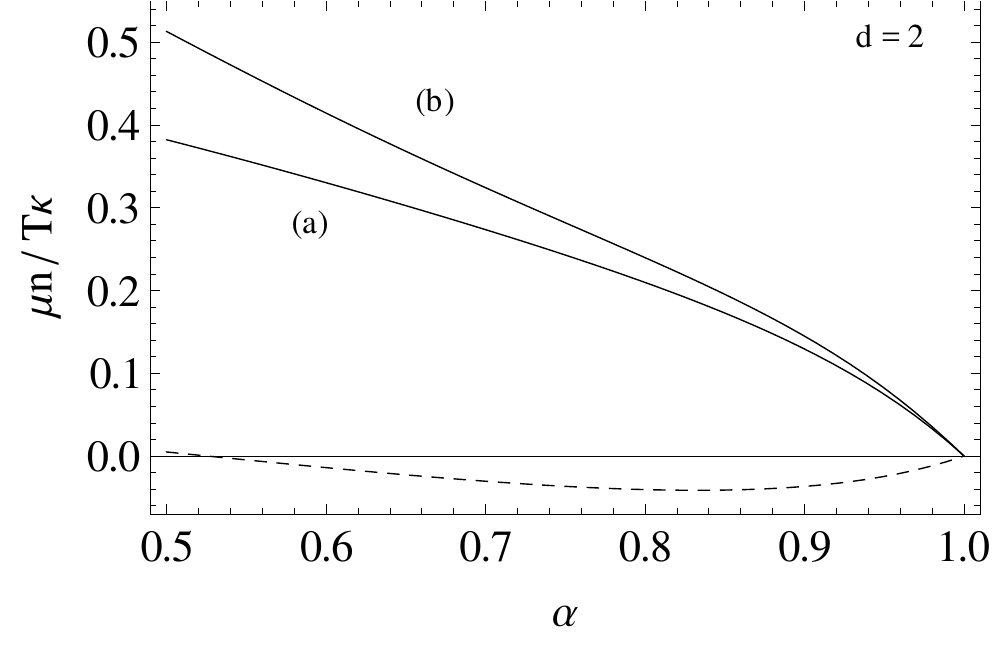}
  \includegraphics[width=0.75 \columnwidth,angle=0]{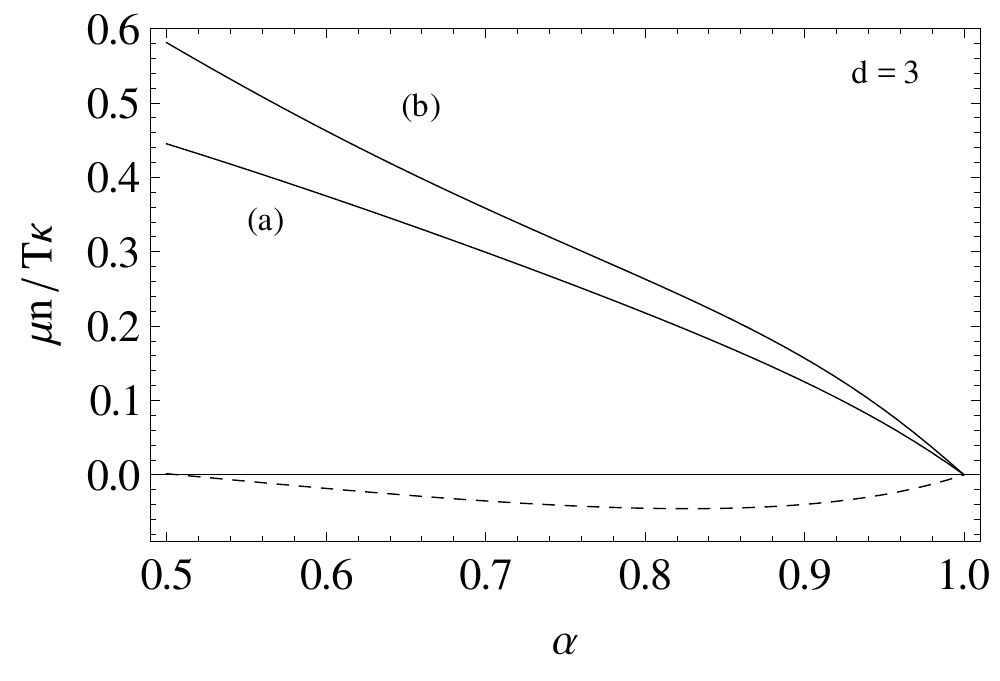}
  \caption{Plot of the dimensionless quantity $n\mu/T\kappa$ versus the coefficient of restitution $\alpha$ for hard disks ($d=2$) with $m=1$, $\sigma=0.01$, $\gamma_\text{b}=T_\text{b}=1$ and two different values of the solid volume fraction $\phi$: $(a)$ $\phi=0.1$, and $(b)$ $\phi=0.3$. The dashed line corresponds to the results obtained by considering the choice B for $\phi=0.1$. Note that $\mu=0$ in the elastic case ($\alpha=1$).
  \label{fig06chap4}}
\end{figure}

\begin{figure}
  \centering
  \includegraphics[width=0.75 \columnwidth,angle=0]{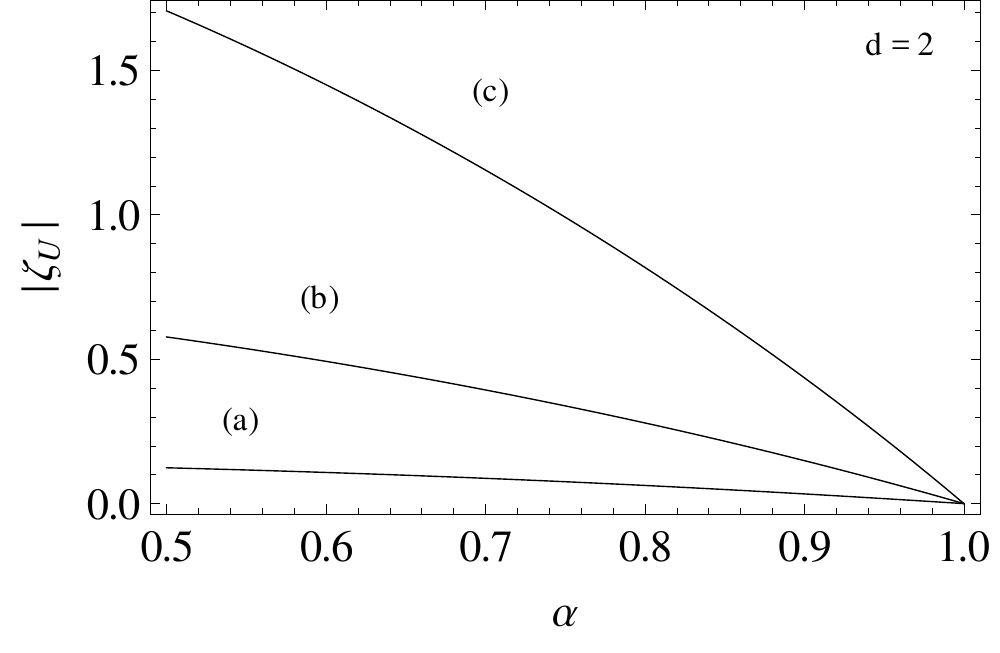}
  \includegraphics[width=0.75 \columnwidth,angle=0]{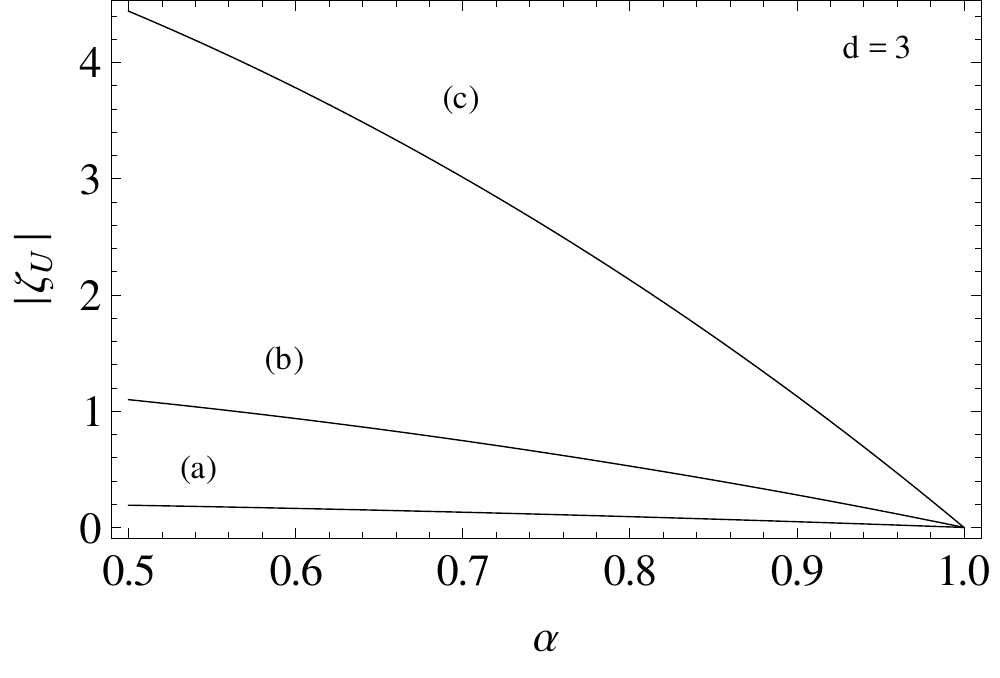}
  \caption{Plot of the magnitude of the first-order contribution $\zeta_U$ to the cooling rate versus the coefficient of restitution $\alpha$ %for hard disks ($d=2$) 
with $m=1$, $\sigma=0.01$, $\gamma_\text{b}=T_\text{b}=1$ and three different values of the solid volume fraction $\phi$: $(a)$ $\phi=0.1$, $(b)$ $\phi=0.3$, and $(c)$ $\phi=0.5$. Note that $\zeta_U=0$ in the elastic case ($\alpha=1$).
  \label{fig07chap4}}
\end{figure}

Now we consider the $\alpha$-dependence of the transport coefficient $\mu$ and the first-order contribution $\zeta_U$ to the cooling rate. Given that both coefficients vanish in the elastic limit, they were also neglected in previous studies for heated granular fluids \cite{NETP99,GSVP11-1}. To assess the impact of the term $-\mu \nabla n$ in the heat flux, the reduced coefficient $\mu n/(T\kappa)$ is plotted in Fig.\ \ref{fig06chap4} versus the coefficient of restitution for two different values of the volume fraction $\phi$ in the case of the choice A. 
Given that the derivatives $(\partial a_2/\partial\xi^*)$, $(\partial a_2/\partial\theta)$ and $(\partial a_2/\partial\chi)$ are in general very small, for the sake of simplicity the contributions proportional to those derivatives have been neglected in the evaluation of $\mu$ in Fig.\ \ref{fig06chap4}.

The results derived for $\mu$ by using the choice B are also plotted for comparison in the case $\phi=0.1$. We observe that the coefficient $\mu$ is negative in the case of the choice B, although its magnitude is practically zero. This drawback ($\mu \leq 0$) of choice B is not present in the case of the choice A since $\mu$ is always positive for any value of $\alpha$ and $\phi$, similarly to what happens in the \emph{undriven} case \cite{GD99,L05}. In addition, although the magnitude of $\mu$ is in general smaller than that of the thermal conductivity $\kappa$, we observe that the influence of $\mu$ on the heat transport could not be considered negligible as the degree of dissipation increases. The $\alpha$-dependence of the magnitude of $\zeta_U$ derived from the choice A is plotted in Fig.\ \ref{fig07chap4} for several values of the volume fraction. It is quite apparent that the influence of dissipation on $|\zeta_U|$ is more significant than in the case of $\mu$, specially at large densities. Consequently, the contribution of $\zeta_U$ to the cooling rate should be considered as the rate of dissipation increases.

\subsection{Comparison with stochastic thermostat data}
\begin{figure}[h]
\centering
  \includegraphics[width=0.75 \columnwidth,angle=0]{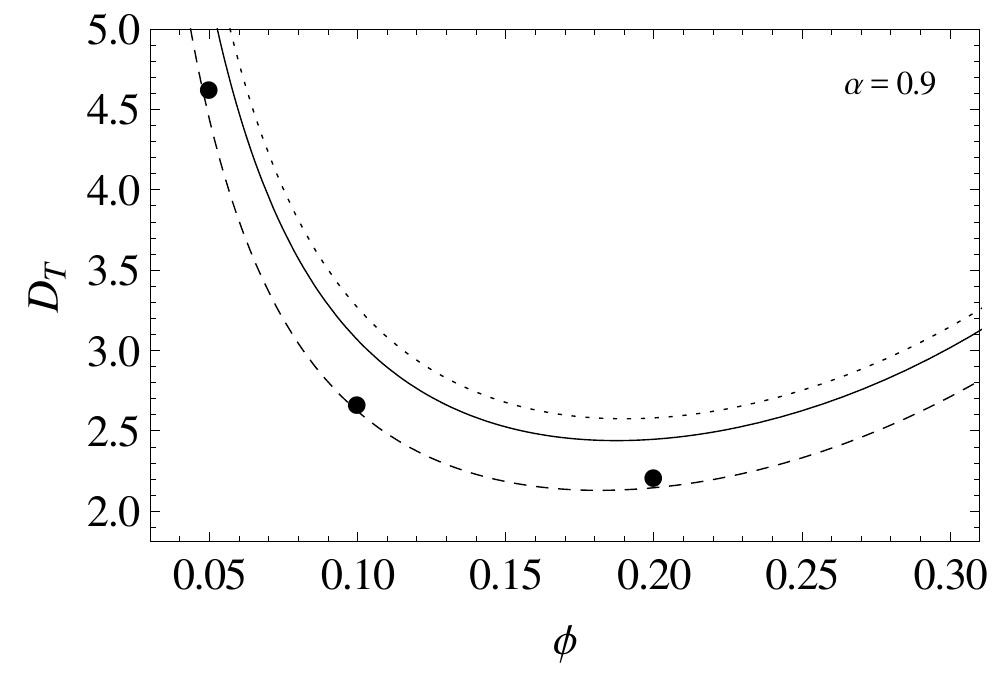}
  \includegraphics[width=0.75 \columnwidth,angle=0]{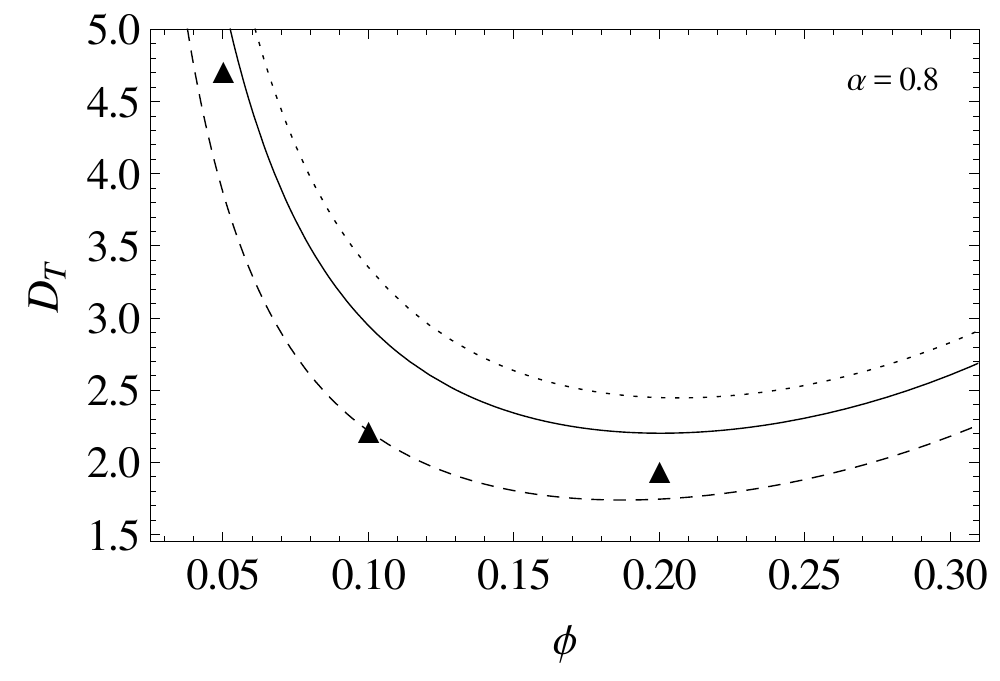}
  \caption{Plot of the thermal diffusivity $D_T$ as a function of the volume fraction $\phi$ for $\alpha=0.9$ and $\alpha=0.8$. Solid and dashed lines are the theoretical predictions for $D_T$ obtained by using Eqs.\ \eqref{eq4.12}--\eqref{eq4.14}  (thermostat A) and Eq.\ \eqref{apD4.24} (thermostat B) respectively whereas dotted line correspond to the preditions for the undriven gas. Circles and triangles are simulation data.}
   \label{fig08chap4}
\end{figure}

In the present Subsection we compare our theoretical predictions in the case of a simple stochastic thermostat, ($\beta=0$), with Molecular Dynamics (MD) simulations carried out by Vollmayr-Lee, Aspelmeier, and Zippelius \cite{VAZ11}. In that paper, time-delayed correlation functions of a homogenous granular fluid of hard spheres ($d=3$) at intermediate volume fractions driven by means of a stochastic external force were determined. In the steady state, the dynamic structure factor $S(k,\omega)$ (being $k$ the wave number and $\omega$ the angular frequency) is measured for several values of the volume fraction $\phi$ and the coefficient of restitution $\alpha$. The corresponding best fit of the simulation results of $S(k,\omega)$ allowed them to identify the thermal diffusivity $D_T$ and the longitudinal viscosity $\nu_l$ coefficients for the smallest values of $k$. As in the simulations performed by Gradenigo \emph{et al.} \cite{GSVP11-1}, this kind of fits require $k$-dependent transport coefficients because they consider wave numbers outside the hydrodynamic regime. As a consequence, their simulation results may not be well-described by our theory, which is based in the limit of $k\rightarrow0$. This can be easily seen in Fig.\ \ref{fig08chap4} where the thermal diffusivity $D_T$ is shown as a function of the volume fraction $\phi$ for $\alpha=0.9$ and $0.8$. For the sake of comparison with those results, the same units are taken here, that is, $m=\sigma=T=1$. Solid and dashed lines are the theoretical predictions for $D_T$ obtained by using Eqs.\ \eqref{eq4.12}--\eqref{eq4.14} (thermostat A) and Eq.\ \eqref{apD4.24} (thermostat B) respectively. 
The dotted line refers to the theoretical prediction for an undriven gas \cite{G11}.
%Lines $(c)$ and $(d)$ are the preditions using the modified Sonine approximations applied to the Revised Enskog Theory (RET) for the driven and undriven gas respectively \cite{G11}. 
Symbols refer to simulation results. 

In general, all theories reproduce the main trends of simulation data. However, at a more quantitative level, we observe that, in principle, thermostat B agrees with simulations better than the one corresponding to thermostat A. As we said before, the discrepancies with our results derived for thermostat A could be explained by the fact that simulation results were obtained for finite wave number values. Thus, an inspection of Table II and Fig.\ 15 in Ref.\ \cite{VAZ11} may help us to figure out that correct results should be nearer to our theoretical predictions with thermostat A as $k\rightarrow0$. %Work along the above lines should be carried out in the future.

\section{Linear stability analysis of the hydrodynamic equations}
\label{sec5chap4}
In this Section we study the stability of the hydrodynamic equations for a dense granular gas with thermostat.
The closed hydrodynamic equations for $n$, ${\bf U}$, and $T$ can be obtained by replacing the constitutive
forms of the pressure tensor \eqref{eq4.6}, the heat flux \eqref{eq4.11}, and the cooling rate \eqref{eq4.18} into the balance equations
\eqref{eqBaln01}--\eqref{eqBalT01}. They are given by
\begin{equation}
\label{75}
D_tn+n\nabla \cdot {\bf U}=0,
\end{equation}
\begin{equation}
D_t U_i + \rho^{-1}\partial_i p=\rho^{-1}\partial_j\left[\eta \left(\partial_iU_j+\partial_jU_i -\frac{2}{3}\delta_{ij}\nabla \cdot {\bf U}\right)+\lambda \delta_{ij}\nabla \cdot {\bf U}\right],
\label{76}
\end{equation}
\begin{eqnarray}
\label{77}
& & \left(D_t+\frac{2\gamma_\text{b}}{m}-\frac{m\xi_\text{b}^2}{T}
+\zeta^{(0)}\right)T+\frac{2}{d n}p\nabla \cdot {\bf U} = \frac{2}{d n}\nabla \cdot \left (\kappa \nabla T + \mu \nabla n\right) +  \nonumber\\
& & + \frac{2}{d n}\left[\eta \left(\partial_iU_j+\partial_jU_i - \frac{2}{d} \delta_{ij}\nabla \cdot {\bf U}\right) + \lambda \delta_{ij}\nabla \cdot {\bf U}\right] \partial_iU_j - T\zeta_U  \nabla\cdot {\bf U}.
\end{eqnarray}

%\end{widetext}
In Eq.\ \eqref{76} we have assumed $\Delta\bt{U}=\bt{U}-\bt{U}_\text{g}=\bt{0}$ for the sake of simplicity.
Note that consistency would require to consider up to second order in the gradients in the expression (\ref{eq4.18}) for the cooling rate, since this is the order of the terms in Eqs.\ \eqref{eq4.6} and (\ref{eq4.11}) coming from the pressure tensor and the heat flux, respectively.
However, it has been shown for a dilute gas that the contributions from the cooling rate
of second order are negligible \cite{BDKS98} as compared with the corresponding contributions from Eqs.\ \eqref{eq4.6} and (\ref{eq4.11}). It is assumed here that the same holds in the dense case \cite{G05}.

The form of the NS Eqs.\ \eqref{75}--\eqref{77} for a driven granular fluid is analogous to that of an ordinary fluid, except for the presence of the external bath parameters $\gamma_\text{b}$ and $\xi_\text{b}^2$, the contributions to the cooling rate $\zeta^{(0)}$ and $\zeta_U$ and the new transport coefficient $\mu$ in the energy balance equation. In addition, as shown in Sec.\ \ref{sec3chap4} and depending on the values of the coefficient of restitution $\alpha$, the transport coefficients are in general different from those obtained for elastic collisions.

Eqs.\ \eqref{75}--\eqref{77} can be linearized around the stationary homogeneous state, where the hydrodynamic fields take the steady values $n_\text{s}\equiv \text{const.}$, $T_\text{s}\equiv \text{const.}$  and ${\bf U}_\text{s}={\bf 0}$.  A linear stability analysis of the hydrodynamic Eqs.\ \eqref{75}--\eqref{77} has also been carried out in Ref.\ \cite{GSVP11-1} but neglecting any dependence of the transport coefficients on inelasticity and assuming that $\mu=\zeta_U=0$. As mentioned in the Introduction, the only impact of inelasticity on the hydrodynamic equations \cite{GSVP11-1} is through the $\alpha$-dependence of the (steady) granular temperature $T_\text{s}$ [see Eq.\ \eqref{eqTs02} with $a_{2,\text{s}}=0$]. Thus, it is worth to assess to what extent the previous theoretical results \cite{GSVP11-1} are indicative of what happens when the correct expressions for the transport coefficients and the cooling rate are considered. This is the main motivation of this Section.

We assume that the deviations $\delta y_{\alpha}({\bf r},t)=y_{\alpha}({\bf r},t)-y_{\text{s} \alpha}(t)$ are small, where $\delta y_{\alpha}({\bf r},t)$ denotes the deviations of $n$, $\mathbf{U}$, and $T$ from their values in the \emph{steady} homogeneous state. To recover previous linear stability results \cite{G05} derived in the undriven case, let us consider the following (reduced) time and space variables:
\begin{equation}
\label{78}
\tau=\frac{1}{2}n_\text{s}\sigma^{d-1}\sqrt{\frac{T_\text{s}}{m}}t, \quad \mathbf{r}'=
\frac{1}{2}n_\text{s}\sigma^{d-1}{\bf r}.
\end{equation}
The dimensionless time scale $\tau$ is a measure of the average number of collisions per particle
in the time interval between $0$ and $t$. The unit length introduced in
the second equality of (\ref{78}) corresponds to the mean free path of gas
particles.

A set of Fourier transformed dimensionless variables is then introduced by
\begin{equation}
\label{79}
\rho_{{\bf k}}(\tau)=\frac{\delta n_{{\bf k}}(\tau)}{n_{\text{s}}}, \quad
{\bf w}_{{\bf k}}(\tau)=\frac{\delta {\bf U}_{{\bf
k}}(\tau)}{\sqrt{T_\text{s}/m}},\quad \theta_{{\bf k}}(\tau)=\frac{\delta
T_{{\bf k}}(\tau)}{T_{\text{s}}},
\end{equation}
where $\delta y_{{\bf k}\alpha}\equiv \{\rho_{{\bf k}},{\bf
w}_{{\bf k}}(\tau), \theta_{{\bf k}}(\tau)\}$ is defined as
\begin{equation}
\label{80}
\delta y_{{\bf k}\alpha}(\tau)=\int d \mathbf{r}'\;
e^{-i{\bf k}\cdot \mathbf{r}'}\delta y_{\alpha}
(\mathbf{r}',\tau).
\end{equation}
Note that in Eq.\ (\ref{80}) the wave vector ${\bf k}$ is dimensionless.

In Fourier space, as expected, Eq.\ \eqref{76} shows that the $d-1$ transverse velocity components
${\bf w}_{{\bf k}\perp}={\bf w}_{{\bf k}}-({\bf w}_{{\bf k}}\cdot
\widehat{{\bf k}})\widehat{{\bf k}}$ (orthogonal to the wave vector ${\bf k}$)
decouple from the other three modes and hence can be obtained more
easily. Their evolution equation can be written as
\begin{equation}
\label{81}
\left(\frac{\partial}{\partial \tau}+\frac{1}{2}\eta^*
k^2\right){\bf w}_{{\bf k}\perp}=0,
\end{equation}
where
\begin{equation}
\label{40.1}
\eta^*=\frac{\eta}{\sigma^{1-d}\sqrt{m T_\text{s}}}.
\end{equation}
The solution to Eq.\ (\ref{81}) is
\begin{equation}
\label{82}
{\bf w}_{{\bf k}\perp}({\bf k}, \tau)={\bf w}_{{\bf k}\perp}(0)\exp\left[\Lambda_\perp(k)\tau\right],
\end{equation}
where
\begin{equation}
\label{82.1}
\Lambda_\perp(k)=-\frac{1}{2}\eta^*k^2.
\end{equation}
%Note that the steady state condition \eqref{20.1} yields $2\gamma^*=\sqrt{2}(\xi^*-\zeta_0^*)$.
Since the (reduced) shear viscosity coefficient $\eta^*$ is positive, then $\Lambda_\perp(k)$ becomes negative for any finite wave number $k$ and so the transversal shear modes of the driven gas are linearly stable. This result contrasts with the ones obtained in the undriven case \cite{G05} where it was shown that the transversal shear modes become \emph{unstable} for values of $k$ smaller than a certain critical wave number.

The remaining (longitudinal) modes correspond to $\rho_{{\bf k}}$, $\theta_{{\bf k}}$, and
the longitudinal velocity component of the velocity field, $w_{{\bf k}||}={\bf w}_{{\bf
k}}\cdot \widehat{{\bf k}}$ (parallel to ${\bf k}$). These modes are coupled and obey the equation
\begin{equation}
\frac{\partial \delta y_{{\bf k}\alpha }(\tau )}{\partial \tau } = M_{\alpha \beta}
\delta y_{{\bf k}\beta }(\tau),
\label{83}
\end{equation}
where $\delta y_{{\bf k}\alpha }(\tau )$ denotes now the set  $\left\{\rho _{{\bf k}},\theta _{{\bf k}},
 w_{{\bf k}||}\right\}$ and $\mathsf{M}$ is the square matrix
%\begin{widetext}
\begin{equation}
{\sf M} = - \left( \begin{array}{ccc}
0 & 0 & i k \\
2\sqrt{2}\zeta_0^*g+\mu^*k^2&\sqrt{2}(\zeta_0^*+2\xi^*)+D_\text{T}^*k^2 & \frac{2}{d}ik(p^*+\frac{d}{2}\zeta_U)\\
ikp^*C_\rho & ikp^* &\nu_l^*k^2
\end{array} \right),
\label{84}
\end{equation}

%\end{widetext}
where
\begin{equation}
\label{84.0}
%\gamma^*=\frac{\gamma_\text{b}}{n_\text{s}\sigma^{d-1}\sqrt{T_\text{s}/m}}, \quad
\zeta_0^*=\frac{\ell \zeta_\text{s}^{(0)}}{\sqrt{2T_\text{s}/m}}, \quad
\xi^*=\frac{m\ell \xi_\text{b}^2}{T_\text{s}\sqrt{2T_\text{s}/m}}
\end{equation}
\begin{equation}
\label{84.1}
p^*=\frac{p_\text{s}}{n_\text{s}T_\text{s}}=1+2^{d-2}(1+\alpha)\chi \phi,
\end{equation}
and
\begin{equation}
\label{84.2}
\nu_l^*= \frac{\rho_\text{s}\nu_{l}}{2\sigma^{1-d}\sqrt{mT_\text{s}}}, \quad
D_\text{T}^*= \frac{n_\text{s}D_\text{T}}{2\sigma^{1-d}\sqrt{T_\text{s}/m}},
\end{equation}
\begin{equation}
\label{84.2.1}
\mu^*=\frac{\rho_\text{s}}{d\sigma^{1-d}T_\text{s}\sqrt{mT_\text{s}}}\mu.
\end{equation}
Here, $\rho_\text{s}=m n_\text{s}$ is the mass density. In the above equations, it is understood that the transport coefficients $\eta$, $\nu_l$, $D_\text{T}$, and $\mu$ are evaluated in the homogeneous steady state. In addition, the quantity $C_\rho(\alpha,\phi)$ appearing in the matrix ${\sf M}$ is given by
%\begin{equation}
%\label{eqgphi} %\label{85}
%g(\phi)=1+\phi\frac{\partial}{\partial \phi}\ln \chi(\phi),
%\end{equation}
\begin{equation}
\label{86}
C_\rho(\alpha,\phi)
= 1+\phi\frac{\partial}{\partial \phi}\ln p^*(\alpha,\phi)
= 1+g(\phi)-\frac{g(\phi)}{1+2^{d-2}(1+\alpha)\phi \chi(\phi)},\nonumber\\
\end{equation}
where in the last equality use has been made of the explicit expression of $p^*$
given by Eq.\ (\ref{84.1}) and $g(\phi)$ is given by Eq.\ \eqref{eqgphi}. If one assumes $\mu^*=\zeta_U=0$, the matrix \eqref{84} agrees with the dynamical matrix obtained when the gas is heated by a stochastic thermostat ($\gamma_\text{b}=0$ but $\gamma_\text{b} T_\text{b}=\text{finite}$ and $\zeta_0^*=\xi^*$) \cite{NETP99}.

\begin{figure}[ht]
\centering
  \includegraphics[width=0.75 \columnwidth,angle=0]{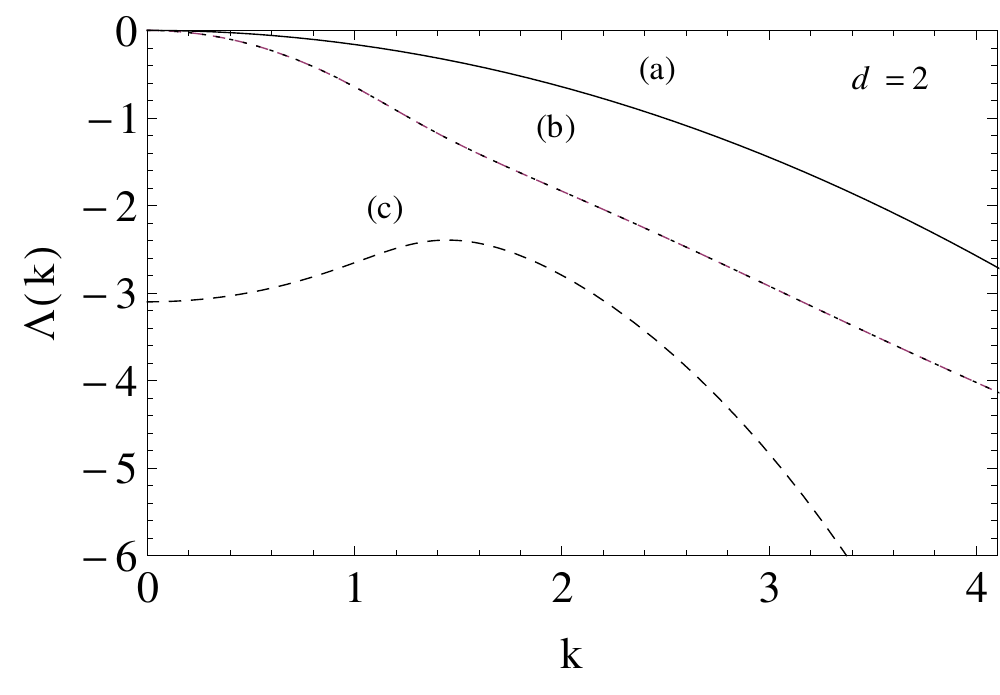}
  \includegraphics[width=0.75 \columnwidth,angle=0]{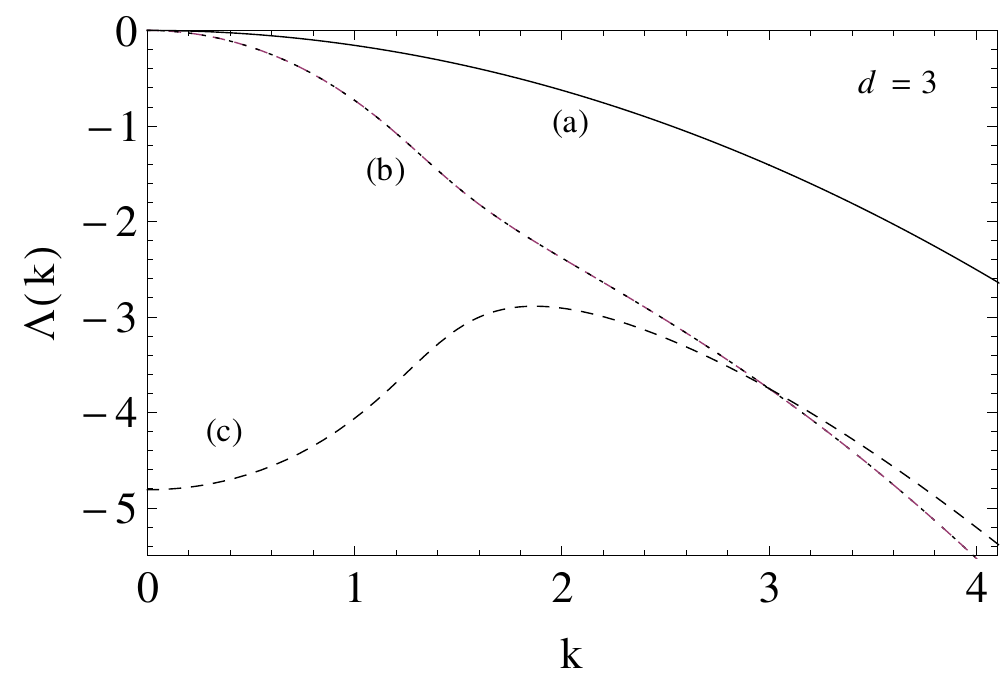}
  \caption{Plot of the dispersion relations for disks ($d=2$) and spheres ($d=3$) with $\sigma=0.01$, $\phi=0.2$ and $\alpha=0.8$. Line $(a)$ corresponds to the $d-1$ degenerate transversal modes while $(b)$ and $(c)$ are the remaining longitudinal modes. Only the real parts of the eigenvalues of the matrix ${\sf M}$ is plotted.}
    \label{fig09chap4}
\end{figure}

The three longitudinal modes have the form $\exp[\Lambda_\ell(k) \tau]$ for $\ell=1,2,3$, where
$\Lambda_\ell(k)$ are the eigenvalues of the matrix ${\sf M}$, namely, they are the solutions of the cubic equation
\begin{equation}
\label{86.1}
\Lambda^3+A(k) \Lambda^2+B(k) \Lambda+C(k)=0,
\end{equation}
where
\begin{equation}
\label{87}
A(k)=\sqrt{2}(\zeta_0^*+2\xi^*)+k^2\left(\nu_l^*+D_\text{T}^*\right),
\end{equation}
\begin{equation}
\label{88}
B(k) = k^4\nu_l^*D_\text{T}^* +k^2\left[p^*C_\rho+p^*\left(\frac{2}{d}p^*+\zeta_U\right)
+\sqrt{2}(\zeta_0^*+2\xi^*)\nu_l^*\right],
\end{equation}
\begin{equation}
\label{89}
C(k) = p^*k^2\left[\sqrt{2}C_\rho\left(\zeta_0^*+2\xi^*\right)
-2\sqrt{2}g\zeta_0^* +\left(C_\rho D_\text{T}^*-\mu^*\right)k^2\right].
\end{equation}
One of the longitudinal modes (the heat mode) could be unstable for $k<k_\text{h}$, where $k_\text{h}$ is obtained from Eq.\ \eqref{86.1} when $\Lambda=0$, namely, $C(k_h)=0$. The result is
\begin{equation}
\label{90}
k_\text{h}^2=\sqrt{2}\frac{2g\zeta_0^*-C_\rho(\zeta_0^*+2\xi^*)}{C_\rho D_\text{T}^*-\mu^*}.
\end{equation}
On the other hand, an analysis of the dependence of $k_\text{h}^2$ on the coefficient of restitution $\alpha$ and the volume fraction $\phi$ shows that  $k_\text{h}^2<0$ for any value of $\alpha$ and $\phi$. Thus, there are no physical values of $k_\text{h}$ for which the heat mode becomes unstable. Consequently, \emph{all} the eigenvalues of the dynamical matrix ${\sf M}$ have a \emph{negative} real part and no instabilities are found due to the presence of the external bath. This conclusion agrees with the results obtained in Refs.\ \cite{NETP99} and \cite{GSVP11-1} for driven granular fluids. To illustrate this behaviour Fig.\ \ref{fig09chap4} shows the dependence of the real parts of the eigenvalues of the matrix $\mathsf{M}$ on the wave number $k$. It appears that $Re\,\Lambda\leq0$ for any value of $k$.

In summary, the results obtained here including the complete $\alpha$-dependence of the transport coefficients show no new surprises relative to the earlier works \cite{NETP99,GSVP11-1}, by considering the elastic Enskog expressions for the above coefficients. Of course, the quantitative forms for the dispersion relations can be quite different in both (elastic and inelastic) approaches since the impact of dissipation on the transport coefficients and the cooling rate is significant and so, their functional forms differ appreciably from their elastic forms.

%The eigenvalues of the matrix ${\sf M}$ for a granular fluid with $\alpha=0.6$ and $\phi=0.5$, as obtained from Eq.\ \eqref{82} and the solutions of the cubic Eq.\ \eqref{86.1}, are plotted in Fig.\ \ref{fig7}. Only the real part (propagating modes) of the solutions to Eq.\ \eqref{86.1} is represented.

\section{Summary and Discussion}

In this Chapter, we have determined the transport coefficients of a granular fluid driven by a stochastic bath with friction. The results have been obtained within the framework of the (inelastic) Enskog kinetic theory and they are expected to apply over a wide range of densities. Our goal is not only academic since, from a practical point of view, many of the simulations reported \cite{PLMV98,PLMV99,CLH00,PEU02,PBL02,VPBTW06,FAZ09,SVGP10,KSZ10,PGGSV12,NETP99,VAZ11,GSVP11-1} for flowing granular materials have used external driving forces to fluidize the system. For this reason, it would be convenient to provide the corresponding expressions of the transport coefficients when the granular fluid is heated by a thermostat. In fact, due to the lack of the above expressions, in most of the cases it is assumed that the forms of the transport coefficients of the driven granular fluid are the same as those given by the \emph{elastic} Enskog theory \cite{FERZIGER72}. However, as expected from previous theoretical works \cite{GM02,G11}, the present results show again that the expressions for the transport coefficients clearly differ from those obtained for ordinary fluids so that, one should use the true inelastic Enskog coefficients to analyze granular flows driven by thermostats.

The transport processes considered are those for a driven fluid with small spatial gradients of the hydrodynamic fields. In this situation, the Enskog equation has been solved by means of the CE method \cite{CHAPMAN70} up to the first order in the spatial gradients. Since these gradients have been assumed to be independent of the coefficient of restitution, although the corresponding hydrodynamic equations restrict their applicability to first order in gradients, the transport coefficients appearing in these equations are valid \emph{a priori} for an arbitrary degree of dissipation.

An important but subtle point is the generalization of the driving external forces (which are mainly used in homogeneous situations) to \emph{non-homogeneous} states. This is a crucial step since one has to consider situations close to steady homogeneous states to determine the transport coefficients from the CE expansion. Although the above generalization is a matter of choice, it has important implications in the final expressions of the transport coefficients. For simplicity, in previous works on heated granular gases \cite{GM02,G11} it was assumed that the external driving force has the same form as in the homogeneous case, except that their parameters are local quantities. As a consequence, the parameters of the force are chosen to impose a stationary temperature in the zeroth-order solution (i.e., $\partial_t^{(0)}T=0$). However, for general small perturbations around the steady homogeneous state, it is expected that the density and temperature are specified separately in the local reference state $f^{(0)}$ and so, the temperature cannot be stationary at any point of the system (i.e., $\partial_t^{(0)}T\neq 0$). This choice is more general than the previous one ($\partial_t^{(0)}T=0$) and has the advantage of a simpler implementation on computer simulations since the parameters of the driven external force are constant, even for non-homogeneous states.

The fact that $\partial_t^{(0)}T\neq 0$ gives rise to conceptual and practical difficulties not present in the case of the choice B. One of them is that evaluation of the complete nonlinear dependence of the transport coefficients on dissipation requires in principle the analysis of the hydrodynamic behavior of the \emph{unsteady} reference state. This involves the corresponding numerical integration of the differential equations obeying the velocity moments of the zeroth-order distribution $f^{(0)}$ [see for instance, Eq.\ \eqref{eqa2inhomo} for the fourth degree moment $a_2$ of $f^{(0)}$]. This is quite an intricate problem. However, given that here we are interested in the evaluation of the momentum and heat fluxes in the first order of the deviations from the steady reference state, the transport coefficients must be determined to zeroth order in the deviations. As a consequence, the steady-state condition \eqref{eqTs01} applies and the transport coefficients and the cooling rate can be defined in terms of the hydrodynamic fields in the steady state. Explicit expressions for these quantities have been obtained after considering the leading terms in a Sonine polynomial expansion. These explicit forms have been displayed in Sec.\ \ref{sec3chap4} and Appendix \ref{AppendixD} for the choices A and B, respectively. More specifically, in the case of the choice A, the bulk $\lambda$ and shear $\eta$ viscosities are given by Eqs.\ \eqref{eq4.7} and \eqref{eq4.9}, respectively, the thermal conductivity $\kappa$ is given by Eqs.\ \eqref{eq4.12} and \eqref{eq4.14}, the coefficient $\mu$ is given by Eqs.\ \eqref{eq4.16} and \eqref{eq4.17} and the cooling rate $\zeta$ is defined by Eqs.\ \eqref{eq4.18}--\eqref{eq4.23}. All these expressions clearly show the complex dependence of the set $\left\{\lambda,\eta, \kappa, \mu, \zeta\right\}$ on the granular temperature $T$, the coefficient of restitution $\alpha$, the solid volume fraction $\phi$ and the model parameter $\xi_\text{b}^2$. In the case of the choice B, our results show that the expressions of $\lambda$ and $\eta$ are the same as those obtained from the choice A but the forms of $\kappa$ and $\mu$ are different [they are given by Eqs.\ \eqref{apD4.24} and \eqref{apD4.25}, respectively]. An important drawback of the results derived from the choice B is that the coefficient $\mu$ can be negative (see Fig.\ \ref{fig06chap4}), although its magnitude is very small.

A comparison with recent Langevin dynamics simulations \cite{GSVP11-1} carried out for a granular fluid driven also by a stochastic bath with friction has been made in Sec.\ \ref{sec4chap4}. The comparison has been displayed in Fig.\ \ref{fig03chap4} for the kinematic viscosity $\nu$, Fig.\ \ref{fig04chap4} for the longitudinal viscosity $\nu_l$ and Fig.\ \ref{fig05chap4} for the thermal diffusivity $D_T$. It is quite apparent that while the predictions of the driven kinetic theory compares very well with simulation data for $\nu$ in a wide range of densities, some discrepancies appear in the cases of $\nu_l$ and $D_T$ as the gas becomes denser. Surprisingly, in the case of $D_T$, the comparison agrees better when one uses the elastic form of $D_T$ in the more inelastic system ($\alpha=0.6$) studied. We think that this disagreement is in part due to the fact that while the simulation data have been obtained for small but \emph{finite} values of the wave number $k$, the Enskog expressions for the transport coefficients only strictly apply in the limit $k\to 0$. Moreover, given that these discrepancies appear at sufficiently high densities, it could also reflect the limitations of the Enskog equation (which is based on the molecular chaos hypothesis) as the granular fluid becomes denser.

For the sake of completeness we have compared our predictions of the thermal diffusivity for hard spheres under the action of a simple stochastic thermostat (without friction) with recent MD simulations carried out by measuring the dynamic structure factor and fitting their numerical data for the dynamical correlations of the longitudinal modes \cite{VAZ11}. It is quite apparent that the results derived from the choice B agree better with simulation data than those obtained from the choice A. Although this discrepancy could be seen as an important drawback of our theory it is neccessary to emphasize that our theoretical predictions (both for choices A and B) only hold in the limit of small gradients, that is, when $k\rightarrow0$ and the simulations were carried out for finite values of $k$. More simulation data for small values of $k$ are needed to asses the accuracy of our theoretical results. %More research is required in this point. %Nevertheless, it must be taken into account that the thermal diffusivity obtained from DM simulation by Vollmayr-Lee \emph{et al.} is still $k$-dependent and, by inspection of the results 

With these new expressions for the momentum and heat fluxes and the cooling rate, a closed set of hydrodynamic equations for situations close to homogeneous steady states has been derived. A stability analysis of these linearized hydrodynamic equations with respect to the homogeneous steady state has been carried out to identify the conditions for stability in terms of dissipation. Our results show that the driven homogeneous state is \emph{stable} for any value of dissipation at sufficiently long wavelengths. This conclusion agrees with previous findings \cite{NETP99,GSVP11-1} obtained by using the \emph{elastic} expressions of the transport coefficients.

\chapter{Navier-Stokes transport coefficients for driven inelastic Maxwell models}
\label{Chapter5}

\blfootnote{The results obtained in this Chapter have been published in M.G. Chamorro, F. Vega Reyes and V. Garz\'o, \emph{J. Stat. Mech.}, P06008 (2014) \cite{CGV14}} 
 
\lhead{Chapter 5. \emph{Navier-Stokes transport coefficients for driven inelastic Maxwell models}}

\section{Introduction}
\label{sec1chap5}
As shown in Chapter \ref{Chapter4}, the determination of the transport coefficients of an inelastic driven granular gas involves the evaluation of certain collision integrals [see Eqs.\ \eqref{apC3}, \eqref{apC8} and \eqref{apC14.2bis}].
However, these integrals cannot be exactly calculated because the collision rate is proportional to the magnitude of the relative velocity of the two colliding pairs. % As for elastic collisions \cite{MAXWELL1867}, 
A possible way of circumventing the above technical difficulty, as for the elastic collisions case \cite{MAXWELL1867}, while keeping the structure of the Boltzmann collision operator is to consider the inelastic Maxwell model (IMM) for a low-density granular gas. For Maxwell particles the collision rate is independent of the relative velocity and thanks to this the collisional moments of the Boltzmann operator for IMM can be exactly obtained without the explicit knowledge of the velocity distribution function \cite{GS07}. Inelastic Maxwell models were introduced in granular literature \cite{BCG00, CCG00, BK00} as an altenative to IHS for assessing in a clean way the influence of dissipation on the dynamic properties of dilute granular gases.

As seen early, IMM share with dilute elastic Maxwell molecules the property that the collision rate is independent of the relative velocity of the two colliding particles but, on the other hand, obeys the collision rules of IHS. Although IMM do not represent any physical microscopic interaction potential, the cost of sacrificing physical realism can be compensated by the availability of exact analytical results. This fact has stimulated the use of IMM in the past few years [see for instance Refs.\ \cite{BCG00, CCG00, BK00, C01, EB02-1, EB02-2, EB02-3, BMP02, BK02-1, KB02, BK02-2, MP02, BC03, BCT03, SE03, BG06, ETB06-1, ETB06-2, BTE07, SGV09, GT11, GT12-1, SG12}). %\cite{IMM}
In addition, inelastic particle collisions can be introduced in the framework of the Boltzmann equation at the level of the cross section, without any reference to a specific interaction potential \cite{E81}. On the other hand, apart from its academic interest, it is worthwhile remarking that experiments for magnetic two-dimensional grains with dipolar interactions in air are well described by IMM \cite{KSSAOB05}.

The goal of this Chapter is to re-examine the problem studied in Chapter \ref{Chapter4} by considering the Boltzmann equation for IMM. The use of this model allows us to determine the expressions of the NS transport coefficients of a driven granular gas without taking any additional and sometimes uncontrolled approximations.

Moreover, the comparison between the results obtained from IMM with those derived from IHS can be used again as a test to assess the reliability of IMM as a prototype model for charaterizing real granular flows. Previous comparisons have shown a mild qualitative agreement in the freely cooling case \cite{S03,GA05} while the agreement between IMM and IHS significantly increases for low order velocity moments in the case of driven states (for instance, the simple shear flow problem) \cite{G03,G07,SG07,GT10}.

As we will show below, the explicit dependence of the transport coefficients of a driven granular gas on the parameters of the system requires in general to solve numerically a set of nonlinear differential equations. However, those equations become simple alebraic equations when steady states are considered. An interesting consequence of using IMM instead of IHS is that a velocity moment of order $k$ of the Boltzmann collision operator only involves moments of order less or equal than $k$. %Thus, the Boltzmann collision moments can be evaluated without the explicit knowledge of the distribution function \cite{GS07}. 
This property allows to get exact solutions for the Boltzmann equation and justifies the interest in IMM in the last years \cite{GS11}. In this Chapter the exact forms of the shear viscosity $\eta$, the thermal conductivity $\kappa$ and the transport coefficient $\mu$ 
% that relates the heat flux with the density gradient 
are determined as a function of the coefficent of restitution $\alpha$ and the thermostat parameters. As in the study of IHS case of the previous Chapter, the expressions of the transport coefficients are obtained by solving the Boltzmann equation for IMM up to first order in the spatial gradients by means of the CE expansion \cite{CHAPMAN70}. Again it has been taken into account that the zeroth-order distribution function $f^{(0)}$ is not in general a stationary distribution since the collisional cooling cannot be compensated locally by the energy supplied by the thermostat. Such energy unbalance introduces new contributions to the transport coefficents which not were considered in previous works \cite{GM02,G11} where local steady state was assumed at zeroth-order.

\section{Inelastic Maxwell Models}
\label{sec2chap5}
IMM are the starting point in this Chapter to determine the NS transport coefficients of a granular inelastic gas driven by stochastic and friction forces. %This problem was solved in the previous Chapter considering the IHS model by mean of Chapman-Enskog method. 
The Boltzmann equation for a driven granular gas modeled as a Maxwell gas of inelastic particles is given by Eq.\ \eqref{eqEB}, namely,
\begin{equation}
  \label{eq5.1}
\partial_{t}f + \mathbf{v} \cdot \mathbf{\nabla}f 
- \frac{\gamma_\text{b}}{m}\frac{\partial }{\partial{\bf v}}\cdot{\bf V}f 
- \frac{\gamma_\text{b}}{m} \Delta \mathbf{U} \cdot \frac{\partial }{\partial{\bf v}}f 
- \frac{1}{2}\xi_\text{b}^2\frac{\partial^2}{\partial v^2}f
= J_\text{IMM}\left[f,f\right],
%= J_\text{IMM}\left[\bt{v}_1|f,f\right],
\end{equation}
where the collision operator $J_\text{IMM}$ is
\begin{equation}
  \label{eq5.2} %  \label{2.2}
  J_\text{IMM}\left[f,f\right]=\frac{\nu(\bt{r},t)}{n(\bt{r},t) \Omega_d}\int\,d\bt{v}_2\int\,d\widehat{\bs{\sigma}}[\alpha^{-1}f(\bt{r},\bt{v}'_1,t)f(\bt{r},\bt{v}'_2,t) - f(\bt{r},\bt{v}_1,t)f(\bt{r},\bt{v}_2,t)].
\end{equation} 
Here $\Omega_d=2\pi^{d/2}/\Gamma(d/2)$ is the total solid angle in $d$ dimensions and the collision frequency $\nu(\bt{r},t)$ is independent of velocity but depends on space and time through its dependence on density and temperature. This frequency can be considered as a free parameter of the model that can be chosen to optimize the agreement with some property of interest of the original Boltzmann equation for IHS.
Moreover, in order to capture in an effective way the velocity dependence of the original IHS collision rate, one usually assumes that the IMM collision rate is proportional to $T^q$ with $q=\frac{1}{2}$. We take $q$ as a generalized exponent so that different values of $q$ can be used to mimic different potentials. As in previous works on IMM \cite{G07,SG07,GT10}, we will assume that $\nu \propto n T^q$, with $q \geq 0$. The case $q=0$ is closer to the original Maxwell model of elastic particles while the case $q=\frac{1}{2}$ is closer to hard spheres. Thus, the collision frequency can be written as \cite{GS11}
\begin{equation}
\label{eq5.3} %\label{2.4} 
\nu=A n T^q,
\end{equation}
where the value of the quantity $A$ will be chosen later.

The macroscopic balance equations for density, momentum, and energy follow directly from Eq.\ \eqref{eq5.1} by multiplying with $1$, $m{\bf v}$, and $\frac{1}{2}mv^2$ and integrating over ${\bf v}$. The result is the same as in Chapter \ref{Chapter2}, Eqs.\ \eqref{eqBaln01}--\eqref{eqBalT01}, where the pressure tensor $\mathsf{P}$ and the heat flux ${\bf q}$ are given by Eqs.\ \eqref{eqPk} and \eqref{eqqk}, respectively. The cooling rate  $\zeta_\text{IMM}$ is defined by Eq.\ \eqref{eqzeta}.
%\begin{equation}
%\label{2.7} D_{t}n+n\nabla \cdot {\bf U}=0\;,
%\end{equation}
%\begin{equation}
%\label{2.8} D_{t}U_i+(mn)^{-1}\nabla_j P_{ij}=0\;,
%\end{equation}
%\begin{equation}
%\label{2.9} D_{t}T+\frac{2}{dn}\left(\nabla \cdot {\bf q}+P_{ij}\nabla_j U_i\right) =-\frac{2 T}{m}\gamma_\text{b}+m \xi_\text{b}^2 -\zeta_\text{IMM} T\;.
%\end{equation}
%Here, $D_{t}=\partial _{t}+{\bf U}\cdot \nabla$ and the microscopic expressions for the pressure tensor ${\sf P}$, the heat flux ${\bf q}$, and the cooling rate $\zeta_\text{IMM}$ are given, respectively, by:
%\begin{equation}
%{\sf P}({\bf r}, t)=\int \dd{\bf v}\,m{\bf V}{\bf V}\,f({\bf r},{\bfv},t),
%\label{2.10}
%\end{equation}
%\begin{equation}
%{\bf q}({\bf r}, t)=\int \dd{\bf v}\,\frac{1}{2}m V^{2}{\bf V}\,f({\bf r},{\bf v},t), \label{2.11}
%\end{equation}
%\begin{equation}
%\label{eq5.4} \zeta_\text{IMM}({\bf r}, t)=-\frac{1}{dn({\bf r},t)T({\bf r},t)}\int\, \dd{\bf v} \; m\; V^2\; J_\text{IMM}\left[f,f\right].
%\end{equation}

Note that the balance equations apply regardless of the details of the interaction model considered. The influence of the collision model appears through the $\alpha$-dependence of the cooling rate and of the momentum and heat fluxes.

One of the main advantages of using IMM instead of IHS is that the collisional moments of $J_\text{IMM}$ can be \emph{exactly} computed. %evaluated in terms of the moments of the distribution $f$, without the explicit knowledge of the latter \cite{TRUESDELL80}. More explicitly, the collisional moments of order $k$ are given as a bilinear combination of moments of order $k'$ and $k''$ with $0\leq k'+k''\leq k$. In particular, the collisional moments involved in the calculation of the momentum and heat fluxes as well as in the fourth cumulant are given by \cite{S03,GS07}:
The first few collisional moments of $J_\text{IMM}$ are provided in Appendix \ref{AppendixE}. In particular the cooling rate $\zeta_\text{IMM}$ can be determined by taking the trace in Eq.\ \eqref{eq5.5}. It is exactly given by \cite{S03}
\begin{equation}
\label{eq5.13.1} \zeta_\text{IMM}=\frac{1-\alpha^2}{2d}\nu.
\end{equation}
Note that while in the case of IHS, the cooling rate $\zeta_\text{IMM}$ is also expressed as a functional of the hydrodynamic fields, $\zeta_\text{IMM}$ is just proportional to $\nu$ in the case of IMM.

In order to compare the results derived here for IMM with those obtained \cite{GCV13} for IHS, we now need a criterion to fix the parameter $\nu$ [or the quantity $A$ in Eq.\ \eqref{eq5.3}]. As in previous works on IMM \cite{S03,G03,GA05,G07,GS07,GS11}, an appropriate choice to optimize the agreement with the IHS results is to take $\nu$ under the criterion that the cooling rate of IMM [as given by Eq. \eqref{eq5.13.1}] with $q=\frac{1}{2}$ be the same as the one obtained for IHS of diameter $\sigma$ evaluated in the Maxwellian approximation \cite{GoSh95,NE98}. With this choice, the collision frequency $\nu$ is
\begin{equation}
\label{eq5.13.2}
\nu=\frac{d+2}{2}\nu_0,
\end{equation}
where
\begin{equation}
\label{eq5.13.3}
\nu_0=\frac{4\Omega_d}{\sqrt{\pi}(d+2)}n \sigma^{d-1} \sqrt{\frac{T}{m}}.
\end{equation}
The collision frequency $\nu_0$ is the one associated with the NS shear viscosity of an ordinary gas ($\al=1$) of both Maxwell molecules and hard spheres, i.e., $\eta_0=p/\nu_0$.

\section{Homogeneous steady states}
\label{sec3chap5}

Before analyzing non-homogeneous states, it is quite convenient first to study the homogeneous problem. In this case, the density $n$ is constant, the flow velocity vanishes and the temperature $T(t)$ is spatially uniform. Consequently, the Boltzmann equation \eqref{eq5.1} becomes
\begin{equation}
\label{eq5.14}
\partial_{t}f-\frac{\gamma_\text{b}}{m} \frac{\partial}{\partial{\bf v}}\cdot {\bf v} f-\frac{1}{2}\xi_\text{b}^2\frac{\partial^2}{\partial v^2}f=J_\text{IMM}[f,f].
\end{equation}
Since the heat flux vanishes and the pressure tensor is diagonal ($P_{ij}=p\delta_{ij}$, where $p=nT$ for a dilute granular gas), then the energy balance equation \eqref{eqBalT01} reads simply
\begin{equation}
\label{eq5.15}
\partial_tT=-\frac{2 T}{m}\gamma_\text{b} +m \xi_\text{b}^2-\zeta_\text{IMM} \,T.
\end{equation}

Following the same steps as in Chapter \ref{Chapter3}, Eq.\ \eqref{eq5.14} can be written in terms of the scaled steady distribution function $\varphi_\text{s}(\bt{c},\xi_\text{s}^*)$. The result is
\begin{equation}
\label{eq5.16}
\frac{1}{2}(\zeta_\text{IMM,s}^* - \xi_s^*) \frac{\partial}{\partial\mathbf{c}} \cdot \mathbf{c} \varphi_\text{s}- \frac{1}{4}\xi_s^* \frac{\partial^2}{\partial c^2} \varphi_\text{s} = J^*[\varphi_\text{s},\varphi_\text{s}],
\end{equation}
where $\zeta_\text{IMM,s}^* \equiv \zeta_\text{IMM,s}/\nu_\text{s}=(1-\al^2)/2d$, $\xi_\text{s}^*=m\xi_\text{b}^2/T_\text{s}\nu_\text{s}$, $J^*[\varphi_\text{s},\varphi_\text{s}]\equiv v_0^d J_\text{IMM}[f_\text{s},f_\text{s}]/(n_\text{s} \nu_\text{s})$ and $\nu_\text{s}=A n_\text{s} T_\text{s}^q$. Here, as before, the subindex $s$ means that all the quantities are evaluated in the steady state.

As already noted in Chapter \ref{Chapter3}, the scaled distribution $\varphi_\text{s}$ depends on the granular temperature through the scaled velocity $\mathbf{c}$ and \emph{also} through the (reduced) noise strength $\xi_s^*$.%On the contrary, in the homogeneous cooling state and in the case of only one thermostat force, the dependence of $\varphi_\text{s}$ is only encoded by the single parameter $\mathbf{c}$ \cite{S03}. In dimensionless form, Eq.\ \eqref{3.6} can be written as

In reduced units, the steady state condition $(\partial_t T=0)$ yields
\beq
\label{eq5.17}
2\gamma_\text{s}^*=\xi_\text{s}^*-\zeta_\text{IMM,s}^*,
\eeq
where $\gamma_\text{s}^*\equiv \gamma_\text{b}/(m \nu_\text{s})$. Since $\gamma_\text{s}^*$ is positive definite, then Eq.\ \eqref{eq5.17} requires that $\xi_\text{s}^*\geq \zeta_\text{IMM,s}^*$. Thus, at a given value of $\al$, there is a minimum threshold value $\xi_\text{th}^*(\al)=\zeta_\text{IMM,s}^*$ needed to achieve a steady state. In particular, for spheres ($d=3$), the smallest value of $\xi_\text{th}^*(\al)$ is $1/6$ (which corresponds to $\al=0$) while the smallest value of $\xi_\text{th}^*(\al)$ for disks ($d=2$) is $1/4$.

%%%%%%%%%%%%%
%%%%%%%%%%%%%
\begin{figure}[h]
  \centering
  \includegraphics[width=0.75 \columnwidth,angle=0]{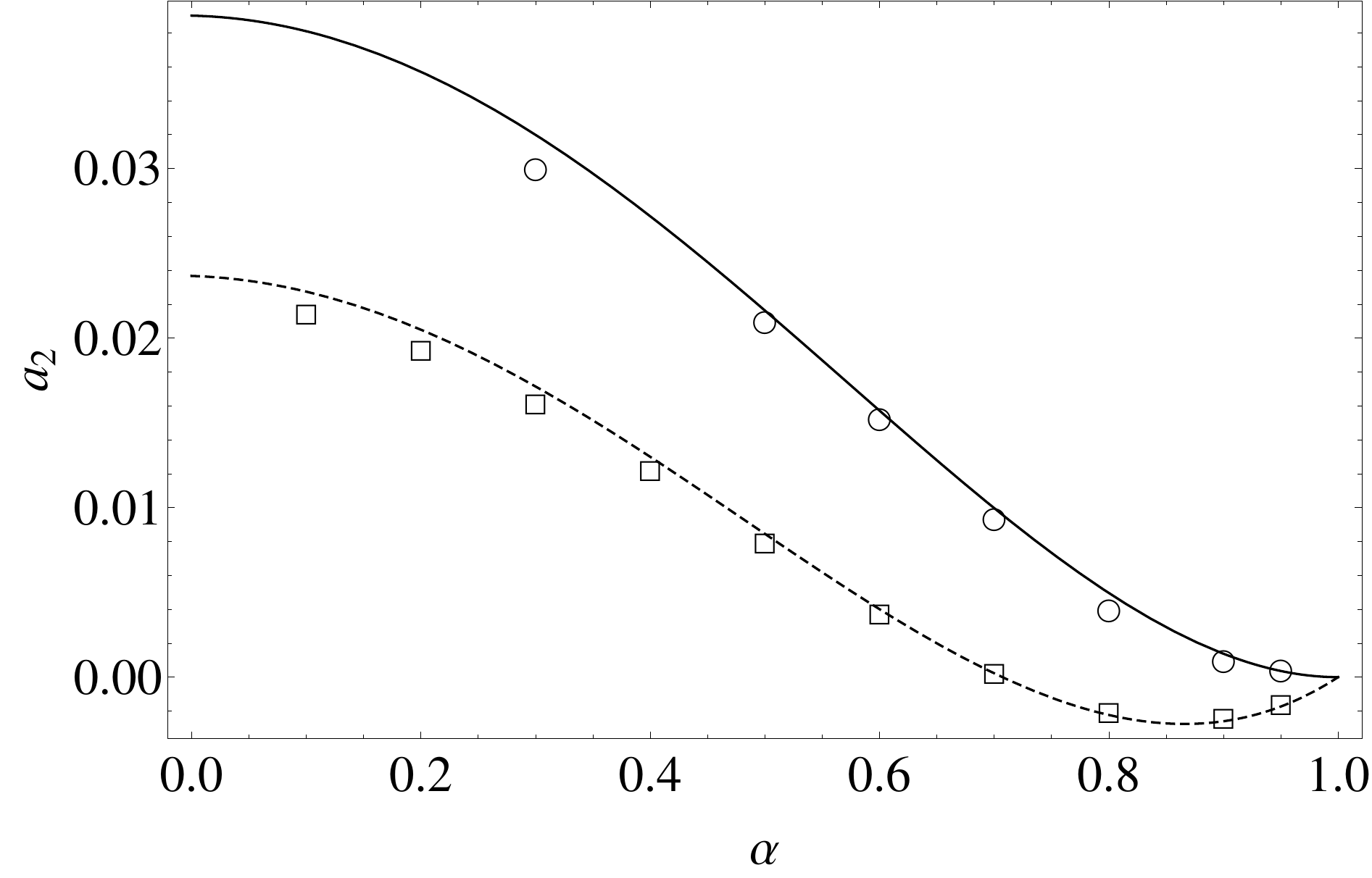}
\caption{The \emph{steady} fourth-cumulant $a_{2,\text{s}}$ as a function of the coefficient of restitution for a three-dimensional system ($d=3$) for $\xi_\text{s}^*=0.62$. The solid and dashed lines are the results obtained for IMM and IHS, respectively. The symbols (circles for IMM and squares for IHS) refer to the Monte Carlo simulation results. \label{fig01chap5}}
\end{figure}
%%%%%%%%%%%%%%%
%%%%%%%%%%%%%%%

In the case of elastic collisions ($\al=1$), $\zeta_\text{s}^*=0$ and the solution to Eq.\ \eqref{eq5.16} is the Maxwellian distribution $\varphi_\text{M}(c)=\pi^{-d/2}e^{-c^2}$. On the other hand, if $\al\neq 1$, then $\zeta_\text{s}^*\neq 0$ and as for IHS the solution to Eq.\ \eqref{eq5.16} is not exactly known. As said in Chapter \ref{Chapter3}, an indirect information of the deviation of $\varphi_\text{s}(c)$ from its Maxwellian form $\varphi_\text{M}(c)$ is given by the kurtosis or fourth-cumulant $a_\text{2,s}$ defined as
\begin{equation}
\label{eq5.18}
a_{2,\text{s}}=\frac{4}{d(d+2)}\langle c^4\rangle-1,
\end{equation}
where $\langle c^k\rangle$ is defined in Eq.\ \eqref{eqcaverage}. 
%\begin{equation}
%\label{eq5.19}
%\langle c^k\rangle=\int\; \dd{\bf c}\; c^k \varphi_s(c).
%\end{equation}
In order to determine $a_{2,\text{s}}$, we multiply Eq.\ \eqref{eq5.16} by $c^4$ and integrate over velocity. The result is
\begin{equation}
\label{eq5.20}
2(\zeta_\text{IMM,s}^*-\xi_s^*)\left(
1+a_{2,\text{s}}\right)+2\xi_s^*=\left(
1+a_{2,\text{s}}\right)\nu_{4|0}^*-\frac{d}{(d+2)}\lambda_1^*,
\end{equation}
where $\nu_{4|0}^*\equiv \nu_{4|0}/\nu_\text{s}$, $\lambda_1^*\equiv \lambda_1/\nu_\text{s}$ and use has been made of Eq.\ \eqref{eq5.7}. The solution to Eq.\ \eqref{eq5.16} is
\begin{eqnarray}
\label{eq5.21}
a_{2,\text{s}}&=&\frac{2\zeta_\text{IMM,s}^*-\nu_{4|0}^*+\frac{d}{d+2}\lambda_1^*}{\nu_{4|0}^*-2(\zeta_\text{IMM,s}^*-\xi_s^*) } \nonumber\\
&=& \frac{6(1-\al^2)^2} {4d-7+   8(d-1)\al +  (2+4d-3\al^2)\al^2 + 16d(d+2)\xi_\text{s}^*}. %%\frac{6(1-\al^2)^2}{4d-7+3\al(2-\al)+16d(d+2)\xi_\text{s}^*}.
\end{eqnarray}

%where the explicit forms of $\zeta_\text{IMM,s}^*$ and $\nu_{4|0}^*$ have been considered. 
In the absence of friction ($\gamma_\text{b}=0$), the steady state condition \eqref{eq5.17} becomes $\xi_\text{s}^*=\zeta_\text{IMM,s}^*$ and Eq.\ \eqref{eq5.21} yields back the results of the theory of a driven granular gas heated only by the stochastic thermostat \cite{S03}, i.e.,
\beq
\label{eq5.22}
a_{2,\text{s}}=\frac{6(1-\al)^2(1+\al)}{12d+9-\alpha(4d+17)+3\alpha^2(1-\alpha)}.
\eeq
Moreover, when $\xi_\text{s}^*=0$, Eq.\ \eqref{eq5.21} is consistent with the one obtained for IMM in the freely cooling case \cite{S03}.

Fig.\ \ref{fig01chap5} shows the steady value of the fourth-cumulant $a_{2,\text{s}}$ versus the coefficient of restitution $\al$ for a three-dimensional system. The theoretical results derived here for IMM given by Eq.\ \eqref{eq5.21} and in Ref.\ \cite{CVG13} [see Eq.\ \eqref{apG7}] for IHS are compared with those obtained by numerically solving the Boltzmann equation for IMM and IHS by means of the DSMC method \cite{BIRD94}. The parameters of the simulations 
%are $m=1$, $\sigma=??$, $\xi_\text{b}^2=??$ and $T_\text{b}=??$ that yields 
for IMM and IHS have been chosen to get $\xi_\text{s}^*=0.62$ in the steady state. It appears that the homogeneous state of IMM deviates from the gaussian distribution $\varphi_\text{M}(c)$ (which corresponds to $a_2=0$) slightly more than the homogeneous state for IHS.
%is seen that the homogeneous state of IHS deviates from the Maxwellian distribution $\varphi_\text{M}(c)$ (which corresponds to $a_2=0$) more than the homogeneous state of IHS. 
This behavior contrasts with the results obtained in the freely cooling case \cite{S03} where the magnitude of $a_\text{2,s}$ for IMM is much larger than that of IHS. As expected, the simulation data for IMM show an excellent agrement with the exact result \eqref{eq5.21}, even for quite small values of $\al$.

\section{Chapman-Enskog method for states close to homogeneous steady states}
\label{sec4chap5}

Let us slightly disturb the homogeneous steady state by small spatial perturbations. In this case, the momentum and heat fluxes are not zero and their corresponding transport coefficients can be identified. The evaluation of these coefficients as functions of the coefficient of restitution and the parameters of the external force is the main goal of the present Section.

As long as the spatial gradients keep small, the Boltzmann equation \eqref{eq5.1} may be solved by means of the CE method \cite{CHAPMAN70} adapted to inelastic collisions. 
%Here, Eq.\ \eqref{eq5.1} will be solved up to the first order in spatial gradients. 
Since the procedure to obtain the kinetic equations for the zeroth- and first-order approximations $f^{(0)}$ and $f^{(1)}$ is similar to those made in Chapter \ref{Chapter4} for IHS, only some intermediate steps will be displayed in this Section.

\subsection{Zeroth-order approximation}

%Following the same steps as in Chapter \ref{Chapter4}, inserting Eqs.\ \eqref{eqfexpand} and \eqref{eqJexpand} into Eq. \eqref{eq5.1} and arranging terms of the same order in the parameter $\epsilon$ we can calculate the different approximations to $f$ separately. 

The zeroth-order approximation $f^{(0)}$ obeys the kinetic equation:
%The kinetic equation to the zeroth-order in the expansion for the distribution $f^{(0)}$ is:
\beq
-\left(\frac{2}{m}\gamma_\text{b} -\frac{m}{T} \xi_\text{b}^2+\zeta_\text{IMM}\right) T\frac{\partial f^{(0)}}{\partial T}-\frac{\gamma_\text{b}}{m}
\frac{\partial}{\partial{\bf v}}\cdot {\bf V}
f^{(0)}-\frac{1}{2}\xi_\text{b}^2\frac{\partial^2}{\partial v^2}f^{(0)}=J_\text{IMM}[f^{(0)},f^{(0)}].
\label{eq5.23}
\eeq
As already noted in the case of IHS in Chapter \ref{Chapter4}, %\cite{GCV13}, 
since density and temperature are specified separately in the \emph{local} reference state $f^{(0)}$, the collisional cooling and the action of the thermostats do not in general cancel each other at all points in the system. Thus, $\partial_t^{(0)} T \neq 0$ and $f^{(0)}$ depends on time through its dependence on the temperature.

In addition, although $\gamma_\text{b}$ and $\xi_\text{b}^2$ can be considered in general as independent parameters, we will assume, analogously to the procedure for IHS in Chapter \ref{Chapter4}, %\cite{GCV13}, 
that both parameters are related by Eq.\ \eqref{eqgamG}.
%\begin{equation}
%\label{eq5.24}
%\gamma_\text{b}=\beta \frac{m^2\xi_\text{b}^2} {T_\text{b}},
%\end{equation}
%where $\beta$ is an \emph{arbitrary} constant and $T_\text{b}$ is the bath temperature. When $\beta=0$ (or equivalently, when $\gamma_\text{b}=0$ but $\gamma_\text{b}T_\text{b}\equiv \text{finite}$) our thermostat reduces to the usual stochastic thermostat \cite{WM96,GMT12} while the choice $\beta=\frac{1}{2}$ reduces to the conventional Fokker-Planck model for ordinary gases \cite{RESIBOIS77, KAMPEN81, H03, GSVP11-1}. Although our results apply for any value of $\beta$, we will consider the latter two values of $\beta$ for practical applications. Eq.\ \eqref{eq5.24} can be rewritten as
In dimensionless form, Eq.\ \eqref{eqgamG} can be written as
\begin{equation}
\label{eq5.25}  %\label{4.11}
\gamma^*=\beta T^* \xi^*=\theta \xi^{*q/(1+q)},
\end{equation}
where $\gamma^*=\gamma_\text{b}/m\nu$, $T^*\equiv T/T_\text{b}$ and
\begin{equation}
\label{eq5.26}
\theta\equiv \beta \left(\frac{m \xi_\text{b}^2}{A n T_\text{b}^{1+q}}\right)^{1/(1+q)}.
\end{equation}
Upon writing Eq.\ \eqref{eq5.25}, use has been made of the identity $\beta T^*=\theta/\xi^{*1/(1+q)}$, where
\begin{equation}
\label{eq5.27}
\xi^*\equiv \frac{m\xi_\text{b}^2}{T \nu(T)}=\frac{m\xi_\text{b}^2}{A n T^{q+1}}.
\end{equation}

Under \emph{unsteady} state, dimensional analysis requires that the zeroth-order distribution $f^{(0)}(\mathbf{r}, \mathbf{v}, t)$ has the scaled form \eqref{eqf0scaled} [once one uses the relation \eqref{eq5.25}], namely
\beq
\label{eq5.28}
f^{(0)}({\bf r}, {\bf v}, t)=n({\bf r},t) v_0({\bf r},t)^{-d}\varphi\left(\mathbf{c}, \theta, \xi^*\right),
\eeq
where now $\mathbf{c}\equiv \mathbf{V}/v_0$, $\mathbf{V}={\bf v}-{\bf U}$ being the peculiar velocity. The temperature dependence of the reduced distribution  $\varphi$ is encoded by the dimensionless velocity $\mathbf{c}$ and the (reduced) noise strength $\xi^*$. Consequently, according to Eq.\ \eqref{eq5.28}, one gets
\begin{equation}
\label{eq5.29}
T\frac{\partial}{\partial T} f^{(0)} =-\frac{1}{2}\frac{\partial}{\partial \mathbf{V}} \cdot \mathbf{V} f^{(0)} -(1+q) \xi^* \frac{\partial}{\partial \xi^*} f^{(0)},
\end{equation}
and the scaled distribution $\varphi$ obeys the kinetic equation
\begin{equation}
\label{eq5.30} 
(1+q)\left[ \left( 2\beta T^* -1 \right) \xi^* + \zeta^* \right] \xi^*
\frac{\partial \varphi}{\partial \xi^*}+
\frac{1}{2}(\zeta^* - \xi^*) \frac{\partial}{\partial\mathbf{c}} \cdot \mathbf{c}\varphi
- \frac{1}{4}\xi^* \frac{\partial^2 \varphi}{\partial c^2}
= J_\text{IMM}^*[\varphi,\varphi],
\end{equation}
where use has been made of the identity \eqref{eq5.25}.

% red: creo que el parrafo de abajo ya se ha dicho para el estado homogeneo, y obviamente tambien es valido para el estado inhomogeneo
%As for IHS \cite{GCV13}, the form of the distribution $\varphi(c,\theta,\xi^*)$ is not \emph{exactly} known. A convenient and useful way of characterizing $\varphi(c,\theta,\xi^*)$ in the range of low to intermediate velocities is through its first cumulants $a_p$. In particular, t

An implicit expression of the fourth-cumulant $a_2(\xi^*)$ [defined by Eq.\ \eqref{eq5.18}] can be obtained for unsteady states by multiplying both sides of Eq.\ \eqref{eq5.30} by $c^4$ and integrating over velocity. The result is
\beq
\label{eq5.30.1} %\label{4.16}
(1+q)\left[ \left( 2\beta T^* -1 \right) \xi^* +\zeta^*\right]\xi^*
\frac{\partial a_2}{\partial\xi^*}=\frac{d}{d+2}\lambda_1^*+(1+a_2)(2\zeta^*-\nu_{4|0}^*) -2\xi^* a_2.
\eeq

% red: esta frase de abajo tambien se podria quitar y quedaria mas resumido, sin perder, creo, significado
%In the steady state, Eq.\ \eqref{3.9.1} applies and thus, the left hand side of Eq.\ \eqref{eq5.30.1} vanishes. In this case, the analytical solution to Eq.\ \eqref{eq5.30.1} is given by Eq.\ \eqref{3.13}. On the other hand, in the unsteady state,

In Eq.\ \eqref{eq5.30.1}, the function $a_2(\xi^*)$ must be obtained numerically. As we will show later, evaluation of the transport coefficients in the steady state requires the knowledge of the derivatives
$\partial a_2/\partial \xi^*$ and $\partial a_2/\partial \theta$ in this state. The explicit expressions of these derivatives are given by Eqs. \eqref{apG8} and \eqref{apG9} of Appendix \ref{AppendixG}.

\subsection{First-order approximation}
The velocity distribution function $f^{(1)}$ verifies the kinetic equation
%\begin{eqnarray}
%\label{eq5.31} %\label{4.21}
%\left(\partial_{t}^{(0)}+{\cal L}\right)f^{(1)}-\frac{\gamma_\text{b}}{m}
%\frac{\partial}{\partial {\bf v}}\cdot {\bf V}f^{(1)}-\frac{1}{2}\xi_\text{b}^2\frac{\partial^2}{\partial v^2}f^{(1)}= \nonumber \\
%-\left(\partial_{t}^{(1)}+{\bf v}\cdot \nabla \right)f^{(0)} + \frac{\gamma_\text{b}}{m}\Delta {\bf U}\cdot\frac{\partial f^{(0)}}{\partial {\bf V}} ,
%\end{eqnarray}
%Inserting the macroscopic balance equations to first order in gradientes \eqref{apB3}--\eqref{apB5} into Eq.\ \eqref{eq5.31} leads to
%The macroscopic balance equations \eqref{2.7}--\eqref{2.9} to first order in the gradients are
%\begin{equation}
%\label{eq5.33}
%D_t^{(1)}n=-n\nabla\cdot {\bf U},\quad
%D_t^{(1)}U_i=-(mn)^{-1}\nabla_i p,
%\end{equation}
%\begin{equation}
%\label{eq5.34}
%D_t^{(1)}T=-\frac{2p}{dn}\nabla\cdot {\bf U},
%\end{equation}
%where $D_t^{(1)}\equiv \partial_t^{(1)}+{\bf U}\cdot \nabla$ and $p=nT$ is the hydrostatic pressure. Use of Eqs.\ (\ref{eq5.33}) in Eq.\ (\ref{eq5.31}) leads to
\beqa
\label{eq5.35} %\label{4.25}
\left(\partial_{t}^{(0)}+{\cal L}\right)f^{(1)}-\frac{\gamma_\text{b}}{m}
\frac{\partial}{\partial {\bf v}}\cdot {\bf V}
f^{(1)}-\frac{1}{2}\xi_\text{b}^2\frac{\partial^2}{\partial v^2}f^{(1)}
= {\bf A}\cdot \nabla \ln T+{\bf B}\cdot \nabla \ln n \nonumber\\
+C_{ij}\frac{1}{2}\left( \nabla_{i}U_{j}+\nabla_{j}U_{i}-\frac{2}{d}\delta _{ij}\nabla \cdot\mathbf{U} \right)+D \nabla \cdot\mathbf{U},
\eeqa
where ${\cal L}$ is the linearized Boltzmann collision operator
\begin{equation}
\label{eq5.32}%\label{4.22}
{\cal L}f^{(1)}=-\left(J_\text{IMM}[f^{(0)},f^{(1)}]+J_\text{IMM}[f^{(1)},f^{(0)}]\right),
\end{equation}
and
\begin{equation}
{\bf A}\left( \mathbf{V}\right)=-\mathbf{V}T\frac{\partial f^{(0)}}{\partial T}
-\frac{p}{\rho }\frac{\partial f^{(0)}}{\partial \mathbf{V}},  \label{eq5.36}
\end{equation}
\beq
{\bf B}\left(\mathbf{V}\right)= -{\bf V}n\frac{\partial f^{(0)}}{\partial n}-\frac{p}{\rho}
\frac{\partial f^{(0)}}{\partial \mathbf{V}},  \label{eq5.37}
\eeq
\begin{equation}
\label{eq5.38}
C_{ij}\left(
\mathbf{V}\right)=V_i\frac{\partial f^{(0)}}{\partial V_j},
\end{equation}
\beq
D=\frac{1}{d}\frac{\partial}{\partial \mathbf{V}}\cdot (\mathbf{V}
f^{(0)})+\frac{2}{d}T\frac{\partial f^{(0)}}{\partial T}-f^{(0)}+n\frac{\partial f^{(0)}}{\partial n}.   \label{eq5.39}%\label{4.29}
\eeq
In Eq.\ \eqref{eq5.36}, $T\partial f^{(0)}/\partial T$ is given by Eq.\ \eqref{eq5.29} while, according to Eqs.\ \eqref{eq5.25}--\eqref{eq5.28}, the term $n\frac{\partial f^{(0)}}{\partial n}$ can be more explicitly written as
\begin{equation}
\label{eq5.40}
n\frac{\partial f^{(0)}}{\partial n}=f^{(0)}-\xi^*\frac{\partial f^{(0)}}{\partial \xi^*}
-\frac{\theta}{1+q}\frac{\partial f^{(0)}}{\partial \theta}.
\end{equation}
It is worth noticing that for $q=\frac{1}{2}$, Eqs.\ \eqref{eq5.35}--\eqref{eq5.39} have the same structure as that of the Boltzmann equation for IHS \cite{GCV13}. The only difference between both models lies in the explicit form of the linearized operator ${\cal L}$.

\section{Transport coefficients}
\label{sec5chap5}

The relevant transport coefficients can be identified from the expressions of the first-order contributions to the pressure tensor
\beq
\label{eq5.41} %\label{5.1}
{\sf P}^{(1)}=\int\; \dd \mathbf{v}\; m \mathbf{V} \mathbf{V} f^{(1)}(\mathbf{V}),
\eeq
and the heat flux vector
\beq
\label{eq5.42} %\label{5.2}
\mathbf{q}^{(1)}=\int\; \dd \mathbf{v}\; \frac{m}{2} V^2 \mathbf{V} f^{(1)}(\mathbf{V}).
\eeq

The evaluation of the above fluxes has been worked out in Appendix \ref{AppendixF}. Only the final results are presented in this Section. As expected, the pressure tensor $P_{ij}^{(1)}$ is given by
\begin{equation}
\label{eq5.42.1} %\label{5.2.1}
P_{ij}^{(1)}=-\eta\left( \nabla_{i}U_{j}+\nabla_{j
}U_{i}-\frac{2}{d}\delta _{ij}\nabla \cdot
\mathbf{U} \right),
\end{equation}
while the heat flux $\mathbf{q}^{(1)}$ is
\beq
\label{eq5.42.2} %\label{5.2.2}
\mathbf{q}^{(1)}=-\kappa \nabla T-\mu \nabla n.
\eeq
%Here, $\eta$ is the shear viscosity coefficient, $\kappa$ is the thermal conductivity coefficient and $\mu$ is a new transport coefficient not present for ordinary gases. 
As for IHS, the transport coefficients $\eta$, $\kappa$ and $\mu$  can be written in the form
\beq
\label{eq5.43} %\label{5.3}
\eta=\eta_0 \eta^*, \quad \kappa=\kappa_0 \kappa^*, \quad \mu=\frac{\kappa_0 T}{n}\mu^*,
\eeq
where $\eta_0=(d+2)\frac{p}{2\nu}$ and $\kappa_0=\frac{d(d+2)}{2(d-1)}\frac{\eta_0}{m}$ 
%\begin{equation}
%\label{eq5.43.1}
%  \eta_0=(d+2)\frac{p}{2\nu}
%\end{equation}
%and 
%\begin{equation}
%\label{eq5.43.2}
%\kappa_0=\frac{d(d+2)}{2(d-1)}\frac{\eta_0}{m}  
%\end{equation}
are the shear viscosity and thermal conductivity coefficients, respectively, of a dilute ordinary gas. The reduced coefficients $\eta^*$, $\kappa^*$ and $\mu^*$ depend on temperature through its dependence on the (reduced) noise strength $\xi^*$. They verify the following first-order differential equations:

\beq
\label{eq5.44} %\label{5.4}
\Lambda^*\left[(1-q)\eta^*-(1+q)\xi^*\frac{\partial \eta^*}{\partial \xi^*}\right]+
\left(\nu_{0|2}^*+2\gamma^*\right)\eta^*=\frac{2}{d+2},
\eeq
\beqa
\label{eq5.45} %\label{5.5}
\Lambda^*\left[(1-q)\kappa^*-(1+q)\xi^*\frac{\partial \kappa^*}{\partial \xi^*}\right]+\left(\Lambda^*-\xi^*-q \zeta^*+\nu_{2|1}^*+3\gamma^*\right)\kappa^* 
\nonumber\\
=\frac{2(d-1)}{d(d+2)}\left[1+2a_2-
(1+q)\xi^*\frac{\partial a_2}{\partial \xi^*}\right],
\eeqa

\beqa
\label{eq5.46} %\label{5.6}
\Lambda^*\left[(2-q)\mu^*-(1+q)\xi^*\frac{\partial \mu^*}{\partial \xi^*}\right]+\left(\nu_{2|1}^*+3\gamma^*\right)\mu^*
\nonumber\\
=\zeta^*\kappa^*
+\frac{2(d-1)}{d(d+2)}\left(a_2-\frac{\theta}{1+q}\frac{\partial a_2}{\partial \theta}-\xi^*\frac{\partial a_2}{\partial \xi^*}\right).
\eeqa
Here,
\beq
\label{eq5.47} %\label{5.7}
\Lambda^*=\xi^*-2\gamma^*-\zeta^*,
\eeq
$\nu_{0|2}^*\equiv \nu_{0|2}/\nu$ and $\nu_{2|1}^*\equiv \nu_{2|1}/\nu$, where $\nu_{0|2}$ and $\nu_{2|1}$ are defined by Eqs.\ \eqref{eq5.8} and \eqref{eq5.9}, respectively.

Apart from the transport coefficients (which are directly related to the second- and third-degree velocity moments of the first order distribution function $f^{(1)}$), another interesting velocity moment of $f^{(1)}$ corresponds to its fourth degree isotropic moment defined as
\beq
\label{eq5.48} %\label{5.8}
e_D=\frac{1}{2d(d+2)}\frac{m^2}{nT^2}\int\; \dd\mathbf{v}\; V^4 f^{(1)}.
\eeq
In dimensionless form, the coefficient $e_D$ is given by
\beq
\label{eq5.49} %\label{5.9}
e_D=e_D^* \nu^{-1} \nabla \cdot \mathbf{U},
\eeq
where $e_D^*$ is the solution of the first-order differential equation 
\beqa
\label{eq5.50} %\label{5.10}
\Lambda^*\left[(2-q)e_D^*-(1+q)\xi^*\frac{\partial e_D^*}{\partial \xi^*}\right]+\left(\nu_{4|0}^*+4\gamma^*\right)e_D^* &=& -\frac{2(1+q)+d}{2d}\xi^*\frac{\partial a_2}{\partial \xi^*} \nonumber\\
&-&\frac{1}{2}\frac{\theta}{1+q}\frac{\partial a_2}{\partial \theta}.
\eeqa
Here, $\nu_{4|0}^*\equiv \nu_{4|0}/\nu$ where $\nu_{4|0}$ is defined by Eq.\ \eqref{eq5.10}.

In the elastic limit ($\al=1$), $\zeta_\text{s}^*=0$, $a_{2,\text{s}}=0$, $\gamma_\text{s}^*=\xi_\text{s}^*/2$, $\nu_{0|2}^*=2/(d+2)$, and $\nu_{2|1}^*=2(d-1)/d(d+2)$. In this case, $\mu_\text{s}^*=e_D^*=0$ and the coefficients $\eta_\text{s}^*$ and $\kappa_\text{s}^*$ become, respectively,
\beq
\label{eq5.51} %\label{5.19}
\eta_\text{s}^*\to \eta_\text{s,0}^*=\frac{1}{1+\frac{d+2}{2}\xi_\text{s}^*},\qquad
\kappa_\text{0}^*\to \kappa_\text{s,0}^*=\frac{1}{1+\frac{d(d+2)}{4(d-1)}\xi_\text{s}^*}.
\eeq

An interesting limit case is the freely cooling gas ($\gamma^*=\xi^*=0$). In this case, $\Lambda^*=-\zeta^*$ and Eq.\ \eqref{eq5.50} gives $e_D^*=0$. In addition, the solution to Eqs.\ \eqref{eq5.44}--\eqref{eq5.46} can be written as
\beq
\label{eq5.52} %\label{5.11}
\eta^*=\frac{2}{d+2}\frac{1}{\nu_{0|2}^*-(1-q)\zeta^*},
\eeq
\beq
\label{eq5.53} %\label{5.12}
\kappa^*=\frac{2(d-1)}{d(d+2)}\frac{1+2a_2}{\nu_{2|1}^*-2\zeta^*},
\eeq
\beq
\label{eq5.54} %\label{5.13}
\mu^*=\frac{\kappa^*}{1+2a_2}\frac{\zeta^*+\nu_{2|1}^*a_2}{\nu_{2|1}^*-\left(2-q\right)\zeta^*}.
\eeq
When $q=\frac{1}{2}$, Eqs.\ \eqref{eq5.52}--\eqref{eq5.54} agree with those previously derived \cite{S03} for an undriven granular gas of IMM.

Apart from the above two situations (elastic collisions and undriven granular gas), the evaluation of the transport coefficients ($\eta^*$, $\kappa^*$, $\mu^*$, and $e_D^*$ ) for the general case of unsteady states requires to solve the differential equations \eqref{eq5.44}--\eqref{eq5.46} and \eqref{eq5.50}.
However, even for the simplest model ($q=0$), it is not possible to obtain an analytical solution to this system of equations, except in the steady state limit. For the steady state ($\Lambda^*=0$), we need still to evaluate the derivatives $\partial a_2/\partial\xi^*$ and $\partial a_2/\partial\theta$. The steady state expressions of these derivatives may be easily deduced, as we will show, from the simplified steady state form of Eq. \eqref{eq5.30.1}. We present the results for steady states in the next Subsection.

\subsection{Transport coefficients under steady state}

Under steady state ($\Lambda^*=0$), the set of differential equations \eqref{eq5.44}--\eqref{eq5.46} and \eqref{eq5.50} becomes a simple set of algebraic equations whose solution is
\beq
\label{eq5.55} %\label{5.14}
\eta_\text{s}^*=\frac{2}{d+2}\frac{1}{\nu_{0|2}^*+2\gamma_\text{s}^*},
\eeq
\beq
\label{eq5.56} %\label{5.15}
\kappa_\text{s}^*=\frac{2(d-1)}{d(d+2)}\frac{1+2 a_{2,\text{s}}-(1+q)\xi_\text{s}^* \left(\frac{\partial a_2}{\partial \xi^*}\right)_{\text{s}}}{\nu_{2|1}^*+\frac{1}{2}\xi_\text{s}^*-\left(q+\frac{3}{2}\right)
\zeta_\text{s}^*},
\eeq
\beq
\label{eq5.57} %\label{5.16}
\mu_\text{s}^*=\frac{
\zeta_\text{s}^*\kappa_\text{s}^*+\frac{2(d-1)}{d(d+2)}\left[
a_{2,\text{s}}-\frac{\theta_\text{s}}{1+q}\left(\frac{\partial a_2}{\partial \theta}\right)_\text{s}-\xi_\text{s}^*\left(\frac{\partial a_2}{\partial \xi^*}\right)_\text{s}\right]}
{\nu_{2|1}^*+3\gamma_\text{s}^*},
\eeq
\beq
\label{eq5.58} %\label{5.17}
e_D^*=-\frac{\frac{2(1+q)+d}{2d}\xi_\text{s}^*\left(\frac{\partial a_2}{\partial \xi^*}\right)_\text{s}
+\frac{1}{2}\frac{\theta_\text{s}}{1+q}\left(\frac{\partial a_2}{\partial \theta}\right)_\text{s}}
{\nu_{4|0}^*+4\gamma_\text{s}^*},
\eeq
where $\gamma_\text{s}^*=(\xi_\text{s}^*-\zeta_\text{s}^*)/2$ and
\beq
\label{eq5.59} %\label{5.18}
\theta_\text{s}=\frac{\xi_\text{s}^*-\zeta_\text{s}^*}{2}\xi_\text{s}^{*q/(1+q)}.
\eeq

The derivatives $(\partial a_2/\partial \xi^*)_\text{s}$ and $(\partial a_2/\partial \theta)_\text{s}$ appearing in Eqs.\ \eqref{eq5.55}--\eqref{eq5.58} can be easily obtained from Eq. \eqref{eq5.30.1}.  According to Eq.\ \eqref{eq5.30.1}, the derivative $\partial a_2/\partial \xi^*$ is given by
\beq
\label{eq5.60} %\label{4.16.1}
\frac{\partial a_2}{\partial \xi^*}=\frac{\frac{d}{d+2}\lambda_1^*+(1+a_2)(2\zeta^*-\nu_{4|0}^*) -2\xi^* a_2}
{(1+q)\xi^*\left[ \left( 2\beta T^* -1 \right) \xi^* +\zeta^*\right]}.
\eeq
In the steady state, the numerator and denominator of Eq.\ \eqref{eq5.60} vanish so that, the quantity $\partial a_2/\partial \xi^*$ becomes indeterminate. As in the case of IHS, this problem can be solved by applying l'Hopital's rule. The final result is

\begin{equation}
\label{eq5.61} %\label{4.17}
\left(\frac{\partial a_2}{\partial \xi^*}\right)_\text{s} = \frac{a_{2,\text{s}}}{ \zeta_\text{s}^* - \frac{\nu_{4|0^*}}{2} -q\xi_\text{s}^*\beta T_\text{s}^*-\frac{1-q}{2}\xi_\text{s}^*}.% \right)^{-1}.
\end{equation}

Upon deriving Eq.\ \eqref{eq5.61}, use has been made of the identity
\begin{equation}
\label{eq5.62} %\label{4.18}
\frac{\partial}{\partial \xi^*}\left[\left( 2\beta T^* -1 \right) \xi^*\right]=\frac{2q}{1+q}\beta T^*-1.
\end{equation}
The other derivative $\partial a_2/\partial \theta$ may be also obtained after taking the derivative on both sides of Eq.\ \eqref{eq5.30.1} with respect to $\theta$ and then taking the steady-state limit. After some algebra, one gets
\begin{equation}
\label{eq5.63} %\label{4.19}
\left(\frac{\partial a_2}{\partial \theta}\right)_{\text{s}}=(1+q)\frac{\xi^{*\frac{1+2q}{1+q}}}{\zeta_\text{s}^* -\frac{1}{2} \nu_{4|0^*}-\xi^*}\left(\frac{\partial a_2}{\partial \xi^*}\right)_\text{s},
\end{equation}
where use has been made of the result
\begin{equation}
\label{eq5.64} %\label{4.20}
\frac{\partial}{\partial \theta}(2\beta T^*)=\frac{\partial}{\partial \theta}
\frac{2\theta}{\xi^{*1/(1+q)}}=\frac{2}{\xi^{*1/(1+q)}}.
\end{equation}

\begin{figure}
  \centering
  \includegraphics[width=0.75 \columnwidth,angle=0]{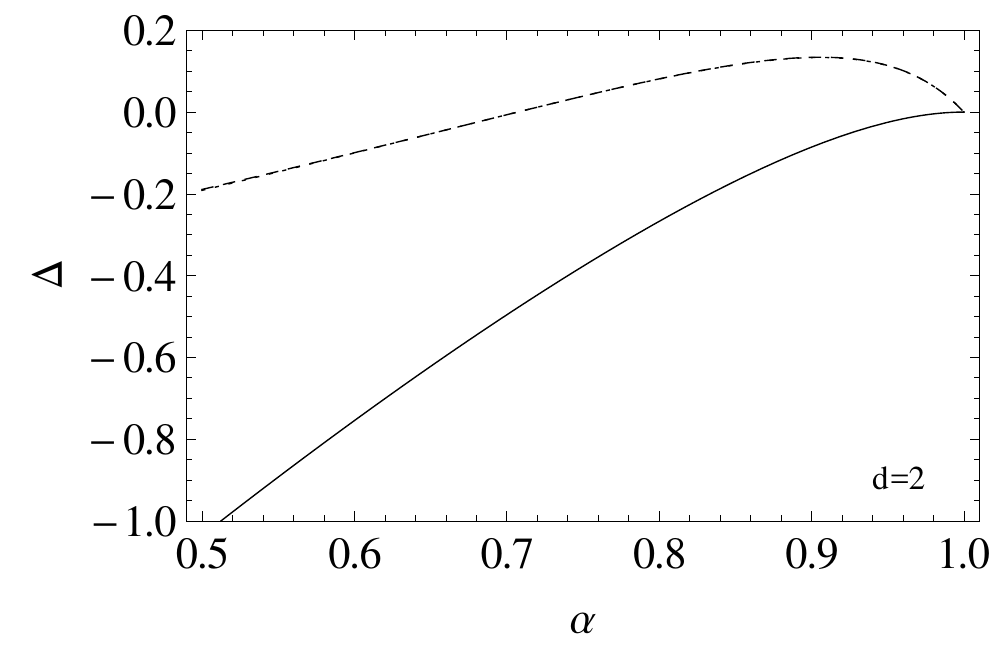}
  \includegraphics[width=0.75 \columnwidth,angle=0]{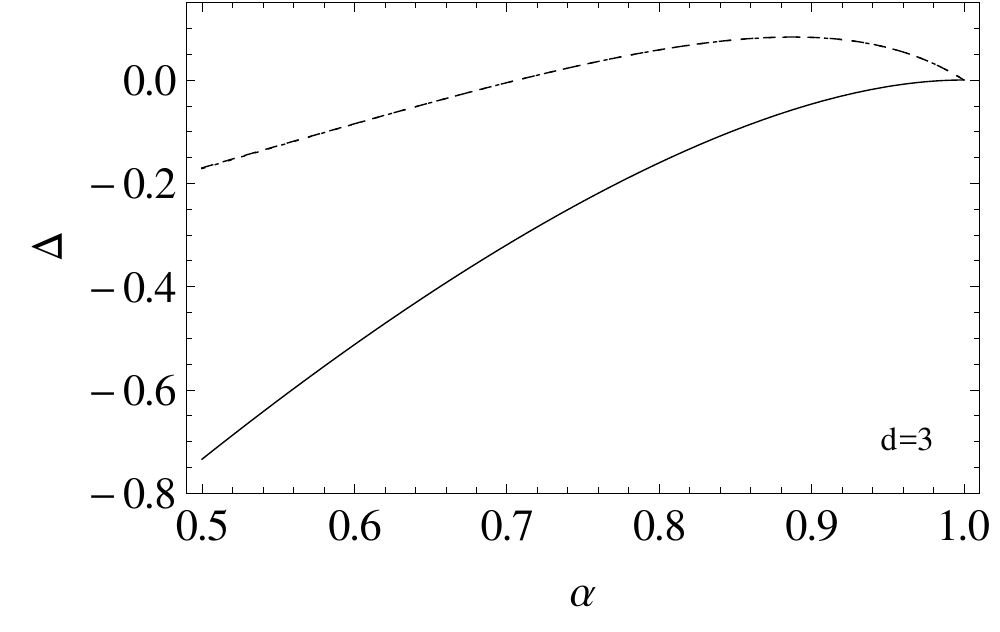}
\caption{Plot of the derivative $\Delta\equiv \left(\frac{\partial a_2}{\partial \xi^*}\right)_\text{s}$ versus the coefficient of restitution $\alpha$ for the stochastic thermostat ($\xi_\text{s}^*=\zeta_\text{s}^*$) for disks ($d=2$) and spheres ($d=3$). The solid lines are the results given by Eq.\ \eqref{eq5.61} for $q=\frac{1}{2}$ while the  dashed lines are the results obtained for IHS.% in Refs.\ \cite{GCV13} and \cite{GMT12}, respectively.
\label{fig02chap5}}
\end{figure}

Fig.\ \ref{fig02chap5} shows the dependence of the derivative $\Delta\equiv \left(\frac{\partial a_2}{\partial \xi^*}\right)_\text{s}$ on the coefficient of restitution $\alpha$ when the gas is heated by the stochastic thermostat ($\beta=0$ and $\xi_\text{s}^*=\zeta_\text{s}^*$). The results obtained from Eq.\ \eqref{eq5.61} when $q=\frac{1}{2}$ are compared with those derived for IHS \cite{GCV13, GMT12}.
% in Refs.\ \cite{GCV13} and \cite{GMT12} by using two different methods. First, it is quite apparent that the results obtained for IHS are practically indistinguishable, showing that the expressions of $\Delta$ obtained in Refs. \cite{GCV13} and \cite{GMT12} are consistent with each other. When comparing IHS and IMM, 
We observe that the discrepancies between both interaction models are small for not too strong dissipation (say for instance $\al \gtrsim 0.8$), although they increase as the coefficient of restitution decreases.

\begin{figure}
  \centering
  \includegraphics[width=0.75 \columnwidth,angle=0]{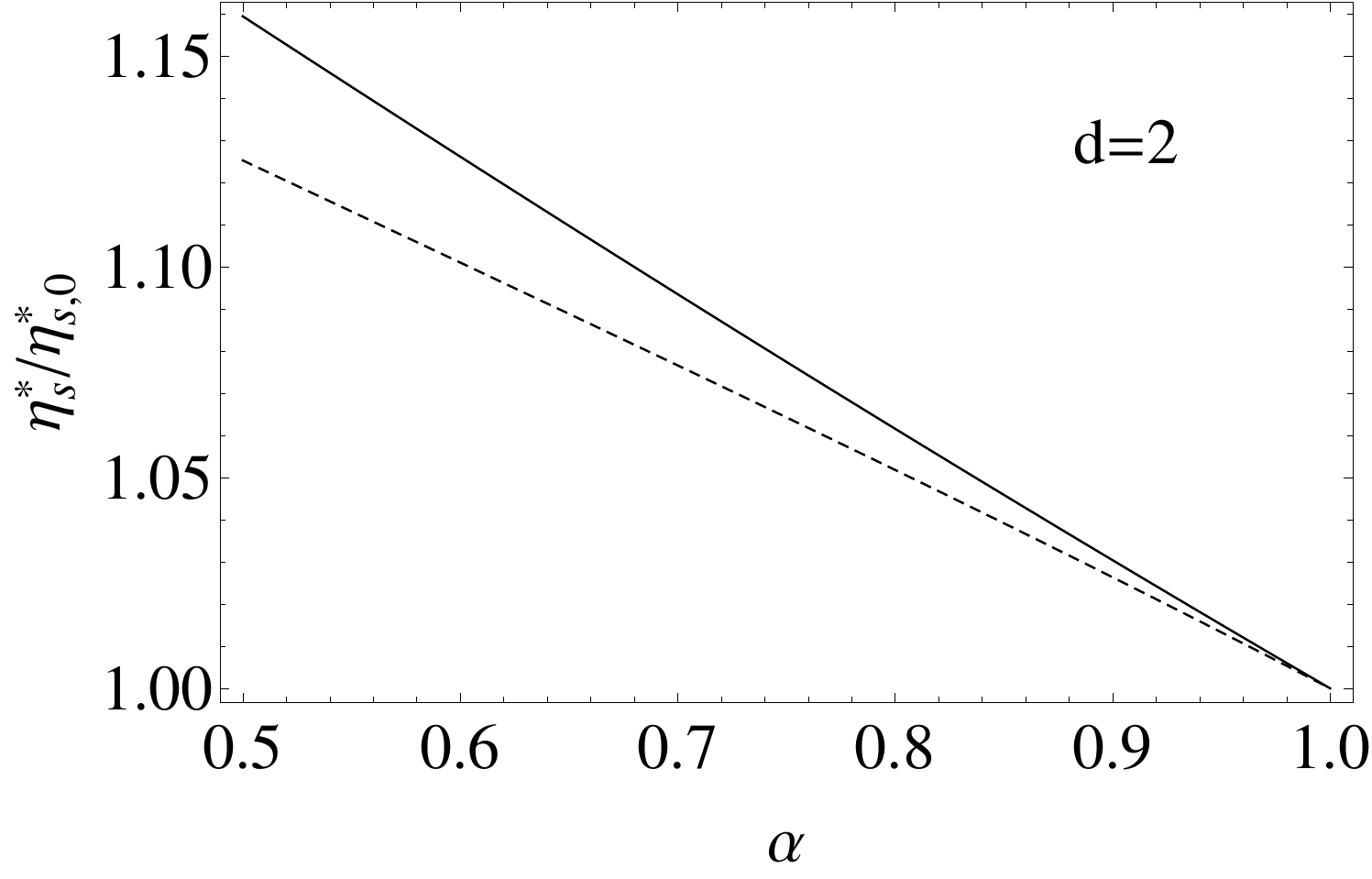}
  \includegraphics[width=0.75 \columnwidth,angle=0]{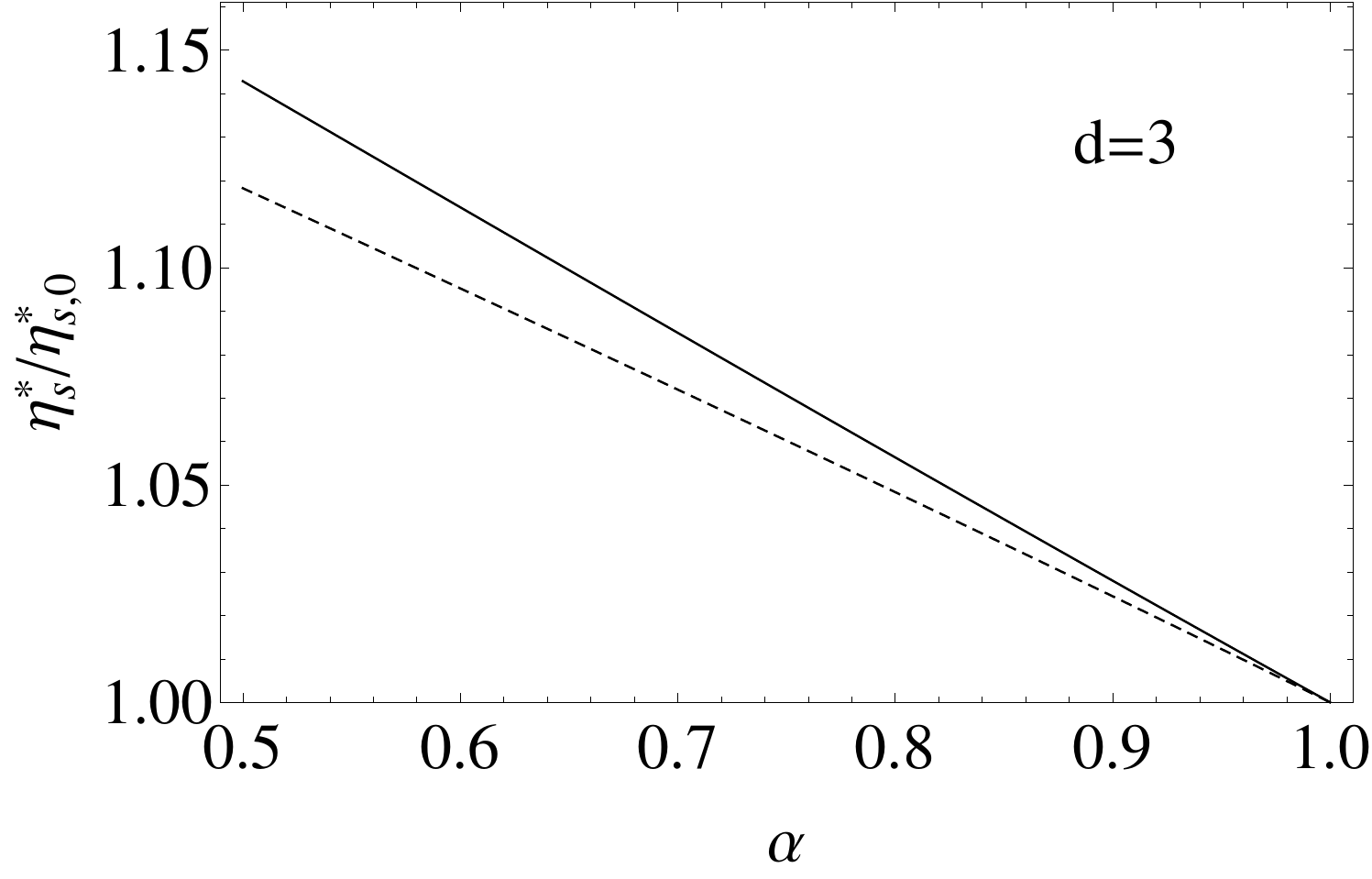}
\caption{Plot of the reduced shear viscosity $\eta_\text{s}^*/\eta_\text{s,0}^*$ as a function of the coefficient of restitution $\al$ for $\beta=\frac{1}{2}$ in the case of a two- and three-dimensional system of IMM with $q=\frac{1}{2}$ (solid lines) and IHS (dashed lines). The value of the (reduced) noise strength is $\xi_\text{s}^*=1$.
\label{fig03chap5}}
\end{figure}

\begin{figure}
  \centering
  \includegraphics[width=0.75 \columnwidth,angle=0]{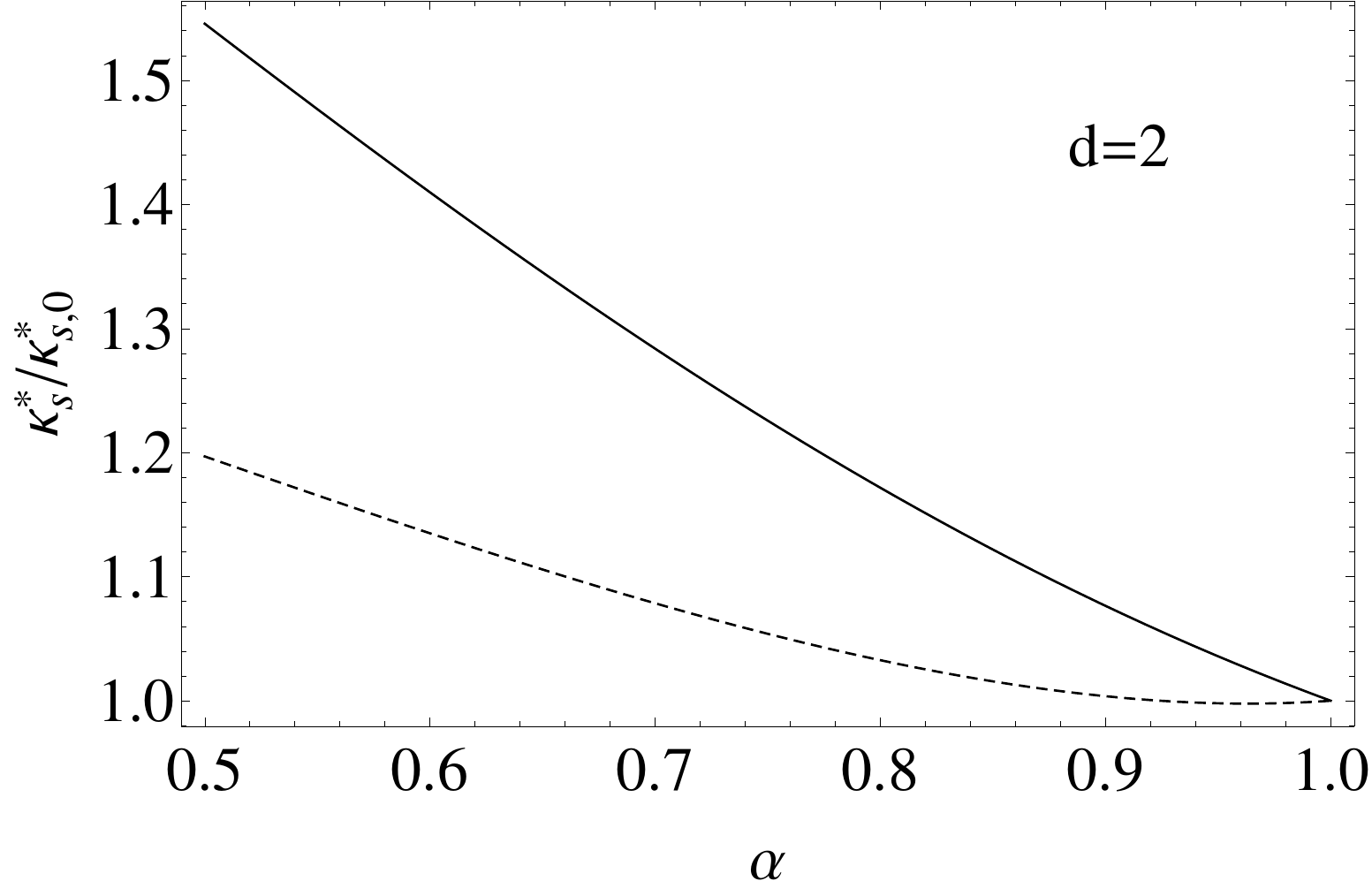}
  \includegraphics[width=0.75 \columnwidth,angle=0]{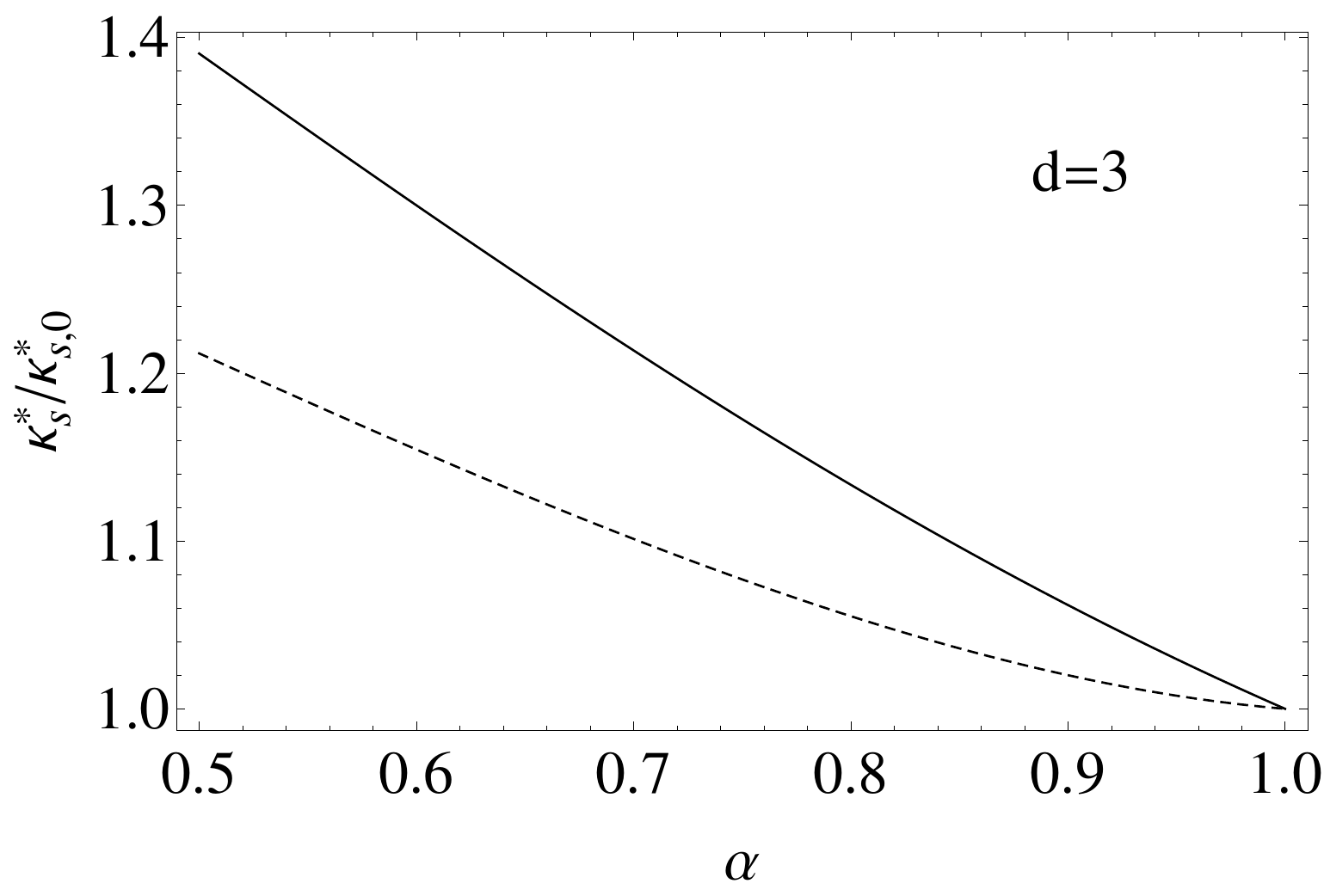}
\caption{The same as in Fig.\ \ref{fig03chap5} for the reduced thermal conductivity $\kappa_\text{s}^*/\kappa_\text{s,0}^*$.
\label{fig04chap5}}
\end{figure}

\begin{figure}
  \centering
  \includegraphics[width=0.75 \columnwidth,angle=0]{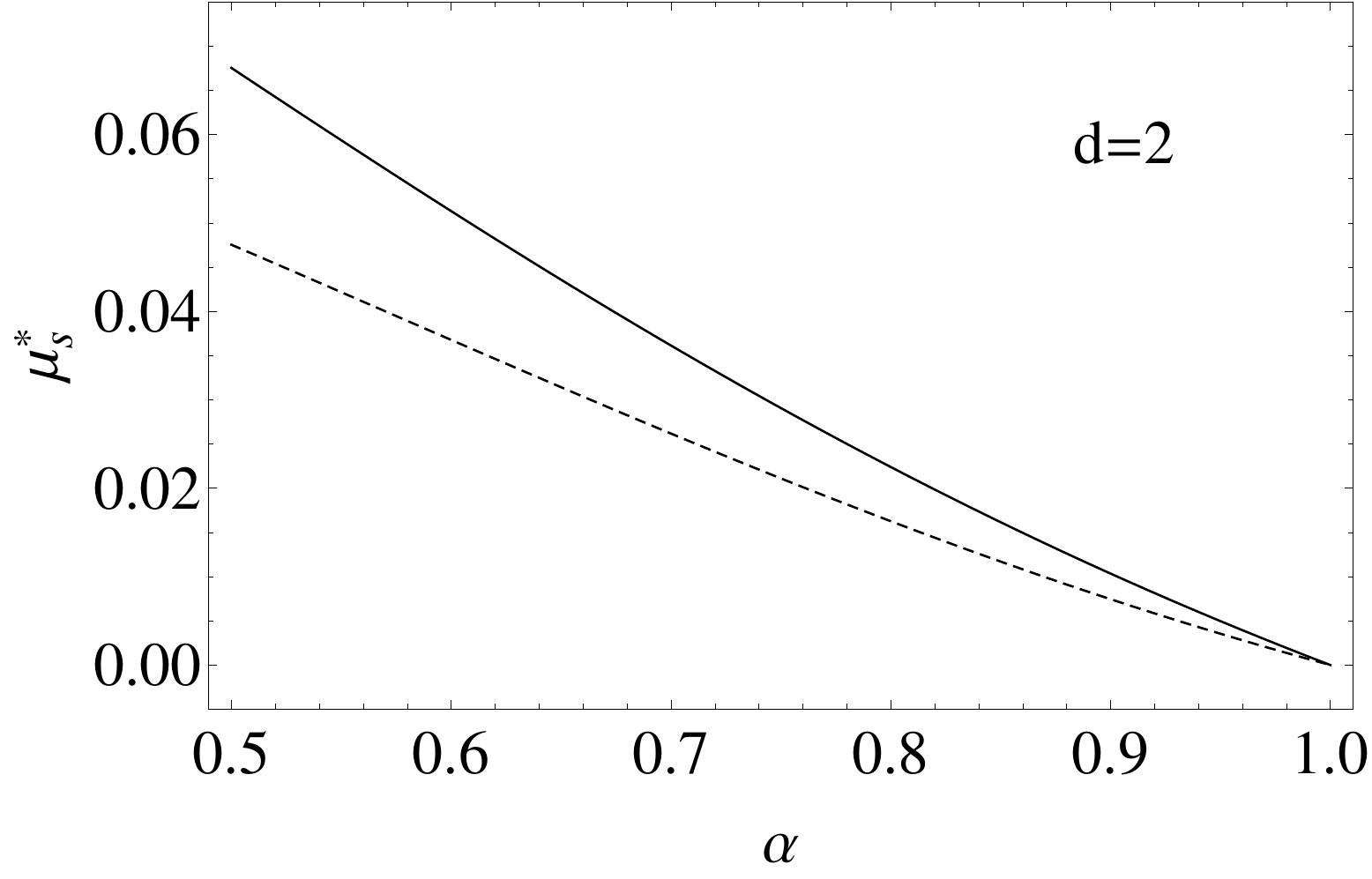}
  \includegraphics[width=0.75 \columnwidth,angle=0]{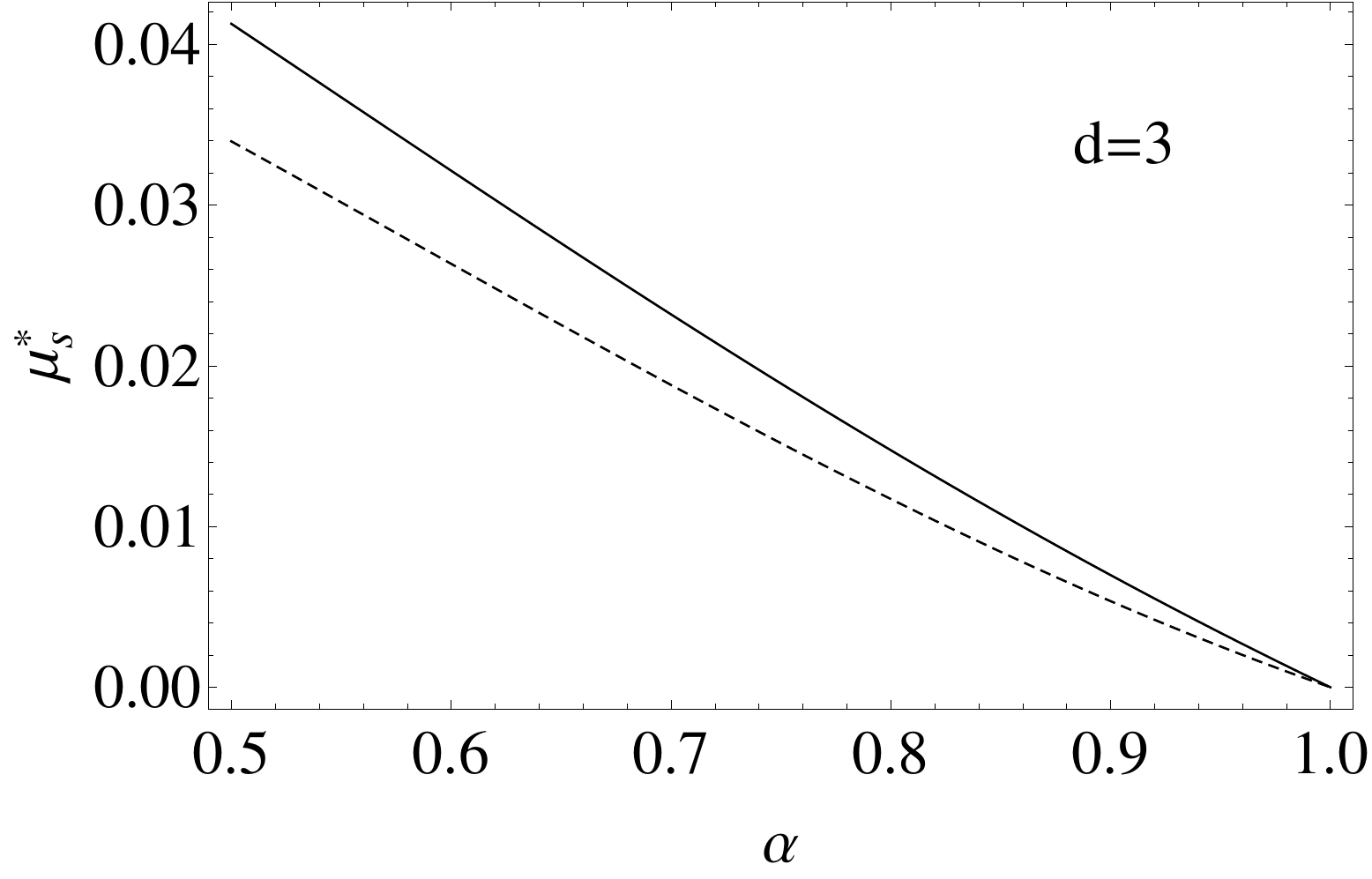}
  \caption{The same as in Fig.\ \ref{fig03chap5} for the reduced coefficient $\mu_\text{s}^*= n\mu_\text{s}/\kappa_0 T$.  \label{fig05chap5}}
\end{figure}

\begin{figure}
  \centering
  \includegraphics[width=0.75 \columnwidth,angle=0]{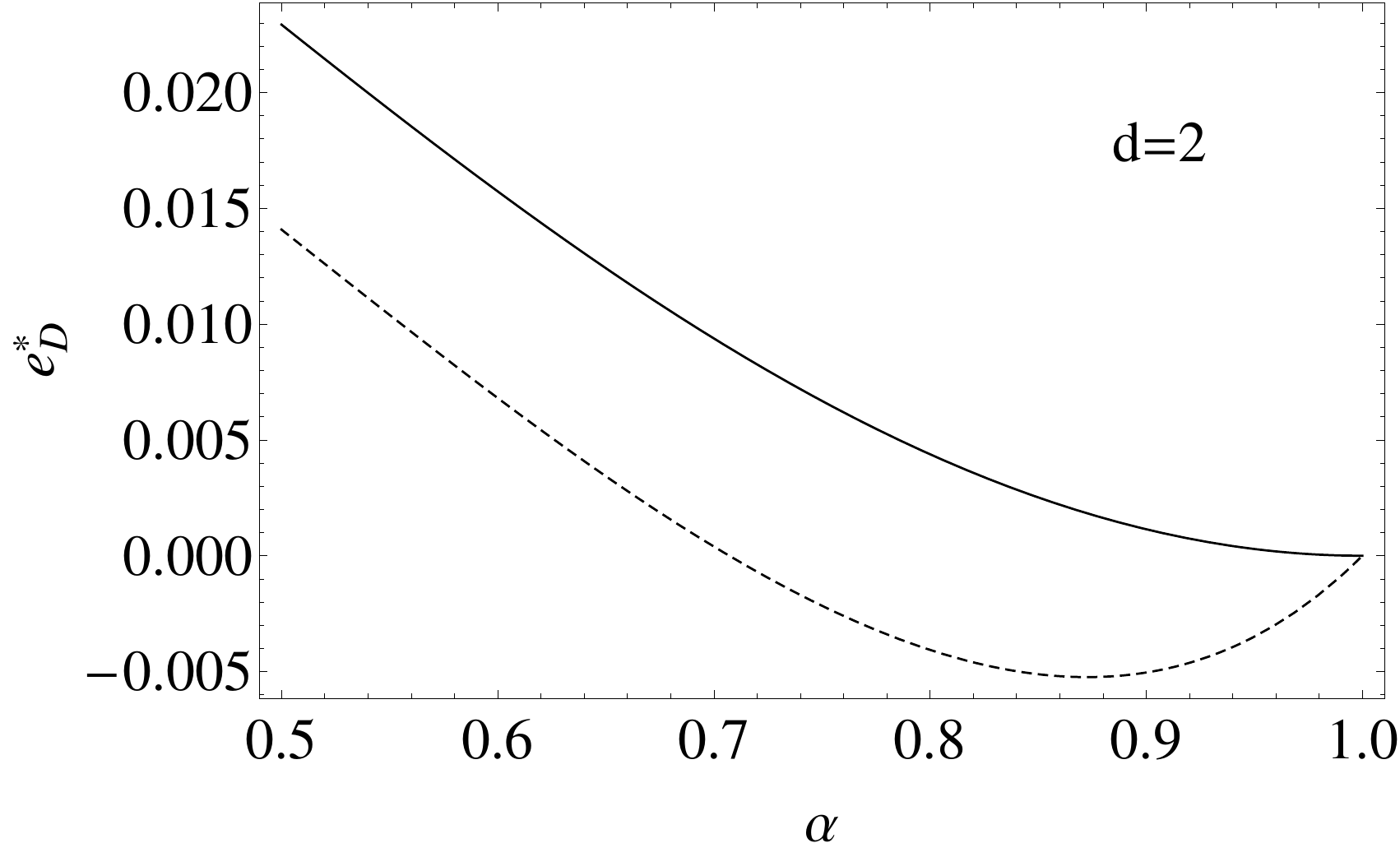}
  \includegraphics[width=0.75 \columnwidth,angle=0]{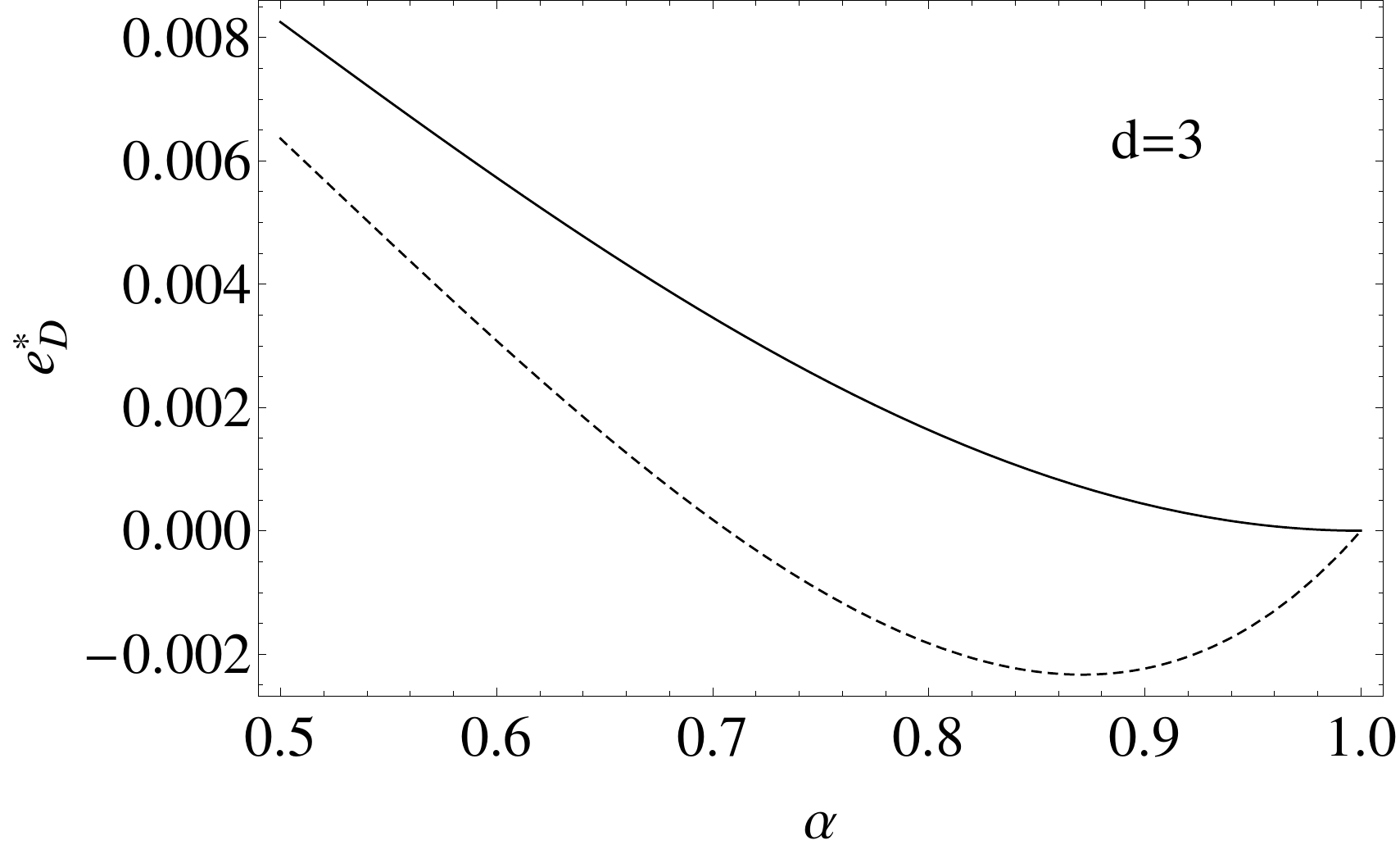}
  \caption{The same as in Fig.\ \ref{fig03chap5} for the reduced coefficient $e_D^*$.  \label{fig06chap5}}
\end{figure}

\subsection{Comparison with the steady state transport coefficients for IHS}
\label{sec6}

The expressions of the transport coefficients of a driven granular gas of IHS at moderate densities have been obtained in Chapter \ref{Chapter4}. 
%The transport coefficients of IHS have been derived in Ref.\ \cite{GCV13} in the first Sonine approximation for a driven granular dense gas. 
For the sake of completeness, the forms of the reduced coefficients $\eta_\text{s}^*$, $\kappa_\text{s}^*$, $\mu_\text{s}^*$ and $e_D^*$ for a low-density gas ($\phi=0$) are listed in Appendix \ref{AppendixG}.

Figs.\ \ref{fig03chap5}--\ref{fig06chap5} show the $\al$-dependence of the reduced transport coefficients $\eta_\text{s}^*/\eta_\text{s,0}^*$, $\kappa_\text{s}^*/\kappa_\text{s,0}^*$, $\mu_\text{s}^*$, and $e_D^*$, respectively, for $\xi_\text{s}^*=1$. Here, since we are mainly interested in analyzing the influence of dissipation on transport, the shear viscosity and thermal conductivity coefficients have been reduced with respect to their corresponding elastic values  $\eta_\text{s,0}^*$ and $\kappa_\text{s,0}^*$, respectively. Note that the coefficients $\mu_\text{s}^*$ and $e_D^*$ vanish for elastic collisions. In addition, we have taken $\beta=\frac{1}{2}$ and the Maxwell model with the power $q=\frac{1}{2}$. This latter choice is closer to IHS.

We observe that in general the qualitative dependence of the transport coefficients on dissipation of IHS is well captured by IMM. The shear viscosity (as expected because the same behavior is observed in analogous systems \cite{S03}) increases with inelasticity. However, this increase is faster for IMM. The (reduced) thermal conductivity of IHS presents a non-monotonic dependence with dissipation, since first it decreases as $\al$ decreases in the region of weak dissipation, reaches a minimum and then, the ratio $\kappa_\text{s}^*/\kappa_\text{s,0}^*$ increases with inelasticity. This behavior differs from the one observed for IMM where $\kappa_\text{s}^*/\kappa_\text{s,0}^*$ always increases with inelasticity. With respect to the new transport coefficient $\mu_\text{s}^*$ (not present for elastic collisions), both interaction models predict that this coefficient is much smaller than the thermal conductivity so that, the impact of the term $-\mu \nabla n$ on the heat flux $\mathbf{q}^{(1)}$ is much smaller than the Fourier's law term $-\kappa \nabla T$. Notice also that the quantitative differences between the NS transport coefficients of IMM and IHS transport coefficients increase with inelasticity, especially in the two-dimensional case. However, and compared to freely cooling granular gases \cite{S03}, these quantitative differences between both models are much less important for driven systems. Therefore, we think the results in this Chapter are particularly useful also for studying the transport properties of the analogous IHS driven system.

\section{Summary and Discussion}
\label{sec6chap5}

Calculation of transport coefficients in driven granular gases from the Boltzmann equation for IHS is a quite difficult problem. In particular, we need to compute three different collision integrals to get the explicit forms of the NS transport coefficients. However, given that these integrals cannot be exactly evaluated, one usually considers the leading terms in a Sonine polynomial expansion of the velocity distribution function (first-Sonine approximation) to estimate them \cite{CHAPMAN70}.  In spite of the simplicity of this approach, the corresponding expressions of the NS transport coefficients compare in general quite well with computer simulations. On the other hand, it could be desirable to introduce interaction models more tractable analytically than IHS that were also capable of capturing the most important properties of the latter (at least within the domain of velocities near thermal velocity).

Based on the experience of elastic particles, a possible alternative that may overcome the technical difficulties embodied in the Boltzmann collision operator of IHS is to consider IMM. In the Boltzmann equation for IMM, the collision rate of the underlying system of IHS is replaced by an effective collision rate independent of the relative velocity of the two colliding particles. This property allows us to evaluate \emph{exactly} the velocity moments of the Boltzmann collision operator without the explicit knowledge of the velocity distribution function.

In this Chapter, the expressions of the transport coefficients of an inelastic Maxwell gas driven by a stochastic bath with friction have been obtained. %This type of thermostat (used in a number of works by other authors \cite{PLMV98,PLMV99,MTC07,VPV08,PV09,GSVP11-1}) % \cite{andrea} 
%is proposed to model the effect of the interstitial fluid on the dynamic properties of grains. 
As noted in the Introduction of this Chapter, the evaluation of the transport coefficients of IMM is an interesting problem by itself since it allows to understand in a clean way the influence of collisional dissipation on transport properties. In addition, the comparison between the \emph{exact} results for IMM with those obtained for IHS by using approximate analytical methods allows us to gauge the degree of reliability of IMM  for the description of granular flows. Here, we have accomplished this comparison with the results  for IHS derived in Chapter \ref{Chapter4} by using the same type of thermostat.

The NS transport coefficients have been obtained by solving the Boltzmann equation for IMM by means of the CE expansion up to first order in the spatial gradients. As noted in the previous Chapter for IHS, depending on the process, collisional cooling will not be necessarily balanced at all points in the system by the thermostat and/or external forces from the boundaries. As a consequence, the zeroth-order solution $f^{(0)}$ depends on time through its dependence on the granular temperature. The fact that  $\partial_t^{(0)}T \neq 0$ gives rise to conceptual and mathematical difficulties not present in previous works \cite{S03,GM02} where the parameters of the force were chosen to impose a steady temperature in the reference state $f^{(0)}$. In particular, we would need to solve numerically (which we have not done in the present work) a set of coupled first-order differential equations [see Eqs. \eqref{eq5.44}--\eqref{eq5.46}], in order to obtain the dependence of the transport coefficients on dissipation and the thermostat forces parameters. This technical difficulty is present even in the simplest Maxwell model where the collision frequency $\nu$ is independent of temperature $T$ [i.e., when $q=0$ in Eq.\ \eqref{eq5.3}]. Thus, the steady state conditions and analytical expressions of all transport coefficients have been considered. The steady state expressions are given by Eq.\ \eqref{eq5.55} for the (dimensionless) shear viscosity $\eta^*$, Eq.\ \eqref{eq5.56} for the (dimensionless) thermal conductivity $\kappa^*$, Eq.\ \eqref{eq5.57} for the coefficient $\mu^*$ and Eq.\ \eqref{eq5.58} for the first-order contribution $e_D^*$ to the fourth-cumulant. The three first coefficients provide the momentum and heat fluxes in the first order of the spatial gradients.

As in previous works \cite{S03,G07,SG07}, the collision frequency $\nu$ appearing in the Boltzmann equation for IMM [see Eq.\ \eqref{eq5.2}] has been chosen to reproduce the cooling rate $\zeta$ of IHS (evaluated in the Maxwellian approximation). With this choice, the comparison between IMM and IHS (see Figs.\ \ref{fig03chap5}--\ref{fig06chap5} for $d=2$ and 3) shows that IMM reproduces qualitatively well the trends observed for IHS, even for strong dissipation. On the other hand, at a more quantitative level, discrepancies between both interaction models increase with inelasticity, especially in the case of hard disks ($d=2$). In any case, the results found in this work contrast with those obtained in the freely cooling case \cite{S03} where IMM and IHS exhibit much more significant differences. Thus, the reliability of IMM as a prototype model for granular flows can be considered more robust in driven states than in the case of undriven states. This conclusion agrees with the results derived in the case of the simple shear flow problem \cite{G03} and more complex shear-induced laminar flows \cite{SGV09}. %In this context, the search for exact solutions for driven IMM, and comparison with computer simulations or experiments, can be considered as an interesting problem in the near future.
         % IMM
\chapter{Non-Newtonian hydrodynamics for a dilute granular suspension under uniform shear flow}
\label{Chapter6}
\blfootnote{The results obtained in this Chapter have been published in M.G. Chamorro, F. Vega Reyes and V. Garz\'o, \emph{Phys. Rev. E}, \textbf{92}:052205 (2015) \cite{CVG15}} 
 
\lhead{Chapter 6. \emph{Non-Newtonian hydrodynamics for a dilute granular suspension under uniform shear flow}}

\section{Introduction}
\label{sec1chap6}

As mentioned in Chapter \ref{Chapter1}, granular matter can be quite usually found in nature surrounded by an interstitial fluid (air, water...) in the form of a gas--solid suspension.
At the level of kinetic theory, the description of granular suspensions is an intricate problem since it involves two phases (solid particles and interstitial fluid) and hence, one would need to solve a set of two coupled kinetic equations for each one of the velocity distribution functions of the different phases. However, due to the mathematical difficulties embodied in this approach and in order to gain some insight into this problem, a classical model for dilute gas--solid flows is to consider a single Boltzmann equation for the solid particles where the influence of the surrounding fluid on them is modeled by means of an effective external force \cite{B74, KH01, K90}.
This has been the approach considered in the model introduced in previous Chapters to determine the dynamic properties of a driven granular gas.
%This will be the approach considered in the present paper.

Moreover, in the study of granular suspensions usually only simple states have been considered, due to the inherent complexity of the system. For instance, in Ref.\ \cite{GTSH12} the NS transport coefficients of monodisperse gas--solid flows at \emph{moderate} densities were obtained by solving a model based on the Enskog kinetic equation by means of the application of the CE method \cite{CHAPMAN70} around the HCS. The external force $\mathbf{F}_\text{ext}$ proposed in Ref.\ \cite{GTSH12} to model the effect of the fluid phase on grains is composed by three different terms: (i) a term proportional to the difference between the mean flow velocities of solid $\mathbf{U}$ and gas $\mathbf{U}_g$ phases, (ii) a drag force $\mathbf{F}_\text{drag}$ proportional to the velocity of particles and (iii) a stochastic force $\mathbf{F}_\text{st}$ accounting for particle neighbor effects (Langevin model). 

As said before, the model introduced in Ref.\ \cite{GTSH12} is quite similar to the one introduced by Gradenigo \emph{et al.} \cite{GSVP11-1} for driven granular fluids. In the case that $\mathbf{U}=\mathbf{U}_g$, the coefficient associated with the stochastic force vanishes and only the drag force interaction $\mathbf{F}_\text{drag}$ remains, namely, mean drag and neighbor effects disappear in the suspension model of Ref.\ \cite{GTSH12}. The above drag force model has been also recently considered in different papers \cite{WZLH09,H13,HT13,SMMD13,WGZS14} to study the shear rheology of frictional hard-sphere suspensions.

Nevertheless, although there exist cases where the hydrodynamics of granular gases are Newtonian \cite{LBLG00, PGVP16}, the ranges of interest fall frequently beyond Newtonian hydrodynamics since the strength of the spatial gradients is large in most situations of practical interest (for example, in steady states). This is essentially due to the coupling between collisional dissipation and spatial gradients that under steady states usually yields moderately large spatial gradients \cite{G03,SGD04,VU09}.
In these steady states, a hydrodynamic description is still valid but with constitutive equations more complex than the NS ones \cite{VSG10,VSG13}. A very neat example of this is the simple or uniform shear flow (USF) \cite{C90}, that except in the quasi-elastic limit, is essentially non-Newtonian. It is characterized by a linear velocity field (that is $\partial U_x/\partial y\equiv a =\text{const}$), zero heat flux, constant density $n$ and constant temperature $T$. In particular, in the USF state the presence of shearing induces anisotropies in the pressure tensor $P_{ij}$, namely, nonzero shear stress $P_{xy}$ and normal stress differences $P_{xx}-P_{yy}\neq0$ and $P_{yy}-P_{zz}\neq0$. In addition, in the case of granular suspensions, it may be assumed \cite{TK95,SMTK96} that $\mathbf{U}=\mathbf{U}_g$ and so, $\mathbf{F}_\text{ext}=\mathbf{F}_\text{drag}$ in the model proposed in Ref.\ \cite{GTSH12}.

%As said before, in the steady USF state, the system departs from equilibrium as dissipation increases and consequently, the results derived in Ref.\ \cite{GTSH12} for the NS hydrodynamics are expected to be reliable in the USF problem only for nearly elastic particles and/or very small values of the (scaled) parameter characterizing the drag force.

%The (steady) USF  state is a particular case (with null heat flux) of a flow class (with constant heat flux) for granular gases that shares common rheological properties \cite{VSG10,VGS11,VSG13}. The fact that for this flow class there is a balance between energy input (viscous heating term) and cooling terms (namely, those arising from dissipative particle-particle collisions and from the viscous drag force) in the bulk of the fluid simplifies the non-Newtonian hydrodynamic equations for this kind of flow.

A detailed study of simple shear flows of granular suspensions at finite Stokes numbers was carried out  by Tsao and Koch \cite{TK95} and Sangani \emph{et al.} \cite{SMTK96}. In both of these works, and like in the model used in Ref.\ \cite{GTSH12}, suspension dynamics is dominated by the drag exerted by the fluid (external drag force) and the solid-body collisions between the particles. In the first paper \cite{TK95}, the authors considered a \emph{dilute} gas--solid suspension of \emph{elastic} particles, thus neglecting the important effect of inelasticity in macroscopic particles. Inelasticity and excluded volume effects (moderated densities) were only considered in the second paper \cite{SMTK96} of the series. Moreover, in the first reference \cite{TK95} (elastic collisions), Tsao and Koch solved the Boltzmann kinetic equation by means of a Grad's moment method approach \cite{G49} where the collisional moment $\Lambda_{ij}$ of the momentum transfer [see Eq.\ \eqref{eq6.20} for its definition] was evaluated by retaining \emph{all} the quadratic terms in the pressure tensor $P_{ij}$ (nonlinear Grad's solution). However, for practical applications, in their actual theoretical results only the term proportional to the shear stress $P_{xy}^2$ was retained in the nonlinear contributions to $\Lambda_{ij}$, (see Eqs.\ (3.14a,b) of \cite{TK95}). Sangani \emph{et al.} \cite{SMTK96} solved first the Enskog kinetic equation by means of Grad's moment method but only \emph{linear} terms in the shear rate and the pressure tensor (linear Grad's solution) were retained in their calculation of $\Lambda_{ij}$ (see Eq.\ (4.21) of \cite{SMTK96}). Some discrepancies were observed in the very dilute regime for the normal stress differences. In particular, their linear Grad's solution yields $P_{yy}=P_{zz}$ (see Eq.\ (4.33) of \cite{SMTK96}) which clearly disagrees with simulation results \cite{SMTK96}.

%The question arises then as to whether, and if so to what extent, the conclusions drawn by Sangani \emph{et al.} \cite{SMTK96} for dilute \emph{granular} suspensions ($\al \neq 1$) may be altered when nonlinear terms in the pressure tensor are considered in the application of Grad's moment method to the Boltzmann equation.}

The objective of this Chapter is to offer a complete study of the USF state for \emph{dilute} granular suspensions where the effect of the fluid phase on grains is taken into account by the presence of an external drag force in the kinetic equation. For the accomplishment of this task, we propose three different approaches: two of them are theoretical and the third one is computational. In the first theoretical approach, the Boltzmann equation is solved by Grad's moment method where both inelasticity and at the same time all of the non-linear terms in shear rate and stress tensor are retained in our expression of the collisional moment $\Lambda_{ij}$. Thus, as we will see, new interesting properties of the suspension arise from this refinement. For instance, we have been able to detect the influence of both viscous friction and inelasticity on the normal stress difference $P_{yy}-P_{zz}$. In this sense, our theory  generalizes previous analyses \cite{TK95,SMTK96}, these being  recovered when the appropriate simplifications are applied to it.

Apart from Grad's moment method, we also consider a second theoretical approach based on the derivation of an \emph{exact} solution to a simplified model kinetic equation \cite{BDS99} for the sheared granular suspension. This will allow us to determine all the velocity moments of the velocity distribution function as well as the explicit form of the latter in terms of the shear rate $a$, the friction coefficient $\gamma_\text{b}$ characterizing the drag force and the coefficient of restitution $\alpha$. In particular, the rheological properties derived from the model kinetic solution are the same as those obtained in linear Grad's solution to the Boltzmann equation.

As a third route and to gauge the accuracy of the previous analytical results, we numerically solve the Boltzmann equation for the granular suspension by means of the DSMC method. 
In this (exact) numerical solution the grain-grain collisions in the context of the hard sphere collision model have been taken into account.
As we will see, the comparison between theory and simulation shows that both (approximate) solutions give in general accurate results even for conditions of quite strong inelasticity (say for instance, $\alpha \gtrsim 0.5$). Moreover, the theoretical predictions for $P_{yy}$ and $P_{zz}$ obtained from our nonlinear Grad's solution agree very well with simulations (see Fig.\ \ref{fig4chap6}), showing the improvement of our theory with respect to the previous analysis of Sangani \emph{et al.} \cite{SMTK96}. On the other hand, the agreement between theory and simulation becomes worse as the (scaled) friction coefficient $\gamma^*$ increases. This means that our theory of rapidly sheared granular flows becomes more reliable as the effects of the inelastic particle collisions dominate over viscous effects.

%The plan of the paper is as follows. In Sec.\ \ref{sec2}, the Boltzmann equation of inelastic hard spheres driven by an external drag force is introduced and the USF problem for granular suspensions is presented. The analytical results derived in the paper are provided in Sec. \ref{sec3} whereas some technical details on the DSMC method used here are briefly described in Sec.\ \ref{sec4}. Section \ref{sec5} deals with the comparison between theory and simulation results. Finally, the paper is closed in Sec. \ref{sec5} with a brief discussion on the results reported in the present contribution.

\section{Description of the system}
\label{sec2chap6}

\subsection{Boltzmann kinetic equation for granular suspensions}

	 Let us consider a set of solid particles of mass $m$ and diameter $\sigma$ immersed in a viscous gas. As we already commented, for big enough particles (typical size $\lesssim 1\,\mu\mathrm{m}$), collisions between particles carry a partial loss of their kinetic energy. Thus, the solid particles can be modeled as a gas of smooth hard spheres (or disks, for two-dimensional systems) with inelastic collisions. %The inelasticity of collisions is characterized by a (positive) \emph{constant} coefficient of normal restitution $0\leq \al \leq 1$, where $\alpha=1$ stands for completely elastic collisions and $\alpha=0$ for completely inelastic collisions \cite{SG98,BDKS98,G03}.
	 
	 	In the dilute limit, the corresponding Langevin equation describing the gas--solid interaction force can be greatly simplified \cite{B74,B74b}. There are several experimental results on the dynamics of dilute particle systems immersed in a gas flow that validate this kind of approach. For instance, this type of system was analyzed in early experimental studies where the corresponding flow properties were carefully measured \cite{TMS84}. These experimental results were later used for validation of a hydrodynamic theory of a granular suspension immersed in gas flow, allowing for characterization of the relevance of grains collisions in the hydrodynamic behavior of the turbulent suspension \cite{LMJ91}. It has been shown more recently, in experiments, that the turbulent gas-grain interaction can also be described by a Langevin equation with a stochastic force that has the form of a white noise, much in the same way as in classic studies at lower Reynolds number \cite{BB88}. Therefore, under the above conditions one can consider the generalized Langevin model for the instantaneous acceleration on a suspended grain given by Eq.\ \eqref{2.2}.
%\begin{equation}
%		 m\frac{\mathrm{d}\mathbf{v}}{dt}=-\beta(\mathbf{U}-\mathbf{U}_g)-\gamma_\text{b}\cdot\mathbf{V}+\mathbf{F}_{\mathrm{st}},
%		\label{eq6.1} %\label{langevin}
%\end{equation}
%where $\mathbf{F}_\text{st}$ is a stochastic force with the properties given in Chapter \ref{Chapter2} by Eq.\ \eqref{noise}. %\cite{BB88}

%The model described by Eq.\ \eqref{eq6.1}  has been recently proposed in Ref.\ \cite{GTSH12} for monodisperse gas-solid flows at moderate density. Although the coefficients %$\beta$, $\gamma_\text{b}$, and $\xi^2_\text{b}$ appearing in Eqs.\ %\eqref{eq6.1} and  \eqref{noise}, respectively, are in general tensors, in the case of a dilute suspension they may be simplified as scalars \cite{BB88}. Those coefficients are associated with the instantaneous gas-solid force \cite{GTSH12}. As we said in the Introduction of this Chapter, the first term on the right-hand side of Eq.\ \eqref{eq6.1} represents the portion of the drag term arising from the mean motion of particle and solid phase; the second term is traced to fluctuations in particle velocity (relative to its mean value) and finally the third term is a stochastic model for the change in particle momentum due to shear stress and pressure contributions at the particle surface that arise from the fluid velocity and pressure disturbances caused by neighbor particles.

As mentioned before, this model has been recently proposed in Ref.\ \cite{GTSH12} for monodisperse gas-solid flows at moderate density. Although the coefficients, $\xi^2_\text{b}$, and $\gamma_\text{b}$ appearing in Eqs.\ \eqref{noise} and \eqref{eqFdrag}, respectively, are in general tensors, in the case of a dilute suspension they may be simplified as scalars \cite{BB88}. Those coefficients are associated with the instantaneous gas-solid force. 
%As we said in Chapter \ref{chapter2}, the external force is composed by two contributions [see Eq.\ \eqref{2.2}]. In the case presented here. As we said in the Introduction of this Chapter,
The first term is a stochastic model for the change in particle momentum due to shear stress and pressure contributions at the particle surface that arise from the fluid velocity and pressure disturbances caused by neighbor particles while 
the second term represents the portion of the drag term arising from the mean motion of particle and solid phase and the fluctuations in particle velocity (relative to its mean value).

According to the model proposed in Ref.\ \cite{GTSH12}, at low Reynolds number, the expressions of $\gamma_\text{b}$ and $\xi^2_\text{b}$ for dilute suspensions of hard spheres ($d=3$) are, respectively, \cite{GTSH12}
\begin{equation}
\label{eq6.3} %\label{2.5}
\xi^2_\text{b}=\frac{1}{6\sqrt{\pi}}\frac{\sigma |\Delta {\bf U}|^2}{\tau^2\sqrt{\frac{T}{m}}},
\end{equation}
\begin{equation}
\label{eq6.2} %\label{2.4}
\gamma_\text{b}=\frac{m}{\tau}R_\mathrm{diss}(\phi),
\end{equation}
where $\tau=m/(3\pi \mu_g \sigma)$ is the characteristic time scale over which the velocity of a particle of mass $m$ and diameter $\sigma$ relaxes due to viscous forces, $\mu_g$ being the gas viscosity. Moreover,  $\phi=(\pi/6)n\sigma^3$ is the solid volume fraction for spheres, and 
\beq
\label{eq6.4} %\label{Rdiss}
R_\text{diss}(\phi)=1+3\sqrt\frac{\phi}{2}.
\eeq
%and $\Delta \mathbf{U}=\mathbf{U}-\mathbf{U}_g$.

In the low-density regime the one-particle distribution function $f(\mathbf{r}, \mathbf{v},t)$ provides complete information on the state of the system. %In the case of an external force composed by the three terms appearing in Eq.\ \eqref{eq6.1}, 
The corresponding Boltzmann kinetic equation for dilute granular suspensions is \cite{GTSH12}
\begin{equation}
\label{eq6.5} %\label{2.6}
\partial_{t}f+\mathbf{v}\cdot \mathbf{\nabla}f-\frac{\gamma_b}{m}\Delta {\bf U}\cdot
\frac{\partial f}{\partial {\bf V}}-\frac{\gamma_\text{b}}{m} \frac{\partial}{\partial
{\bf V}}\cdot {\bf V} f -\frac{1}{2}\xi^2_\text{b}\frac{\partial^2}{\partial V^2}f=J\left[f,f\right],
\end{equation}
where the Boltzmann collision operator $J\left[{\bf v}|f,f\right]$ is defined by Eq.\ \eqref{eqJE2}.% with $\chi=1$.

Note that in the suspension model defined by Eq.\ \eqref{eq6.5} the form of the Boltzmann collision operator $J[f,f]$ is the same as for a dry granular gas and hence, the collision dynamics does not contain any gas--phase parameter. As it has been previously discussed in several papers \cite{TK95, SMTK96, WKL03}, the above assumption requires that the mean--free time between collisions is much less than the time taken by the fluid forces (viscous relaxation time) to significantly affect the motion of solid particles. Thus, the suspension model \eqref{eq6.5} is expected to describe situations where the stresses exerted by the interstitial fluid on particles are sufficiently small that they have a weak influence on the dynamics of grains. However, as the density of fluid increases (liquid flows), the above assumption could be not reliable and hence one should take into account the presence of fluid into the binary collisions event.

\subsection{Steady base state: the uniform shear flow}
\label{USFstate}

Let us assume now that the suspension is under \emph{steady} USF. As we said before, this state is macroscopically defined by a zero heat flux, constant density $n$ and temperature $T$ and the mean velocity $\mathbf{U}$ is
\beq
\label{eq6.6} %\label{2.15}
U_i=a_{ij}r_j, \quad a_{ij}=a\delta_{ix}\delta_{jy},
\eeq
where $a$ is the constant shear rate. The USF state appears as spatially uniform when one refers the velocity of particles  to the Lagrangian frame moving with the flow velocity $\bt{U}$, namely, $f(\bt{r},\bt{v})\equiv f(\bt{V})$, where $V_i=v_i-a_{ij}r_j$ is the peculiar velocity \cite{DSBR86}.

In addition, as usual in uniform sheared suspensions \cite{TK95,SMTK96}, the average velocity of particles follows the velocity of the fluid phase and so, $\mathbf{U}=\mathbf{U}_g$. In this case, $\Delta {\bf U}=\textbf{0}$ and according to Eq.\ \eqref{eq6.3}, $\xi^2_\text{b}=0$. Thus, the steady Boltzmann equation \eqref{eq6.5} becomes
\beq
\label{eq6.7} %\label{2.16}
-aV_y\frac{\partial f}{\partial V_x}-\frac{\gamma_\text{b}}{m} \frac{\partial}{\partial
{\bf V}}\cdot {\bf V} f =J[\mathbf{V}|f,f].
\eeq
%In Eq.\ \eqref{eq6.7} we use the USF property of spatial uniformity when the Boltzmann equation is expressed in terms of the peculiar velocity $V_i=v_i-a_{ij}r_j$ \cite{GARZO03}. 
We note that the Boltzmann equation \eqref{eq6.7} is equivalent to the one employed by Tsao and Koch \cite{TK95} (in the case of elastic collisions) and Sangani \emph{et al.} \cite{SMTK96} for granular suspensions.

In the USF problem, the heat flux vanishes ($\mathbf{q}=\mathbf{0}$) and the only relevant balance equation is that of the temperature \eqref{eqBalT01}. In the steady state and for the geometry of the USF, Eq.\ \eqref{eqBalT01} reads
\beq
\label{eq6.8} %\label{2.17}
-\frac{2}{d\, n} P_{xy} a=\frac{2\,T}{m}\gamma_\text{b} + \zeta\, T,
\eeq
where $P_{xy}$ and $\zeta$ are defined by Eqs.\ \eqref{eqPk} and \eqref{eqzeta}, respectively.
%\beq
%\label{cooling_int}
%\zeta=-\frac{m}{d\, n\, T}\int\; \dd \mathbf{v}\, V^2 J[\mathbf{v}|f,f].
%\eeq

Eq.\ \eqref{eq6.8} implies that in the steady state the viscous heating term ($-a\,P_{xy}>0$) is exactly compensated by the cooling terms arising from collisional dissipation ($\zeta\, T$) and viscous friction ($\gamma_\text{b}\, T/m$) \cite{VU09}. As a consequence, for a given shear rate $a$, the (steady) temperature $T$ is a function of the friction coefficient $\gamma_\text{b}$ and the coefficient of restitution $\alpha$. Note that in contrast to what happens for \emph{dry} granular gases ($\gamma_\text{b}=0$), a steady state is still possible for suspensions when the particle collisions are elastic ($\al=1$ and so, $\zeta=0$). Moreover, the balance equation \eqref{eq6.8} also holds for flows with \emph{uniform} heat flux (the so-called LTu class of Couette flows) \cite{VSG10,VGS11,VSG13} with no friction ($\gamma_\text{b}=0$). %For this class of flows, the physical meaning of Eq.\ \eqref{eq6.8} is that there is an \emph{exact} balance at \emph{every} point of the system between the heating (coming from viscosity) and cooling (coming from inelasticity and friction) terms.

The USF state is in general non-Newtonian. This can be characterized by the introduction of generalized transport coefficients measuring the departure of transport coefficients from their NS forms. First, we define a non-Newtonian shear viscosity coefficient $\eta(a, \gamma_\text{b}, \al)$ by
\beq
\label{eq6.9} %\label{2.18}
P_{xy}=-\eta(a, \gamma_\text{b}, \al)a.
\eeq
In addition, while $P_{xx}=P_{yy}=P_{zz}=nT$ in the NS hydrodynamic order, normal stress differences are expected to appear in the USF state ($P_{xx}\neq P_{yy} \neq P_{zz})$. We are interested here in determining the (reduced) shear stress $P_{xy}^*$ and the (reduced) normal or diagonal elements $P_{xx}^*$, $P_{yy}^*$ and $P_{zz}^*$, where $P_{ij}^*\equiv P_{ij}/p$ and $p=nT$ is the hydrostatic pressure. With respect to the cooling rate $\zeta$ (which vanishes for elastic collisions \cite{TK95}), since this quantity is a scalar, its most general form is
\beq
\label{eq6.10} %\label{zeta}
\zeta=\zeta_0+\zeta_{2} a^2+\cdots.
\eeq
The zeroth-order contribution to the cooling rate $\zeta_0$ was derived in Chapter\ \ref{Chapter4}. For a dilute gas ($\phi=0$), it is given by
\beq
\label{eq6.11} %\label{2.18.1}
\zeta_0=\frac{d+2}{4d}\left(1-\alpha^2\right)\nu_0,
\end{equation}
where $\nu_0$ is an effective collision frequency of hard spheres defined by \eqref{eq5.13.3}.
%\begin{equation}
%\label{eq6.12} %\label{nu}
%\nu_0=\frac{8}{d+2}\frac{\pi^{(d-1)/2}}{\Gamma(d/2)}n\sigma^{d-1}\sqrt{\frac{T}{m}}.
%\end{equation}
For hard spheres ($d=3$), Eq.\ \eqref{eq6.11} is consistent with the results derived by Sangani \emph{et al.} \cite{SMTK96} in the dilute limit (solid volume fraction $\phi=0$). On the other hand, given that the latter theory \cite{SMTK96} only retains linear terms in the pressure tensor in the evaluation of the collisional moment $\Lambda_{ij}$ [defined in Eq.\ \eqref{eq6.20}], then $\zeta_2=0$. We calculate the second-order contribution $\zeta_2$ to the cooling rate in Sec.\ \ref{sec3chap6}. %To the best of our knowledge, this contribution has not yet been computed in previous works on granular sheared suspensions.

Eq.\ \eqref{eq6.8} can be rewritten in dimensionless form when one takes into account Eq.\ \eqref{eq6.9}:
\begin{equation}
\label{eq6.13} %\label{USF}	
\frac{2}{d}\eta^*a^{*2} = 2\gamma^*+\zeta^*,
\end{equation}
where $\eta^*\equiv \eta/\eta_0$, $a^*\equiv a/\nu_0$, $\gamma^*\equiv \gamma_\text{b}/(m \nu_0)$ and $\zeta^*\equiv \zeta/\nu_0$. We recall that $\eta_0=p/\nu_0$ is the NS shear viscosity of a dilute (elastic) gas. Since $\eta^*$ and $\zeta^*$ are expected to be functions of the (reduced) shear rate $a^*$, the (reduced) friction coefficient $\gamma^*$ and the coefficient of restitution $\al$, Eq.\ \eqref{eq6.13} establishes a relation between $a^*$, $\gamma^*$ and $\al$ and hence, only \emph{two} of them can be independent. Here, we will take $\gamma^*$ and $\al$ as the relevant (dimensionless) parameters measuring the departure of the system from equilibrium.

Before closing this Subsection, it is instructive to display the results derived for the granular suspension in the NS domain (small values of the shear rate). In this regime, the normal stress differences are zero and the form of the shear viscosity coefficient is \cite{GTSH12}
\beq
\label{eq6.14} %\label{2.19}
\eta_\text{NS}=\frac{nT}{\nu_\eta-\frac{1}{2}\left(\zeta_0-\frac{2}{m}\gamma_\text{b}\right)},
\eeq
where $\zeta_0$ is given by Eq.\ \eqref{eq6.11} and the collision frequency $\nu_\eta$ is defined by Eq.\ \eqref{eq4.10} with $a_2=0$ and $\chi=1$.

%\cite{BDKS98}
%\begin{equation}
%\label{eq6.15} %\label{2.21}
%\nu_\eta=\frac{3\nu_0}{4d}\left(1-\alpha+\frac{2}{3}d\right)(1+\alpha).
%\end{equation}
%In Eqs.\ \eqref{eq6.11}, \eqref{eq6.14} and \eqref{eq6.15}, for the sake of simplicity, we have neglected non-gaussian corrections (proportional to the fourth cumulant) to $\zeta_0$, $\eta$ and $\nu_\eta$, respectively.

\subsection{Characteristic time scales and dimensionless numbers}
\label{time_scales}

As it is known, in general there is more than one independent reduced length or time scale in a real flow problem (and, thus, more than one independent Knudsen number \cite{BIRD94}). This feature was analyzed in Ref.\ \cite{VU09} in the context of granular gases. Thus, let us analyze the dimensionless energy balance equation \eqref{eq6.13}. It contains three \emph{homogeneous} terms, each one of them stands for the inverse of the three relevant (dimensionless) time scales of the USF problem: the first term is proportional to the (reduced) shear rate $a^*$ that, according to its definition, is the shearing rate time scale (let us call it $\tau_\text{shear}$); the second term is proportional to $\gamma^*$, thus setting the drag friction time scale ($\tau_\text{drag}$); and finally, the third one, $\zeta^*$ comes from the inelastic cooling characteristic time scale ($\tau_\text{inelastic}$).

A relevant dimensionless number in fluid suspensions is the Stokes number $\text{St}$ \cite{B74}. As in previous works \cite{TK95,SMTK96}, it is defined as the relation between the inertia of suspended particles and the viscous drag characteristic time scale :
\beq
\label{eq6.16} %\label{2.22}
\text{St}=\frac{m a}{3\pi \sigma \mu_g},
\eeq
where we recall that $\mu_g$ is the gas viscosity. According to Eq.\ \eqref{eq6.2}, $\text{St}$ can be easily expressed in terms of $\gamma^*$ and $a^*$ as
\begin{equation}
\label{eq6.17} %\label{Stokes}
\mathrm{St}=\frac{a^*}{\gamma^*/R_\text{diss}},
\end{equation}	
where $R_\text{diss}=1$ for dilute suspensions ($\phi=0$). %Note that the Stokes number is a relevant parameter in fluid suspensions \cite{B74} since it measures the competition between the shearing and viscous friction mechanisms ($a^*$ and $\gamma^*$) on its rheological properties.

\begin{figure}[h]
\centering
\includegraphics[width=0.75 \columnwidth,angle=0]{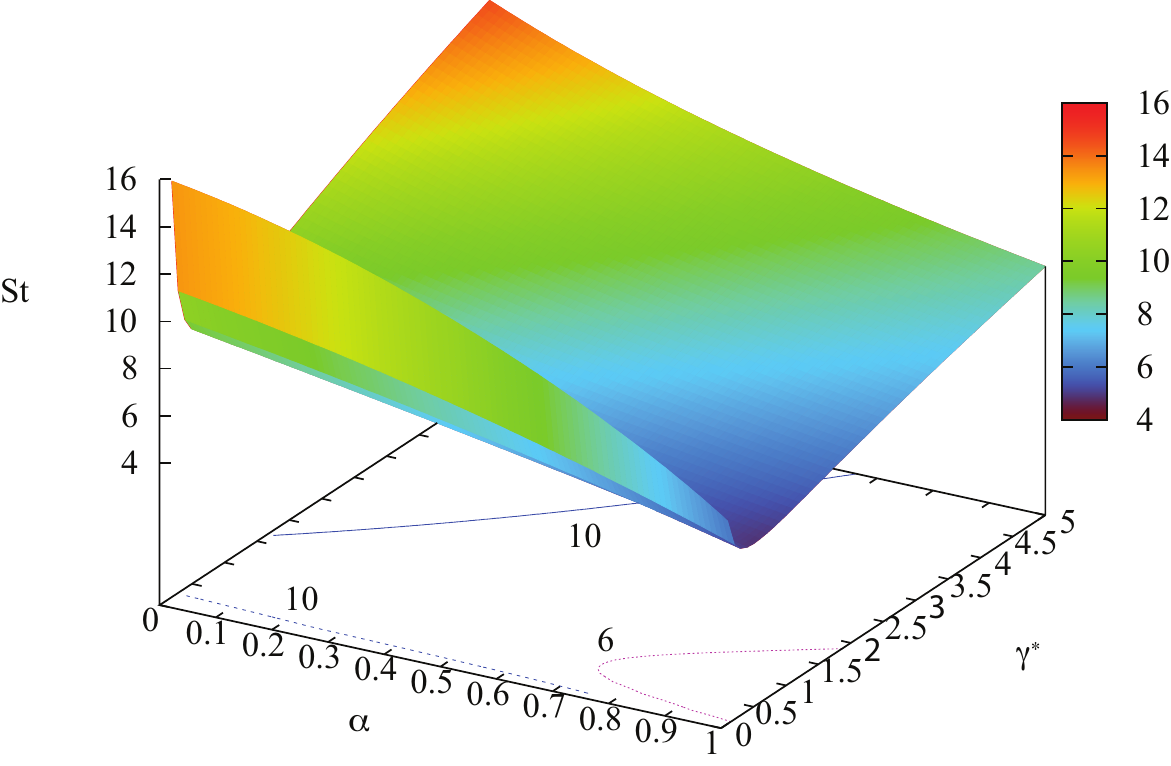}
\caption{$\mathrm{St}(\alpha,\gamma^*)$ surface for a dilute suspension of granular particles. The contours for $\mathrm{St}=6, 10$ have been marked in the $\mathrm{St}=0$ plane. \label{fig1chap6}} % \label{surface}}
\end{figure}

\begin{figure}[h]
\centering
\includegraphics[width=0.75 \columnwidth,angle=0]{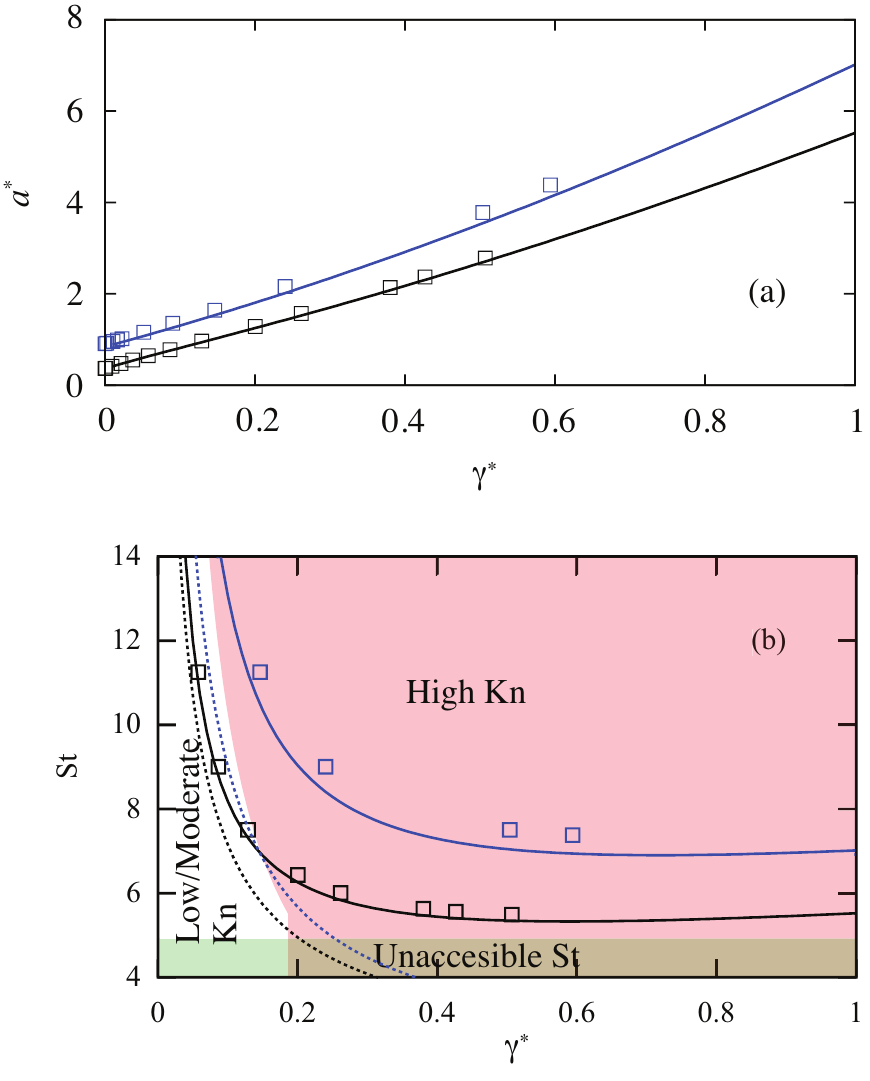}
\caption{Scheme of the flow regimes as they result from the relation \eqref{eq6.13} between the (reduced) shear rate $a^*$, the (reduced) friction coefficient $\gamma^*$ and the Stokes number $\text{St}$ for a dilute granular suspension under USF. Blue (symbols and lines) stands for the case $\alpha=0.5$  and black (symbols and lines) stands for the case $\alpha=0.9$. The solid lines correspond to the results derived from Grad's moment method while the dashed lines refer to the NS predictions. Panel (a): Reduced shear rate $a^*$ vs. $\gamma^*$. Panel (b) Stokes number $\mathrm{St}$ vs. $\gamma^*$. In this panel the three regions commented in the text have been marked: a high Knudsen number region to the right of the panel (in pale red); a low/moderate Knudsen number region (in white) and finally, in the lower part of the panel, the forbidden small $\mathrm{St}$ region (green) may be found.  \label{fig2chap6}}  %\label{class}}
\end{figure}

Since the reduced time scales ($\tau_\text{shear}$, $\tau_\text{drag}$, and $\tau_\text{inelastic}$) have been scaled with the inverse collision frequency $\nu_0^{-1}$, they may be regarded also as the characteristic Knudsen numbers ($\mathrm{Kn}$) of the system \cite{VU09}. For this reason, it is a necessary precondition for a NS hydrodynamic description of the problem (valid only for small enough spatial gradients), that all of them are small. In other words, as soon as one of them (just one) is close to one or higher, the NS approximation is expected to fail \cite{VSG13}.

However, as said before, for the case of the USF regime only two of the relevant Knudsen numbers are actually independent since they are related through Eq.\ \eqref{eq6.13}. For this reason, we additionally need to explore the relation between $\tau_\text{shear}$, $\tau_\text{drag}$ and $\tau_\text{inelastic}$ in order to analyze the limits of a NS description for the granular suspension under USF. For this, the reduced energy balance equation \eqref{eq6.13} can be written in a perhaps more meaningful way for granular suspensions as a function of the Stokes number $\mathrm{St}$, namely,
\begin{equation}
\label{eq6.18} %\label{LTu2}
-\frac{2}{d}\eta^*a^{*} +2\mathrm{St}^{-1}+\frac{\zeta^*}{a^*}=0.
\end{equation}
Once the (scaled) non-Newtonian shear viscosity $\eta^*$ and the (scaled) cooling rate $\zeta^*$ are given in terms of both $\alpha$ and $\gamma^*$, one can obtain the (scaled) shear rate $a^*$ (or equivalently, the reduced temperature $T^*\equiv \nu_0^2/a^2= a^{*-2}$) by solving the energy equation \eqref{eq6.18}. This yields a cubic equation for $T^{*1/2}$ and has therefore three roots. A detailed study of the behavior of these roots has been previously made by Tsao and Koch \cite{TK95} for elastic suspensions and by Sangani \emph{et al.} \cite{SMTK96} for inelastic systems. The analysis shows that in general only one root is real at high values of the Stokes number while the other two are zero and negative (\emph{unphysical} solution). We focus now on the physical solution with positive temperature (that corresponds to the \textit{ignited} state of \cite{SMTK96}) by using the more general nonlinear Grad's solution derived in Sec.\ \ref{sec3chap6}.

In Fig.\ \ref{fig1chap6} we plot first the surface $\mathrm{St}(\alpha,\gamma^*)$ verifying Eq.\ \eqref{eq6.18}. According to Fig.\ \ref{fig1chap6}, it is quite evident that it is not possible to reach a null value of the Stokes number. This is consistent with the energy equation \eqref{eq6.18} since the latter value would imply $\text{St}^{-1}\to \infty$ and so, a balance between the different effects would not be possible. Fig.\ \ref{fig2chap6} is the representation of two constant $\alpha$ curves of this surface, as obtained from the nonlinear Grad's solution (solid lines), explained in Section \ref{sec3achap6}, and DSMC (symbols) for $d=3$ (spheres). The NS prediction for $\text{St}(\gamma^*)$ obtained from Eqs.\ \eqref{eq6.13} and \eqref{eq6.14} is also plotted (dashed lines) for the sake of comparison. In Fig.\ \ref{fig2chap6} (b) we have marked with different colors three different regions: white stands for the region with $a^* \lesssim 1$, where the NS description is expected to apply (or in other words, where non-Newtonian corrections to rheological properties would not be significant), whereas red stands for the region where the NS approximation is expected to fail ($a^*\gtrsim 1$). The inelastic time scale $\tau_\text{inelastic}$ would keep small as long as we do not represent too large inelasticity values. The drag time scale $\tau_\text{drag}$ (or equivalently $\gamma^*$) is represented here only below 1. Thus, the only concern would be tracking small enough values of $\tau_\text{shear}$ (or equivalently $a^*$) values. For this reason, the moderate to large $\mathrm{Kn}$ regions in Fig.\ \ref{fig2chap6}(b) are separated by the curve that follows from the value $\gamma^*(\alpha,a^*=1)$ extracted from Eq.\ \eqref{eq6.18}

The dark green region denotes the low $\mathrm{St}$ region that is not accessible for hydrodynamics (negative solutions for $T^{*1/2}$). As we can see in both panels (a) and (b), the agreement between Grad's solution (which takes into account non-Newtonian corrections to the shear viscosity) and simulations is excellent as long as keep in the small $\mathrm{Kn}$ region (both $\gamma^*<1$ and $a^*<1$). 

The accuracy of Grad's solution extends deep inside the large $\mathrm{Kn}$ region, specially for lower inelasticities (note the black curve and symbols in the pale red region of Fig.\ \ref{fig1chap6}). On the other hand, as expected, the NS prediction exhibits significant discrepancies with simulations when $\text{Kn}\gg 1$. 

Please note that, although this is somewhat masked in the small range of values of $\gamma^*$ considered in Fig.\ \ref{fig2chap6}(b) the Stokes number $\mathrm{St}$ is always a bivalued function of the (scaled) friction coefficient $\gamma^*$, as can be clearly seen in Fig.\ \ref{fig1chap6}. Also notice from Fig.\ \ref{fig1chap6} that $\mathrm{St}$ always has a minimum with respect to $\gamma^*$ (at a given value of $\alpha$), although for scale reasons it is not very noticeable in Fig.\ \ref{fig2chap6}.

It is important to recall again that the need for more complex constitutive equations (namely, those provided by Grad's moment method) is not a signal of a breakdown of hydrodynamics \cite{G03,TG98}, only a failure of the NS approximation \cite{DB99,VU09}. %Also, let us note as an  important feature not described previously that $\mathrm{St}(\gamma^*)$ has two roots for each $\mathrm{St}$ value, as we can see in Fig.\ \ref{fig1chap6}.

\section{Theoretical approaches}
\label{sec3chap6}

\subsection{Grad's moment method of the Boltzmann equation}
\label{sec3achap6}

We are interested here in obtaining the explicit forms of the relevant elements of the (scaled) pressure tensor $P_{ij}^*$ for a dilute granular suspension in terms of $a^*$, $\gamma^*$ and $\al$. To get it, we multiply both sides of Eq.\ \eqref{eq6.7} by $mV_iV_j$ and integrate over velocity. The result is
\beq
\label{eq6.19} %\label{3.13}
a_{ik}P_{kj}+a_{jk}P_{ki}+\frac{2\gamma_\text{b}}{m} P_{ij}=\Lambda_{ij},
\eeq
where
\beq
\label{eq6.20} %\label{3.14}
\Lambda_{ij}\equiv \int\; \dd \mathbf{V}\; m V_iV_j J[\mathbf{V}|f,f],
\eeq
and we recall that $a_{ij}=a\delta_{ix}\delta_{jy}$. The exact expression of the collision integral $\Lambda_{ij}$ is not known, even in the elastic case. However, a good estimate can be expected by using Grad's approximation \cite{G49} where the exact distribution function $f$ is replaced by
\begin{equation}
\label{eq6.21} %\label{3.1}
f(\mathbf{V})\to f_\text{M}(\mathbf{V}) \left(1 +\frac{m}{2nT^2}V_iV_j \Pi_{ij}\right).
\end{equation}
Here, $f_M(\mathbf{V}$) is the Maxwellian distribution defined by Eq.\ \eqref{eqfMaxwell}
%\begin{equation}
%\label{eq6.22} %\label{3.11}
%f_\text{M}(\mathbf{V})=n\left(\frac{m}{2\pi T}\right)^{d/2}e^{-mV^2/2T}
%\end{equation}
%is the (local) equilibrium distribution function  and
and
\begin{equation}
\label{eq6.23} %\label{3.12}
\Pi_{ij}=P_{ij}-p\delta_{ij}
\end{equation}
is the traceless part of the pressure tensor. Upon writing the distribution function \eqref{eq6.21} we have taken into account that the heat flux is zero in the USF and we have also neglected the contribution of the fourth-degree velocity moment to $f$. This contribution has been considered in Ref.\ \cite{G13} for the calculation of the NS transport coefficients of a granular fluid at moderate densities.

The collisional moment $\Lambda_{ij}$ can be determined when Eq.\ \eqref{eq6.21} is inserted into Eq.\ \eqref{eq6.20}. After some algebra (see Appendix \ref{AppendixH} for details), we obtain the expression of $\Lambda_{ij}$ for inelastic hard spheres ($d=3$) given by
\beq
\label{eq6.24} %\label{3.15}
\Lambda_{ij}=-p\nu_0 (1+\al)\left[\frac{5}{12}(1-\al)\delta_{ij}+\frac{3-\al}{4}
\left(\Pi_{ij}^*+\frac{1}{14}\Pi_{ik}^*\Pi_{kj}^*
\right)-\frac{5+3\al}{672}\Pi_{k\ell}^*\Pi_{k\ell}^*\delta_{ij}\right],
\eeq
where $\Pi_{ij}^*\equiv \Pi_{ij}/p$. In the case of inelastic hard disks ($d=2$), the expression of $\Lambda_{ij}$ is
\beq
\label{eq6.25} %\label{3.16}
\Lambda_{ij}=-p\nu_0 \frac{1+\al}{2}\left[(1-\al)\delta_{ij}+\frac{7-3\al}{4}\Pi_{ij}^*+\frac{3}{64}(1-\al) \Pi_{k\ell}^*\Pi_{k\ell}^*\delta_{ij}\right].
\eeq
As we noted before, $\Lambda_{ij}$ has been evaluated by retaining \emph{all} the quadratic terms in the tensor $\Pi_{ij}^*$. In particular, Eq.\ \eqref{eq6.24} reduces to the simpler expression obtained by Sangani \emph{et al.} \cite{SMTK96} for $d=3$ if we suppress  the quadratic terms in $\Pi_{ij}^*$. Also, if we particularize Eq.\ \eqref{eq6.24} for $\al=1$, we obtain
\beq
\label{eq6.26} %\label{3.15.1}
\Lambda_{ij}=-p \nu_0 \left[\Pi_{ij}^*+\frac{1}{14}\left( \Pi_{ik}^*\Pi_{kj}^*-\frac{1}{3}\Pi_{k\ell}^*\Pi_{k\ell}^*\delta_{ij}\right)\right],
\eeq
and hence the expression of $\Lambda_{ij}$ derived by Tsao and Koch \cite{TK95} for the special case of perfectly elastic particles (see Eq.\ (3.7) of \cite{TK95}) is recovered. Thus, the expression \eqref{eq6.24} for the collisional moment $\Lambda_{ij}$ for inelastic hard spheres is more general and includes the results derived in previous works as particular cases.

In addition, we have also checked that the expression \eqref{eq6.24} agrees with a previous and independent derivation of $\Lambda_{ij}$ for inelastic hard spheres \cite{GT12}. This shows the consistency of our nonlinear Grad's solution.

The nonlinear contribution $\zeta_2$ to the cooling rate [defined by Eq.\ \eqref{eqzeta}] can be obtained for spheres and disks from Eqs.\ \eqref{eq6.24} and \eqref{eq6.25}, respectively. The final expressions for the dimensionless cooling rate $\zeta^*$ are
\beq
\label{eq6.27} %\label{3.17}
\zeta_\text{spheres}^*=\frac{5}{12}(1-\al^2)\left(1+\frac{1}{40}\Pi_{k\ell}^*\Pi_{k\ell}^*\right),
\eeq
\beq
\label{eq6.28} %\label{3.18}
\zeta_\text{disks}^*=\frac{(1-\al^2)}{2}\left(1+\frac{3}{64}\Pi_{k\ell}^*\Pi_{k\ell}^*\right).
\eeq
%Here again, this is a more general and accurate expression of the cooling rate for dilute granular suspensions. Of course, for elastic collisions ($\al=1$), we recover the limit $\zeta^*=0$ \cite{TK95}. Moreover in the linear in $\Pi_{ij}^*$ approach, $\zeta^*\to (5/12)(1-\al^2)$ for spheres, which agrees with the previous results \cite{SMTK96}.

The knowledge of the collisional moment $\Lambda_{ij}$ allows us to get the explicit form of the relevant elements of the pressure tensor $P_{ij}^*$. Their forms are provided in Appendix \ref{AppendixH}.

%%%%%%%%%%%%%%%%%%%%%%%%%%%%%%%%%%%%%%%%%%%%%%%%%%%%%%%%%%%%%%%%%%%%%%%%%%%%%%%%%%%%%%%%%%%%%%

\subsection{BGK-type kinetic model of the Boltzmann equation}
\label{sec3bchap6}

Now we consider the results derived for the USF from a BGK-type kinetic model of the Boltzmann equation \cite{BDS99}. In the USF problem, the steady kinetic model for the granular suspension described by the Boltzmann equation \eqref{eq6.7} becomes
\beq
\label{eq6.29} %\label{3.26}
-aV_y\frac{\partial f}{\partial V_x}-\frac{\gamma_\text{b}}{m} \frac{\partial}{\partial
{\bf V}}\cdot {\bf V} f =-\psi(\alpha)\nu_0 \left(f-f_\text{M}\right)+\frac{\zeta_0}{2}\frac{\partial}{\partial{\bf V}}\cdot {\bf V} f,
\eeq
where $\nu_0$ is the effective collision frequency defined by Eq.\ \eqref{eq5.13.3}, $f_\text{M}$ is given by Eq.\ \eqref{eqfMaxwell}, $\zeta_0$ is defined by Eq.\ \eqref{eq6.11}, and $\psi(\alpha)$ is a free parameter of the model chosen to optimize the agreement with the Boltzmann results.

One of the main advantages of using the kinetic model \eqref{eq6.29} instead of the Boltzmann equation is that it lends itself to get an exact solution. The knowledge of the form of $f(\mathbf{V})$ allows us to determine \emph{all} its velocity moments. The explicit forms of the distribution function $f(\mathbf{V})$ as well as its moments are provided in Appendix \ref{AppendixI}. In particular, the relevant elements of the pressure tensor are given by
\beq
\label{eq6.30} %\label{3.39}
\Pi_{yy}^*=\Pi_{zz}^*=-\frac{2\widetilde{\epsilon}}{1+2\widetilde{\epsilon}},
\quad \Pi_{xy}^*=-\frac{\widetilde{a}}{(1+2\widetilde{\epsilon})^2},
\eeq
where the (dimensionless) shear rate $\widetilde{a}$ obeys the equation
\beq
\label{eq6.31} %\label{3.40}
\widetilde{a}^2=d\widetilde{\epsilon}(1+2\widetilde{\epsilon})^2.
\eeq
Here, $\widetilde{a}\equiv a^*/\psi$, $\widetilde{\zeta}\equiv \zeta^*/\psi$, $\widetilde{\epsilon}\equiv \widetilde{\gamma}+\widetilde{\zeta}/2$, and  $\widetilde{\gamma}\equiv \gamma^*/\psi$. The expressions \eqref{eq6.30} and \eqref{eq6.31} are fully equivalent to linear Grad's predictions \eqref{apH15}-\eqref{apH17}, except that $\psi$ is replaced by $\beta$.

%%%%%%%%%%%%%%%%%%%%%%%%%%%%%%%%%%%%%%%%%%%%%%

\section{Numerical solutions: DSMC method}
\label{sec4chap6}

As we said in the Introduction, the third method consists in obtaining a numerical solution to the Boltzmann equation \eqref{eq6.7} by means of the DSMC method \cite{BIRD94} applied to inelastic hard spheres. More concretely, the algorithm we used is analogous to the one employed in Ref.\ \cite{MGSB99} where the USF state becomes homogeneous in the frame moving with the flow velocity $\mathbf{U}$. Here, we have simply added the drag force coming from the interaction between the solid particles and the surrounding interstitial fluid. The initial state is the same for all simulations, namely, Gaussian velocity distributions with homogeneous density and temperature. We have observed in most of the cases that, after a relatively short transient, a steady state is reached. In this state, the relevant quantities of the USF problem (nonzero elements of the pressure tensor, the kurtosis and the velocity distribution function) are measured.
%Since the base of the algorithm has been explained in detail in previous papers \cite{MGSB99,CVG13}, we skip here these details and only comment that
 
We have performed systematic simulation series for two different situations: (i) by varying the (scaled) friction coefficient $\gamma^*$ at a given value of $\alpha$ and, conversely, (ii) by varying the coefficient of restitution $\alpha$ at a given value of $\gamma^*$. In addition, the series corresponding to varying $\gamma^*$ have been employed for graphs with varying the Stokes number $\mathrm{St}$ also.

%The use of the DSMC method is convenient since it is considered as an accurate method of solving the Boltzmann equation. Here, the DSMC results can be considered as a clean way to assess the degree of reliability of the theoretical descriptions we developed (Grad's moment method and BGK-type kinetic model). This is what we do, along with presentation of the results, in the following Section.

%%%%%%%%%%%%%%%%%%%%%%%%%%%%%%%%%%%

\section{Results}
\label{sec5chap6}

We devote this Section to direct comparative presentation of the results obtained from all three independent routes we have followed for this Chapter. Although the theoretical expressions apply for spheres and disks, for the sake of brevity we present only results for the physical case of a three-dimensional system ($d=3$). Given that the computational algorithm can be easily adapted to disks, a comparison between theory and simulation for $d=2$ could be also performed.

\subsection{Dilute granular suspensions}
\label{sec5achap6}

\begin{figure}[h]
\centering
\includegraphics[width=0.75 \columnwidth,angle=0]{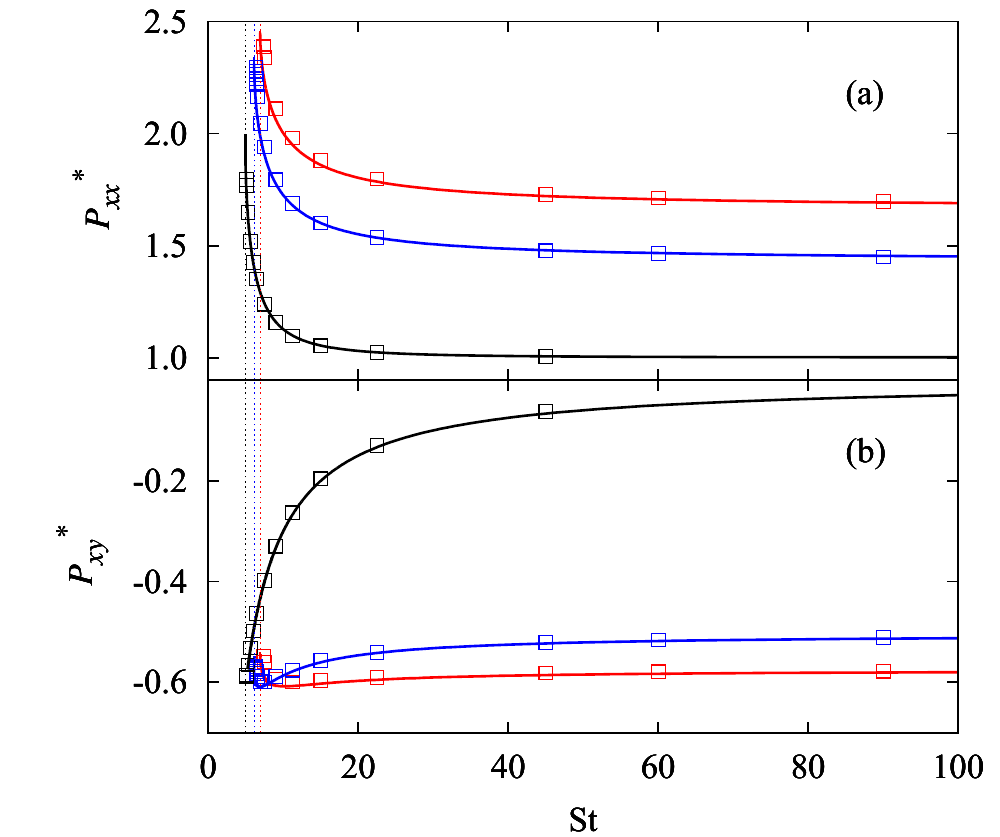}
\caption{Dependence of the (reduced) elements of the pressure tensor $P_{xx}^*$ (panel (a)) and $P_{xy}^*$ (panel (b)) on the Stokes number $\text{St}$ for several values of the coefficient of restitution $\al$: $\alpha=1$ (black), $\alpha=0.7$ (blue) and $\alpha=0.5$ (red). The solid lines are the theoretical results obtained from nonlinear Grad's solution while the symbols refer to the results obtained from DSMC. We have marked as vertical dotted lines the minimum allowed value for the Stokes number $\mathrm{St}$.\label{fig3chap6} } %\label{figPij}
\end{figure}

\begin{figure}[h]
\centering
\includegraphics[width=0.75 \columnwidth,angle=0]{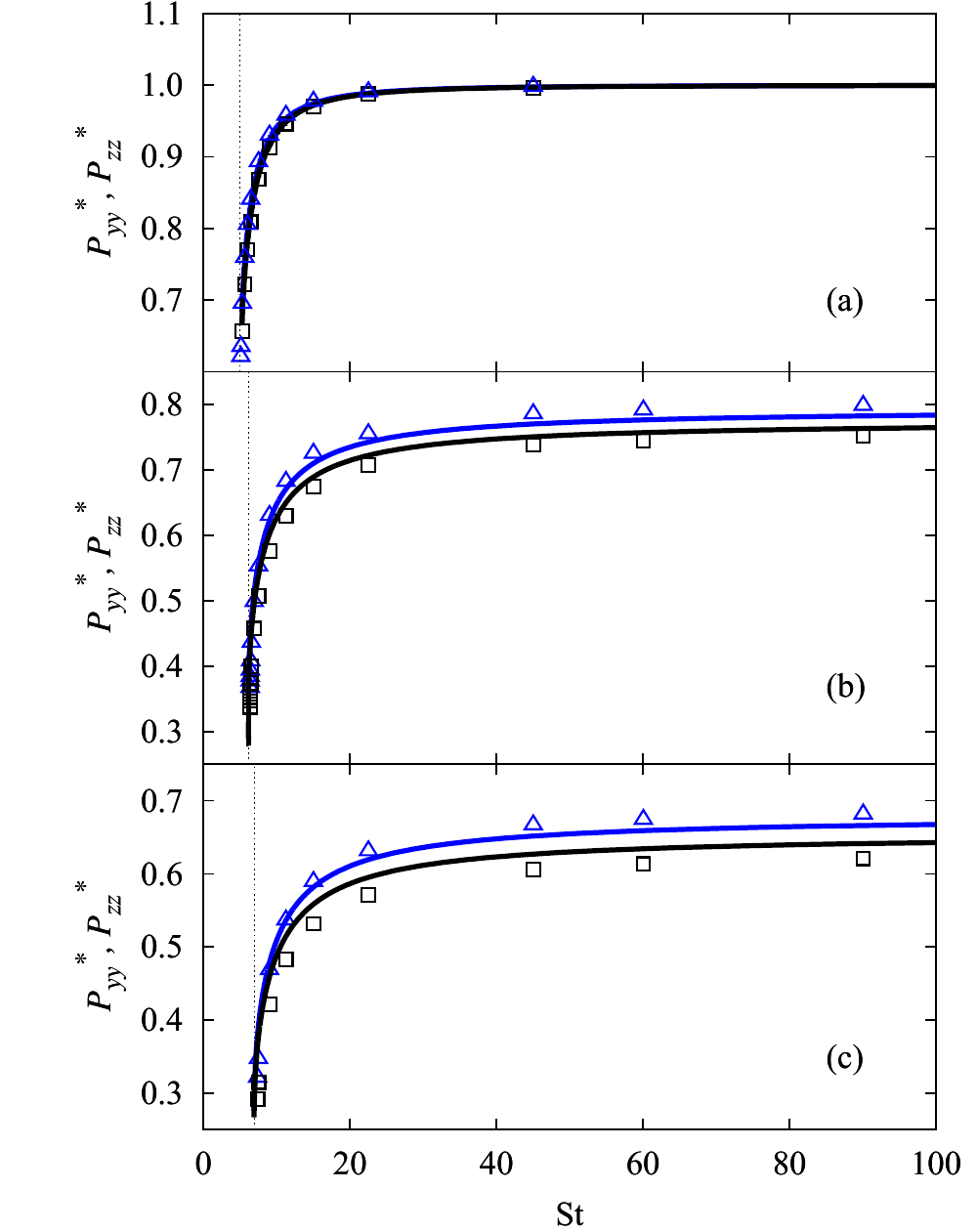}
\caption{Dependence of the (reduced) diagonal elements of the pressure tensor $P_{yy}^*$ (black lines and squares) and $P_{zz}^*$ (blue lines and triangles) on the Stokes number $\text{St}$ for several values of the coefficient of restitution $\al$: $\alpha=1$ (a), $\alpha=0.7$ (b) and $\alpha=0.5$ (c). The solid lines are the theoretical results obtained from nonlinear Grad's solution while the symbols refer to the results obtained from DSMC. As in Fig.\ \ref{fig3chap6}, we have marked as vertical dotted lines the minimum allowed value of the Stokes number $\mathrm{St}$ for each value of $\al$.
\label{fig4chap6}} %\label{figPii}
\end{figure}
\begin{figure}[h]
\centering
\includegraphics[width=0.75 \columnwidth,angle=0]{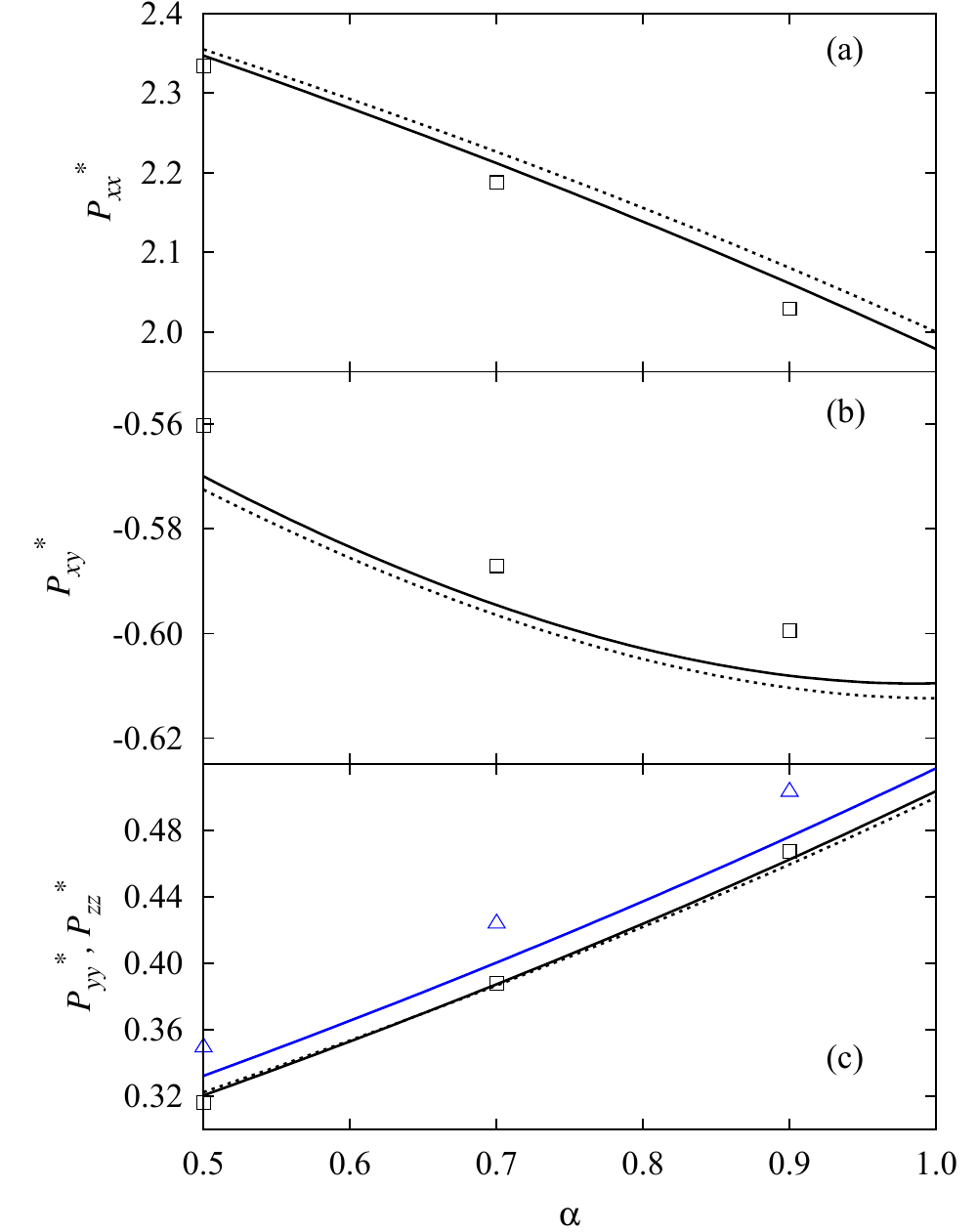}
\caption{Plot of the (reduced) nonzero elements of the pressure tensor $P_{xx}^*$ (panel a), $P_{xy}^*$ (panel b), $P_{yy}^*$ and $P_{zz}^*$ (panel c) as functions of the coefficient of restitution $\al$ for  $\gamma^*=0.5$. The solid and dotted lines correspond to the results obtained from nonlinear and linear Grad's solution, respectively. Symbols refer to DSMC. In the panel (c), the blue solid line and triangles are for the element $P_{zz}^*$ while the black solid line and squares are for the element $P_{yy}^*$. Note that linear Grad's solution (dotted line) yields $P_{yy}^*=P_{zz}^*$.
\label{fig5chap6}} %\label{figPijalfa}
\end{figure}

Fig.\ \ref{fig3chap6} shows the dependence of the (reduced) elements $P_{xx}^*$ and $P_{xy}^*$ of the pressure tensor on the Stokes number $\text{St}$. Here, we have performed simulation series by varying the (reduced) friction coefficient $\gamma^*$ (or equivalently, $\text{St}$)  for three different values of the coefficient of restitution: $\alpha=1$ (elastic case), $\al=0.7$ and $\al=0.5$. Recall that the diagonal elements of the pressure tensor are related through $P^*_{xx}+P^*_{yy}+(d-2)P^*_{zz}=d$. In this graph, only the predictions given by the so-called nonlinear Grad's solution are plotted. The results obtained from linear Grad's solution are practically indistinguishable from the latter ones for the cases considered in this plot. The comparison between theory (solid lines) and computer simulations (symbols) shows an excellent agreement for all values of the Stokes number represented here, independently of the degree of inelasticity of collisions in the granular gas.

As noted in the Introduction, one of the drawbacks of linear Grad's solution is that it yields $P_{yy}^*=P_{zz}^*$ and hence, the second viscometric function (proportional to $P_{yy}^*-P_{zz}^*$ \cite{BAH87}) vanishes. This failure of linear Grad's solution is also present at moderate densities (see Eq.\ (4.33) of \cite{SMTK96}). Fig.\ \ref{fig4chap6} shows the dependence of the normal elements $P_{yy}^*$ and $P_{zz}^*$ on the Stokes number $\text{St}$ as obtained from the DSMC method (symbols) and nonlinear Grad's solution. It is quite apparent that both simulations and theory show that $P_{zz}^*>P_{yy}^*$. This is specially relevant in granular suspensions since we have two different sink terms ($\gamma^*$ and $\zeta^*$) in the energy balance equation \eqref{eq6.13}. And thus, the non-Newtonian effects like $P_{yy}^*\neq P_{zz}^*$ are expected to be stronger. The balance of these two terms with the viscous heating term ($\eta^* a^{*2}$) requires high shear rates as can be seen in Fig.\ \ref{fig2chap6}. We observe in Fig.\ \ref{fig4chap6} that our theory captures quantitatively well the tendency of $P^*_{yy}$ (the diagonal element of the pressure tensor in the direction of shear flow) to become smaller than $P^*_{zz}$, this tendency being stronger as inelasticity increases (and disappearing completely in the elastic limit $\alpha=1$). It is also apparent that the dependence of both $P_{zz}^*$ and $P_{yy}^*$ on the Stokes number is qualitatively well captured by the nonlinear Grad's solution, even for strong collisional dissipation. Finally, regarding rheology and as a complement of Figs.\ \ref{fig3chap6} and \ref{fig4chap6}, Fig.\ \ref{fig5chap6} shows the $\al$-dependence of the relevant elements of the pressure tensor at a given value of the (scaled) friction coefficient $\gamma^*$. Since the value of $\gamma^*$ is relatively high ($\gamma^*=0.5$), the results presented in Fig.\ \ref{fig5chap6} can be considered as a stringent test for both linear and nonlinear Grad's solutions. Although the linear Grad's solution exhibits a reasonably good agreement with DSMC data, we see that the nonlinear Grad's solution mitigates in part the discrepancies observed by using the linear approach since the former theory correctly predicts the trend of the normal stress difference $P_{zz}^*-P_{yy}^*$ and also improves the agreement with simulations for the elements $P_{xx}^*$ and $P_{xy}^*$. On the other hand, since the system is quite far from equilibrium, there are still quantitative discrepancies between the nonlinear theory and simulations.

\begin{figure}[h]
\centering
\includegraphics[width=0.75 \columnwidth,angle=0]{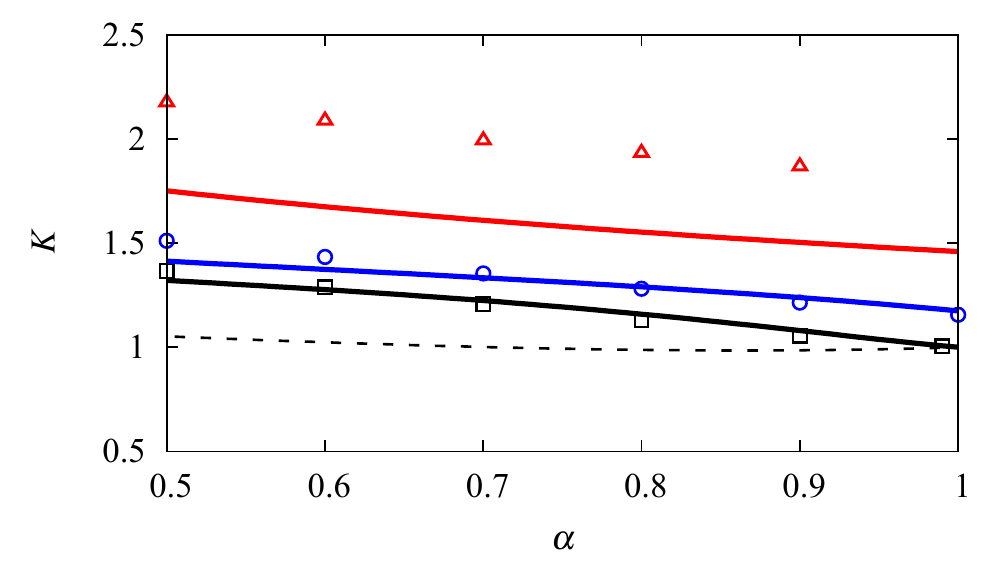}
\caption{Plot of the kurtosis $K\equiv \langle V^4 \rangle/\langle V^4 \rangle_0$ versus the coefficient of restitution $\al$ for three different values of the (reduced) friction coefficient $\gamma^*$: $\gamma^*=0$ (black line and squares), $\gamma^*=0.1$ (blue line and circles) and $\gamma^*=0.5$ (red line and triangles). The solid lines correspond to the results obtained from the BGK-type model while symbols refer to DSMC results. The dashed line is the result obtained in Ref.\ \cite{GTSH12} for the homogeneous cooling state \label{fig6chap6}.} %\label{kurtosis}
\end{figure}

Next, we present results for the kurtosis or fourth order cumulant $K\equiv \langle V^4 \rangle/\langle V^4 \rangle_0$ where
\beq
\label{eq6.32} %\label{K}
\langle V^k \rangle=\frac{1}{n}\int\; \dd \mathbf{V} V^k f(\mathbf{V}),
\eeq
and
\beq
\label{eq6.33} %\label{K0}
\langle V^k \rangle_0=\frac{1}{n}\int\; \dd \mathbf{V} V^k f_\text{M}(\mathbf{V}).
\eeq

The dependence of the kurtosis on both $\gamma^*$ and $\al$ can be easily obtained from the results derived from the BGK-type kinetic model [see Eq.\ \eqref{3.38} for the BGK velocity moments].
Note that $\langle V^k \rangle=\langle V^k \rangle_0$ if one uses Grad's distribution \eqref{eq6.21}, which is a failure of Grad's solution since $K$ is clearly different from 1. Fig.\ \ref{fig6chap6} shows the dependence of $K$ on the coefficient of restitution $\al$ for hard spheres ($d=3$) and three different values of the (reduced) friction coefficient $\gamma^*$: $\gamma^*=0$ (dry granular gas),  $\gamma^*=0.1$ and $\gamma^*=0.5$. In the case of elastic collisions ($\al=1$), $K=1$ only for $\gamma^*=0$ since in this case the system is at equilibrium ($f=f_\text{M}$). We have also included the result obtained in Ref.\ \cite{GTSH12} in the HCS, which is independent of $\gamma^*$. It is important to remark first that the simulation results obtained independently here for $\gamma^*=0$ in Fig.\ \ref{fig6chap6} are consistent with those previously reported for a sheared granular gas with no interstitial fluid \cite{AS05}. For low values of $\gamma^*$, we see that the agreement between theory and simulation is very good in the full range of values of inelasticities represented here. This shows again the reliability of the BGK model to capture the main trends observed in granular suspensions. On the other hand, the agreement is only qualitative for relatively high values of the friction coefficient $\gamma^*$ since the BGK results clearly underestimate the value of the kurtosis given by computer simulations. These discrepancies between the BGK-type model and DSMC for the fourth-degree velocity moment in non-Newtonian states are not surprising since the above kinetic model does not intend to mimic the behavior of the \emph{true} distribution function beyond the thermal velocity region. As expected, it is apparent that the prediction for $K$ in the homogeneous state differs clearly from the one obtained in the DSMC simulations at $\gamma^*=0$.

\begin{figure}[h]
\centering
\includegraphics[width=0.75 \columnwidth,angle=0]{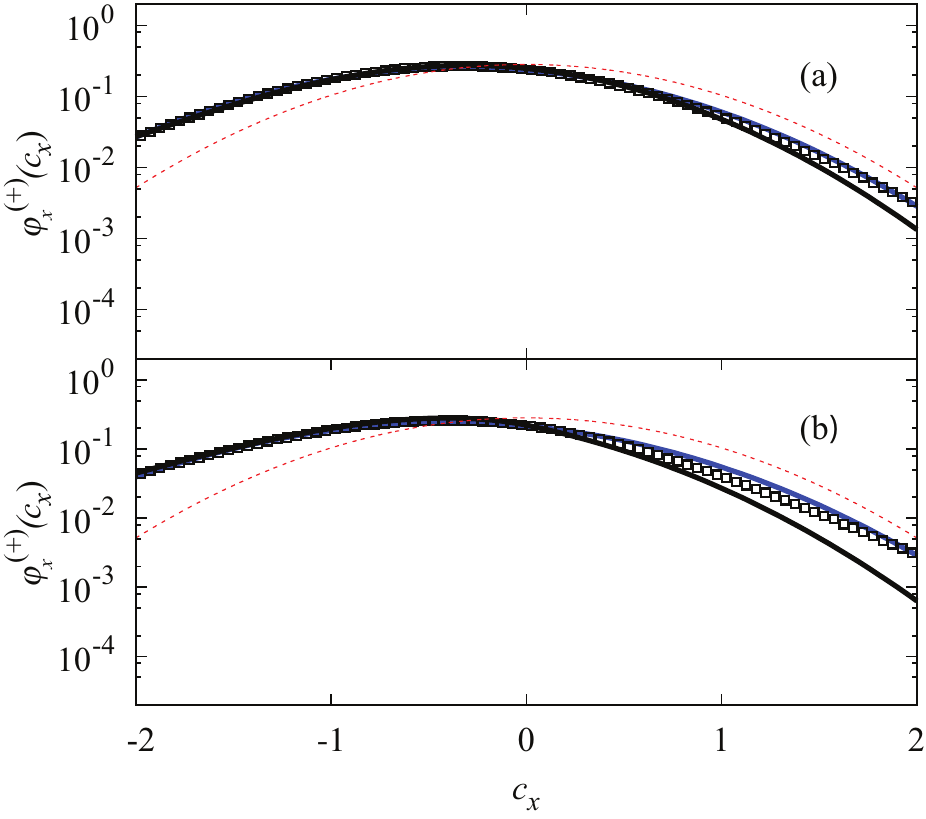}
\caption{Logarithmic plots of the marginal distribution function $\varphi_x^{(+)}(c_x)$, as defined in Eq.\ \eqref{phix}. Two cases are represented here: (a) $\alpha=0.9$, $\gamma^*=0.1$ and (b) $\alpha=0.5$, $\gamma^*=0.1$. The black and blue solid lines are the theoretical results derived from the BGK model and the ME formalism, respectively, while the symbols represent the simulation results. The red dotted lines are the (local) equilibrium distributions.  \label{fig7chap6}}%    \label{fxre}}
\end{figure}
\begin{figure}[h]
\centering
\includegraphics[width=0.75 \columnwidth,angle=0]{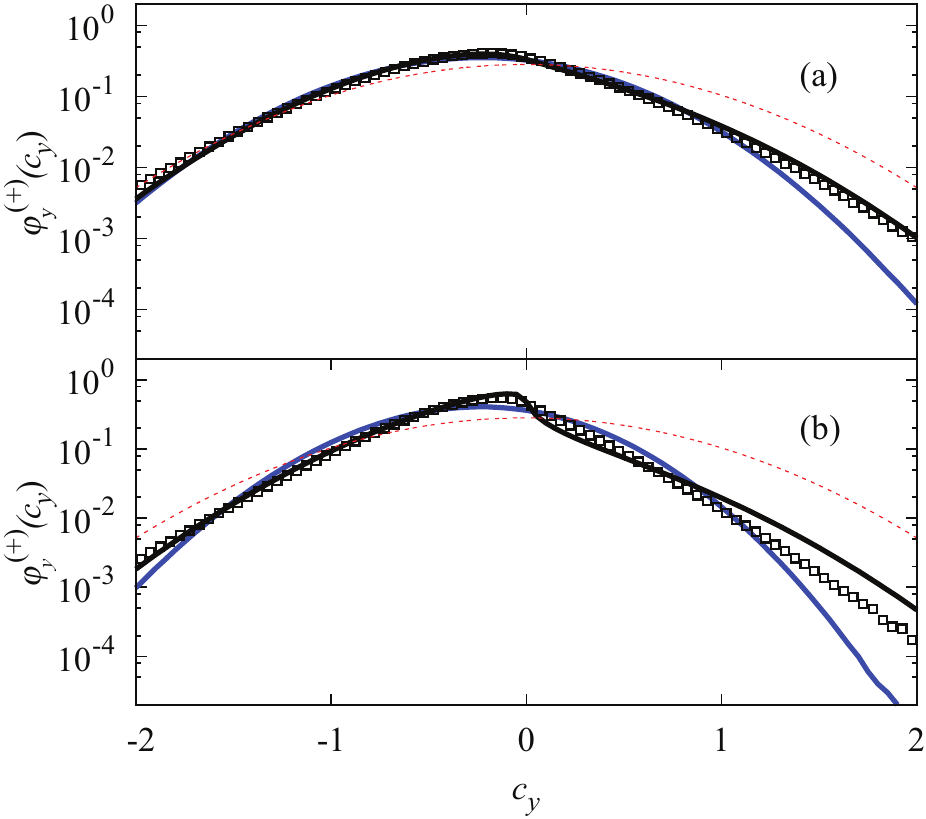}
\caption{Logarithmic plots of the marginal distribution function $\varphi_y^{(+)}(c_y)$, as defined in Eq.\ \eqref{phiy}. Two cases are represented here: (a) $\alpha=0.9$, $\gamma^*=0.1$ and (b) $\alpha=0.5$, $\gamma^*=0.1$. The black and blue solid lines are the theoretical results derived from the BGK model and the ME formalism, respectively, while the symbols represent the simulation results. The red dotted lines are the (local) equilibrium distributions. \label{fig8chap6}}% \label{fyre}}
\end{figure}

Apart from the rheological properties and the high velocity moments, the solution to the BGK-type model provides the explicit form of the velocity distribution function $f(\mathbf{V})$. Figs.\ \ref{fig7chap6} and \ref{fig8chap6} show the marginal distributions $\varphi_x^{(+)}(c_x)$ [defined by Eq.\ \eqref{phix}] and $\varphi_y^{(+)}(c_y)$ [defined by Eq.\ \eqref{phiy}], respectively, for $\gamma^*=0.1$ and two different values of the coefficient of restitution $\al$: $\al=0.9$ (moderate inelasticity) and $\al=0.5$ (strong inelasticity). The black solid lines are the results derived from the BGK model and the symbols represent DSMC. For the sake of completeness, it is interesting to use the Maximum-Entropy (ME) formalism \cite{BM91} to construct the distribution maximizing the functional
\beq
\label{eq6.34} %\label{ME}
-\int\; \dd \mathbf{V}\; f(\mathbf{V})\; \ln f(\mathbf{V}),
\eeq
subjected to the constraints of reproducing the density $n$ and the pressure tensor $\mathsf{P}$. In the three-dimensional case, this yields
\beq
\label{eq6.35} %\label{ME.1}
f(\mathbf{V})=n \pi^{-3/2} \det\left( \mathbf{Q} \right)^{1/2} \exp\left(-\mathbf{V}\cdot \mathbf{Q} \cdot \mathbf{V} \right),
\eeq
where $\mathbf{Q}\equiv \frac{1}{2}m n \mathsf{P}^{-1}$. The ME approximation was employed by Jenkins and Richman \cite{JR88} in order to determine the kinetic contributions to the pressure tensor in a sheared granular fluid of hard disks. Moreover, in Figs.\ \ref{fig7chap6} and \ref{fig8chap6}, as a reference the (local) equilibrium distributions (red dotted lines) are also represented. Although not shown in Figs.\ \ref{fig7chap6} and \ref{fig8chap6}, Grad's distribution \eqref{eq6.21} could lead to unphysical (negative) values of the marginal
distributions $\varphi_x^{(+)}(c_x)$ and $\varphi_y^{(+)}(c_y)$ for large velocities. This is again a drawback of Grad's solution not shared by the BGK's since the latter is always positive definite for any range of velocities considered. Regarding the comparison between the different results, since the (reduced) shear rate is not small [see for instance, Fig.\ \ref{fig2chap6} for $\al=0.5$ and $\gamma^*=0.1$], we observe that the distortion from the Gaussian distribution is quite apparent in the three different approaches (BGK, ME and DSMC). Two anisotropic features of the USF state are seen. First, the functions $\varphi_x^{(+)}(c_x)$ and $\varphi_y^{(+)}(c_y)$ are asymmetric since $\varphi_x^{(+)}(|c_x|)<\varphi_x^{(+)}(-|c_x|)$ and $\varphi_y^{(+)}(|c_y|)<\varphi_y^{(+)}(-|c_y|)$. This is a physical effect induced by shearing since the shear stress $P_{xy}^*<0$. The second feature is the non-Newtonian property $\varphi_x^{(+)}(c_x)<\varphi_y^{(+)}(c_y)$. In fact, the marginal distribution $\varphi_x^{(+)}(c_x)$ is thicker than $\varphi_y^{(+)}(c_y)$, in consistency with the result $P_{xx}^*-P_{yy}^*>0$. The above two effects are more pronounced for $\al=0.5$ than for $\al=0.9$. With respect to the comparison between theory and simulation, we observe that in general the agreement between theoretical predictions (the BGK model and the ME formalism) and simulation data is excellent in the region of thermal velocities ($|c_i|\sim 1$). It is also apparent that while the ME approach compares better with simulations than the BGK results for the distribution $\varphi_x^{(+)}(c_x)$, the opposite happens for the distribution $\varphi_y^{(+)}(c_y)$. In particular, in the case of $\al=0.9$ the BGK model (the ME formalism) yields an excellent agreement with DSMC over the complete range of velocities studied for the distribution $\varphi_y^{(+)}(c_y)$ [$\varphi_x^{(+)}(c_x)$]. On the other hand, for larger velocities and strong collisional dissipation, there are quantitative discrepancies between theoretical predictions and simulations.

\subsection{Granular suspensions at moderate densities}
\label{sec5bchap6}

Although the main results of this Chapter follow from to the study of sheared granular dilute suspensions described by the Boltzmann kinetic equation (which strictly applies for zero volume fraction $\phi=0$), it is interesting to extend them to the case of \emph{moderately} dense suspensions. For this regime of densities, the inelastic Enskog kinetic equation \cite{BDS97} describes the time evolution of the velocity distribution function. %Although the Enskog equation neglects velocity correlations among particles that are about to collide, it retains the spatial correlations arising from volume exclusion effects and thus is expected to be applicable for moderate densities.

Application of Grad's moment method to the Enskog equation follows similar steps as those made before for the Boltzmann equation (see Subsection \ref{sec3achap6}). On the other hand, the application of this method to dense systems is much more intricate than for dilute granular suspensions and hence, one has to consider additional approximations such as to neglect terms that are quadratic or higher order in the pressure tensor and/or the shear rate. The Enskog kinetic equation has been recently solved \cite{G13} by means of Grad's moment method to determine the NS transport coefficients of a $d$-dimensional dry granular fluid. The forms of the kinetic $P_{ij}^k$ and collisional $P_{ij}^c$ contributions to the pressure tensor $P_{ij}=P_{ij}^k+P_{ij}^c$ can be easily obtained in the USF problem when one takes into account the results derived in Ref.\ \cite{G13}.
%These results are displayed in Appendix \ref{AppendixJ}, for completion. 
In particular, in the case of hard spheres ($d=3$) our analytical results agree with those reported by Sangani \emph{et al.} \cite{SMTK96} by using previous results derived by Jenkins and Richman \cite{JR85} from the classical Grad's moment method.

To compare with the dynamic simulations performed in Ref.\ \cite{SMTK96} for hard spheres ($d=3$), it is convenient to introduce the reduced (steady) shear viscosity
\beq
\label{4.3}
\mu_s=-\frac{4P_{xy}}{\rho_s \phi \sigma^2 a^2},
\eeq
and the (steady) granular temperature
\beq
\label{4.4}
\theta_s=\frac{4T}{m\sigma^2 a^2},
\eeq
where $\rho_s=6\,m/(\pi \sigma^3)$ is the mass density of a particle. The relation between $\mu_s$ and $\theta_s$ with $a^*$ and $P_{xy}^*$ is
\beq
\label{4.5}
\mu_s=-\frac{25\pi}{2304}\frac{P_{xy}^*}{\phi^2 a^{*2}},
\eeq
\beq
\label{4.6}
\theta_s=\frac{25\pi}{2304}\frac{1}{\phi^2 a^{*2}},
\eeq
where we recall that $\phi=(\pi/6)n\sigma^3$ is the volume fraction for spheres.

\begin{figure}
\centering
\includegraphics[width=0.75 \columnwidth,angle=0]{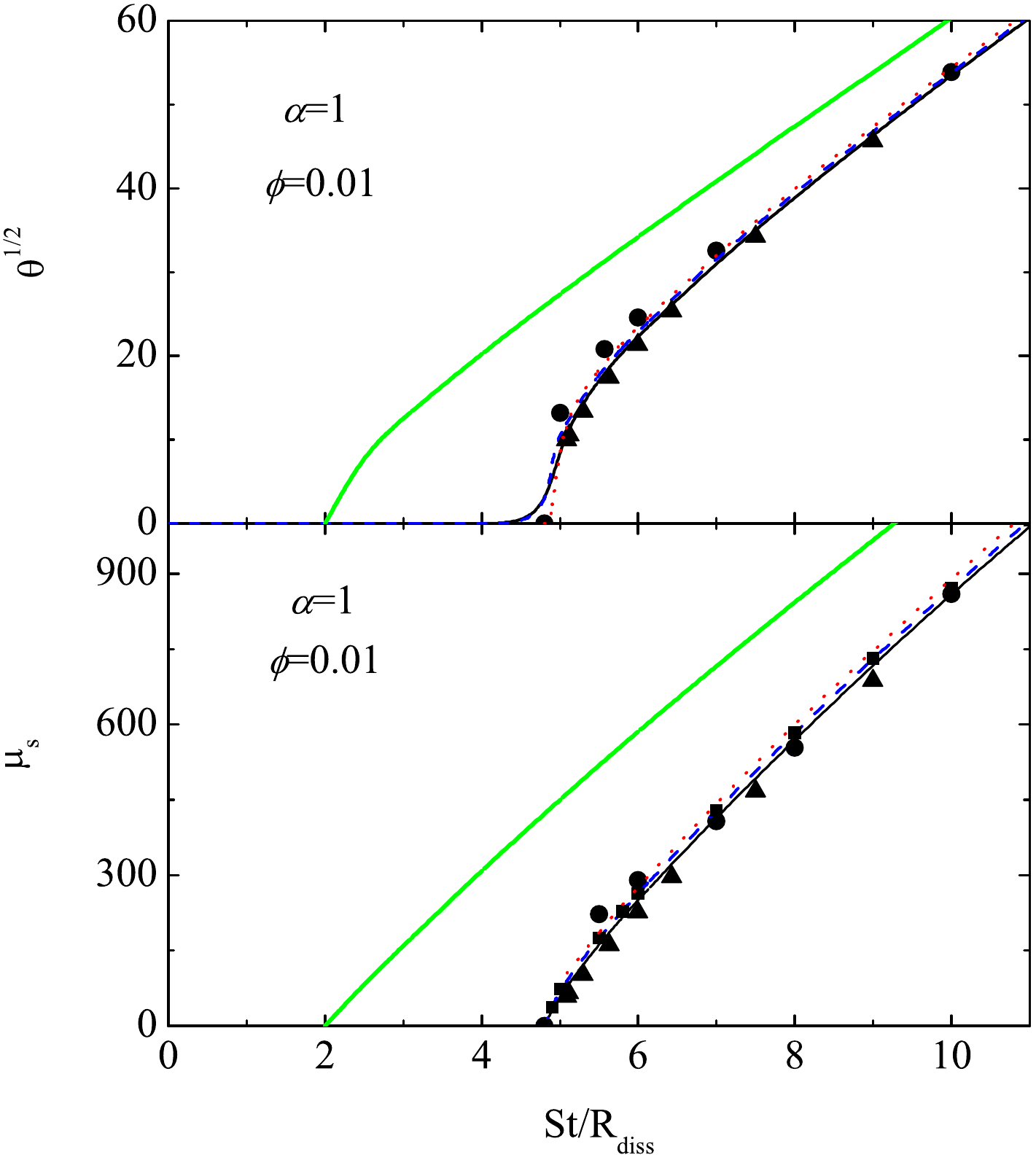}
\caption{Plot of the reduced steady shear viscosity $\mu_s$ and the square root of the steady granular temperature $\theta^{1/2}$ as a function of $\text{St}/R_\text{diss}$ in the case of hard spheres ($d=3$) with $\al=1$ and $\phi=0.01$. The solid black lines are the Grad's solution (including nonlinear contributions) to the Boltzmann equation, the dashed (blue) lines correspond to the BGK results (which coincide with those obtained from the linear Grad's solution) and the dotted (red) lines refer to the results obtained from the Enskog equation by applying (linear) Grad's method. The circles are the simulation results obtained by Sangani \emph{et al.} \cite{SMTK96} while the triangles correspond to the DSMC carried out in this work. The green solid lines are the predictions obtained from the NS hydrodynamic equations derived in Ref.\ \protect\cite[]{GTSH12}.\label{fig9chap6}}
\end{figure}

The shear viscosity $\mu_s$ and the square root of temperature $\sqrt{\theta}$ are plotted in Fig.\ \ref{fig9chap6} as functions of $\text{St}/R_\text{diss}$ for hard spheres with $\al=1$ with a solid volume fraction $\phi=0.01$ (very dilute system). Here, we consider the theoretical predictions provided by Grad's solution (including nonlinear contributions in the pressure tensor) to the Boltzmann equation, the exact results of the BGK equation \eqref{eq6.29} with the choice of the free parameter $\psi(\al)=(1+\al)(2+\al)/6$ (which coincide with the results obtained from the linear Grad's solution) and the results obtained from the Enskog equation by applying the linear Grad's moment method. Symbols are the simulation results obtained by Sangani \emph{et al.} \cite{SMTK96} (circles) and those obtained here by the DSMC method (triangles). %In the case of the Enskog results [Eqs.\ \eqref{3.49}, \eqref{3.50}, \eqref{3.52} and \eqref{3.53}], we have chosen for the pair correlation function $\chi(\phi)$ the Carnahan-Starling approximation  given by Eq.\ \eqref{chi3d}.% \cite{CS69}.
%\beq
%\label{4.7}
%g(\phi)=\frac{1-\frac{1}{2}\phi}{(1-\phi)^3}.
%\eeq

We observe first that dynamic simulations \cite{SMTK96} and DSMC results are consistent among themselves in the
range of values of the Stokes number explored. It is also important to recall that the (nonlinear) Grad's solution to the Boltzmann equation predicts the extinction of the hydrodynamic solution at $\text{St}/R_\text{diss}\simeq 5$ while the prediction of (linear) Grad's solution to the Enskog equation and the exact BGK solution is for $\text{St}/R_\text{diss}\simeq 4.8$. Fig.\ \ref{fig9chap6} shows clearly an excellent agreement between \emph{all} the theoretical predictions and both simulation methods. In fact, as expected the Boltzmann results are practically indistinguishable from the Enskog ones showing that the density corrections to the rheological properties are very small for this volume fraction ($\phi=0.01$). It is also important to remark the reliability of the BGK model to capture the main trends of the true Boltzmann kinetic equation. Moreover, given that the suspension is far away from equilibrium, as expected the NS description fails to describe the dependence of $\mu_s$ and $\theta$ on the Stokes number.

\begin{figure}
\includegraphics[width=0.75 \columnwidth,angle=0]{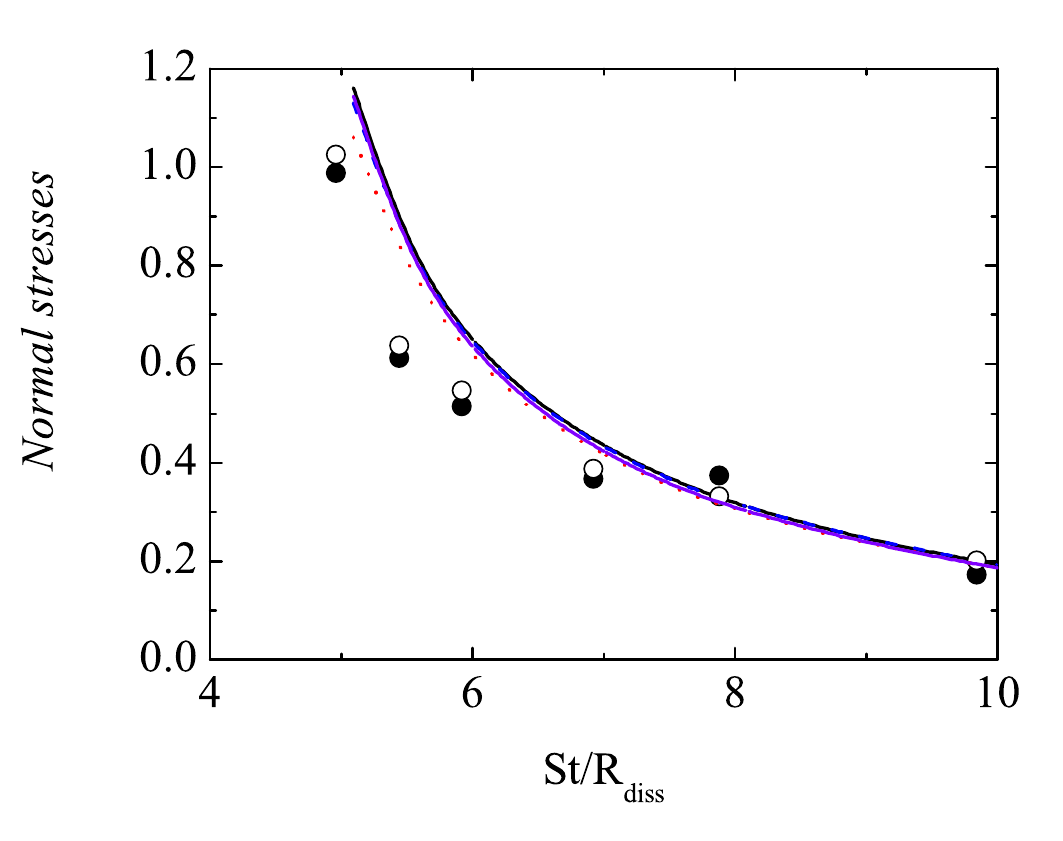}
\caption{Plot of the normal stress differences $P_{xx}^*-P_{yy}^*$ and $P_{xx}^*-P_{zz}^*$ as a function of $\text{St}/R_\text{diss}$ in the case of hard spheres ($d=3$) with $\al=1$ and $\phi=0.01$. The solid lines are the Grad's solution (including nonlinear contributions) to the Boltzmann equation for $P_{xx}^*-P_{yy}^*$ (black line) and $P_{xx}^*-P_{zz}^*$ (violet line), the dashed (blue) line corresponds to the BGK results (which coincide with those obtained from the linear Grad's solution) and the dotted (red) line refers to the results obtained from the Enskog equation by applying (linear) Grad's method. The black and empty circles are the simulation results obtained by Sangani \emph{et al.} \cite{SMTK96} for $P_{xx}^*-P_{yy}^*$ and $P_{xx}^*-P_{zz}^*$, respectively.  \label{fig10chap6}}
\end{figure}

Concerning the normal stress differences, Fig.\ \ref{fig10chap6} shows $P_{xx}^*-P_{yy}^*$ and $P_{xx}^*-P_{zz}^*$ versus $\text{St}/R_\text{diss}$. Except the nonlinear Grad's solution to the Boltzmann equation, all the other theories only predict normal stress differences in the plane of shear flow ($P_{xx}^*\neq P_{yy}^*=P_{zz}^*$). We observe that the simulations of Ref.\ \cite{SMTK96} also show that there is anisotropy in the plane perpendicular to the flow velocity ($P_{yy}^* < P_{zz}^*$), in accordance with nonlinear Grad's theory. For small Stokes numbers, although the different theories overestimate the simulation results, linear Grad's solution to the Enskog equation slightly compares better with simulations of \cite{SMTK96} than the more sophisticated nonlinear Grad's solution to the Boltzmann equation. Based on the good agreement found in Subsection \ref{sec5achap6} for the diagonal elements of the pressure tensor when the volume fraction is strictly zero, we think that the disagreement between the nonlinear Grad's solution and simulations for $\phi=0.01$ is due essentially to the (small) density corrections to the above elements which are of course not accounted for in the Boltzmann equation.

\begin{figure}[h]
\centering
\includegraphics[width=0.75 \columnwidth,angle=0]{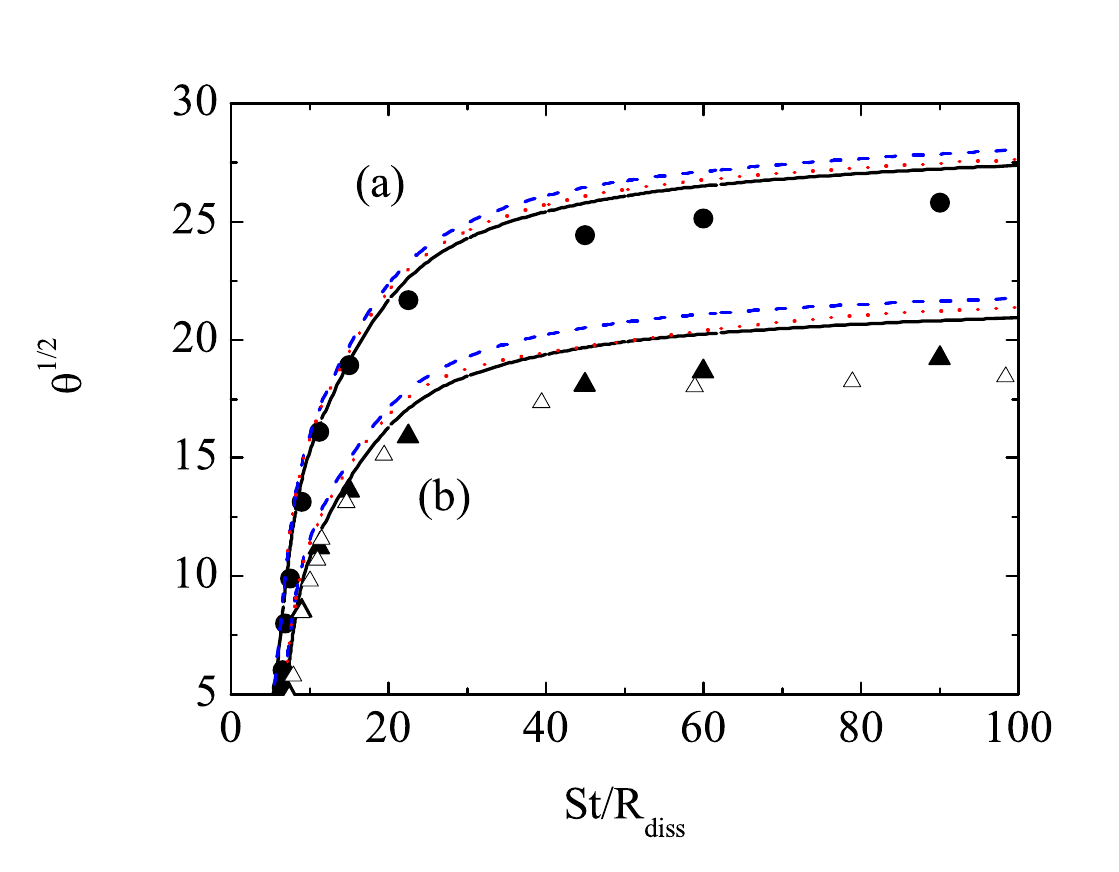}
\caption{Plot of the square root of the steady granular temperature $\theta^{1/2}$ as a function of $\text{St}/R_\text{diss}$ in the case of hard spheres ($d=3$) for $\phi=0.01$. Two different values of the coefficient of restitution have been considered: $\al=0.7$ (a) $\al=0.5$ (b). The solid line is the Grad's solution (including nonlinear contributions) to the Boltzmann equation, the dashed (blue) line corresponds to the BGK results (which coincide with those obtained from the linear Grad's solution) and the dotted (red) line refers to the results obtained by Sangani \emph{et al.} \cite{SMTK96} from the Enskog equation by applying (linear) Grad's method. The black circles and triangles are the simulation results obtained here by means of the DSMC method for $\al=0.7$ and $\al=0.5$, respectively, while the empty triangles are the results obtained in Ref.\ \cite{SMTK96}.  \label{fig11chap6}} %  \label{fig5}}
\end{figure}

%Finally, it is quite interesting to compare the dynamic simulation results reported in Ref.\ \cite{SMTK96} in the case of very dilute suspensions ($\phi=0.01$) with those carried here by means of the DSMC method. To do it, we introduce the reduced (steady) shear viscosity
%\beq
%\label{4.3}
%\mu_s = -\frac{4P_{xy}}{\rho_s\phi\sigma^2 a^2} = -\frac{25\pi}{2304}\frac{P^*_{xy}}{\phi^2 a^{*2}},
%\eeq
%and the (steady) granular temperature $\theta$ as
%\beq
%\label{4.4}
%\theta=\frac{4T}{m\sigma^2 a^2}=\frac{25\pi}{2304}\frac{1}{\phi^2 a^{*2}},
%\eeq
%where we recall that $\phi=(\pi/6)n\sigma^3$ is the volume fraction for spheres. 

Finally, Fig.\ \ref{fig11chap6} shows $\sqrt{\theta}$ versus $\text{St}/R_\text{diss}$ for two different values of the coefficient of restitution: $\al=0.7$ and $\al=0.5$.  We have considered the DSMC performed here for $\al=0.7$ and $\al=0.5$ and those made in Ref.\ \cite{SMTK96} in the case $\al=0.5$. In addition, we have also included the theoretical results derived in \cite{SMTK96} from the Enskog equation.  We observe first that the dynamic simulations for finite Stokes number and the DSMC results are consistent among themselves in the range of values of $\text{St}/R_\text{diss}$ explored. This good agreement gives support to the applicability of the model for dilute granular suspensions introduced in Eq.\ \eqref{eq6.7}. It is also apparent that the performance of nonlinear Grad's theory for the (steady) temperature is slightly better than the remaining theories. 
Note also that the agreement between theory and computer simulations improves as we approach the dry granular limit $\text{St}/R_\text{diss}\rightarrow\infty$. Thus, at $\alpha=0.7$, for instance, the discrepancies between nonlinear Grad's theory and DSMC results for $\text{St}/R_\text{diss}=11.3,\, 22.5,\, 45,\, 60,$ and $90$ are about $8.5\%,\, 6.4\%,\, 5.8\%,\, 5.5\%$ and $5.3\%,$ respectively, while at $\alpha=0.5$ the discrepancies are about $14\%,\, 10\%,\, 9\%,\, 8.6\%$ and $8.5\%,$ respectively. This shows again that our Grad's solution compares quite well with simulations for not too large values of the (scaled) friction coefficient $\gamma^*$ (or equivalently, for large values of the Stokes number $\text{St}$).

\section{Summary and Discussion}
\label{sec6chap6}

In this Chapter, we have presented a complete and comprehensive theoretical description of the non-Newtonian transport properties of a dilute granular suspension under USF in the framework of the (inelastic) Boltzmann equation. The influence of the interstitial fluid on the dynamic properties of grains has been modeled via a viscous drag force proportional to the particle velocity. This type of external force has been recently employed in different works on gas-solid flows \cite{WZLH09,H13,HT13,SMMD13,WGZS14}. The study performed here has been both theoretical and computational. In the theory part, we have presented results from two different approaches: Grad's moment method and a BGK-type kinetic model used previously in other granular flow problems and now applied specifically to the model of granular suspensions. In contrast to previous works in granular sheared suspensions \cite{SMTK96}, we have included in Grad's solution quadratic terms in the pressure tensor $P_{ij}$ in the collisional moment $\Lambda_{ij}$ associated with the momentum transport (nonlinear Grad's solution). This allows us to evaluate the normal stress differences in the plane normal to the laminar flow (namely, the normal stress difference $P_{yy}^*-P_{zz}^*$) and of course, one obtains more accurate expressions of the non-Newtonian transport properties. The inclusion of quadratic terms in $P_{ij}$ in the evaluation of $\Lambda_{ij}$ was already considered by Tsao and Koch \cite{TK95} in an analogous system but only in the limit of perfectly elastic collisions ($\al=1$).

%Therefore, for strictly granular particles (i.e., beyond the elastic limit) this is the first time that, to the best of our knowledge, the difference $P_{yy}^*-P_{zz}^*$ has been analytically detected and evaluated in a theory of sheared \emph{granular} suspensions. %This is one of the most relevant achievements of the present contribution. 
Moreover, the development of the corresponding BGK-type model for the dilute granular suspension under uniform shear has allowed us also to formally compute all velocity moments as well as the velocity distribution function of the suspension.

Additionally, to gauge the accuracy of the above theoretical approaches, we have presented simulation results (DSMC method applied to the inelastic Boltzmann equation). The comparison between theory and DSMC has been done by varying both the (scaled) friction coefficient $\gamma^*$ (or equivalently, the Stokes number $\text{St})$ characterizing the magnitude of the drag force and the coefficient of restitution $\al$ characterizing the inelasticity of collisions. The agreement for the reduced shear rate [see Fig.\ \ref{fig2chap6} (a)] and the elements of the pressure tensor [see Figs.\ \ref{fig3chap6} and \ref{fig4chap6}] between DSMC and both theoretical solutions is excellent (especially in the case of nonlinear Grad's solution) for not too large value values of $\gamma^*$. As the magnitude of the friction coefficient increases the agreement between Grad's solution and simulations decreases [cf. Fig.\ \ref{fig5chap6}], although being the discrepancies smaller than 6\%. This good performance of Grad's method has been also observed for monodisperse dry granular gases for Couette flow sustaining a uniform heat flux \cite{VSG10,VGS11,VSG13} and also in the case of granular binary mixtures under USF \cite{MG02,L04}. Regarding high velocity moments, we also obtain good agreement for the kurtosis $K$ since the BGK results compare very well with simulations for not too large values of $\gamma^*$ [cf. Fig.\ \ref{fig6chap6}]. Finally, as expected, the BGK model reproduces very well the behavior of the marginal distributions $\varphi_x^{(+)}(c_x)$ and $\varphi_y^{(+)}(c_y)$ in the region of thermal velocities [see Figs.\ \ref{fig7chap6} and \ref{fig8chap6}], although they quantitatively disagree with simulations for higher velocities especially for strong collisional dissipation.

%Finally, it is also important to remark that the objective of this Chapter has been to set a non-linear hydrodynamic theory for the USF, state that as we know is necessarily non-Newtonian \cite{VSG10}, as a starting point for the deployment of a more comprehensive and systematic theory for more complex flows in this kind of system. In this context, we expect in the near future to extend the present results to other related flows such as the so-called LTu flows \cite{VSG10,VGS11} (i.e., the more general case of uniform but non-null heat flux) or to the more general class of Couette flows \cite{VSG13}. We want also to carry out further studies on the more realistic case of multicomponent granular suspensions where problems like segregation can be addressed. Work along these lines is underway.
         % Suspensions
\chapter{Conclusions and Outlooks}

\label{Chapter7}

\lhead{Chapter 7. \emph{Conclusions and Outlooks}}

%----------------------------------------------------------------------------------------

\section{Conclusions} 

In this work transport properties of a granular gas homogeneously driven by a thermostat composed by two different external forces (stochastic heating and drag friction) have been studied. Such forces compensate for the inelastic cooling due to the binary collisions (inelastic hard spheres, IHS) and the system is maintained in a non-equilibrium steady state. The Enskog kinetic equation has been considered as the starting point and DSMC simulations have been carried out to assess the validity of the theoretical predictions.

Firstly, the velocity distribution function of a homogeneous system has been characterized through its first non-zero velocity moments, that is, the Sonine coefficients $a_2$ and $a_3$ defined in terms of  the fourth and sixth-degree velocity moments of the scaled velocity distribution function $\varphi$.
%The new feature found here is the dependence of $\varphi$ on both the reduced noise strength $\xi^*$ [defined in Eq.\ \eqref{eqxistar}] and the granular temperature $T$ through the scaled velocity $\bt{c}$ and also through $\xi^*$. 
The new feature found here is the dependence of $\varphi$ on the granular temperature T through both the scaled velocity $\mathbf{c}$ and the (reduced) strength noise $\xi^*$ [defined in Eq.\ \eqref{eqxistar}].
The simulation results have confirmed the above dependence (see Figs.\ \ref{fig06chap3} and \ref{fig07chap3}).

%Se ha caracterizado el comportamiento de la función de distribución escalada $\varphi$ a través de sus primeros momentos. Más concretamente a través del segundo y tercer coeficiente ($a_2$, $a_3$) de Sonine. Dado que estos coeficientes obedecen una jerarquía de ecuaciones no pueden ser calculados de forma exacta y hay que truncar la expansión de Sonine para obtener una estimación aproximada. En este trabajo se ha realizado dos aproximaciones diferentes (Aproximaciones I y II) para obtener las expresiones explícitas de $a_2$ y $a_3$ en función de la dimensionalidad, del coeficiente de restitución normal $\alpha$ y del parámetro del termostato $\xi^*$, siendo la Aproximación II más compleja que la Aproximación I. La comparación con los resultados numéricos medidos mediante DSMC muestra un buen acuerdo en general incluso para grandes inelasticidades. Como se esperaba, la Aproximación II predice resultados más cercanos, aunque ligeramente, a la simulación que la I. Teniendo esto en cuenta, se puede concluir que usando la Aproximación I se obtiene un razonable compromiso entre simplicidad y precisión y es la que se ha usado en los cálculos posteriores.

The Navier-Stokes transport coefficients have been obtained by solving the Enskog kinetic equation by means of the Chapman-Enskog expansion. Heat and momentum fluxes have been determined in first-order deviations of the hydrodynamic fields from their homogeneous steady state. 

An important point here is the generalization of the driving external forces (which are mainly used in homogeneous situations) to \emph{non-homogeneous} states. The choice of a general external driving force that homogeneously acts over the whole system cannot compensate locally the collisional cooling and the local reference state $f^{(0)}$ depends on time through the non-steady temperature. The above choice has the adventage of a simpler implementation on computer simulations but introduces some difficulties in the analytical study.
However, given that the knowledge of the complete time-dependence of the transport coefficients is a formidable task and we are interested in the evaluation of the momentum and heat fluxes in the first order of the expansion, then the transport coefficients must be determined to zeroth order in the deviations.
%in the first order of the deviations from the steady reference state, the transport coefficients must be determined to zeroth order in the deviations. 
As a consequence, the steady-state condition \eqref{eqTs01} applies and the transport coefficients and the cooling rate can be defined in terms of the hydrodynamic fields in the steady state. Our theoretical predictions of the driven kinetic theory compares reasonably well with Langevin dynamics simulations in a wide range of densities. However, some discrepancies appear in the cases of $\nu_l$ and $D_T$ as the gas becomes denser. 

The stability of the linearized hydrodynamic equations with respect to the homogeneous steady state with the new expressions for the momentum and heat fluxes and the cooling rate has been analyzed in order to identify the conditions for stability in terms of dissipation. Results presented here show that the driven homogeneous state is \emph{stable} for any value of dissipation at sufficiently long wavelengths.

The transport coefficients have also been obtained by solving the Boltzmann equation for inelastic Maxwell models (IMM) by means of the Chapman-Enskog method. The comparison between IMM and IHS shows that IMM reproduces qualitatively well the trends observed for IHS, even for strong dissipation. At a more quantitative level, discrepancies between both interaction models increase with inelasticity. In any case, the results found in this work contrast with those obtained in the freely cooling case \cite{S03} where IMM and IHS exhibit much more significant differences. Thus, the reliability of IMM as a prototype model for granular flows can be considered more robust in driven states than in the case of undriven states.

Finally, a complete theoretical description of the non-Newtonian transport properties of a dilute granular suspension under USF in the framework of the (inelastic) Boltzmann equation has been presented. In that case, 
the effect of the interstitial fluid on the solid particles has been modeled via 
%the complex hydrodynamical perturbations of the interstitial fluid on the particles have been modeled via 
a viscous drag force proportional to the particle velocity and a stochastic force accounting for particle neighbor effects (Langevin model) proportional to the difference between the mean velocity of solid and interstitial fluid respectively. 
In the USF, the mean flow velocity of gas phase follows the mean flow velocity of solid particles and hence, only the drag force and the collisional cooling compensate for the viscous heating due to shearing work. 
%That difference has been taken as zero and only drag force and collisional cooling remain to compensate the shear heating. 
Here we have used three different but complementary routes: Grad's moment method, BGK-type kinetic model and DSMC computer simulation.

We have included in Grad's solution quadratic terms in the pressure tensor $P_{ij}$ in the collisional moment $\Lambda_{ij}$ associated with the momentum transport (nonlinear Grad's solution). This allows us to evaluate the normal stress differences in the plane normal to the laminar flow (namely, the normal stress difference $P_{yy}^*-P_{zz}^*$) and of course, one obtains more accurate expressions of the non-Newtonian transport properties. 

In the case of inelastic collisions and to the best of our knowledge, this is the first time that 
%For strictly granular particles this is the first time that, to the best of our knowledge, 
the difference $P_{yy}^*-P_{zz}^*$ has been analytically detected and evaluated in a theory of sheared \emph{granular} suspensions. This is one of the most relevant achievements of the present work. 

The agreement between DSMC and both theoretical solutions for the reduced shear rate and the elements of the pressure tensor is excellent (especially in the case of nonlinear Grad's solution).
This shows the accuracy of our analytical results that allows us to describe in detail the flow dynamics of the granular sheared suspension. 
We also obtain good agreement between DSMC and BGK kinetic model for the kurtosis and the behavior of the marginal distributions in the region of thermal velocities.

\section{Outlooks}

An interesting point is the usefulness of the theoretical results derived in this work to modelize the experiments performed by using boundary driven conditions. As usual in computer simulations, in this work we have fluidized the system by means of a thermostat composed by a friction term which mimics the presence of an interstitial fluid and a stochastic force that could model the effect of a vibrating wall. The main advantage of using this type of driving mechanism is the possibility of making theoretical progress. In addition, although the relationship of the last external force with real vibrating walls is not clear to date, some theoretical results (see, for instance Fig.\ 2 of Ref.\ \cite{G09}) obtained for the temperature ratio of a granular impurity immersed in a granular gas heated by the stochastic thermostat compare quite well with MD simulations of shaken mixtures \cite{SUKSS06}. This agreement could stimulate the use of this simple stochastic driving for qualitative comparisons with experimental results. On the other hand, more comparisons between kinetic theory results for heated granular gases and computer simulations performed in realistic vibrating beds are needed before qualitative conclusions can be drawn.

%It is important to remark that one of the objective of this work has been to set a non-linear hydrodynamic theory for the USF, state that as we know is necessarily non-Newtonian \cite{VSG10}, as a starting point for the deployment of a more comprehensive and systematic theory for more complex flows in this kind of system. 

One of the objectives of the present work has been to determine the non-Newtonian transport properties of a granular suspension under USF. This study can be considered as the starting point for the deployment of a more comprehensive and systematic theory for more complex flows. 
In this context, we expect in the near future to extend the present results to other related flows such as the so-called LTu flows \cite{VSG10,VGS11} (i.e., the more general case of uniform but non-null heat flux) or to the more general class of Couette flows \cite{VSG13}. 

We want also to carry out further studies on the more realistic case of multicomponent granular suspensions where problems like segregation can be addressed. %Work along this line is underway.

Another interesting point here is that the non-linear solution to the Boltzmann equation proposed in Chapter \ref{Chapter6} predicts the existence of a second branch in the curve for the Stokes number $St$ for high values of the friction coefficient $\gamma_b$ (see Fig.\ \ref{fig1chap6}) not present in previous works \cite{TK95,SMTK96}. This second branch, which lies for high Knudsen numbers, could not be reached numerically due to the limitations of our uniform DSMC simulations. In order to test this branch, new non-uniform DSMC or MD simulation will be carried out in the future.

\addtocontents{toc}{\vspace{2em}} % Add a gap in the Contents, for aesthetics
\appendix % Cue to tell LaTeX that the following 'chapters' are Appendices
% Include the appendices of the thesis as separate files from the Appendices folder

% Appendix A
\chapter{Expressions for $A_{i}$, $B_{i}$, and $C_{i}$} % Main appendix title
\label{AppendixA} % For referencing this appendix elsewhere, use \ref{AppendixA}
\lhead{Appendix A. \emph{Expressions for $A_i$, $B_i$, and $C_i$}}

\section{Expressions for $A_i$, $B_i$, and $C_i$}
In this Appendix we provide the explicit expressions of the coefficients $A_i$, $B_i$, and $C_i$ as functions of $d$ and $\alpha$. They are given by \cite{NE98,BP06,BP06err}
\begin{equation}
\label{apA1}
A_0=K(1-\alpha^2),\quad A_2=\frac{3K}{16}(1-\alpha^2),\quad 
A_3=\frac{K}{64}(1-\alpha^2),
\end{equation}
\begin{equation}
\label{apA2}
B_0=K(1 -\alpha^2) \left(d + \frac{3}{2} + \alpha^2\right),
\end{equation}
\begin{equation}
\label{apA3}
B_2=K(1+\alpha)\left[d -1+\frac{3}{32} (1 - \alpha) (10 d + 39 + 10 \alpha^2) 
\right],
\end{equation}
\begin{equation}
\label{apA4}
B_3=-\frac{K}{128}(1+\alpha)\left[(1 - \alpha) (97 + 10 \alpha^2) +2(d -
1)(21 - 5 \alpha)\right],
\end{equation}
\begin{equation}
\label{apA5}
C_0= \frac{3K}{4} (1 - \alpha^2) \left[d^2 +
    \frac{19}{4}+(d + \alpha^2) (5 + 2 \alpha^2)\right],
\end{equation}
\begin{equation}
\label{apA6}
C_2=\frac{3K}{256} (1 - \alpha^2) \left[1289 + 172 d^2+
     4 (d + \alpha^2) (311 + 70 \alpha^2) \right] + \frac{3}{4} {\lambda},
\end{equation}
\begin{equation}
\label{apA7}
C_3= -\frac{3K}{1024} (1 - \alpha^2) \left[2537 + 236 d^2+
     4 (d + \alpha^2) (583 + 70 \alpha^2)\right] -
  \frac{9}{16} \lambda,
\end{equation}
where
\begin{equation}
\label{apA8}
K\equiv \frac{\pi^{(d-1)/2}}{\sqrt{2}\Gamma(d/2)},\quad
\lambda\equiv K(1 + \alpha) \left[(d - \alpha) (3 +
       4 \alpha^2) + 2 (d^2 - \alpha)\right].
\end{equation}
%%%%%%%%%%%%%%%%%%%%%%%%%%%%%%%%%%%%%%
%%%%%%%%%%%%%%%%%%%%%%%%%%%%%%%%%%%%%%
\section{Approximations I and II}
The forms of the Sonine coefficients $a_2$ and $a_3$ by using Approximations I and II are displayed in this Appendix. Let us start by considering Approximation I. In this case, we neglect $a_3$ versus $a_2$ in equation \eqref{eqJerarquia01} and so, one gets
\begin{equation}
\label{apA9}
\left[B_2-(d+2)(A_0+A_2)+\frac{d(d+2)}{2}\xi^*\right]a_2=(d+2)A_0-B_0,
\end{equation}
whose solution is
\begin{eqnarray}
\label{apA10}
a_2^{(I)}(\alpha,\xi^*)&=&\frac{(d+2)A_0-B_0}{B_2-(d+2)(A_0+A_2)+\frac{d(d+2)}{2}\xi^*}\nonumber\\
&=&\frac{16(1-\alpha)(1-2\alpha^2)}{9+24d-\alpha(41-8d)+30(1-\alpha)\alpha^2
+G_\text{d}\frac{\xi^*}{(1+\alpha)}},
\end{eqnarray}
where in the last step use has been made of the explicit expressions of $A_0$, $A_2$, $B_0$, and $B_2$. Here,
\begin{equation}
\label{apA11}
G_\text{d}=16\sqrt{2}d(d+2)\frac{\Gamma(d/2)}{\pi^{(d-1)/2}}.
\end{equation}

Once $a_2$ is determined, we can use equation\ \eqref{eqJerarquia02} to express $a_3$ in terms of $a_2$. The result can be written as
\begin{equation}
\label{apA12}
a_3^{(I)}(\alpha,\xi^*)=F\left(\alpha,a_2^{(I)}(\alpha),\xi^*\right),
\end{equation}
where
\begin{equation}
\label{apA13} %b4
F\left(\alpha,a_2,\xi^*\right)\equiv\frac{\frac{3}{4}(d+2)(d+4)A_0-C_0-\left[C_2+\frac{3}{4}(d+2)(d+4)
(d\xi^*-3A_0-A_2)\right]a_2}{C_3-\frac{3}{4}(d+2)(d+4)\left(A_3-A_0+\frac{d}{2}\xi^*\right)}.
\end{equation}
%\subsection{Approximation II}

In Approximation II, $a_3$ is formally treated as being of the same order of magnitude as $a_2$ and so, Eqs. \eqref{eqJerarquia01} and \eqref{eqJerarquia02} become a linear set of two coupled equations for $a_2$ and $a_3$. The problem is algebraically more involved as in  Approximation I. The form of $a_2^{(II)}$ is given by Eq.\ \eqref{eqa2II} where 
\begin{eqnarray}
\label{apA14}
M(\alpha,\xi^*)&\equiv&\left[C_3-\frac{3}{4}(d+2)(d+4)\left(A_3-A_0+\frac{d}{2}\xi^*\right)\right]\left[(d+2)A_0-B_0\right]\nonumber\\
&&-\left[B_3-(d+2)A_3\right]\left[\frac{3}{4}(d+2)(d+4)A_0-C_0\right],
\end{eqnarray}
and
\begin{eqnarray}
\label{apA15}
N(\alpha,\xi^*)&\equiv& \left[B_2-(d+2)(A_0+A_2)+\frac{d(d+2)}{2}\xi^*\right]\nonumber\\
&\times&\left[C_3-\frac{3}{4}(d+2)(d+4)\left(A_3-A_0+\frac{d}{2}\xi^*\right)\right]\\
&-&\left[B_3-(d+2)A_3\right]\left[C_2+\frac{3}{4}(d+2)(d+4)(d\xi^*-3A_0-A_2)\right].\nonumber
\end{eqnarray}
The corresponding result for $a_3^{(II)}$ in Approximation II has the same form as for Approximation I except that it now relies on $a_2^{(II)}$, i.e,
\begin{equation}
\label{apA16}
a_3^{(II)}(\alpha,\xi^*)=F\left(\alpha,a_2^{(II)}(\alpha),\xi^*\right).
\end{equation}
 % del Chapter 3
%%--------------------------------------------
\chapter{First-order approximation} 

\label{AppendixB} 

\lhead{Appendix B. \emph{First-order approximation}}
The application of the Chapman-Enskog method up to the first-order approximation follows similar mathematical steps as those made before in the undriven case \cite{GD99, L05}. Up to first order in the expansion, the velocity distribution function $f^{(1)}$ obeys the kinetic equation

\begin{eqnarray}
    \label{apB1}
  \left(\partial_t^{(0)}+\mathcal{L}\right)f^{(1)}-\frac{\gamma_\text{b}}{m}\frac{\partial}{\partial \bt{v}}\cdot \bt{V} f^{(1)}  - \frac{1}{2}\xi_\text{b}^2\frac{\partial^2}{\partial v^2}f^{(1)}  = -\left(\partial_t^{(1)}+\bt{v}\cdot\nabla\right)f^{(0)} \nonumber \\
+\frac{\gamma_\text{b}}{m}\Delta \bt{U}\cdot\frac{\partial f^{(0)}}{\partial \bt{V}} - J_E^{(1)}[f],
\end{eqnarray}
where $J_E^{(1)}[f]$ is the first-order contribution to the expansion of the Enskog collision operator and $\mathcal{L}$ is the linear operator defined as:
\begin{equation}
  \label{apB2}
  \mathcal{L}f^{(1)}=-\left(J_E^{(0)}[f^{(0)},f^{(1)}]+J_E^{(0)}[f^{(1)},f^{(0)}]\right).
\end{equation}
The macroscopic balance equations to first order in gradients are
\begin{eqnarray}
  \label{apB3}
  &(\partial_t^{(1)} + \bt{U}\cdot\nabla)n& = - n\nabla\cdot\bt{U}, \\
  &(\partial_t^{(1)} + \bt{U}\cdot\nabla)U_i& = - (mn)^{-1}\nabla_i p - \frac{\gamma_\text{b}}{m} \Delta \mathbf{U}, \\
  &(\partial_t^{(1)} + \bt{U}\cdot\nabla)T&   = - \frac{2p}{dn}\nabla\cdot \bt{U}-\zeta^{(1)}T,
\end{eqnarray}
where $\zeta^{(1)}$ is the first-order contribution to the cooling rate. Introducing these balance equations into Eq.\ \eqref{apB1} and with the expression for $J_E^{(1)}[f]$ given in \cite{GD99} for the undriven case, Eq.\ \eqref{apB1} can be rewritten as
\begin{eqnarray}
  \label{apB4}
  \left(\partial_t^{(0)}+\mathcal{L}\right)f^{(1)}-\frac{\gamma_\text{b}}{m}\frac{\partial}{\partial \bt{v}}\cdot \bt{V} f^{(1)} 
  - \frac{1}{2}\xi_\text{b}^2\frac{\partial^2}{\partial v^2}f^{(1)}  = \nonumber \\
  \bt{A}\cdot\nabla \ln T + \bt{B}\cdot\nabla \ln n + C_{ij}\frac{1}{2}\left(\partial_i U_j + \partial_j U_i - \frac{2}{d}\delta_{ij}\nabla\cdot\bt{U}\right) + D\nabla\cdot\bf{U},
\end{eqnarray}
where
\begin{equation}
  \label{apB5}
  \bt{A}(\bt{V})=-\bt{V}T \frac{\partial f^{(0)}}{\partial T} - \frac{p}{\rho}\frac{\partial f^{(0)}}{\partial \bt{V}} - \boldsymbol{\mathcal{K}}\left[T \frac{\partial f^{(0)}}{\partial T}\right],
\end{equation}
\begin{equation}
  \label{apB6}  
  \bt{B}(\bt{V}) = -\bt{V} n \frac{\partial f^{(0)}}{\partial n} - \frac{p}{\rho}\left(1+\phi\frac{\partial}{\partial\phi}\ln p^*\right)\frac{\partial f^{(0)}}{\partial\bt{V}} - \boldsymbol{\mathcal{K}}  \left[ n \frac{\partial f^{(0)}}{\partial n}\right] - \frac{1}{2}\phi\frac{\partial}{\partial \phi}\ln\chi\, \boldsymbol{\mathcal{K}}[f^{(0)}],
\end{equation}
\begin{equation}
  \label{apB7}
  C_{ij}(\bt{V}) = V_i \frac{\partial f^{(0)}}{\partial V_j}  + \mathcal{K}_i\left[\frac{\partial f^{(0)}}{\partial V_j}\right],
\end{equation}
\begin{equation}
   \label{apB8}
  D = \frac{1}{d}\frac{\partial}{\partial\bt{V}}\cdot(\bt{V}f^{(0)}) + \left(\zeta_U +\frac{2}{d}p^*\right)T\frac{\partial f^{(0)}}{\partial T} - f^{(0)} + n\frac{\partial f^{(0)}}{\partial n} + \frac{2}{d}\mathcal{K}_i[\frac{\partial f^{(0)}}{\partial V_i}].
\end{equation}
Here, $p^*$ is the steady hydrodynamic pressure
\begin{equation}
  \label{apB9}
  p^*=\frac{p_\text{s}}{n_\text{s} T_\text{s}}=1+2^{d-2}(1+\alpha)\chi\phi,
\end{equation}
$\mathcal{K}_i$ is the operator defined as
\begin{eqnarray}
    \label{apB10}
  \mathcal{K}_i[\bt{V}|X] &=& 
  \sigma^d \chi \int\;d\bt{v}_2 \int\;d\widehat{\bs{\sigma}} \Theta(\widehat{\bs{\sigma}}\cdot\bt{g}_{12})(\widehat{\bs{\sigma}}\cdot\bt{g}_{12})\widehat\sigma_i [\alpha^{-2} f^{(0)}(\bt{V}_1'') X(\bt{V}_2'') \nonumber\\
  &+& f^{(0)}(\bt{V}_1) X(\bt{V}_2)],
\end{eqnarray}
where $\bt{V}_1''=\bt{V}_1 - \frac{1}{2}(1+\alpha^{-1})(\widehat{\bs{\sigma}}\cdot\bt{g}_{12})\widehat{\bs{\sigma}}$, $\bt{V}_2''=\bt{V}_2 - \frac{1}{2}(1+\alpha^{-1})(\widehat{\bs{\sigma}}\cdot\bt{g}_{12})\widehat{\bs{\sigma}}$, $\bt{g}_{12}=\bt{V}_1-\bt{V}_2$, and $\zeta_U$ is the first-order contribution to the cooling rate defined by Eq.\ \eqref{eq4.18}. In addition, upon deriving Eqs.\ \eqref{apB4}--\eqref{apB8} use has been made of the spherical symmetry of $f^{(0)}$ which allows us to write the tensor derivative of the flow field $\partial_i U_j$ in terms of its independent trace and traceless parts:
\begin{eqnarray}
    \label{apB12}
  V_i\frac{\partial f^{(0)}}{\partial V_j}\partial_i U_j &=&   V_i\frac{\partial f^{(0)}}{\partial V_j} \frac{1}{2}(\partial_i U_j+\partial_j U_i) \nonumber \\
  &=&  V_i\frac{\partial f^{(0)}}{\partial V_j} \frac{1}{2}(\partial_i U_j+\partial_j U_i - \frac{2}{d}\delta_{ij}\nabla\cdot\bt{U}) + \nonumber \\
  &+&\frac{1}{d}\bt{V}\cdot\frac{\partial f^{(0)}}{\partial\bt{V}}\nabla\cdot\bt{U}
\end{eqnarray}
and a similar analysis of the contribution from $\mathcal{K}_i\left[\partial f^{(0)}/\partial V_j\right]$.

Furthermore, due to the complex dependence of the distribution function $f^{(0)}$ on the density through $\chi$ and $\xi^*$, the derivative $\partial f^{(0)}/\partial n$ has now two new terms not present in previous works:
\begin{equation}
  \label{apB13}
  n \frac{\partial  f^{(0)}}{\partial n} = f^{(0)} - \xi^*\frac{\partial f^{(0)}}{\partial \xi^*} - \frac{2}{3}\theta\frac{\partial f^{(0)}}{\partial \theta}  + \phi\frac{\partial\chi}{\partial\phi}\frac{\partial f^{(0)}}{\partial\chi}.
\end{equation}

The solution of Eq.\ \eqref{apB1} can be written in the form:
\begin{eqnarray}
  \label{apB14}
  &f^{(1)}&=\bs{\mathcal{A}}(\bt{V})\cdot\nabla \ln T + \bs{\mathcal{B}}(\bt{V}) \cdot\nabla \ln n \nonumber \\
  && + \mathcal{C}_{ij}(\bt{V}) \frac{1}{2}\left(\partial_i U_j + \partial_j U_i - \frac{2}{d}\delta_{ij}\nabla\cdot\bt{U}\right) + \mathcal{D}(\bt{V}) \nabla\cdot\bf{U},
\end{eqnarray}
where $\bs{\mathcal{A}}$, $\bs{\mathcal{B}}$, $\mathcal{C}_{ij}$ and $\mathcal{D}$ are unknown functions of the peculiar velocity. Introducing this first-order distribution function \eqref{apB14} in \eqref{apB4} and since the gradients of the hydrodynamic fields are all independent, a set of independent, linear, non-homogeneous integral equations for each unknown coefficient is obtained:
\begin{eqnarray}
  \label{apB15}
  -\left(\frac{2\gamma_\text{b}}{m}-\frac{m\xi_\text{b}^2}{T}+\zeta^{(0)}\right)T\frac{\partial \bs{\mathcal{A}}}{\partial T}  
  - \left[ \frac{m\xi_\text{b}^2}{T}\left(1+\frac{3}{2}\frac{\partial \zeta^*_0}{\partial\xi^*}\right) + \frac{1}{2}\zeta^{(0)}\right]  \bs{\mathcal{A}} - \nonumber\\
  - \frac{\gamma_\text{b}}{m}\frac{\partial}{\partial\bt{v}}\cdot\bt{V} \bs{\mathcal{A}} 
  - \frac{1}{2}\xi_\text{b}^2\frac{\partial^2}{\partial v^2} \bs{\mathcal{A}}  + \mathcal{L} \bs{\mathcal{A}}  = \bt{A},
\end{eqnarray}
\begin{eqnarray}
  \label{apB16}
  - \left(\frac{2\gamma_\text{b}}{m}-\frac{m\xi_\text{b}^2}{T}+\zeta^{(0)}\right)T\frac{\partial \bs{\mathcal{B}}}{\partial T} 
  - \frac{\gamma_\text{b}}{m}\frac{\partial}{\partial\bt{v}}\cdot\bt{V} \bs{\mathcal{B}} 
  - \frac{1}{2}\xi_\text{b}^2\frac{\partial^2}{\partial v^2}\bs{\mathcal{B}} + \mathcal{L}\bs{\mathcal{B}}= \nonumber\\
  = \bt{B} + \zeta^{(0)}g(\phi)\bs{\mathcal{A}} + \left[ \phi\frac{\partial\chi}{\partial\phi}\frac{\partial}{\partial\chi}\left(\frac{\zeta^{(0)}}{\chi}\right)-\xi^*\frac{\partial\zeta^{(0)}}{\partial\xi^*}
  - \frac{2}{3}\theta\frac{\partial \zeta^{(0)}}{\partial \theta} \right]\bs{\mathcal{A}},
\end{eqnarray}
\begin{eqnarray}
  \label{apB17}
  - \left(\frac{2\gamma_\text{b}}{m}-\frac{m\xi_\text{b}^2}{T}+\zeta^{(0)}\right)T\frac{\partial \mathcal{C}_{ij}}{\partial T} 
  - \frac{\gamma_\text{b}}{m}\frac{\partial}{\partial\bt{v}}\cdot\bt{V} \mathcal{C}_{ij}
  - \frac{1}{2}\xi_\text{b}^2\frac{\partial^2}{\partial v^2} \mathcal{C}_{ij} + \mathcal{L} \mathcal{C}_{ij}
  =  C_{ij},
\end{eqnarray}
\begin{eqnarray}
  \label{apB18}
  - \left(\frac{2\gamma_\text{b}}{m}-\frac{m\xi_\text{b}^2}{T}+\zeta^{(0)}\right)T\frac{\partial \mathcal{D}}{\partial T} 
  - \frac{\gamma_\text{b}}{m}\frac{\partial}{\partial\bt{v}}\cdot\bt{V} \mathcal{D}
  - \frac{1}{2}\xi_\text{b}^2\frac{\partial^2}{\partial v^2} \mathcal{D} + \mathcal{L} \mathcal{D}
  =  D,
\end{eqnarray}
where use has been made of the result
\begin{eqnarray}
  \label{apB19}
  \partial_t^{(0)}\nabla\ln T = \nabla \partial_t^{(0)} \ln T = 
  \nabla\left(\frac{m\xi_\text{b}^2}{T}-\frac{2\gamma_\text{b}}{m}-\zeta^{(0)}\right) = \nonumber\\
  \left[ -\zeta^{(0)}g(\phi) - \left( \phi\frac{\partial\chi}{\partial\phi}\frac{\partial}{\partial\chi}\left(\frac{\zeta^{(0)}}{\chi}\right)-\xi^*\frac{\partial\zeta^{(0)}}{\partial\xi^*} -\frac{2}{3}\theta\frac{\partial \zeta^{(0)}}{\partial \theta}\right)\right]\nabla\ln n -\nonumber\\
   - \left[\frac{m\xi_\text{b}^2}{T}\left(1-\frac{3}{2}\frac{\partial\zeta^*_0}{\partial\xi^*}\right) +\frac{1}{2}\zeta^{(0)}\right]\nabla\ln T.
\end{eqnarray}

In the first order of the deviations from the steady state, we only need to know the transport coefficients to zeroth order in the deviations (steady state conditions). This means that the term
$$
\frac{2\gamma_\text{b}}{m}-\frac{m\xi_\text{b}^2}{T} +\zeta^{(0)}
$$ appearing in the left-hand side of Eqs.\ \eqref{apB13}--\eqref{apB16} vanishes. The differential equations for the transport coefficients thus become simple coupled algebraic equations. They are given by Eqs.\ \eqref{eq4.2}--\eqref{eq4.5}.
 % del Chapter 4
\chapter{Kinetic contributions to the transport coefficients} 

\label{AppendixC} 

\lhead{Appendix C. \emph{Kinetic contributions to the transport coefficients}}
In this Appendix we determine from Eqs.\ \eqref{eq4.2}--\eqref{eq4.5} the kinetic contributions to the transport coefficients $\eta$, $\kappa$ and $\mu$ as well as the first order contribution $\zeta_U$ to the cooling rate. Given that all these coefficients are evaluated in the steady state, the subscript $s$ appearing along the main text will be omitted in this Appendix for the sake of brevity.

We start with the kinetic contribution $\eta_\text{k}$ to the shear viscosity $\eta$. It is defined as: 
\begin{equation}
\label{apC1}
\eta_k=-\frac{1}{(d-1)(d+2)}\int\; d{\bf v} D_{ij} \mathcal{C}_{ij}({\bf V}),
\end{equation}
where
\begin{equation}
\label{apC1.1}
D_{ij}=m\left(V_iV_j-\frac{1}{d}V^2\delta_{ij}\right).
 \end{equation}
To obtain $\eta_k$, we multiply Eq.\ (\ref{eq4.4}) by $D_{ij}$ and integrate over velocity. The result is
\begin{equation}
\label{apC2}
\left(\frac{2\gamma_\text{b}}{m}+\nu_\eta\right)\eta_k = nT-\frac{1}{(d-1)(d+2)}\int\; d{\bf V} D_{ij}({\bf V}) \mathcal{K}_i\left[\frac{\partial}{\partial V_j}f^{(0)}\right],
\end{equation}
where
\begin{equation}
\label{apC3}
\nu_\eta=\frac{\int d{\bf v} D_{ij}({\bf V}){\cal L}{\cal C}_{ij}({\bf V})}
{\int d{\bf v}D_{ij}({\bf V}){\cal C}_{ij}({\bf V})}.
\end{equation}
The collision integral of the right hand side of Eq.\ (\ref{apC2}) has been evaluated in previous works \cite{GD99,L05} and the result is:
\begin{equation}
\label{apC3.1}
\int\; d{\bf V}\; D_{ij}({\bf V}) \mathcal{K}_i\left[\frac{\partial f^{(0)}}{\partial V_j}
\right]=2^{d-2}(d-1)nT\chi \phi(1+\alpha)(1-3\alpha).
\end{equation}
Thus, the kinetic part $\eta_k$ can be written as
\begin{equation}
\label{apC4}
\eta_k=\frac{nT}{\nu_\eta+\frac{2\gamma_\text{b}}{m}}
\left[1-\frac{2^{d-2}}{d+2}(1+\alpha)
(1-3 \alpha)\phi \chi \right].
\end{equation}
In order to get an explicit expression for $\eta_k$, one has to evaluate the collision frequency $\nu_\eta$. It can be determined by considering the leading terms in a Sonine polynomial expansion of the function $\mathcal{C}_{ij}({\bf V})$. Here, we have considered a recent modified version of the standard method \cite{GSM07,GVM09} that yields good agreement with computer simulations even for quite strong values of dissipation \cite{MSG07}. The expression of $\nu_\eta$ is given by Eq.\ \eqref{eq4.10}. The final form \eqref{eq4.9} of the shear viscosity $\eta$ is obtained when one takes into account the relation \eqref{eqgamG}.

The kinetic parts $\kappa_k$ and $\mu_k$ of the transport coefficients characterizing the heat flux are defined, respectively, as
\begin{equation}
\label{apC5}
\kappa_k = -\frac{1}{dT}\int \, d{\bf v} {\bf S}({\bf V})\cdot \boldsymbol{\mathcal{A}} ({\bf V}),
\end{equation}
\begin{equation}
\label{apC5.1}
\mu_k=-\frac{1}{dn}\int\, d{\bf v} {\bf S}({\bf V})\cdot \boldsymbol{\mathcal{B}} ({\bf V}),
\end{equation}
where
\begin{equation}
\label{apC6}
{\bf S}({\bf V})=\left(\frac{m}{2}V^2-\frac{d+2}{2}T\right){\bf V}.
\end{equation}
We obtain first the kinetic part $\kappa_k$. It is obtained by multiplying Eq.\ (\ref{eq4.2})
by ${\bf S}({\bf V})$ and integrating over ${\bf V}$. The result is

%\begin{widetext}

\begin{equation}
\label{apC7}
\left[\nu_\kappa+ \frac{1}{2}\frac{m\xi_\text{b}^2}{T} \left(1+3\frac{\partial \zeta_0^*}{\partial \xi^*}\right)-2\zeta^{(0)}\right]\kappa_k
=-\frac{1}{dT}\int\; d{\bf V} {\bf S}({\bf V})\cdot {\bf A},
\end{equation}
where
\begin{equation}
\label{apC8}
\nu_\kappa=\frac{\int d{\bf v} {\bf S}({\bf V})\cdot {\cal L}\boldsymbol{\mathcal{A}}({\bf V})}
{\int d{\bf v}{\bf S}({\bf V})\boldsymbol{\mathcal{A}}({\bf V})}.
\end{equation}

The right hand side of Eq.\ \eqref{apC7} is given by
\begin{eqnarray}
\label{apC9}
-\frac{1}{d T}\int\; d{\bf V}{\bf S}({\bf V})\cdot {\bf A} &=& \frac{1}{d T}\left\{\frac{d(d+2)}{2m}nT^2\left(1+2a_2-\frac{3}{2}\xi^*\frac{\partial a_2}{\partial \xi^*}\right)\right. 
\nonumber\\
& &\left. -\frac{1}{2}\int\; d{\bf V}{\bf S}({\bf V})\cdot \boldsymbol{\mathcal{K}}\left[\frac{\partial}{\partial \mathbf{V}}\cdot \left( \mathbf{V} f^{(0)}\right) \right]\right.
\nonumber\\
& & \left. -\frac{3}{2}\xi^* \int\; d{\bf V}{\bf S}({\bf V})\cdot \boldsymbol{\mathcal{K}}\left[\frac{\partial f^{(0)}}{\partial \xi^*}\right]\right\}
\nonumber\\
&=& \frac{1}{d T}\left\{\frac{d(d+2)}{2m}nT^2\left(1+2a_2-\frac{3}{2}\xi^*\frac{\partial a_2}{\partial \xi^*}\right)\right. 
\nonumber\\
& &\left. -\frac{1}{2}\int\; d{\bf V}{\bf S}({\bf V})\cdot \boldsymbol{\mathcal{K}}\left[\frac{\partial}{\partial \mathbf{V}}\cdot \left( \mathbf{V} f^{(0)}\right) \right]\right.
\nonumber\\
& & \left. -\frac{3}{2}\xi^*\frac{\partial \ln a_2}{\partial \xi^*}\frac{\partial}{\partial a_2}\int\; d{\bf V}{\bf S}({\bf V})\cdot \boldsymbol{\mathcal{K}}\left[f^{(0)}-f_\text{M}\right]\right\},
\nonumber\\
\end{eqnarray}
where use has been made of Eq.\ \eqref{eqTtime}. The last two terms on the right hand side of Eq.\ (\ref{apC9}) can be evaluated more explicitly and the result is %\cite{G12}
\begin{equation}
\label{apC10}
\int\; d{\bf V}{\bf S}\cdot
\boldsymbol{\mathcal{K}}\left[\frac{\partial}{\partial \mathbf{V}}\cdot \left( \mathbf{V}f^{(0)}\right) \right] 
= -\frac{3}{8} 2^{d} d \frac{n T^2}{m} \chi \phi (1+\alpha)^2 \left[2\alpha-1+a_2(1+\alpha)\right],
\end{equation}
\begin{equation}
\label{apC11}
\int\; d{\bf V}{\bf S}\cdot
\boldsymbol{\mathcal{K}}\left[f^{(0)}-f_\text{M}\right] = 
\frac{3}{32} 2^{d} d \frac{n T^2}{m} \chi \phi (1+\alpha)^3 a_2.
\end{equation}
%\begin{equation}
%\label{apC12}
%\int\; d{\bf V}{\bf V}\cdot
%\boldsymbol{\mathcal{K}}\left[f^{(0)}\right]=2^{d-1} d \frac{n T}{m} \chi \phi (1+\alpha),
%\end{equation}
With the above results, the kinetic part $\kappa_k$ can be finally written as
%\begin{widetext}
\begin{eqnarray}
\label{apC13}
\kappa_k&=&\kappa_0\nu_0\frac{d-1}{d}\left(\nu_\kappa+\frac{1}{2}\frac{m\xi_\text{b}^2}{T} \left(1+3\frac{\partial \zeta_0^*}{\partial \xi^*}\right)-2\zeta^{(0)}\right)^{-1}
\nonumber\\
&\times& \left\{1+2a_2-\frac{3}{2}\xi^*\frac{\partial a_2}{\partial \xi^*}+3\frac{2^{d-3}}{d+2}\phi \chi(1+\alpha)^2\right.
\nonumber\\
&\times& \left. \left[2\alpha-1 + a_2(1+\alpha)-\frac{3}{8} (1+\alpha)\xi^*\frac{\partial a_2}{\partial \xi^*}\right]\right\},
\end{eqnarray}
where $\kappa_0$ is the low density value of the thermal conductivity of an elastic gas (defined by Eq.\ \eqref{eq4.13}).

The expression \eqref{apC13} for $\kappa_k$ is still \emph{exact}. In order to get an explicit expression for $\kappa_k$, one considers the form \eqref{eqZetas} for $\zeta^{(0)}$ and evaluates $\nu_\kappa$ by considering again the leading terms in a Sonine polynomial expansion of $\boldsymbol{\mathcal{A}}({\bf V})$. With these approaches, one gets the expression \eqref{eq4.15} for $\nu_\kappa$ while
\begin{equation}
\label{apC14.1}
\frac{\partial \zeta_0^*}{\partial \xi^*}=\zeta_\text{M} \frac{\partial a_2}{\partial \xi^*},
\end{equation}
where $\zeta_\text{M}$ is defined by Eq.\ \eqref{eq4.14.1}. Use of Eq.\ \eqref{apC14.1} in Eq.\ \eqref{apC13} gives the final result.

%The evaluation of the coefficient $\mu$ is quite intricate since it involves the derivatives $\partial a_2/\partial \xi^*$ and $\partial a_2/\partial \chi$. Since both derivatives are in general very small, for the sake of simplicity we will neglect contributions proportional to those derivatives in the calculation of the coefficient $\mu$. With this approximation, in particular the term $n\left(\partial f^{(0)}/\partial n\right)$ appearing in Eq.\ \eqref{apB6} becomes simply $n\left(\partial f^{(0)}/\partial n\right)=f^{(0)}$. 
In order to determine $\mu_k$, we multiply Eq.\ \eqref{eq4.3} by ${\bf S}({\bf V})$ and integrate over velocity to achieve:
\begin{eqnarray}
\label{apC14.2}
\left[\nu_\mu-\frac{3}{2}\left(\zeta_\text{s}^{(0)}-\frac{m\xi_\text{b}^2}{T_\text{s}}\right)\right]\mu_k
&=& -\frac{1}{d\, n}\int\; d{\bf V}{\bf S}({\bf V})\cdot \left\{\zeta ^{(0)} \left(1+\phi\frac{\partial  }{\partial \phi}\ln \chi \right)\boldsymbol{\mathcal{A}}\right.\nonumber\\
&&+\left. \left[\phi \frac{\partial \chi}{\partial \phi}\frac{\partial}{\partial \chi}\left(\frac{\zeta^{(0)}}{\chi}\right) - \xi^*\frac{\partial \zeta^{(0)}}{\partial \xi^*} - \frac{2}{3}\theta \frac{\partial \zeta^{(0)}}{\partial \theta} \right] \boldsymbol{\mathcal{A}}+{\bf B}\right\}  \nonumber\\
&=&\frac{T}{n}\left[  \zeta^{(0)}\left(1+\phi\frac{\partial  }{\partial \phi}\ln\chi \right)+\phi \frac{\partial \chi}{\partial \phi}\frac{\partial}{\partial \chi}\left(\frac{\zeta^{(0)}}{\chi}\right) - \xi^*\frac{\partial \zeta^{(0)}}{\partial \xi^*} \right.\nonumber\\
&&-\left. \frac{2}{3}\theta \frac{\partial \zeta^{(0)}}{\partial \theta} \right]\kappa_k-\frac{1}{d\, n}\int\; d{\bf V}{\bf S}({\bf V})\cdot {\bf B},
\end{eqnarray}
where
\begin{equation}
\label{apC14.2bis}
\nu_\mu=\frac{\int d{\bf v} {\bf S}({\bf V})\cdot {\cal L}\boldsymbol{\mathcal{B}}({\bf V})}
{\int d{\bf v}{\bf S}({\bf V})\cdot \boldsymbol{\mathcal{B}}({\bf V})}.
\end{equation}

The last term on the right hand side of Eq.\ \eqref{apC14.2} is
\begin{eqnarray}
\label{apC14.3} 
-\frac{1}{dn}\int\;d{\bf V}{\bf S}\cdot {\bf B} 
&=& \frac{d+2}{2}\frac{T^2}{m}\left(a_2-\xi^*\frac{\partial a_2}{\partial \xi^*} - \frac{2}{3}\theta\frac{\partial a_2}{\partial \theta}+\phi \frac{\partial \chi}{\partial \phi}\frac{\partial a_2}{\partial \chi}\right)
\nonumber\\
&-& \frac{3}{32}2^{d}\frac{T^2}{m}\chi \phi (1+\alpha)^3\left(\xi^*\frac{\partial a_2}{\partial \xi^*} + \frac{2}{3}\theta\frac{\partial a_2}{\partial \theta}-\phi \frac{\partial \chi}{\partial \phi} \frac{\partial a_2}{\partial \chi}\right)
\nonumber\\
&+& \frac{1}{dn} \left(1+\frac{1}{2}\phi\frac{\partial}{\partial \phi}\ln \chi\right)\int\;d{\bf V}{\bf S} \cdot
\boldsymbol{\mathcal{K}}\left[f^{(0)}\right],
\end{eqnarray}
where use has been made of teh result \eqref{apC11}. The last term in Eq.\ \eqref{apC14.3} is given by
\begin{equation}
  \label{apC14.4}
  \int\;d{\bf V}{\bf S}\cdot \boldsymbol{\mathcal{K}}\left[f^{(0)}\right] = \frac{3}{8}2^d d\frac{n T^2}{m}\chi\phi(1+\alpha)\left[\alpha(\alpha-1)+\frac{a_2}{6}(10+2d-3\alpha+3\alpha^2)\right]
\end{equation}

The final expression of $\mu_k$ is obtained from Eq.\ \eqref{apC14.2} when one substitutes Eq.\ \eqref{apC11} into Eq.\ \eqref{apC14.2}. However, this expression is not explicit unless one knows the collision frequency $\nu_\mu$. To determine it, one takes the leading terms in a Sonine polynomial expansion of $\boldsymbol{\mathcal{B}}({\bf V})$ and gets $\nu_\mu=\nu_\kappa$. This finally yields Eq.\ \eqref{eq4.17} for $\mu_k$.

We consider finally the first-order contribution $\zeta_U$ to the cooling rate. This coefficient is given by Eq.\ \eqref{eq4.19}, where
\begin{equation}
  \label{apC15}
  \zeta_{11}=\frac{1}{2nT}\frac{\pi ^{(d-1)/2}}{d\Gamma \left( \frac{d+3}{2} \right)}
\sigma ^{d-1}\chi m (1-\alpha^{2})\int d\mathbf{V}_{1}\,\int d \mathbf{V}_{2}\,g^{3}f^{(0)}(\mathbf{V}_{1})\mathcal{D}(\mathbf{V}_{2}), 
\end{equation}
and the unknown $\mathcal{D}$ verifies the integral in Eq.\ \eqref{eq4.5}. An approximate solution to Eq.\ \eqref{eq4.5} can be obtained by taking the Sonine approximation
\begin{equation}
\mathcal{D}(\mathbf{V})\rightarrow e_{D}f_{M}(\mathbf{V})F( \mathbf{V}), \label{apC16}
\end{equation}
where
\begin{equation}
F(\mathbf{V})=\left( \frac{m}{2T}\right) ^{2}V^{4}-\frac{d+2}{2}
\frac{m}{T}V^{2}+\frac{d(d+2)}{4}, \label{apC17}
\end{equation}
and the coefficient $e_D$ is
\begin{equation}
e_{D}=\frac{2}{d(d+2)}\frac{1}{n}\int \;d\mathbf{V}\;\mathcal{D}(
\mathbf{V})F(\mathbf{V}). \label{apC18}
\end{equation}
Substitution of Eq.\ (\ref{apC16}) into Eq.\ (\ref{apC15}) gives
\begin{equation}
\label{apC19}
\zeta_{11}=\frac{3(d+2)}{32d}\chi (1-\alpha^2)
\left(1+\frac{3}{32}a_2\right)\nu_0 e_D.
\end{equation}
The coefficient $e_D$ is determined by substituting Eq.\ (\ref{apC16}) into the integral
equation (\ref{eq4.5}), multiplying by $F({\bf V})$ and integrating over ${\bf V}$. After some algebra one gets the
expression \eqref{eq4.21} for $\zeta_{11}$. Here, for the sake of simplicity, we have neglected the contributions proportional to the derivatives $\partial a_2/\partial \chi$ and $\partial a_2/\partial \theta$.
 % del Chapter 4

\chapter{Expressions for choice B} % Main appendix title

\label{AppendixD} % For referencing this appendix elsewhere, use \ref{AppendixA}

\lhead{Appendix D. \emph{Expressions for choice B}} % This is for the header on each page - perhaps a shortened title

In this Appendix we display the expressions for the NS transport coefficients $\eta$, $\lambda$, $\kappa$, and $\mu$ by using the choice B defined by the condition $\partial_t^{(0)}T=0$. The application of the Chapman-Enskog method to this case follows similar mathematical steps as those made for the choice A ($\partial_t^{(0)}T\neq 0$). The results show that the expressions of $\eta$ and $\lambda$ are the same as those obtained for the choice A
[see Eqs.\ \eqref{eq4.7}--\eqref{eq4.9}]. However, the forms of $\kappa$ and $\mu$ are different since they
are given by Eqs.\ \eqref{eq4.12} and \eqref{eq4.16}, respectively, but their corresponding kinetic contributions are
\begin{equation}
\label{apD4.24}
\kappa_k=\frac{d-1}{d}\frac{\kappa_0\nu_0}{\nu_\kappa+\frac{3\beta m}{T_\text{b}}\xi_\text{b}^2}
\left(1+2a_{2,\text{s}}+3\frac{2^{d-3}}{d+2}\phi \chi(1+\alpha)^2\left[2\alpha-1+a_{2,\text{s}}(1+\alpha)\right]\right)
\end{equation}
and
\begin{eqnarray}
\label{apD4.25}
\mu_k &=& \frac{\kappa_0\nu_0T_\text{s}}{\nu_\kappa+\frac{3\beta m}{T_\text{b}}\xi_\text{b}^2}\left\{\frac{d-1}{d}a_2+3\frac{2^{d-2}(d-1)}{d(d+2)}\phi\chi(1+\alpha)\left(1+\frac{1}{2}\phi\partial_\phi\ln\chi\right)\right.\nonumber\\
&&\times\left.\left[\alpha(\alpha-1) \frac{a_{2,\text{s}}}{6}(10+2d-3\alpha+3\alpha^2)\right]\right\},
\end{eqnarray}

where the collision frequency $\nu_\kappa$ is defined by Eq.\ \eqref{eq4.15}.
 % del Chapter 4
%%---------------------------------------------
% Appendix Template
\chapter{Collisional moments of $J_\text{IMM}[f,f]$} % Main appendix title
\label{AppendixE} % Change X to a consecutive letter; for referencing this appendix elsewhere, use \ref{AppendixX}
\lhead{Appendix E. \emph{Collisional moments of $J_\text{IMM}[f,f]$}} % Change X to a consecutive letter; this is 

%\chapter{Inelastic Maxwell Model} % Main appendix title
%\label{AppendixE} % Change X to a consecutive letter; for referencing this appendix elsewhere, use \ref{AppendixX}
%\lhead{Appendix E. \emph{Inelastic Maxwell Model}} % Change X to a consecutive letter; this is for the header on each page - perhaps a shortened title
%\section{Collisional moments of $J_\text{IMM}[f,f]$}

As said in the Introduction of Chapter \ref{Chapter5}, one of the advantages of the Boltzmann equation for Maxwell models (both elastic and inelastic) is that the collisional moments of the operator $J_\text{IMM}[f,f]$ can be \emph{exactly} evaluated in terms of the moments of the distribution $f$, without the explicit knowledge of the latter \cite{TRUESDELL80}. More explicitly, the collisional moments of order $k$ are given as a bilinear combination of moments of order $k'$ and $k''$ with $0\leq k'+k''\leq k$. In particular, the collisional moments involved in the calculation of the momentum and heat fluxes as well as in the fourth cumulant are given by \cite{S03,GS07}
\begin{equation}
\label{eq5.5}
\int\dd\mathbf{v}\; m\; V_iV_j\; J_\text{IMM}[f,f]=-\nu_{0|2}\left(P_{ij}-p\delta_{ij}\right)-\nu_{2|0} p \delta_{ij},
\end{equation}
\begin{equation}
\label{eq5.6}
\int\dd\mathbf{v}\; \frac{m}{2}\;V^2\;\mathbf{V}\, J_\text{IMM}[f,f]=-\nu_{2|1}\mathbf{q},
\end{equation}
\begin{equation}
\label{eq5.7} %\label{2.17}
\int\dd\mathbf{v}\; \;V^4\; J_\text{IMM}[f,f]=-\nu_{4|0}\langle V^4 \rangle+\lambda_1 d^2\frac{pT}{m^2}-
\frac{\lambda_2}{nm^{2}}\left(P_{ij}-p\delta_{ij}\right)\left(P_{ji}-p\delta_{ij}\right),
\end{equation}
where $p=nT$ is the hydrostatic pressure,
\beq
\nuzt=\frac{(1+\al)(d+1-\al)}{d(d+2)}\nu, \quad \nu_{2|0}=\frac{1-\alpha^2}{2d}\nu,
\label{eq5.8} %\label{2.18}
\eeq

\beq
\nuto=\frac{(1+\al)\left[5d+4-\al(d+8)\right]}{4d(d+2)}\nu,
\label{eq5.9}
\eeq
\beq
\nufz=\frac{(1+\al)\left[12d+9-\alpha(4d+17)+3\alpha^2-3\alpha^3\right]}{8d(d+2)}\nu,
\label{eq5.10}
\eeq
\beq
\lambda_1=\frac{(1+\al)^2\left(4d-1-6\al+3\al^2\right)}{8d^2}\nu,
\label{eq5.11}
\eeq
\beq
\lambda_2=\frac{(1+\al)^2\left(1+6\al-3\al^2\right)}{4d(d+2)}\nu.
\label{eq5.12}
\eeq
Here we have introduced the fourth-degree isotropic velocity moment
\beq
\label{eq5.13}
\langle V^4 \rangle=\int\; \dd \mathbf{v}\; V^4\; f(\mathbf{v}).
\eeq
 % del Chapter 5
% Appendix Template

\chapter{First-order contributions to the fluxes for Inelastic Maxwell Models}

\label{AppendixF} 

\lhead{Appendix F. \emph{First-order contributions to the fluxes in the Inelastic Maxwell Model}} 

In this Appendix we determine the first-order contributions to the momentum and heat fluxes for a driven granular gas of IMM. Let us consider each flux separately. The first order contribution to the pressure tensor $P_{ij}^{(1)}$is defined by Eq.\ \eqref{eq5.41}. To obtain it, we multiply both sides of Eq.\ \eqref{eq5.35} by $m V_i V_j$ and integrate over $\mathbf{v}$. The result is
\beq
\label{apF1} % \label{apE1}
\partial_t^{(0)}P_{ij}^{(1)}+\nu_{0|2}P_{ij}^{(1)}+\frac{2\gamma_\text{b}}{m} P_{ij}^{(1)}=-p
\left( \nabla_{i}U_{j}+\nabla_{j}U_{i}-\frac{2}{d}\delta_{ij}\nabla \cdot
\mathbf{U} \right).
\eeq
Upon writing Eq.\ \eqref{apF1}, use has been made of the result
\beq
\label{apF2} %\label{apE2}
\int \dd \mathbf{v}\; m V_i V_j {\cal L}f^{(1)}=\nu_{0|2} P_{ij}^{(1)},
\eeq
where $\nu_{0|2}$ is given by Eq.\ \eqref{eq5.7}. The solution to Eq.\ \eqref{apF1} can be written in the form \eqref{eq5.42.1}, where the shear viscosity coefficient $\eta$ obeys the time dependent equation
\beq
\label{apF4} %\label{apE4}
\partial_t^{(0)}\eta+\left(\nu_{0|2}+\frac{2\gamma_\text{b}}{m}\right)\eta=p.
\eeq
The shear viscosity can be written in the form \eqref{eq5.43} where $\eta^*$ is a dimensionless function of the reduced noise strength $\xi^*$ [or the reduced drag parameter $\gamma^*$  through Eq.\ \eqref{eq5.25}] and the coefficient of restitution $\al$. Thus,
\beq
\label{apF5} %\label{apE5}
\partial_t^{(0)}\eta=(T\frac{\partial \eta}{\partial T})(\partial_t^{(0)} \ln T)=\Lambda T\frac{\partial }{\partial T} (\eta_0 \eta^*)=
\Lambda \left[(1-q)\eta-(1+q)\eta_0\xi^*\frac{\partial \eta^*}{\partial \xi^*}\right],
\eeq
where
\beq
\label{apF6} %\label{apE6}
\Lambda \equiv \frac{m\xi_b^2}{T}-\frac{2\gamma_b}{m} -\zeta.
\eeq
Equation \eqref{eq5.44} for $\eta^*$ can be easily obtained when one takes into account the relation \eqref{apF5} in Eq.\ \eqref{apF4}.

The first order contribution to the heat flux is defined by Eq.\ \eqref{eq5.42}. As in the case of the pressure tensor, to obtain $\mathbf{q}^{(1)}$ we multiply both sides of Eq.\ \eqref{eq5.35} by $\frac{m}{2} V^2 \mathbf{V}$ and integrate over $\mathbf{v}$. After some algebra, one gets
\beqa
\label{apF9} %\label{apE9}
&&\partial_t^{(0)}\mathbf{q}^{(1)}+\left(\nu_{2|1}+\frac{3\gamma_\text{b}}{m}\right)\mathbf{q}^{(1)} =
-\frac{d+2}{2}\frac{p}{m}\left[1+2a_2-(1+q)\xi^*\frac{\partial a_2}{\partial \xi^*}\right]\nabla T \nonumber\\
& &-\frac{d+2}{2}\frac{T^2}{m}\left(a_2-\frac{\theta}{1+q}\frac{\partial a_2}{\partial \theta}-\xi^*\frac{\partial a_2}{\partial \xi^*}\right)\nabla n.
\eeqa
Upon writing Eq.\ \eqref{apF9}, the following results have been used:
\beq
\label{apF10.1} %\label{apE10.1}
\int \dd \mathbf{v}\; \frac{m}{2} V^2 \mathbf{V} {\cal L}f^{(1)}=\nu_{2|1} \mathbf{q}^{(1)},
\eeq
\beqa
\label{apF10} %\label{apE10}
\int\; \dd \mathbf{v}\; \frac{m}{2}V^2 V_i A_j(\mathbf{V})&=&
-\frac{d+2}{2}\frac{p T}{m}\delta_{ij}\left(1+2a_2+T\frac{\partial a_2}{\partial T} \right) \nonumber\\
&=& -\frac{d+2}{2}\frac{p T}{m}\delta_{ij}\left(1+2a_2-
(1+q)\xi^*\frac{\partial a_2}{\partial \xi^*}\right),
\eeqa
\beqa
\label{apF11} %\label{apE11}
\int\; \dd \mathbf{v}\; \frac{m}{2}V^2 V_i B_j(\mathbf{V})&=&-\frac{d+2}{2}\frac{p T}{m}\delta_{ij}\left(a_2+n\frac{\partial a_2}{\partial n} \right) \nonumber\\
&=&-\frac{d+2}{2}\frac{p T}{m}\delta_{ij}\left(a_2-\frac{\theta}{1+q}\frac{\partial a_2}{\partial \theta}-\xi^*\frac{\partial a_2}{\partial \xi^*}\right).
\eeqa
In Eq.\ \eqref{apF10.1}, $\nu_{2|1}$ is defined by Eq.\ \eqref{eq5.8}. The solution to Eq.\ \eqref{apF9} is given by Eq.\ \eqref{eq5.42.2}, where the transport coefficients $\kappa$ and $\mu$ can be written in the form \eqref{eq5.43}. Since the (reduced) coefficients $\kappa^*$ and $\mu^*$ depend on $T$ through their dependence on $\xi^*$, then
\beq
\label{apF14} %\label{apE14}
\partial_t^{(0)}\kappa=(T\frac{\partial \kappa}{\partial T})(\partial_t^{(0)} \ln T)=\Lambda T \frac{\partial}{\partial T} (\kappa_0 \kappa^*)=
\Lambda \left[(1-q)\kappa-(1+q)\kappa_0\xi^*\frac{\partial \kappa^*}{\partial \xi^*}\right],
\eeq
\beq
\label{apF15} %\label{apE15}
\partial_t^{(0)}\mu=(T \frac{\partial \mu}{\partial T})(\partial_t^{(0)} \ln T)=\Lambda T \frac{\partial}{\partial T} \left(\frac{\kappa_0 T}{n} \mu^*\right)=\Lambda \left[(2-q)\mu-(1+q)\frac{\kappa_0 T}{n} \xi^*\frac{\partial \mu^*}{\partial \xi^*}\right].
\eeq
Moreover, there are also contributions to Eq.\ \eqref{apF9} coming from the term
\beq
\label{apF16} %\label{apE16}
\nabla \partial_t^{(0)} T=\left(\Lambda-\frac{m\xi_\text{b}^2}{T}-q \zeta\right) \nabla T-\frac{T \zeta}{n}\nabla n.
\eeq
The corresponding differential equations for $\kappa^*$ and $\mu^*$ can be obtained when one takes into account the constitutive form \eqref{eq5.42} and the relations  \eqref{apF14}--\eqref{apF16} in Eq.\ \eqref{apF9}. These equations are given by Eq.\ \eqref{eq5.45} for $\kappa^*$ and Eq.\ \eqref{eq5.46} for $\mu^*$.

%\subsection{Fourth cumulant}

We consider finally the isotropic fourth degree moment \eqref{eq5.48}. Since $e_\text{D}$ is a scalar, it can be only coupled to the divergence of flow velocity $\nabla \cdot \mathbf{U}$:
\beq
\label{apF20} %\label{apE20}
e_D=e_D^* \nu^{-1} \nabla \cdot \mathbf{U}.
\eeq
In order to determine the (reduced) coefficient $e_D^*$, we multiply both sides of Eq.\ \eqref{4.1} by $V^4$ and integrate over velocity. After some algebra one arrives to Eq.\ \eqref{eq5.50} where use has been made of the partial result
\beq
\label{apF23} %\label{apE23}
\int\; \dd \mathbf{v}\; V^4 D(\mathbf{V})=d(d+2)\frac{nT^2}{m^2}\left(\frac{2(1+q)+d}{d}\xi^*\frac{\partial a_2}{\partial \xi^*}+\frac{\theta}{1+q}\frac{\partial a_2}{\partial \theta}\right).
\eeq
 % del Chapter 5
% Appendix Template

\chapter{Transport coefficients for IHS in the steady state in the low-density limit} 

\label{AppendixG} 

\lhead{Appendix G. \emph{Transport coefficients for IHS in the steady state \\ in the low-density limit}}

%\section{Transport coefficients for IHS in the steady state}

%In this Appendix we display the explicit expressions of the transport coefficients obtained in Ref.\ \cite{GCV13} for a moderately dense gas by considering the leading terms in a Sonine polynomial expansion. In the low-density limit, 

The expressions of the NS transport coefficents obtained in Chapter\ \ref{Chapter4} for a moderately dense gas of IHS are displayed in this Appendix in the low-density limit $(\phi=0)$. In this case the forms of the dimensionless coefficients $\eta_\text{s}^*$, $\kappa_\text{s}^*$, and $\mu_\text{s}^*$ for IHS in the steady state are given, respectively, by
\beq
\label{apG1} % \label{apE24}
\eta_\text{s}^*=\frac{2}{d+2}\frac{1}{\nu_\eta^*+2\gamma_\text{s}^*},
\eeq
\beq
\label{apG2} % \label{apE25}
\kappa_\text{s}^*=\frac{2(d-1)}{d(d+2)}\frac{1+2 a_{2,\text{s}}-\frac{3}{2}\xi_\text{s}^*\Delta_\xi}{\nu_\kappa^*+
\frac{\xi_\text{s}^*}{2}\left[1+\frac{9}{32d}(1-\al^2)\Delta_\xi\right]-2\zeta_\text{s}^*},
\eeq
\beq
\label{apG3} % \label{apE26}
\mu_\text{s}^*=\frac{
\kappa_\text{s}^*\left[\zeta_\text{s}^*-\frac{3(1-\al^2)}{32d}\left(\theta_\text{s}\Delta_\theta+
\xi_\text{s}^*\Delta_\xi\right)\right]
+\frac{2(d-1)}{d(d+2)}\left[
a_{2,\text{s}}-\theta_\text{s} \Delta_\theta-
\xi_\text{s}^*\Delta_\xi\right]}
{\nu_{\kappa}^*+3\gamma_\text{s}^*},
\eeq
where
\beq
\label{apG4} % \label{apE27}
\nu_\eta^*=\frac{3-3\al+2d}{2d(d+2)}(1+\al)\left(1+\frac{7}{16}a_{2,\text{s}}\right),
\eeq
\beq
\label{apG5} % \label{apE28}
\nu_\kappa^*=\frac{2}{d(d+2)}(1+\al)\left[\frac{d-1}{2}+\frac{3}{16}(d+8)(1-\alpha)
+\frac{296+217d-3(160+11d)\alpha}{256}a_{2,\text{s}}\right],
\eeq
\beq
\label{apG6} % \label{apE29}
\zeta_\text{s}^*=\frac{1-\al^2}{2d}\left(1+\frac{3}{16}a_{2,\text{s}}\right),
\eeq
and
\begin{equation}
\label{apG7} % \label{apE30}
a_{2,\text{s}}=\frac{16(1-\alpha)(1-2\alpha^2)}{9+24d-\alpha(41-8d)+30(1-\alpha)\alpha^2
+\frac{64d(d+2)}{1+\alpha}\xi_\text{s}^*}.
\end{equation}

In addition, the quantities $\Delta_\xi$ and $\Delta_\theta$ are related to the derivatives $(\partial a_2/\partial \xi^*)_\text{s}$ and $(\partial a_2/\partial \theta)_\text{s}$, respectively. The derivative $(\partial a_2/\partial \xi^*)_\text{s}$ obeys the quadratic equation \eqref{a7}. They are given by
%However, given that the magnitude of this derivative is in general quite small, one can neglect the nonlinear term  $(\partial a_2/\partial \xi^*)^2$ in this quadratic equation and get an explicit expression for this derivative. In this approximation, the quantities $\Delta_\xi$ and $\Delta_\theta$ can be written as
\beq
\label{apG8} % \label{apE31}
\Delta_\xi=\frac{ a_{2,\text{s}}}{\frac{19}{32d}(1-\al^2)-\frac{1+\left(\frac{2^{5/2}}{d+2}\right)^{2/3}\theta_\text{s}\xi_\text{s}^{*-2/3} }{4}\xi_\text{s}^*-\frac{1-\al^2}{2d(d+2)}
\left[\frac{3}{32}(10d+39+10\al^2)+\frac{d-1}{1-\al}\right]},
\eeq
\beq
\label{apG9} % \label{apE32}
\Delta_\theta=\frac{\left(\frac{2^{5}}{(d+2)^2}\right)^{1/3}\xi_\text{s}^{*4/3}\Delta_\xi}{\frac{3}{16d}(1-\al^2)\left(1+
a_{2,\text{s}}-\frac{3}{4}\xi_\text{s}^*\Delta_\xi\right)+2(\zeta_\text{s}^*-\xi_\text{s}^*)-\frac{1-\al^2}{d(d+2)}
\left[\frac{3}{32}(10d+39+10\al^2)+\frac{d-1}{1-\al}\right]},
\eeq
where
\beq
\label{apG10} % \label{apE33}
\theta_\text{s}=\frac{\xi_\text{s}^*-\zeta_\text{s}^*}{2}\xi_\text{s}^{*1/3}.
\eeq

Upon deriving Eq.\ \eqref{apG8}, the nonlinear term $\left(\partial a_2/\partial \xi^*\right)^2_\text{s}$ has been neglected in Eq.\ \eqref{a7} for the sake of simplicity.

Finally, the coefficient $e_D^*$ can be written as
\beq
\label{apG11} % \label{apE34}
e_D^*=-\frac{\frac{d+3}{2d}\xi_\text{s}^*\Delta_\xi }{\nu_{\gamma}^*+4\gamma_\text{s}^*},
\eeq
where
\beq
\label{apG12} % \label{apE35}
\nu_\gamma^*=-\frac{2}{96(d+2)}(1+\al)\left[30\alpha^3-30\alpha^2+(105+24 d) \alpha-56d-73\right].
\eeq
 % del Chapter 5
%%---------------------------------------------
% Appendix Template

\chapter{Results from Grad's moment method. Rheological properties} % Main appendix title

\label{AppendixH} % Change X to a consecutive letter; for referencing this appendix elsewhere, use \ref{AppendixX}

\lhead{Appendix H. \emph{Results from Grad's moment method. Rheological properties}} % Change X to a consecutive letter; this is for the header on each page - perhaps a shortened title

In this Appendix we provide the approximate results obtained from Grad's moment method. First, we evaluate the collisional moment $\Lambda_{ij}$ defined in Eq.\ \eqref{eq6.20} by using Grad's approximation \eqref{eq6.21}. Before considering the trial distribution function \eqref{eq6.21}, the collision integral $\Lambda_{ij}$ can be written as
\beqa
\label{apH1} %\label{a1}
\Lambda_{ij}=
m\sigma^{d-1}\int\; \dd\mathbf{V}_1\dd\mathbf{V}_2\;f(\mathbf{V}_1)f(\mathbf{V}_2)\int \dd \widehat{\boldsymbol {\sigma}}
\Theta(\widehat{\boldsymbol {\sigma}}\cdot \mathbf{g})(\widehat{\boldsymbol {\sigma }}\cdot \mathbf{g})\left(V_{1i}''V_{1j}''-V_{1j}V_{1j}\right),
\nonumber\\
\eeqa
where $\mathbf{g}\equiv\mathbf{g}_{12}=\mathbf{V}_1-\mathbf{V}_2$ is the relative velocity and
\begin{equation}
\label{apH2} %\label{a2}
\mathbf{V}_1''=\mathbf{V}_1-\frac{1+\alpha}{2}(\widehat{\boldsymbol {\sigma}}\cdot \mathbf{g})\widehat{\boldsymbol {\sigma }}.
\end{equation}
Using Eq.\ \eqref{apH2}, Eq.\ \eqref{apH1} becomes
\begin{eqnarray}
\label{apH3} %\label{a3}
\Lambda_{ij}&=&m\sigma^{d-1}\int\; \dd\mathbf{V}_1\dd\mathbf{V}_2\;f(\mathbf{V}_1)f(\mathbf{V}_2)
\int \dd \widehat{\boldsymbol {\sigma}}\Theta(\widehat{\boldsymbol {\sigma}}\cdot \mathbf{g})\left[\left(\frac{1+\alpha}{2}\right)^2(\widehat{\boldsymbol {\sigma }}\cdot \mathbf{g})^3
\widehat{\sigma}_i\widehat{\sigma}_j\right.\nonumber\\
&-& \left.\frac{1+\alpha}{2}(\widehat{\boldsymbol {\sigma }}\cdot \mathbf{g})^2\left(
\widehat{\sigma}_j V_{1i}+\widehat{\sigma}_iV_{1j}\right)\right].
\end{eqnarray}
To perform the angular integrations, we need the results \cite{NE98}
\begin{equation}
\label{apH4} %\label{a4}
\int\; \dd \widehat{\boldsymbol {\sigma}}\Theta(\widehat{\boldsymbol {\sigma}}\cdot \mathbf{g})
(\widehat{\boldsymbol {\sigma}}\cdot \mathbf{g})^n=\beta_n g^n,
\end{equation}
\begin{equation}
\label{apH5} %\label{a5}
\int\; \dd \widehat{\boldsymbol {\sigma}}\Theta(\widehat{\boldsymbol {\sigma}}\cdot \mathbf{g})
(\widehat{\boldsymbol {\sigma}}\cdot \mathbf{g})^n \widehat{\boldsymbol {\sigma}}=\beta_{n+1} g^{n-1}\mathbf{g},
\end{equation}
\begin{equation}
\label{apH6} %\label{a6}
\int\; \dd \widehat{\boldsymbol {\sigma}}\Theta(\widehat{\boldsymbol {\sigma}}\cdot \mathbf{g})
(\widehat{\boldsymbol {\sigma}}\cdot \mathbf{g})^n \widehat{\boldsymbol {\sigma}}\widehat{\boldsymbol {\sigma}}=\frac{\beta_{n}}{n+d} g^{n-2}\left(n\mathbf{g}\mathbf{g}+g^2 \mathbb{1} \right),
\end{equation}
where $\mathbb{1}$ is the unit tensor and
\begin{equation}
\label{apH7} %\label{a7}
\beta_n=\pi^{(d-1)/2}\frac{\Gamma\left((n+1)/2\right)}
{\Gamma\left((n+d)/2\right)}.
\end{equation}
Taking into account these integrals, the integration over $\widehat{\boldsymbol {\sigma}}$ in Eq.\ \eqref{apH3} yields
\beqa
\label{apH8} %\label{a8}
\Lambda_{ij}&=&-m\sigma^{d-1}\beta_3\frac{1+\alpha}{2}
\int \,\dd{\bf v}_{1}\,\int \,\dd{\bf v}_{2}f({\bf V}_{1})f({\bf V}_{2})g\nonumber\\
&\times&\left[g_i G_j+g_j G_i
+\frac{2d+3-3\alpha}{2(d+3)}g_i g_{j}-\frac{1+\alpha}{2(d+3)}g^2\delta_{ij}\right],
\nonumber\\
\eeqa
where ${\bf G}=({\bf V}_1+{\bf V}_2)/2$ is the center of mass velocity.

The expression \eqref{apH8} is still \emph{exact}. However, to compute \eqref{apH8} one has to replace the true $f(\mathbf{V})$ by its Grad's approximation \eqref{eq6.21}. The result is
\beq
\label{apH9} %\label{a9}
\Lambda_{ij}=-p n\sigma^{d-1}\sqrt{\frac{2T}{m}}(1+\al)\beta_3 I_{ij},
\eeq
where $I_{ij}$ is the dimensionless quantity
\beqa
\label{apH10} %\label{a10}
I_{ij} &=& 
\pi^{-d}\int \,\dd{\bf c}_{1}\,\int \,d{\bf c}_{2} e^{-(c_1^2+c_2^2)}
\left[(c_{1\mu}c_{1\lambda} + c_{2\mu}c_{2\lambda})\Pi_{\mu\lambda}^*+c_{1\lambda}c_{1\mu}c_{2\gamma}c_{2\nu}\Pi_{\mu\lambda}^*\Pi_{\gamma\nu}^*\right]\nonumber\\
&\times&g^*\left[g_i^* G_j^*+g_j^* G_i^* + \frac{2d+3-3\alpha}{2(d+3)}g_i^* g_{j}^*-\frac{1+\alpha}{2(d+3)}g^{*2}\delta_{ij}\right].
\eeqa
Here, $\mathbf{c}_i=\mathbf{v}_i/v_0$, $\mathbf{g}^*=\mathbf{g}/v_0$, $\mathbf{G}^*=\mathbf{G}/v_0$, $\Pi_{ij}^*=\Pi_{ij}/p$, and $v_0=\sqrt{2T/m}$ is the thermal velocity. The Gaussian integrals involved in the calculation of $I_{ij}$ can be easily computed by considering $\mathbf{g}^*$ and $\mathbf{G}^*$ as integration variables instead of $\mathbf{c}_1$ and $\mathbf{c}_2$. The corresponding integrals can be done quite efficiently by using a computer package of symbolic calculation. Here, we have used MATHEMATICA \cite{W96b}. The final expressions of $\Lambda_{ij}$  are given by Eq.\ \eqref{eq6.24} for $d=3$ and Eq.\ \eqref{eq6.25} for $d=2$.

Once the collisional moment $\Lambda_{ij}$ is known, the hierarchy \eqref{eq6.19} can be solved. According to the geometry of USF, the only non-zero elements of the pressure tensor are the off-diagonal element $P_{xy}=P_{yx}$ (shear stress) and the diagonal elements $P_{kk}$ ($k=x,y$ and also $z$, if $d=3$). The equations defining these elements (including the $zz$ element that would only raise if $d=3$) can be easily obtained from Eq.\ \eqref{eq6.19}. They are given by
\beq
\label{apH11} %\label{3.19}
2 a^* \Pi_{xy}^*+2\gamma^*(1+\Pi_{xx}^*)=\Lambda_{xx}^*,
\eeq
\beq
\label{apH12} %\label{3.19.1}
2\gamma^*(1+\Pi_{yy}^*)=\Lambda_{yy}^*,
\eeq
\beq
\label{apH13} %\label{3.20}
a^* (1 + \Pi_{yy}^*) +2\gamma^*\Pi_{xy}^*=\Lambda_{xy}^*,
\eeq
where  $\Lambda_{ij}^*\equiv \Lambda_{ij}/p \nu$. Note that in the physical case $d=3$, $\Pi_{zz}^*$ can be obtained from the constraint $\Pi_{zz}^*=-(\Pi_{xx}^*+\Pi_{yy}^*)$.

The solution to Eqs.\ \eqref{apH11}--\eqref{apH13} gives the elements $\Pi_{xx}^*$, $\Pi_{yy}^*$ and $\Pi_{xy}^*$ as functions of the reduced shear rate $a^*$. Note that $a^*$ is proportional to the square root of the (steady) temperature. In order to close the problem, we need an extra condition to express $a^*$ in terms of $\gamma^*$ and $\al$. This is provided by the energy balance equation \eqref{eq6.8}, whose dimensionless form is
\beq
\label{apH14} %\label{3.22}
-\frac{2}{d}\Pi_{xy}^* a^*=2\gamma^*+\zeta^*,
\eeq
where $\zeta^*$ is defined by Eqs.\ \eqref{eq6.27} and \eqref{eq6.28} for spheres and disks, respectively. Thus, the solution to Eqs.\ \eqref{apH11}--\eqref{apH14} provides the forms of $\Pi_{ij}^*$ in terms of the coefficient of restitution $\al$ and the (dimensionless) friction coefficient $\gamma^*$. On the other hand, given that the collisional moments $\Lambda_{ij}^*$ are nonlinear functions of $\Pi_{ij}^*$, Eqs.\ \eqref{apH11}--\eqref{apH14} must be solved numerically (nonlinear Grad's solution).

An analytical solution to Eqs.\ \eqref{apH11}--\eqref{apH14} can be easily obtained when one only considers linear terms to $\Pi_{ij}^*$ in the expressions \eqref{eq6.24} and \eqref{eq6.25} for $\Lambda_{ij}$. This was the approach considered by Sangani \emph{et al.} \cite{SGD04} to get the kinetic contributions to the pressure tensor at moderate densities. In this linear approximation (linear Grad's solution), the solution to Eqs.\ \eqref{apH11}--\eqref{apH14} can be written as
\beq
\label{apH15} %\label{3.23}
\Pi_{yy}^*=\Pi_{zz}^*=-\frac{\zeta_0^*+2\gamma^*}{\beta+\zeta_0^*+2\gamma^*}, \quad \Pi_{xx}^*=-(d-1)\Pi_{yy}^*,
\eeq
\beq
\label{apH16} %\label{3.24}
\Pi_{xy}^*=-\frac{\beta a^*}{(\beta+\zeta_0^*+2\gamma^*)^2},
\eeq
\beq
\label{apH17} %\label{3.24.1}
a^*=\sqrt{\frac{d (2\gamma^*+\zeta_0^*)}{2\beta}}(\beta+\zeta_0^*+2\gamma^*),
\eeq
where $\zeta_0^*\equiv \zeta_0/\nu_0=\left[(d+2)/4d\right](1-\alpha^2)$ and
\begin{equation}
\label{apH18} %\label{3.25}
\beta=\frac{1+\alpha}{2}\left[1-\frac{d-1}{2d}(1-\alpha)\right].
\end{equation}

In the dry granular case ($\gamma^*=0$), Eqs.\ \eqref{apH15}--\eqref{apH17} are consistent with previous results \cite{SGD04} obtained in the USF problem by using Grad's moment method. In addition, the expressions obtained by Sangani \emph{et al.} \cite{SGD04} agree with Eqs.\ \eqref{apH15}--\eqref{apH17} in the limit of dilute granular suspensions.

\chapter{Results from the BGK-like kinetic model} % Main appendix title

\label{AppendixI} % Change X to a consecutive letter; for referencing this appendix elsewhere, use \ref{AppendixX}

\lhead{Appendix I. \emph{Results from the BGK-like kinetic model}} % Change X to a consecutive letter; this is for the header on each page - perhaps a shortened title

The exact results derived from the BGK-like kinetic model \eqref{eq6.29} are displayed in this Appendix. In terms of the dimensionless quantities $\widetilde{a}$, $\widetilde{\zeta}$ and $\widetilde{\epsilon}$, the BGK equation \eqref{eq6.29} can be rewritten as
\beq
\label{3.29}
\left(1-d\widetilde{\epsilon}-\widetilde{a}V_y \frac{\partial}{\partial V_x}-\widetilde{\epsilon}\mathbf{V}\cdot \frac{\partial}{\partial \mathbf{V}} \right)f(\mathbf{V})=f_\text{M}(\mathbf{V}).
\eeq
The hydrodynamic solution to Eq.\ \eqref{3.29} is
\beqa
\label{3.30}
f(\mathbf{V})&=&\left(1-d\widetilde{\epsilon}-\widetilde{a}V_y \frac{\partial}{\partial V_x}-\widetilde{\epsilon}\mathbf{V}\cdot \frac{\partial}{\partial \mathbf{V}} \right)^{-1}f_\text{M}(\mathbf{V})\nonumber\\
&=&
\int_0^{\infty}\; \dd t e^{-(1-d\widetilde{\epsilon})t}\;
e^{\widetilde{a}tV_y \frac{\partial}{\partial V_x}}\; e^{\widetilde{\epsilon}t\mathbf{V}\cdot \frac{\partial}{\partial \mathbf{V}}}
f_\text{M}(\mathbf{V}).\nonumber\\
\eeqa
The action of the velocity operators $e^{\widetilde{a}tV_y \frac{\partial}{\partial V_x}}$ and
$e^{\widetilde{\epsilon}t\mathbf{V}\cdot \frac{\partial}{\partial \mathbf{V}}}$ on an arbitrary function $g(\mathbf{V})$ is
\beq
\label{3.31}
e^{\widetilde{a}tV_y \frac{\partial}{\partial V_x}}g(\mathbf{V})=g\left(\mathbf{V}+\widetilde{a}tV_y \mathbf{\hat{x}}\right),
\eeq
\beq
\label{3.31.1}
e^{\widetilde{\epsilon}t\mathbf{V}\cdot \frac{\partial}{\partial \mathbf{V}}}g(\mathbf{V})=g\left(e^{\widetilde{\epsilon}t}\mathbf{V}\right).
\eeq
Taking into account these operators, the velocity distribution function $f$ can be written as
\beq
\label{3.32}
f(\mathbf{V})=n \left(\frac{m}{2T}\right)^{d/2}\varphi(\mathbf{c}),
\eeq
where $\mathbf{c}\equiv (m/2T)^{1/2}\mathbf{V}$ and the (scaled) velocity distribution function $\varphi(\mathbf{c})$ is
\beq
\label{3.33}
\varphi(\mathbf{c})=\pi^{-d/2}\int_0^{\infty}\; \dd t\;  e^{-(1-d\widetilde{\epsilon})t}\;
\exp \left[-e^{2\widetilde{\epsilon}t}\left(\mathbf{c}+t \widetilde{\mathbf{a}}\cdot \mathbf{c}\right)^2\right],
\eeq
where we have introduced the tensor $\widetilde{a}_{ij}=\widetilde{a}\delta_{ix}\delta_{jy}$.

Equations \eqref{3.32} and \eqref{3.33} provide the explicit form of the velocity distribution function in terms of the parameter space of the system. The knowledge of $f(\mathbf{V})$ allows us to evaluate its velocity moments. In order to accomplish it, it is convenient to introduce the general velocity moments
\beq
\label{3.34}
M_{k_1,k_2,k_3}=\int\; \dd \mathbf{V}\; V_x^{k_1} V_y^{k_2} V_z^{k_3} f(\mathbf{V}).
\eeq
The only nonvanishing moments correspond to even values of $k_1+k_2$ and $k_3$. Insertion of Eq.\ \eqref{3.33} into Eq.\ \eqref{3.34} yields
\beqa
\label{3.35}
M_{k_1,k_2,k_3}&=&n\left(\frac{2T}{m}\right)^{k/2}\pi^{-d/2} \int_0^{\infty}\; \dd t\; e^{-(1-d\widetilde{\epsilon})t}
\int \dd \mathbf{c}\; c_x^{k_1}c_y^{k_2}c_z^{k_3}\; e^{\widetilde{a} t c_y\partial_{c_x}}\exp\left(-e^{2\widetilde{\epsilon}t}c^2\right)
\nonumber\\
&=&n\left(\frac{2T}{m}\right)^{k/2} \pi^{-d/2}\int_0^{\infty}\; \dd t\; e^{-(1+k\widetilde{\epsilon})t}
\int \dd \mathbf{c}\; (c_x-\widetilde{a}t c_y)^{k_1}c_y^{k_2}c_z^{k_3} e^{-c^2},\nonumber\\
\eeqa
where $k=k_1+k_2+k_3$. It is now convenient to expand the term $(c_x-\widetilde{a}t c_y)^{k_1}$, so that
Eq.\ \eqref{3.35} becomes
\beq
\label{3.36}
M_{k_1,k_2,k_3}=n\left(\frac{2T}{m}\right)^{k/2}
\sum_{q=0}^{k_1}\frac{k_1!}{q!(k_1-q)!}\langle c_x^{k_1-q}c_y^{k_2+q}c_z^{k_3}\rangle_{\text{L}}\int_0^{\infty}\; \dd t\; (-\widetilde{a}t)^q e^{-(1+k\widetilde{\epsilon})t},
\eeq
where
\beq
\label{3.37}
\langle c_x^{k_1}c_y^{k_2}c_z^{k_3}\rangle_{\text{L}}=\pi^{-3/2}\Gamma\left(\frac{k_1+1}{2}\right)
\Gamma\left(\frac{k_2+1}{2}\right)\Gamma\left(\frac{k_3+1}{2}\right)
\eeq
if $k_1$, $k_2$ and $k_3$ are even, being zero otherwise. Finally, after performing the $t$-integration in Eq.\ \eqref{3.36} one achieves the result
\beq
\label{3.38}
M_{k_1,k_2,k_3}=n\left(\frac{2T}{m}\right)^{k/2}
\sum_{q=0}^{k_1}\frac{k_1!}{q!(k_1-q)!}(-\widetilde{a})^q (1+k\widetilde{\epsilon})^{-(1+q)}
\langle c_x^{k_1-q}c_y^{k_2+q}c_z^{k_3}\rangle_{\text{L}}.
\eeq

In order to write more explicitly the form of the (scaled) distribution function $\varphi(\mathbf{V})$, we consider here a three-dimensional system ($d=3$). In this case, the distribution $\varphi$ can be written as
\beq
\label{3.40.1}
\varphi(\mathbf{c})=\pi^{-3/2}\int_0^{\infty}\; \dd t\; e^{-(1-3\widetilde{\epsilon})t}\exp\left[-e^{2\widetilde{\epsilon}t}(c_x+\widetilde{a}tc_y)^2-
e^{2\widetilde{\epsilon}t}c_y^2-e^{2\widetilde{\epsilon}t}c_z^2\right].
\eeq
To illustrate the dependence of $\varphi$ on the parameter space of the problem, it is convenient
to introduce the following \emph{marginal} distributions:
\beq
\label{3.41}
\varphi_x^{(+)}(c_x)=\int_0^{\infty}\; \dd c_y\; \int_{-\infty}^{\infty}\; \dd c_z\; \varphi(\mathbf{c}),
\eeq
\beq
\label{3.43}
\varphi_y^{(+)}(c_y)=\int_0^{\infty}\; \dd c_x\; \int_{-\infty}^{\infty}\; \dd c_z\; \varphi(\mathbf{c}).
\eeq
Their explicit forms can be easily obtained from Eq.\ \eqref{3.40.1}:
\beq
\label{phix}
\varphi_x^{(+)}(c_x)=\frac{1}{2\sqrt{\pi}}\int_0^{\infty}\;\dd t \; \frac{e^{-(1-\widetilde{\epsilon})t}}{\sqrt{1+\widetilde{a}^2t^2}}
\exp\left(-e^{2\widetilde{\epsilon}t}\frac{c_x^2}{1+\widetilde{a}^2t^2}\right)\text{erfc}\left(
e^{\widetilde{\epsilon}t}\frac{\widetilde{a}tc_x}{\sqrt{1+\widetilde{a}^2t^2}}\right),
\eeq
\beq
\label{phiy}
\varphi_y^{(+)}(c_y)=\frac{1}{2\sqrt{\pi}}\int_0^{\infty}\;\dd t \; e^{-(1-\widetilde{\epsilon})t}
\exp\left(-e^{2\widetilde{\epsilon}t}c_y^2\right) \text{erfc}\left(e^{\widetilde{\epsilon}t}\widetilde{a}tc_y\right).
\eeq
In Eqs.\ \eqref{phix} and \eqref{phiy}, $\text{erfc}(x)$ is the complementary error function.

So far, $\psi$ has remained free. Henceforth, to agree with the results derived from linear Grad's solution, we will take $\psi=\beta$, where $\beta$ is defined by Eq.\ \eqref{apH18}.
 % del Chapter 6

\addtocontents{toc}{\vspace{2em}} % Add a gap in the Contents, for aesthetics
\backmatter
%------------------
%	BIBLIOGRAPHY
%------------------
\label{Bibliography}
\lhead{\emph{Bibliography}} % Change the page header to say "Bibliography"
\bibliographystyle{unsrtnat} % Use the "unsrtnat" BibTeX style for formatting the Bibliography
\bibliography{BIBLIOGRAFIA-tesis_moy} % The references (bibliography) information are stored in the file named "Bibliography.bib"

\end{document}